\documentclass[preprint,12pt,numbers]{elsarticle}
\RequirePackage[T1]{fontenc}

\RequirePackage{graphicx}
\RequirePackage{mathptmx}      % use Times fonts if available on your TeX system
\RequirePackage{flushend}
\usepackage{braket}
\usepackage{amsmath}
\usepackage{float}
\usepackage{lineno}
\usepackage{enumitem}%https://www.overleaf.com/project/5cd98a6adbf7c77b862fed66
\usepackage{isotope}
\usepackage[utf8]{inputenc}
\usepackage{gensymb}
\usepackage{hyperref}
\usepackage{footnote}
\usepackage{url}
\usepackage{xcolor}
\usepackage{caption}
\usepackage{subcaption}
\usepackage{mathtools}
\usepackage{siunitx}
\usepackage{lineno}
\cortext[cor1]{Corresponding author: david.milstead@fysik.su.se}
\begin{document}
\begin{frontmatter}

\title{New high-sensitivity searches for neutrons converting into antineutrons and/or sterile neutrons at the European Spallation Source \\
%Version 1.5
}

\begin{abstract}
%suggest to open with high level motivation for broader community
The violation of Baryon Number, $\mathcal{B}$, is an essential ingredient for the preferential creation of matter over antimatter needed to account for the observed baryon asymmetry in the universe. However, such a process has yet to be experimentally observed. 
The HIBEAM/NNBAR %experiment
program is a proposed two-stage experiment at the European Spallation
Source (ESS) to search for baryon number violation. The program will include high-sensitivity searches for processes that violate baryon number by one or two units: free neutron-antineutron oscillation ($n\rightarrow \bar{n}$) via mixing, neutron-antineutron oscillation via regeneration from a sterile neutron state ($n\rightarrow [n',\bar{n}'] \rightarrow \bar{n}$), and neutron disappearance ($n\rightarrow n'$); the effective $\Delta \mathcal{B}=0$ process of neutron regeneration ($n\rightarrow [n',\bar{n}'] \rightarrow n$) is also possible. The program can be used to discover and characterise mixing in the neutron, antineutron, and sterile neutron sectors. The experiment addresses topical open questions such as the origins of baryogenesis, the nature of dark matter, and is sensitive to scales of new physics substantially in excess of those available at colliders. A goal of the program is to open a discovery window to neutron conversion probabilities (sensitivities) by up to \textit{three orders of magnitude} compared with previous searches. The opportunity to make such a leap in sensitivity tests should not be squandered. 
%represents a set of cross-disciplinary experiments with clear %particle
%fundamental physics goals, %seems redundant to me
The experiment pulls together a diverse international team of physicists from the particle (collider and low energy) and nuclear physics communities, while also including specialists in neutronics and magnetics. 
\end{abstract}

\author[8,46]{A. Addazi}
\author[43]{K. Anderson}
\author[65]{S.~Ansell}
\author[52]{K.~S.~Babu}
\author[23]{J. Barrow}
\author[4,5,6]{D.~V.~Baxter}
\author[29]{P.~M.~Bentley}
\author[2,12]{Z.~Berezhiani}
\author[29]{R.~Bevilacqua}
\author[2]{R.~Biondi}
% ...

%%   Affiliations
% ...
\address[2]{INFN, Laboratori Nazionali del Gran Sasso, 67010 Assergi AQ, Italy}
% ...
\address[4]{Department of Physics, Indiana University, 727 E. Third St., Bloomington, IN, USA, 47405}
\address[5]{Indiana University Center for Exploration of Energy \& Matter, Bloomington, IN 47408, USA}
\address[6]{Indiana University Quantum Science and Engineering Center, Bloomington, IN 47408, USA}
% ...
\address[8]{Center for Theoretical Physics, College of Physics Science and Technology, Sichuan University, 610065 Chengdu, China}
% ...
\address[12]{Dipartimento di Scienze Fisiche e Chimiche, Universit\`a di L'Aquila, 67100 Coppito AQ}
% ...
\address[23]{Department of Physics and Astronomy, The University of Tennessee, Knoxville, TN 37996, USA}
% ...
\address[29]{European Spallation Source ERIC, Lund, Sweden}
% ...
\address[43]{Oak Ridge National Laboratory, Oak Ridge, TN 37831, USA}
% ...
\address[46]{INFN sezione Roma Tor Vergata, I-00133 Rome, Italy}
% ...
\address[52]{Department of Physics, Oklahoma State University, Stillwater, OK, 74078, USA}
% ...
\address[65]{MAX IV Laboratory, Box 118 22100 Lund, Sweden}
% ...

\end{frontmatter}

%\tableofcontents
%\input{authors

\section{Introduction}
\label{Sec:intro}
%LJB wordsmith: opening is pretty muted
The observation of baryon number violation (BNV) in a laboratory experiment would be a discovery of fundamental importance to particle physics.  
%Additional early work on $n\rightarrow\bar n$ transitions %includes Refs.~\cite{Chang:1980ey,Kuo:1980ew,Cowsik:1980np,Rao:1%982gt,Rao:1983sd}. 
Within the Standard Model (SM), baryon number, $\mathcal{B}$, is a good global symmetry for tests up to the TeV scale. However, BNV is anticipated. Nonperturbative instanton effects in the SU(2) sector of the SM break $\mathcal{B}$ and total lepton number, $\mathcal{L}$, while conserving $\mathcal{B}-\mathcal{L}$ \cite{tHooft:1976snw}. Although these are negligible at temperatures that are low compared with the electroweak scale of $\mathcal{O}$(100) GeV, they gain dynamic importance via sphaleron processes in the early universe at temperatures of this order \cite{Kuzmin:1987wn,Dolgov:1991fr}. Furthermore, precision tests of the Equivalence Principle~\cite{Smith:1999cr,Schlamminger:2007ht,Cowsik:2018jbq} offer no evidence for a long-range force coupled to baryon number, a key requirement for any hypothetical local gauge symmetry \textit{forbidding} BNV. Most compellingly, according to Sakharov's conditions~\cite{Sakharov:1967dj}, BNV is required to understand the matter-antimatter asymmetry of the universe.

Processes of the neutron transition $n \to \bar n$ ($\Delta \mathcal{B}=2$) into antineutrons~\cite{Kuzmin:1970nx,Mohapatra:1980de,Phillips:2014fgb,Chang:1980ey,Kuo:1980ew,Cowsik:1980np,Rao:1982gt,Rao:1983sd,Caswell:1982qs}, and/or a transition $n\to n'$ ($\Delta \mathcal{B}=1$) into sterile (mirror) neutrons \cite{Berezhiani:2005hv,Berezhiani:2008bc,Grojean:2018fus,Bringmann:2018sbs}, offer unique and comparatively unexplored discovery windows for BNV.  Some early studies of $n - \bar n$ transitions include \cite{Mohapatra:1980de,Mohapatra:1980qe,Chang:1980ey,Kuo:1980ew,Cowsik:1980np,Rao:1982gt,Rao:1983sd}. A recent review is \cite{Phillips:2014fgb}. Neutron conversion processes, at potentially observable rates, are anticipated in scenarios of baryogenesis and dark matter~\cite{Mohapatra:1980qe,Berezhiani:2005hv,Babu:2006xc,Dev:2015uca,Allahverdi:2017edd,Grojean:2018fus}, supersymmetry~\cite{Barbier:2004ez,Calibbi:2016ukt}, extra dimensions~\cite{Nussinov:2001rb,Girmohanta:2019fsx,Girmohanta:2020qfd}, cosmic rays \cite{Berezhiani:2006je,Berezhiani:2011da}  neutrino mass generation mechanisms~\cite{Mohapatra:1980qe,Mohapatra:1980de,Mohapatra:2009wp,Dev:2015uca,Allahverdi:2017edd,Berezhiani:2015afa}, extensions of the Standard Model with certain types of scalar fields \cite{Arnold:2012sd}, and even in oscillations of (anti)atomic matter \cite{Mohapatra:1982xz,Senjanovic:1982np}.

In this Article, a proposed two-stage program of experiments at the European Spallation Source (ESS) is shown which is able to perform high precision searches for neutron conversions in a range of BNV channels, culminating in an ultimate sensitivity increase for $n\rightarrow \bar{n}$ oscillations of three orders of magnitude over that previously attained with free neutrons after a search at the Institut Laue-Langevin (ILL)~\cite{BaldoCeolin:1994jz}. This concept developed from an original proposal for a single $n\rightarrow \bar{n}$ search~\cite{EOInnbar}. As part of the new staged approach, an expanded set of searches together with R\&D for NNBAR is planned. As developmental stepping stones toward the final $n\rightarrow\bar{n}$ NNBAR search, searches with world-leading  experimental sensitivities for neutron conversion phenomenon into a dark (sterile neutron) sector~\cite{Berezhiani:2005hv} will be performed at the HIBEAM stage. Taken together, the HIBEAM/NNBAR program will enable the discovery and characterisation of a mixing sector involving neutrons, antineutrons, and sterile neutrons. Furthermore, by designing and exploiting a flexible and easily interchangeable set of different experimental configurations for sterile neutron searches on a single beamline, multiple potential discoveries across a single experimental apparatus could be supported.

%\footnote{The process $n\rightarrow n' \rightarrow \bar{n}$ is specific to scenarios in which a field in the dark sector affects a sterile neutron and must be balanced by a magnetic field in the visible sector. In this paper, this second order process is not considered to be included in the first order process of $n \rightarrow \bar{n}$, which is taken here to refer to neutron oscillations via mixing, a classic signature requiring a quasi-free neutron which has been searched for previously by a number of experimental collaborations~\cite{Bressi:1989zd,Bressi:1990zx,Fidecaro:1985cm,BaldoCeolin:1994jz}.}). 

The first stage of the program, the {\it High Intensity Baryon Extraction and Measurement} (HIBEAM), will employ the planned fundamental physics beamline ANNI \cite{Soldner:2018ycf} during the first phase of ESS operation, as it \textit{does not} require the planned full beam power to achieve its goals. This stage focuses principally on searches for neutron  conversions to sterile neutrons $n'$ : (i) neutron disappearance ($n\rightarrow [n',\bar{n}']$), (ii) neutron regeneration ($n\rightarrow [n',\bar{n}'] \rightarrow n$), and (iii) 
neutron-antineutron conversion via regeneration from a sterile neutron state ($n\rightarrow [n',\bar{n}'] \rightarrow \bar{n}$)\footnote{To distinguish the two types of searches for neutrons converting to antineutrons conducted in the program,  $n\rightarrow \bar{n}$ corresponds to free neutrons converting into antineutrons which can be parameterised by a single mass mixing term in the Hamiltonian, whereas $n\rightarrow [n',\bar{n}'] \rightarrow \bar{n}$ refers to a two-stage conversion mediated via sterile neutron states, as explained in Sections~\ref{sec:nnbar} and ~\ref{sec:nnbarsterile}, respectively.}. The HIBEAM program will include a sensitivity increase to (i) of an order of magnitude compared with previous experimental work~\cite{Ban:2007tp,Serebrov:2007gw,Altarev:2009tg,Bodek:2009zz,Serebrov:2008hw,Berezhiani:2012rq,Berezhiani:2017jkn}. An early attempt to search for (ii) has resulted in weak and unpublished limits~\cite{Schmidt2007} while (iii) is hitherto unexplored. Stage one acts as a pilot for the second stage of the program, a high-sensitivity search for $n \rightarrow \bar{n}$ via direct mixing. HIBEAM will provide a test platform for detector and neutron transmission technologies, and allow in-situ development of background mitigation techniques. The second stage, NNBAR, will exploit the Large Beam Port (LBP), a unique component of the ESS facility, to search for direct $n\rightarrow \bar{n}$ oscillations. Due to the  substantially higher flux and neutron propagation time compared to that available at other neutron facilities worldwide, as well as advances in neutronics and detector technology since the last search with free neutrons in 1990~\cite{BaldoCeolin:1994jz}; an increase of three orders of magnitude in sensitivity is possible. %In addition to the above the collaboration will take advantage of a neutron test beam at the ESS for detector prototype tests and {\it in situ} background measurements to validate  theoretical estimates. 

This Article is organised as follows. A brief motivation for searches for neutron conversion %$n\rightarrow \bar{n}$, $n\rightarrow n'$,  $n\rightarrow n' \rightarrow n$ and $n\rightarrow n'\rightarrow \bar{n}$
processes is given in Section~\ref{Sec:motivation}, followed by descriptions of the phenomenology of neutron oscillations in Section~\ref{Sec:mixing_formalism}. The results of earlier complementary searches for both free and bound neutrons and the experimental principles underpinning these searches are given in Section~\ref{Sec:previous_searches}. The ESS moderator system, beamlines, and shielding are described in Section~\ref{sec:ess}. The technical design of the ANNI beamline~\cite{Soldner:2018ycf} at which HIBEAM would operate is outlined in Section~\ref{sec:HIBEAMProgram}. The HIBEAM program of searches and their expected sensitivities are described for processes involving sterile-neutron phenomena in Section~\ref{sec:nnprimesearches} and for direct $n\rightarrow \bar{n}$ oscillations in Section~\ref{sec:small_nnbar_exp}. Section~\ref{sec:nnbarsec} then outlines the proposed neutronics for the final-stage NNBAR experiment and its expected sensitivity for $n\rightarrow \bar{n}$ oscillations. A dedicated section on backgrounds to these searches (Section~\ref{Sec:backgrounds}) is also included. Future plans and research directions are then described in Section~\ref{Sec:future}, followed by a summary in Section~\ref{sec:summary}. Simulations of the prototype test set-up to be used in the neutron test beam at the ESS are given in an appendix.
 \section{Motivation for searches for free neutron conversions}\label{Sec:motivation}
 
%LJB  section has a lot of duplicated references in very short spans; also some incorrect refs, need to double check
 
 %LJB  lead with the big picture; is this too fluffy?
 %DM A bit too fluffy
%The mysteries of the origin and makeup of our universe have driven our pursuit of scientific knowledge throughout history. Today we can describe the known particles and interactions with remarkable precision using the Standard Model, yet we still grapple with several fundamental questions. For example, we lack definitive experimental evidence to understand two key topics: the matter/antimatter asymmetry, and the particle nature of dark matter. Searches to clarify these questions remain a high priority for the field of particle physics.
 
%LJB  suggest theory paragraph first
Neutron conversions are unique observables able to probe the new physics which could address the deficiencies of the SM. A number of theoretical arguments motivate their existence, chief among them, arguably, is baryogenesis, a critical but poorly understood area in particle physics~\cite{Babu:2006xc,Mohapatra:1980qe,Berezhiani:2015uya,Dev:2015uca,Allahverdi:2017edd}. Other motivations include the possible existence of observable low scale BNV which can occur in models of extra dimensions~\cite{Nussinov:2001rb}, branes \cite{Dvali:1999gf}, and supersymmetry~\cite{Barbier:2004ez,Dutta:2005af,Calibbi:2016ukt}, as well connections to dark matter~\cite{Berezhiani:2015uya,Dev:2015uca,Allahverdi:2017edd,Dvali:2009ne}, neutrino masses and neutrino mass orderings~\cite{Mohapatra:2009wp,Dev:2015uca,Allahverdi:2017edd,Berezhiani:2015afa}. In this Section, theoretical motivations for the existence of neutron-antineutron and neutron-sterile neutron conversion processes are outlined in detail.
 
In addition to the theoretical arguments described below, it is also important to note that a strictly experimentalist consideration highlights the 
importance of searches for neutron conversions. In such processes, baryon number can be violated independently of other quantities hitherto observed to be conserved. Single nucleon two-body decay searches (e.g., $p\rightarrow \pi^0 e^+$ 
or $p \rightarrow \pi^+ \nu$) always require lepton number violation. 
%to ensure angular momentum conservation. 
Neutron-antineutron transitions also give rise to matter instability via dinucleon decays, and these have been sought in a number of experiments, most recently, Super-Kamiokande~\cite{Abe:2011ky,Gustafson:2015qyo,Sussman:2018ylo,Girmohanta:2019cjm}. However, searches for free neutron conversions offer a theoretically clean and high-precision sensitivity to BNV-only processes.

\subsection{Baryogenesis, dark matter and neutron conversions}% \boldmath{$n\to \bar{n},n'$}}
According to Sakharov \cite{Sakharov:1967dj}, there must be baryon number violating processes to explain the universe's baryon asymmetry as observed today. Early grand unified theories (GUTs) such as SU(5)~\cite{Weinberg:1979bt,Fry:1980bd,Yoshimura:1978ex,Ellis:1978xg} that contained BNV do not provide a good source of baryogenesis. The original baryon asymmetry generated by such models conserves $\mathcal{B}-\mathcal{L}$ and violates $\mathcal{B}+\mathcal{L}$, just as in the Standard Model (SM), and any leftover asymmetry below the unification scale would be erased by electroweak sphaleron interactions.
%%%%%%%%
A more promising class of models attempting to explain the origin of matter are those focused on electroweak baryogenesis, which does not succeed in the SM but could work in some SM extensions (see, e.g., Ref.~\cite{Morrissey:2012db} and references therein). Alternatively, baryogenesis can be generated via leptogenesis~\cite{Fukugita:1986hr}, which utilizes the seesaw mechanism \cite{Weinberg:1979sa,Mohapatra:1979ia} of neutrino masses and allows for an initial lepton asymmetry to be converted into a baryon asymmetry via the sphaleron processes \cite{Kuzmin:1985mm}.  The simplest examples of such models require the baryogenesis scale to be very high, and are very hard to test experimentally. More specific lepto-baryogenesis models include $\nu$MSM \cite{Asaka:2005pn,Asaka:2005an} and co-leptogenesis models via the neutrino interactions with sterile neutrinos from a dark sector  \cite{Bento:2001rc,Bento:2002sj,Berezhiani:2008zza}. 

A subset of weak-scale baryogenesis models have the attractive feature of being experimentally testable.  Explicit UV-complete models featuring post-sphaleron baryogenesis (PSB)~\cite{Babu:2006xc,Babu:2008rq,Babu:2013yca} use interactions that violate baryon number by two units and predict magnitudes of observable phenomena such as $n\to\bar{n}$ oscillation~\cite{Mohapatra:1980qe} periods. %both of which can be experimentally tested. Searches for 
These models also connect the neutrino's Majorana mass to $n\to\bar{n}$ transformations, and present an upper limit for the $n\to\bar{n}$ oscillation time which can be accessible at next-generation facilities like NNBAR at ESS. Other simplified models that could realize PSB with a connection to neutrino masses and dark matter, while simultaneously giving rise to an observable $n\rightarrow\bar{n}$ rate, have been studied in Refs.~\cite{Dev:2015uca,Allahverdi:2017edd}. Scenarios of co-baryogenesis with a dark sector have been discussed in Refs.~\cite{Berezhiani:2005hv,Berezhiani:2018zvs,Bringmann:2018sbs}.
 %~\cite{Bringmann:2018sbs} and $n\rightarrow n' \rightarrow \bar{n}$
These searches % for $n\rightarrow \bar{n}$ transformations are thus highly valuable in elucidating fundamental physical processes pertinent to baryogenesis, 
represent dedicated probes of selection rules ($\Delta \mathcal{B} = 1,\textbf{2}; \Delta \mathcal{L} = 0$), which fulfill a Sakharov condition but have been comparatively overlooked in the program of experiments probing fundamental symmetries and lepto/baryogenesis.

This points to another open question in modern physics: what is the nature of dark matter? The fact that our astronomical observations are not sufficiently well described by the SM is unquestioned. %, but few clues are provided to resolve
In proposing dark matter candidates, the physics community has largely employed a strategy of linking dark matter to other problems in the SM~\cite{Feng:2010gw}. With no conclusive experimental observations of any prospective particles, the number of plausible candidates has only grown, and resolution will require a thorough and comprehensive search utilizing multiple experimental techniques to fully explore the range of possibilities~\cite{Battaglieri:2017aum}. 

The existence of a dark sector, interacting primarily gravitationally with our familiar visible sector, has long been postulated as a means of explaining astronomical data. When such a dark (sterile) sector is assumed to have particles having gauge interactions similar to our own SM interactions, one easily implies the existence of, e.g., sterile neutrinos and sterile baryons which could  
represent asymmetric dark matter induced by a baryon asymmetry in the dark sector. Self-interacting and dissipative characteristics of such dark matter would have interesting astrophysical implications~\cite{Berezhiani:1995am,Mohapatra:2000qx,Berezhiani:2000gw,Foot:2014mia,Essig:2013lka}. In principle, observable portals onto such a sector can occur ia mixing phenomena between any stable or meta-stable electrically neutral particles, allowing for conversion into a dark partner particle. For example, photons may become “dark photons” via kinetic mixing~\cite{Holdom:1985ag}, while neutrinos can oscillate into sterile neutrinos of the dark sector \cite{Berezhiani:1995yi,Berezhiani:1996sz}. The neutron represents another possible generic portal.

One of the simplest examples of a hidden sector is the theory of mirror matter, a dark sector represented by a replica of the SM~(for reviews see \cite{Foot:2014mia,Berezhiani:2003xm,Berezhiani:2005ek}, for a historical overview see \cite{Okun:2006eb}). %LJB  need to replace with appropriate ref
The assumption of this minimal model forms the basis of the phenomenological framework for the sterile neutron transition searches considered in this work, though these searches have some sensitivity to a more generic dark sector. %LJB  sensitivity is not as strong as statement had implied
Forms of $n\rightarrow n'$ transitions have been proposed~\cite{Berezhiani:2018eds, Berezhiani:2018qqw}, to which HIBEAM is sensitive, that can also shed light onto the apparent anomaly present between experimental free cold and ultracold neutron ``beam" and ``bottle" measurements of the neutron lifetime~\cite{nlife:2018wfe}.

\subsection{Exploring the TeV-PeV regime with $n\rightarrow \bar{n}$ searches}\label{sec:tevpev}
Baryon number violation is a generic feature of grand unified theories (GUTs) \cite{Georgi:1974sy} and many other proposed extensions of the SM~\cite{Barbier:2004ez,Dutta:2005af,Calibbi:2016ukt,Nussinov:2001rb,Berezhiani:2015uya,Dev:2015uca,Allahverdi:2017edd}. 

Classic BNV signatures include proton and bound neutron decay, mediated by four-fermion operators, and $n\to \bar n$ oscillations \cite{Kuzmin:1970nx}, mediated by six-quark operators\cite{Glashow:1979nm,Mohapatra:1980qe}. In SM effective field theory, the lowest orders of these operators have mass dimensions $+6$ and $+9$, respectively; also, in supersymmetric models these can take dimensions $+4$ and $+5$. Thus, if there were only one mass scale $M_{BNV}$ characterizing BNV processes, $n\to\bar{n}$ oscillations would be more highly suppressed (like $1/M_{BNV}^5$) compared with proton and bound neutron decay, for which the effective Lagrangian would only involve a suppression by $1/M_{BNV}^2$. However, there is no good reason to assume that BNV processes correspond to a single scale nor is it known which processes Nature has chosen should there be one BNV scale. There are a number of approaches where $n\to\bar{n}$ oscillations are the dominant manifestation of BNV, while proton decay is either absent or suppressed well below experimental limits \cite{Mohapatra:1980de,Mohapatra:1980qe,Nussinov:2001rb,Dolgov:2006ay,Bambi:2006mi}. 
Some early studies of $n\to\bar{n}$ oscillations include~\cite{Mohapatra:1980de,Mohapatra:1980qe,Chang:1980ey,Kuo:1980ew,Cowsik:1980np,Rao:1982gt,Rao:1983sd};  a recent review is \cite{Phillips:2014fgb}.

%In a generic approach to neutron conversions within an effective field theory formalism, dimension $d=9$, six-quark BNV operators mediate $n\rightarrow \bar{n}$ conversions. 
Basic dimensional analysis based on the above considerations implies sensitivity to mass scales of $\mathcal{O}(10)-\mathcal{O}(1000)$~TeV, accessible via a precision $n\rightarrow \bar{n}$ oscillation search such as the experimental program proposed in this Article. Such a scale is substantially in excess of the reach of current or planned colliders. This is a complementary the large volume single nucleon decay experiments which are sensitive to a different set of BNV processes with scales near the GUT energy.      

Examples of models predicting observable $n\rightarrow \bar{n}$ arising from BNV at TeV and PeV scales include  $R$-parity violating supersymmetry scenarios~\cite{Barbier:2004ez,Calibbi:2016ukt} and extra dimensional models. Extra dimensional scenarios arise from the leading candidate for quantum gravity, i.e.~superstring theory~\cite{Schwarz:1982jn}, which predicts extra spatial dimensions beyond the three which are observed. Compactification radii characterizing these extra dimensions might be much larger than the Planck length. A model of this type~\cite{Nussinov:2001rb,Girmohanta:2019fsx} provides an explicit example of how proton decay can be strongly suppressed, while $n\rightarrow\bar{n}$ transformations can occur at levels comparable to existing limits, which will be probed by a new, high sensitivity experiment. This is also true of an extra-dimensional model with a left-right gauge symmetry broken at the scale of $\mathcal{O}(10^3)$ TeV~\cite{Girmohanta:2020qfd}. 

%%JLB: Mention Rabi's triangle of n-nbar/majorana nus/p-decay?
\subsection{The connection of \boldmath{$n\rightarrow\bar{n}$} with neutrino masses and proton decay}
Another topical issue in which the $n\rightarrow \bar{n}$ process may play a role concerns the origin of neutrino mass. There is a symbiosis between $n\rightarrow \bar{n}$ transitions and neutrinoless double $\beta$ decays: they both violate $\mathcal{B}-\mathcal{L}$ (the anomaly-free SM symmetry) by two units and imply Majorana masses, and both processes are connected in unification models, such as the left-right-symmetric model \cite{Senjanovic:1975rk,Senjanovic:1978ev,Mohapatra:1980yp} based on the gauge group $G_{LRS} = {\rm SU}(3)_c \times {\rm SU}(2)_L \times {\rm SU}(2)_R \times {\rm U}(1)_{B-L}$. In these models, the spontaneous symmetry breaking of $G_{LRS}$ to the SM is produced by the vacuum expectation value (VEV) of a Higgs field transforming as $(1,1,3)_2$ under $G_{LRS}$, so that its VEV breaks $\mathcal{B}-\mathcal{L}$ by 2 units. This gives rise to both an operator with $\Delta \mathcal{L}=0$ and $\Delta \mathcal{B}=2$, such as the six-quark operator mediating $n\rightarrow\bar n$ oscillations, and to an
operator such as the bilinear Majorana product of right-handed neutrinos, with $\Delta \mathcal{B}=0$ and $\Delta \mathcal{L}=2$ that is responsible for a seesaw mechanism leading to Majorana neutrino masses.
%%{\bf \color{red} GB I don't understand the sentence above}
Thus in theories with Majorana neutrino masses and quark-lepton unification, it is natural to expect both Majorana neutrinos as well as $n\rightarrow\bar{n}$ transitions~\cite{Mohapatra:1980qe}. In fact, there exist both left-right-symmetric and $SO(10)$ models with observable $n\to\bar{n}$ oscillation where this connection is explicit; there has also been recent work on these connections within $SU(5)$ effective field theory \cite{deGouvea:2014lva}. 

Setting aside specific theories of physics beyond the SM, the sphaleron interaction itself being a nine-quark-three lepton interaction can be written as \\ $QQQQQQ~QQQL~LL$. This implies that if any two of the following processes are seen, then the other should exist:   $n\rightarrow\bar{n}$ transition (the first six quark operator), proton decay (the second 4-fermion operator) and $\Delta \mathcal{L}$ =2, lepton number violation. The last term implies low energy processes such as neutrinoless double beta decay and direct lepton number violation in the form of same sign charged lepton pairs \cite{Keung:1983uu} at hadron colliders, possibly even at the LHC. The neutrinoless double beta decay can  result from the neutrino Majorana mass or the new physics \cite{Mohapatra:1980yp} that leads to same sign dileptons at colliders, and in the left-right symmetric model of neutrino mass there is a deep connection between the two processes~\cite{Tello:2010am}.

In summary, together with the discovery of proton decay, an observation of $n\rightarrow \bar{n}$ oscillations, could, therefore, establish evidence for the Majorana nature of neutrinos and/or probe the theory behind neutrino Majorana mass. Equivalently, discoveries of $n\rightarrow \bar{n}$ oscillations and $\Delta \mathcal{L}$ =2 lepton number violation would imply proton decay. Searches for $n\rightarrow\bar{n}$ oscillations thus play a key and complementary role in a wider experimental program of $\mathcal{B}$ and $\mathcal{L}$ violation searches~\cite{Babu:2014tra}.

\section{Phenomenology of the neutron conversion processes }\label{Sec:mixing_formalism}
That Nature must violate baryon number is a statement that can be made with confidence. However, should Nature have chosen BNV-only processes, then not only does this imply that the channels which are available for high precision study are limited, but also that a BNV signal is \textit{fragile}. Each channel can require special experimental conditions in order for BNV to manifest itself. In this Section, the formalism of neutron conversions and the conditions for a signal to appear are outlined. A description is also given of other relevant phenomenological aspects of the HIBEAM and NNBAR search programs, such as the modelling of the scattering of a neutron off a guide and antineutron-nucleon annihilation on a nucleus.

\subsection{Neutron-antineutron conversions}\label{sec:nnbar}
%%JLB: Rabi, Kaladi and Bhupal--feel free to add some things here, possibly more of the QFT background and associated Feynman diagrams of the process, along with their dynamical relation to the baryon asymmetry. 

In the SM frames the neutron has only the Dirac Mass term 
$m \overline{n} n$ which conserves $\mathcal{B}$. 
However, as mentioned in Section~\ref{sec:tevpev}, $n\rightarrow \bar{n}$ can proceed by effective six-quark (dimension 9) operators. These 
involve light quarks $u$ and $d$ and violate $\mathcal{B}$ by two units,  
\begin{equation}\label{uud-operator}
{\cal O}_{\Delta \mathcal{B} =2} = \frac{1}{{\cal M}^5}(udd)^2 \, + \, {\rm h.c.}
\end{equation}
with ${\cal M}$ being a large cutoff scale originated from new physics 
beyond the Standard Model, can induce a Majorana mass term 
\begin{equation}\label{nnbar-mass}
\frac{\epsilon_{n\bar n}}{2}(n^T C n + \bar n C \bar n^T) = 
\frac{\epsilon_{n\bar n}}{2}(\overline{n_c} n + \overline{n} n_c) 
\end{equation} 
where $C$ is the charge conjugation matrix and $n_c = C\overline{n}^T$ 
stands for the antineutron field.\footnote{Generically these operators 
induce four bilinear terms $\overline{n} n_c$, $\overline{n}\gamma^5 n_c$, $\overline{n_c} n$ and $\overline{n_c}\gamma^5 n$, with complex 
coefficients. However, by proper redefinition of fields, these terms can be reduced to just one combination (\ref{nnbar-mass}) with a real $\epsilon_{n\bar n}$
which is explicitly invariant under transformations of the charge conjigation 
($n \to n_c$) and parity ($n\to i\gamma^0 n$, $n_c\to i\gamma^0 n_c$) 
\cite{Berezhiani:2018xsx,Berezhiani:2018pcp}.} 
Thus, the $n\rightarrow\bar n$ matrix element/mixing mass term $\epsilon_{n\bar n}$ 
depends on the scale of new physics: 
\begin{equation}\label{epsilon-nnbar}
\epsilon_{n\bar n} = \frac{C \Lambda_{\rm QCD}^6}{{\cal M}^5} 
= C \left(\frac{500~{\rm TeV}}{\cal M} \right)^5 
\times 7.7 \cdot 10^{-24}~{\rm eV}\, , 
\end{equation} 
with $C=O(1)$ being the model dependent factor in the determination 
of matrix element $\bra{\bar n} {\cal O}_{\Delta \mathcal{B} =2}\ket{n} $. 
This mixing between the neutron and antineutron fields gives rise to  
the phenomenon of $n\rightarrow\bar n$ oscillation \cite{Kuzmin:1970nx,Mohapatra:1980de}. 
The direct bound on $n\rightarrow\bar n$ oscillation time 
$\epsilon_{n\bar n}^{-1}=\tau_{n\bar n} > 0.86 \times 10^{8}$~s 
\cite{BaldoCeolin:1994jz}, 
i.e. $\epsilon_{n\bar n}< 7.7 \times 10^{-24}$~eV, 
corresponds to ${\cal M} > 500$~TeV or so. 
%$(discussed below in this section) are at the level of $10^{-24}$~eV. 
By improving the experimental sensitivity by two orders of magnitude one could test the new physics above the PeV scale. 
%which is practically unaccessible in direct search with accelerators.
%${\cal M}\sim 1$~PeV or so. 

Conversion of $n\rightarrow \bar{n}$ can be understood  
as the evolution of a beam of initially pure neutrons
\begin{equation}
 \ket{\Psi(t)} =  
 \begin{pmatrix} \psi_n(t) \\ \psi_{\bar n}(t)  \end{pmatrix} = 
 e^{-i \hat{\mathcal{H}} t} \ket{\Psi(t=0)}, \quad\quad 
 \ket{\Psi(t=0)} = \ket{n} =
% \begin{pmatrix} \ket{n} \\ \ket{\bar{n}} \end{pmatrix} =  
\begin{pmatrix} 1 \\ 0 \end{pmatrix},
\end{equation}
described by $2\times 2$ Hamiltonian 
\begin{equation}
\hat{\mathcal{H}}=
\begin{pmatrix}
E_n & {\epsilon}_{n\bar{n}} \\
{\epsilon}_{n\bar{n}}  & E_{\bar{n}},
\end{pmatrix}.
\quad \label{eq:nnbarmixing}
\end{equation}
where $E_n$ and $E_{\bar{n}}$ are the neutron and antineutron energies, respectively. While the neutron and antineutron masses are equal 
by CPT invariance, $E_n$ and $E_{\bar{n}}$ are not generically equal 
due to the environmental effects which differently act on the neutron and antineutron states, as a presence of matter medium or magnetic 
fields, or perhaps some hypothetical fifth forces \cite{Addazi:2016rgo,Babu:2016rwa}.  

The probability to find an antineutron at a time $t$ is given by $P_{n\bar{n}}(t)= | \psi_{\bar n}(t) |^2$, 
or explicitly 
\begin{equation}
\label{eq:nnbarevol}
P_{n\bar{n}}(t)= 
\frac{ \epsilon_{n\bar{n}}^2} { (\Delta E/2)^2 + \epsilon_{n\bar{n}}^2}
\sin^2\big[ t \, \sqrt{ (\Delta E/2)^2+ \epsilon_{n\bar{n}}^2}
\big] \, e^{-t/\tau_n} ,
%{ 4 {\epsilon_{n\bar{n}}^2}    { {\Delta E}^2+4{\epsilon_{n\bar{n}}^2}} %\sin^2(frac{\sqrt{ {\Delta E}^2+4{\epsilon_{n\bar{n}}^2}}{2})
\end{equation}
where ${\Delta E} = {E_n} - E_{\bar{n}}$ and $\tau_n$ denotes the mean life of the free neutron. It thus becomes immediately clear that the
probability of a conversion is suppressed when the energy degeneracy between neutron and antineutron is broken. 
In particular, for free neutrons
suppression occurs  due to the interaction of the magnetic field ($B\simeq 0.5$~G at the Earth) with the neutron and antineutron's magnetic dipole moments ($\vec{\mu}_n=-\vec{\mu}_{\bar{n}}$), 
equivalent to $\Delta E/2 = \vert\vec{\mu}_n\vec{B}\vert\approx 
(B/1~{\rm G})\times 10^{-11}$~eV in Eq.~(\ref{eq:nnbarevol}). 
To prevent significant suppression of $n\rightarrow\bar n$ conversion,  one must maintain so called quasi-free regime ${|\Delta E|} t \ll 1$ 
which can be realized in vacuum in nearly zero magnetic field
\cite{BITTER1985461,SCHMIDT1992569,Davis:2016uyk}.   
In this case Eq.~(\ref{eq:nnbarevol}) reduces to
\begin{equation}
\label{eq:freennbarosc}
 P_{n\bar{n}}(t)=\epsilon_{n\bar{n}}^2t^2
 =\frac{t^2}{\tau^2_{n\bar n}} 
 = \left(\frac{t}{0.1~{\rm s}}\right)^2 
 \left(\frac{10^8\,{\rm s}}{\tau_{n\bar n}}\right)^2 \times 10^{-18},
 \end{equation}  
where $\tau_{n\bar n}= 1/\epsilon_{n\bar n}$ is characteristic oscillation time. Since in real experimental situation the neutron flight time is small, 
$t\sim 0.1$~s or so, the exponential factor related to the neutron decay can be neglected in Eq. (\ref{eq:nnbarevol}).

This necessitates magnetic shielding for searches utilizing a neutron beam~\cite{Bressi:1989zd,Bressi:1990zx,Fidecaro:1985cm,BaldoCeolin:1994jz}. HIBEAM and NNBAR must employ such shielding, as will be discussed in Section~\ref{sec:magnetics}. 

In the experiment \cite{BaldoCeolin:1994jz} 
performed at the ILL, the magnetic field was suppressed below $10$~mG or so and the lower limit 
$\tau_{n\bar n} > 0.86 \times 10^8$~s ($90~\%$ C.L.) was obtained. In turn, this translates into upper limit 
$\epsilon_{n\bar n} < 7.7 \times 10^{-24}$~eV 
which by now remains a strongest limit 
on $n\rightarrow\bar n$ mass mixing obtained with free neutrons. 
The effects of not perfect vacuum (residual gas pressure) on the observation of neutron to antineutron transformation were discussed in the papers ~\cite{Costa:1983wc,BaldoCeolin:1994jz,Kerbikov:2017spv,Gudkov:2019gro}.

    As for bound neutrons in a nucleus, the potential energy difference experienced between a neutron and antineutron in the strong nuclear field (${\Delta E} \sim 10-100$~MeV, depending on nuclei) introduces a suppression of $\sim 10^{-31}$ with respect to the conversion of a free neutron. This of course inhibits the conversion of neutrons bound in nuclei, with sensitive searches only possible with large volume detectors~\cite{Homestake,KGF,NUSEX,IMB,Kamiokande,Frejus,Soudan-2,Abe:2011ky} such as Super-Kamiokande~\cite{Abe:2011ky}, SNO~\cite{Aharmim:2017jna}, DUNE \cite{Abi:2020evt,Barrow:2019viz,Hewes:2017xtr}, or Hyper-Kamiokande \cite{Labarga:2018owv}. The comparatively large number of neutrons permits searches with currently complementary limits. However, event identification is obscured by atmospheric backgrounds, intranuclear scattering of the decay products and other nuclear physics effects.

A limit on $n\to \bar n$ conversion time in a specific nucleus ($T$) can be related to that of a free neutron ($\tau_{n\bar{n}}$) via a nuclear suppression factor, $R\sim 10^{22}s^{-1}$, which can be calculated with phenomenological nuclear models~\cite{Dover:1982wv,Alberico:1982nu,Alberico:1984wk,Dover:1989zz,Alberico:1990ij,Hufner:1998gu,Friedman:2008es,Barrow:2019viz} and  predict quadratic scaling such that
\begin{equation}\label{eq:nuclearsuppression} 
T=R\cdot\tau_{n\bar{n}}^2 \sim 
\left(\frac{10^8\,{\rm s}}{\tau_{n\bar n}}\right)^2 \times 
10^{31}\,{\rm yr}.
\end{equation}
For today, the strongest limit obtained by Super-Kamiokande ~\cite{Abe:2011ky} for Oxygen reads $\tau_{n\bar{n}}> 2.7 \times 10^8$~s, or equivalently 
$\epsilon_{n\bar n} < 2.5 \times 10^{-24}$~eV. Super-Kamiokande has also carried out searches for $\Delta B = -2$ dinucleon decays to specific multi-meson and leptonic final states \cite{Gustafson:2015qyo,Sussman:2018ylo,Girmohanta:2019cjm,Nussinov:2020wri} 
More details on current limits and future sensitivities are in Section~\ref{sec:search-nnbar}. 

Caution is required when comparing limits and sensitivities for free and bound neutron searches. Calculations relating $T$ and $\tau_{n\bar{n}}$ rely on underlying model assumptions, such as a point-like conversion process, while the physics behind $n\rightarrow \bar{n}$ conversion is \textit{a priori} unknown\footnote{This being said, there has been great progress in a broad program of intranuclear suppression factor calculations across many nuclei which show rather remarkable similarity despite their quite disparate theoretical origins\cite{Barrow:2019viz,Oosterhof:2019dlo,Haidenbauer:2019fyd,Dover:1982wv}. One should also note that intranuclear experiments like SNO \cite{Aharmim:2017jna} have chosen specific targets (deuterium) and techniques to minimize contamination from these and other model dependent nuclear effects, including avoiding excessive final state interactions.}. The visibility of a signal in a bound neutron search could therefore be arbitrarily suppressed compared to a free search, or vice-versa. For example, a recently proposed model of low scale BNV contains the possibility of a suppressed (or even {enhanced}) bound neutron conversion probability~\cite{Berezhiani:2015afa}. There can be also some environmental effects which can affect free $n\rightarrow\bar n$ oscillations even if the magnetic field is properly shielded. 
These effects can be related, e.g. with long range fifth-forces 
induced by very light $\mathcal{B}-\mathcal{L}$ baryophotons. Present high-sensitivity limits on such forces \cite{Wagner:2012ui} with Yukawa radius comparable to the Earth radius or to sun-Earth distance still allow significant contribution to the neutron-antineutron energy level splitting, which in fact can be as large as $\Delta E \sim 10^{-11}$~eV or so \cite{Addazi:2016rgo,Babu:2016rwa}. 

Consideration of free and bound neutron searches is thus complementary: neither makes the other redundant, and indeed they require one another to help constrain the underlying physical process.
%confirmation and detailed quantification of any discovery.

Within the framework of an assumed ultraviolet extension of the Standard Model that features $n-\bar n$ transitions, one has a prediction for the coefficients of the various types of six-quark operators in the resultant low-energy effective Lagrangian, and the next step in obtaining a prediction for the $n-\bar n$ transition rate of free neutrons is to estimate the matrix elements of these six-quark operators between $|n\rangle$ and $|\bar n\rangle$ states.  Since the six-quark operators have dimension 6, their matrix elements are of the form $\Lambda_{eff}^6$. The relevant scale is set by the QCD confinement scale, $\Lambda_{QCD} \sim 0.25$ GeV, so one expects, roughly speaking, that the matrix elements are of order 
$\sim \Lambda_{QCD}^6 \simeq 2.4 \times 10^{-4}$ GeV$^6$, and this expectation is borne out by both 
early estimates using the MIT bag model \cite{Rao:1982gt,Rao:1983sd} and recent calculations using 
lattice QCD (LQCD) \cite{WagmanPrivComm,Rinaldi:2018osy,Rinaldi:2019thf,Buchoff:2015qwa,Buchoff:2015wwa}, including approximate assessments of modeling uncertainties. The LQCD results in \cite{Rinaldi:2018osy,Rinaldi:2019thf} indicate that for most operators, the corresponding $\Lambda_{eff}$ is larger, by $\sim 10-40$ \%, than the $\Lambda_{eff}$ characterizing the MIT bag model results, i.e., a factor $\sim 2-8$ for $\Lambda_{eff}^6$, and thus for the matrix elements themselves. This suggests that overall experimental sensitivities may reach higher than previously expected \cite{WagmanPrivComm}. This being said, direct constraint of PSB and its predicted upper bound on $\tau_{n\rightarrow\bar{n}}$ \cite{Babu:2013yca} is slightly different, as this limit is derived not from tree level amplitudes but instead from loop diagrams involving $W$-boson exchange. In \cite{Babu:2013yca}, larger MIT bag-model estimates are used, and so the LQCD matrix element for this particular amplitude appears smaller by some $\sim15\%$ \cite{Rinaldi:2019thf}; this leads to an expectation that the upper limit for $\tau_{n\rightarrow\bar{n}}$ will be shifted slightly up by roughly the same proportion. The community's integration of this new knowledge is continuing, and still more accurate predictions are being actively discussed and developed \cite{BhupalPrivComm}. Similar computational methods may eventually advance peripheral modeling of secondary processes, such as the annihilation itself and  background interactions.

\subsection{Phase shift suppression}\label{sec:bounce-idea}
A key attribute of a traditional free $n\rightarrow \bar{n}$ conversion search is a neutron beam focused onto an annihilation target to minimize interactions with, e.g, a guide wall.  The difference in neutron and antineutron interactions with wall material have been assumed to act as a large potential difference, suppressing the oscillation.  The interaction can be seen as destroying any  wave function component, which effectively ``resets the clock" for the oscillation time measurement. With this assumption, only the neutron's free flight time since last wall interaction contributes to the probability to find an antineutron, necessitating a large area experimental apparatus in practice.

An \textit{almost} free $n\rightarrow \bar{n}$ oscillation search has recently been proposed~\cite{Nesvizhevsky2019,Kerbikov:2018mct} in which one allows slow, cold neutrons (and antineutrons) (with energies of $<10^{-2}$ eV) to reflect from effective $n$/$\bar{n}$ optical mirrors. Although the reflection of $n$/$\bar{n}$ had been considered in the 1980's for ultra cold neutrons (UCNs)~\cite{Kazarnovskii80,Chetyrkin81,Yoshiki89,Yoshiki92} and recently in ~\cite{Kerbikov:2018mct} for proposed experiments to constrain $\tau_{n\rightarrow\bar{n}}$, the  authors now extended this approach to higher energies, namely where nominally cold, initially collimated neutrons can be reflected from neutron guides when their transverse velocities with respect to the wall are similarly very or ultra-cold. Conditions for suppressing the phase difference for $n$ and $\bar{n}$ were studied, and the required low transverse momenta of the $n$/$\bar{n}$ system was quantified, leading to new suggestions for the nuclei composing the reflective guide material. It was shown that, over a broad fraction of phase space, the relative phase shift of the $n$ and $\bar{n}$ wave function components upon reflection can be small, while the probability of coherent reflection of the $n$/$\bar{n}$ system from the guide walls can remain high. The theoretical uncertainties associated with a calculation of the experimental sensitivity, even in the absence of direct measurements of low energy $\bar{n}$ scattering amplitudes, can be small. 

An important consequence would be that the conversion probability depends on the neutron's total flight time, as wall interactions no longer reset the clock. Such an experimental mode relaxes some of the constraints on free $n$ oscillation searches, and in principle allows a much higher sensitivity to be achieved at reduced complexity and costs. The above represents a new idea from within the community which requires a program of simulation and experimental verification. While this doesn't form part of the \textit{current} core plan for the HIBEAM/NNBAR experiment, it is considered as a promising future research direction, and experimental verification of this concept
is under investigation.

\subsection{Neutron-mirror neutron conversions}\label{sec:nnprimeform}
Though the HIBEAM searches are generic in nature, the mixing of $n$ and its sterile twin $n'$ is considered here within the paradigm of a parallel gauge sector in a mirror matter model ~\cite{Berezhiani:2005hv,Berezhiani:2006je}. There are a range of possible conversion processes which can be explored experimentally, motivating a suite of searches outlined in Section~\ref{sec:nnprimesearches}.

In addition to external fields and interactions in the standard sector, the possibility of equivalent fields in the sterile sector must be taken into consideration. There can exist also some hypothetical forces between ordinary and sterile sector particles which can be 
induced e.g. by the photon kinetic mixing with dark photon \cite{Holdom:1985ag,Glashow:1985ud,Carlson:1987si,Gninenko:1994dr}, or by new gauge bosons interacting with particles of both sectors as e.g. common $\mathcal{B}-\mathcal{L}$ gauge bosons \cite{Addazi:2016rgo} or 
common flavor gauge bosons of family symmetry \cite{Berezhiani:1996ii,Berezhiani:2018ill,Belfatto:2019swo}. 
The respective forces can provide portals for direct detection of dark matter components from a parallel sterile sector and give a possibility for identification of their nature \cite{Foot:2004pa,Addazi:2015cua,Cerulli:2017jzz}. 
In addition, flavor gauge bosons can induce mixing between neutral ordinary particles and their sterile partners and induce oscillations e.g. between Kaons and sterile Kaons, conversion 
of muonium into hidden muonium, etc. \cite{Berezhiani:2018ill,Belfatto:2019swo}.%The Hamiltonian for $n\rightarrow n'$ is %given in Eq.~\ref{eq:hamnprime}

The possibility of neutron-mirror neutron mass mixing 
$\alpha_{nn'} \overline{n}n' + {\rm h.c.} $
was proposed in \cite{Berezhiani:2005hv}.
It can be induced by six-fermion effective operators 
$\frac{1}{M^5} (\overline{u} \bar{d} \bar{d}) (u'd'd') $ 
similar to operator (\ref{uud-operator}) but involving 
three ordinary quarks and three quarks of sterile (mirror) sector. 
This mixing violates conservations of both baryon number 
and mirror baryon number  ($ \Delta \mathcal{B}=1$,  $ \Delta \mathcal{B}'=-1$) 
but it conserves the combination $\mathcal{B}+\mathcal{B}'$. The mixing mass 
$\alpha_{nn'}$ can be estimated as 
\begin{equation}\label{alpha-nnprime}
\alpha_{nn'} = \frac{C \Lambda_{\rm QCD}^6}{M^5} 
= C^2 \left(\frac{10~{\rm TeV}}{M} \right)^5 
\times 2.5 \cdot 10^{-15}~{\rm eV}    
\end{equation}
It was shown that no direct, astrophysical or cosmological effects forbid that $n\rightarrow n'$ oscillation time $\tau_{nn'}=1/\alpha_{nn'}$ can be smaller than the neutron decay time, 
and in fact it can be as small as a second. 
So rapid $n\rightarrow n'$ oscillations could have interesting implications for the propagation of ultra-high cosmic rays 
\cite{Berezhiani:2006je,Berezhiani:2011da} or for neutrons from solar flares \cite{Mohapatra:2005ng}. 
Thus, the effective scale $M$ of underlying new 
physics can be of few TeV, with direct implications for the 
search at the LHC and future accelerators. 
Effects of $n\rightarrow n'$ oscillation can 
be directly observed in experiments searching for anomalous neutron disappearance ($n\to n'$) and/or regeneration ($n\to n' \to n$) 
processes\cite{Berezhiani:2005hv}. 
Experimental sensitivities of such searches with ultra-cold and cold 
neutrons were discussed in Refs. \cite{Pokotilovski:2006gq,Berezhiani:2017azg}.

The Hamiltonian for $n\rightarrow n'$ is given in Eq.~\ref{eq:hamnprime}. The presence of a static magnetic moment shifts the $n$ and $n'$s total energies. 
The Hamiltonian is expressed for the general case of neutrons propagating in magnetic fields $B$ (of the standard sector) and $B'$ (of the sterile sector); the former of these is generated by the magnetic poles of the Earth, the latter by hypothetical ionization and flow of gravitationally captured dark material in and around the Earth \cite{Berezhiani:2008bc}. Such an accumulation could occur due to ionized gas clouds of sterile atoms captured 
by the Earth e.g. due to photon--sterile photon kinetic mixing; present experimental and cosmological limits on such mixing 
\cite{Berezhiani:2008gi,Raaijmakers:2019hqj} 
and geophysical limits \cite{Ignatiev:2000yw} still allow 
the presence of a relevant amount of sterile material at the Earth
\cite{Berezhiani:2016ong}. Then a sterile magnetic field can be induced by the drag of dark electrons due to the Earth rotation via mechanism described in \cite{Berezhiani:2013dea} which can  be enhanced through the dynamo effect \cite{Berezhiani:2008bc}. 
In addition to the static magnetic moments, $\vec{\mu}_n$ and $\vec{\mu}_{n'}$, and unlike for the $n\rightarrow \bar{n}$ transition\footnote{ A non-zero TMM between the neutron and antineutron is forbidden by Lorentz invariance then \cite{Berezhiani:2018xsx,Berezhiani:2018pcp}. Moreover, any 
transition  $n \rightarrow \bar{n}\gamma^\ast$ with an external virtual photon connected to a proton  
would destabilize nuclei even in the absence of $n\rightarrow\bar n$ mixing.}, 
transition magnetic moments (TMMs)~\cite{Berezhiani:2018zvs} may also be present in the off-diagonal transitional magnetic moments (TMM) $\vec{\mu}_{nn'}=\kappa \vec{\mu_n}$  between the neutron and sterile neutron term.\footnote{TMMs play a role both in understanding SM processes, e.g. hadronic decays~\cite{Sharma:2010vv}, and the development of BSM physics models, such as those predicting neutrino flavour changing processes~\cite{Broggini:2012df}.}

As analogous contributions to the Hamiltonian, it can be seen that TMMs are quantities which are as fundamental as the more familiar static magnetic moments. The TMMs contribute to the mixing via the interaction of new physics processes with the external magnetic fields, leading to terms $\kappa\vec{\mu}_n \vec{B}$ and $\kappa' \vec{\mu_{n}} \vec{B}'$, where dimensionless parameter $\kappa\ll 1$ measures the magnitudes of the TMM in units of neutron magnetic moment $\mu_n$. 
\begin{equation}
\label{eq:hamnprime}
\hat{\mathcal{H}}=
\begin{pmatrix}
m_n + \vec{\mu}_n \vec{B}  & \alpha_{nn'} + 
\kappa \vec{\mu}_{n} \vec{B} + \kappa' \vec{\mu}_{n} \vec{B}'  \\
\alpha_{nn'} + \kappa \vec{\mu}_{n} \vec{B} + \kappa' \vec{\mu}_{n} \vec{B}' & 
m_{n'} + \vec{\mu}_{n'}\vec{B'}  \\
%{\alpha_{nn'}}+  \kappa {\mu_n\cdot B}+ \kappa' {\mu'\cdot B'}   & m_n^' + %\mu_n'\cdot B'
\end{pmatrix}.
\end{equation}

Beyond removing exponential decay, a number of simplifications can be made to the picture of $n\rightarrow n'$ mixing described above. In the simplest model, it is assumed that $n$ and $n'$ share degeneracies such as $m_n = m_{n'}$, $|\vec{\mu_n}|=|\vec{\mu_n'}|$ and $\kappa=\kappa'$. The magnitude, direction, and time dependence of the mirror magnetic field $\vec{B}'$ is {\it a priori} unknown. 

If only $n\rightarrow n'$ mass mixing $\alpha_{nn'}$ is present, i.e. 
assuming for the moment that $\kappa=0$, 
then probability $n\rightarrow n'$ oscillation at time $t$ 
 is given by \cite{Berezhiani:2008bc,Berezhiani:2012rq}:
\begin{equation} \label{eq:nnprimeevol}
P_{nn'}(t)= \frac{\alpha_{nn'}^2 \cos^2 \frac{\beta}{2} } {(\omega-\omega')^2} \sin^2\left[(\omega-\omega')t \right] +  
\frac{\alpha_{nn'}^2 \sin^2 \frac{\beta}{2} } {(\omega+\omega')^2} \sin^2\left[(\omega+\omega')t \right]
\end{equation}
where $2\omega = | \mu_n B|$ and $2\omega' = | \mu_n B'|$, 
$\beta$ is the angle between the directions of magnetic fields  $\vec{B}$ and $\vec{B}'$,  
and contribution of $\alpha_{nn'}$ in oscillation frequencies is neglected assuming that $\alpha_{nn'} < |\omega -\omega'|$. 
If $|\omega -\omega'|\, t \gg 1$, the oscillations can be 
averaged in time and one obtains
\begin{equation} 
\label{eq:nnprimeevo2}
\overline{P}_{nn'} = \frac{\alpha_{nn'}^2 \cos^2 \frac{\beta}{2} } {2(\omega-\omega')^2} \, + \,  
\frac{\alpha_{nn'}^2 \sin^2 \frac{\beta}{2} } {2(\omega+\omega')^2} .
\end{equation}
In particular, if $B'=0$ (i.e. $\omega'=0$),  from (\ref{eq:nnprimeevol}) the expression
$P_{nn'}(t) < (\alpha_{nn'}/\omega)^2$ is obtained if $\omega t > 1$, 
and $P_{nn'}(t) \approx (\alpha_{nn'} t)^2$ if $\omega t \ll 1$. 

When $B$ getting close to $B'$,  $|\omega - \omega'|$ decreases and
probability $P_{nn'}(t)$ resonantly increases.  
In a quasi-free regime, when $|\omega -\omega'|\, t \gg 1$, it reaches 
the value 
\begin{equation} 
\label{eq:nnprimeevo3}
P_{nn'}(t) \approx\frac12 (\alpha_{nn'} t)^2 \cos^2 \frac{\beta}{2} = \cos^2 \frac{\beta}{2} 
\left(\frac{t} { 0.1\, {\rm s}}\right)^2 
\left(\frac{1\, {\rm s}}{\tau_{nn'}} \right)^2 
\times 5\cdot 10^{-3}
\end{equation}
%\begin{equation}\label{eq:nnprimeevol-X}
%P_{nn'}(t)=A\frac{\sin^2(\sqrt{({\mu_n %B})^2 +4(\alpha_{nn'}+ \kappa {\mu_n %B})^2}\times t)}{2} ,
%\end{equation}
%\begin{equation}A=\frac{ 4(\alpha_{nn'}+ %\kappa {\mu_nB} ) }
%{(\mu_n \cdot B)^2+ 4(\alpha_{nn'}+ \kappa %{\mu_n B})^2},
%\end{equation}
where $\tau=1/\alpha_{nn'}$ is the characteristic $n\rightarrow n'$ oscillation time for free neutrons in a field-free vacuum. Therefore, this leads to  
a situation when $n\rightarrow n'$ oscillation probability non-trivially depends 
on the value (and direction) of magnetic field which effect can 
be observed in experiments searching for anomalous neutron disappearance ($n\to n'$) and regeneration ($n\to n' \to n$) \cite{Berezhiani:2005hv}. Experimental sensitivities of such searches with cold and ultra-cold 
neutrons were discussed in Ref. \cite{Pokotilovski:2006gq}. 

Several dedicated experiments searching for $n\rightarrow n'$ oscillation 
with ultra-cold neutrons (UCN) were performed in last decade \cite{Ban:2007tp,Serebrov:2007gw,Altarev:2009tg,Bodek:2009zz,Serebrov:2008hw,Berezhiani:2012rq,Berezhiani:2017jkn}. Under the hypothesis 
that there is no mirror magnetic field at the Earth, i.e. $B'=0$, the strongest lower limit $\tau_{nn'} > 414$~s (90 \% C.L.) was obtained by comparing the UCN losses in zero ($B< 10^{-3}$~G) and non-zero 
($B = 0.02$~G) magnetic fields \cite{Serebrov:2007gw}. However, this limit becomes invalid in the presence of $B'$. Lower limits 
on $\tau_{nn'}$ and $\tau_{nn'}/\sqrt{\cos\beta}$ in the presence 
of non-vanishing $B'$ following from experiments  \cite{Ban:2007tp,Serebrov:2007gw,Altarev:2009tg,Bodek:2009zz,Serebrov:2008hw,Berezhiani:2012rq,Berezhiani:2017jkn} are summarized in Ref. 
\cite{Berezhiani:2017jkn}. In fact. some experiments show deviations from null-hypothesis which may point towards $\tau_{nn'}\sim 10$~s 
and $B'\sim 0.1$~G. For $B' > 0.5$~G or so, the oscillation time as 
small as 1 second remains allowed \cite{Berezhiani:2017jkn}. 

In the case of TMM induced $n\rightarrow n'$ transition the average oscillation probability becomes  \cite{Berezhiani:2018qqw}:
\begin{equation} 
\label{eq:nnprimeevo-TMM}
\overline{P}_{nn'} = \frac{2\kappa^2 (\omega+\omega')^2 \cos^2 \frac{\beta}{2} } {(\omega-\omega')^2} \, + \,  
\frac{2\kappa^2 (\omega-\omega')^2 \cos^2 \frac{\beta}{2} } {(\omega+\omega')^2} .
\end{equation}
The upper limits on parameter $\kappa$ that can be obtained by data analysis of 
experiments \cite{Ban:2007tp,Serebrov:2007gw,Altarev:2009tg,Bodek:2009zz,Serebrov:2008hw,Berezhiani:2012rq,Berezhiani:2017jkn} are given in Ref. \cite{Berezhiani:2018qqw}. 
 For the limits on the TMM 
$\kappa \mu_n$ obtained from these experiments see \cite{Berezhiani:2018qqw}. In the case when $\alpha_{nn'}$ and $\kappa$ 
are both present, the average probability of $n\rightarrow n'$ transition is 
given just by a sum of terms (\ref{eq:nnprimeevo2}) and (\ref{eq:nnprimeevo-TMM})

\begin{figure}[tb]
 \setlength{\unitlength}{1mm}
 \includegraphics[width=0.48\linewidth, angle=0]{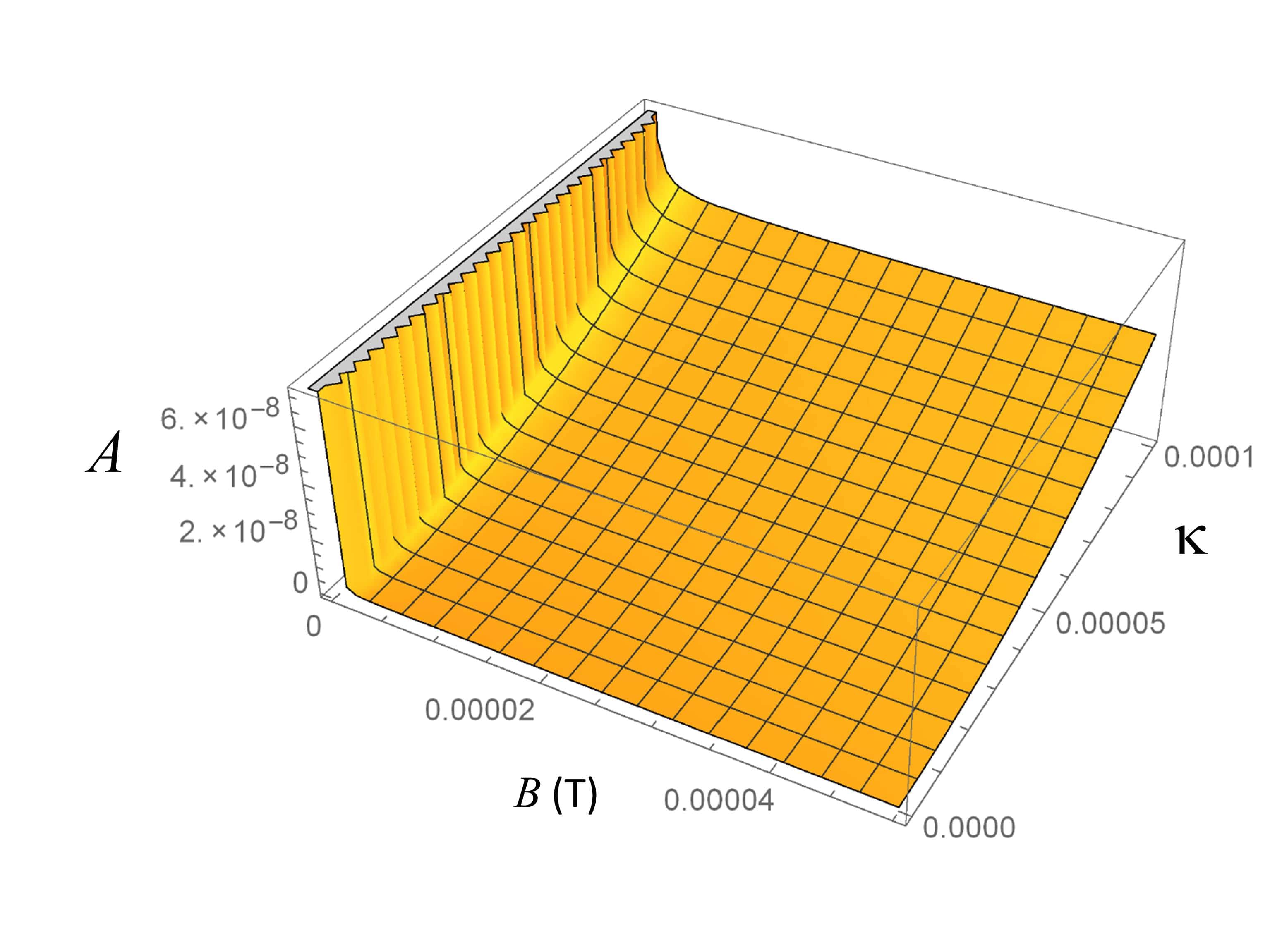}
 \includegraphics[width=0.48\linewidth, angle=0]{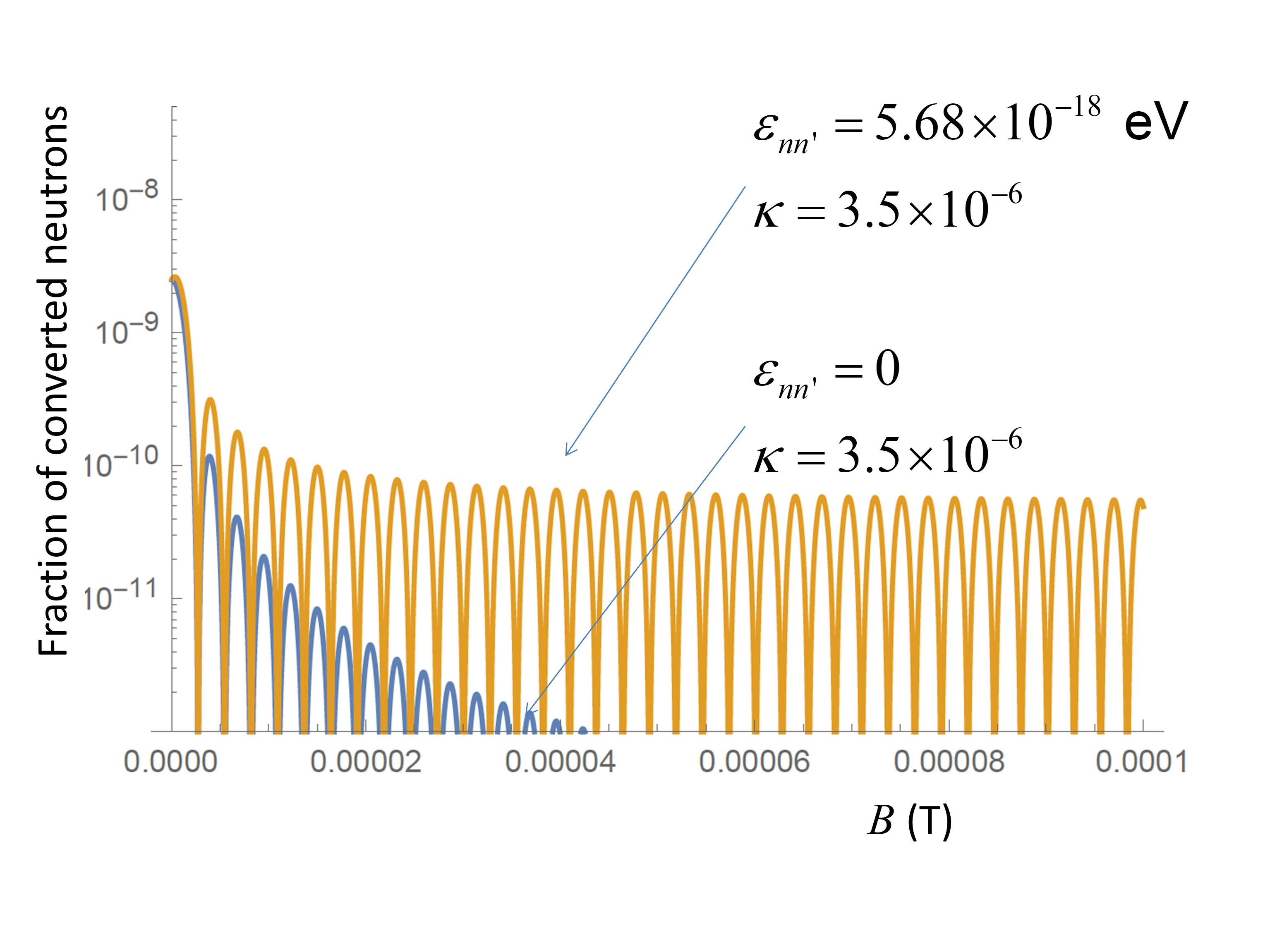}
 \caption{Left: amplitude of the probability function for $n\rightarrow n'$ as a function of $|\vec{B}|$ and $\kappa$ for a value of $\tau=500$~s. Right: fraction of neutrons which have been converted as a function of $|\vec{B}|$ travelling 25 m in a vacuum at a velocity of $1000$~m/s. Predictions are shown for conversions induced by mass mixing and a TMM ($\tau=500$~s and $\kappa=3.5\times 10^{-6}$), and for mass mixing alone ($\tau=500\,$s) alone.}
 \label{fig:amp-vs-deltab-kappa-nprime}
 % last sentence in figure caption does not correspond to what 
 %is written on plots
%  \vspace{2.75cm}
\end{figure}

In the following, let us chose $\vec{B}'$ to be zero,  
%Hereafter, the quantity $\vec{B}$ should then be regarded roughly as the (vectorial) \textit{difference} in magnetic fields between the standard and dark sectors.
%
%Inspection of Eq.~\ref{eq:nnprimeevol} reveals that 
in which case there remain three parameters determining the 
probability of the $n\rightarrow n'$ process: $\alpha_{nn'}$, $\kappa$ and $\vec{B}$. Fig.~\ref{fig:amp-vs-deltab-kappa-nprime} illustrates the interplay between these parameters and their impact oscillation, taking $\alpha_{nn'} = 5.68\times 10^{-18}$~eV ($\tau = 500$~s). For 
only TMM transitions when $\alpha_{nn'} = 0$, 
%$\kappa=3.5\times 10^{-6}$ when  
one has 
%, and a neutron velocity of 1000~m/s\footnote{In a practical experiment neutrons would have a range of velocities. 
%The fine details of the structure, shown for illustrative purposes in  Fig.~\ref{fig:amp-vs-deltab-kappa-nprime}, would be washed out}. The amplitude $A$ is at a maximum for $\vec{B}=0$ ($B-B'=0$) as a consequence of the resonance behaviour expected for $n\rightarrow n'$ mixing. Maximal mixing occurs when the visible and dark sector fields are the same ($\vec{B}=0$), thereby ensuring energy degeneracy for $n$ and $n'$. As seen in Fig.~\ref{fig:amp-vs-deltab-kappa-nprime} (left), with a non-negligible TMM the amplitude does not vanish as $B$ increases but becomes finite and independent of $\vec{B}$, and the time-averaged probability becomes 
\begin{equation}\label{eq:nnprimetimm}
P_{nn'}=2\kappa^2.
\end{equation}
Fig.~\ref{fig:amp-vs-deltab-kappa-nprime} (right) compares the fraction of converted neutrons after having travelled 25~m in a vacuum for the case of a transition magnetic moment term and nonzero mass mixing versus zero mass mixing.

The TMM can also lead to an enhanced $n\rightarrow n'$ transformation in a gas atmosphere due to the creation of a positive Fermi potential along the neutron path~\cite{Berezhiani:2018qqw}. A constant magnetic field $\vec{B}$ in the flight volume can be 
chosen such that for one polarization of neutron it will provide a negative magnetic potential compensating the positive Fermi potential of the gas: $V_F=\vec{\mu} \vec{B}$. Thus, for example, the Fermi potential of air at Normal Temperature and Pressure corresponds to the constant magnetic field of $\sim 10$~G. Continuing to assume that $|B'|=0$, the oscillation Hamiltonian becomes 
\begin{equation}
\mathcal{H}=
\begin{pmatrix}
V_{F}-\mu B & {\alpha}_{nn'} +\kappa \mu B \\
{\alpha}_{nn'} +\kappa \mu B  & 0
\end{pmatrix}.
\quad
\end{equation}\label{eq:compensate1}
With a zero diagonal term (in the resonance) it will correspond to pure oscillation with the probability:
\begin{equation}
P_{nn'}=(\alpha_{nn'}+ \kappa \mu B)^2 t^2
\end{equation}\label{eq:Probc1}
The probability due to the mass mixing term here 
is enhanced by the term due to the TMM that is proportional to field $B$.

%In addition to the sterile sector quantities, there are other notable differences between $n\rightarrow n'$ and $n-\bar{n}$ mixing. A non-zero TMM is possible for the former but forbidden for the latter case due to rotational symmetry. Furthermore, despite its suppression, $n\rightarrow \bar{n}$ in nuclei plays as experimentally significant a role in searches as do free neutrons. Decays due to the process $n\rightarrow n'$ would, however, be unobservable due to backgrounds from non-sterile neutron-induced decays~\cite{Berezhiani:2006je}.

%subsection{Further processes connecting the visible and mirror sectors}
\subsection{Conversions of neutrons to antineutrons via sterile neutrons }\label{sec:nnbarsterile}
Sections~\ref{sec:nnbar} and ~\ref{sec:nnprimeform} address the transformations of $n\rightarrow \bar{n}$ and $n\rightarrow n'$, respectively. However, should a sterile neutron sector exist, processes connecting the visible and sterile sectors need not be restricted to the above processes, as proposed in Refs.~\cite{Berezhiani:2018zvs,Berezhiani:2020nzn}. It is essential to test the full range of conversions between the sectors: $n\rightarrow \{n{'}, \bar{n}{'}\}$,$\bar{n}\rightarrow \{n{'},\bar{n}{'}\}$, 
$n{'}\rightarrow \{n,\bar{n}\}$ and 
$\bar{n}{'}\rightarrow \{n,\bar{n}\}$ 

In principle, a transformation to four states mixed in the $n$, $(n,\bar{n},n{'},\bar{n}{'})$, in free space without any fields
can be described by the symmetric Hamiltonian 
\begin{equation}
    \label{eq:nbarmatrix}
    \hat{\mathcal{H}}=\left(\begin{array}{cccc}
    m_n + \vec{\mu}_n \vec{B} & \varepsilon_{n\bar{n}} & \alpha_{nn'} & \alpha_{n\bar{n}'}  \\
    \varepsilon_{n\bar{n}} & m_n -\vec{\mu}_n \vec{B} & \alpha_{n\bar{n}'}  & \alpha_{nn'}  \\
    \alpha_{nn'} & \alpha_{n\bar{n}'}  & m_{n'} +\vec{\mu}_{n'} \vec{B}' & \varepsilon_{n\bar{n}}  \\
    \alpha_{n\bar{n}'}  & \alpha_{nn'} & \varepsilon_{n\bar{n}} & m_{n'} - \vec{\mu}_{n'} \vec{B}'
    \end{array}\right)
\end{equation}
Here, $\varepsilon_{n\bar{n}}$ is the $n \bar{n}$ Majorana 
mass mixing parameter, and  $\alpha_{nn'}$  and $\alpha_{n\bar{n}'}$
are mass mixing parameters for $n n{'}$ and for $n \bar{n}{'}$ correspondingly. In the follwoing we neglect possible TMM terms between $n,\bar n$ and $n',\bar n'$ states, and assume $m_{n'}=m_n$, 
$\mu_{n'} = \mu_{n}$. 

Thus, in this case, the final state antineutron can be a result of the classical $n\rightarrow \bar{n}$ with mixing mass amplitude $\varepsilon$, and with baryon number change $\Delta \mathcal{B}=-2$ and probability $P_{n\rightarrow \bar{n}}=\varepsilon_{n\bar{n}}^2 t^2$.
It can also arise due to the second order oscillation process: $n\rightarrow n{'}\rightarrow\bar{n}$ with amplitude ($\alpha_{nn'} \alpha_{n\bar{n}'}$) or $n \rightarrow \bar{n}{'}\rightarrow \bar{n}$ with amplitude $(\alpha_{n\bar{n}'}\alpha_{nn'})$ plus interference of all three channels. If $\varepsilon_{n\bar{n}}$ is very small and $\alpha_{nn'}$ and $\alpha_{n\bar{n}'}$ are relatively large, then $n\rightarrow \bar{n}$ could be observed for a non-zero sterile magnetic field. However, neither previous limits on free $n\rightarrow\bar n$ oscillation from experiments in which the magnetic field was suppressed \cite{BaldoCeolin:1994jz},  
nor nuclear stability limits from $n\rightarrow\bar n$ conversion in nuclei \cite{Abe:2011ky} would be valid for this scenario, since a fixed field $\vec{B}$ compensating for the magnetic field in the sterile sector would be needed to allow the full process $n\rightarrow n' \rightarrow \bar{n}$ to proceed. 
In fact, $n\rightarrow\bar n$ conversion in free neutron experiments can emerge as second order process induced by $n\rightarrow n'$ and $n\rightarrow n'$
conversions, with the probability 
$P_{n\bar n}(t)\simeq P_{nn'}(t) P_{n\bar n'}(t)$. 
Existing limits allow the oscillation times $\tau_{nn'}$ 
and $\tau_{n\bar n'}$ to be as small as $1\div 10$~s
(for a summary of present experimental situation see \cite{Berezhiani:2017jkn}). Therefore,
by properly tuning the value of magnetic field $B$ resonantly close to $B'$ (with precision of mG or so) and thus achieving the  quasi-free regime, the probability of 
induced $n\rightarrow\bar n$ oscillation can be rendered as large as 
\begin{equation}
\label{eq:indnnbarosc}
 P_{n\bar{n}}(t)=
 \frac14 \alpha_{n\bar{n}'}^2\alpha_{n\bar{n}'}^2t^4 \sin^2\beta
 %=\frac{t^2}{\tau^2_{n\bar n}} 
 = \frac{\sin^2\beta}{4} \left(\frac{t}{0.1~{\rm s}}\right)^4 
 \left(\frac{10^2\,{\rm s}^2}{\tau_{nn'}\tau_{n\bar n'}}\right)^2 \times 10^{-8}
 \end{equation}  
where $\beta$ is (unknown) angle between the directions of $\vec{B}$ and $\vec{B}'$ \cite{Berezhiani:2020nzn}. Hence, the probability of induced $n\rightarrow\bar n$ transition can be be several orders of magnitude larger than the present sensitivity in direct $n\rightarrow\bar n$ conversion (\ref{eq:freennbarosc}). Once again, for achieving such enhancement, 10 orders of magnitude or perhaps more, the magnetic field should not be suppressed but  one must scan over its values and directions for finding the resonance when magnitudes $B\approx B'$ and angle $\phi$ is non-zero. In addition, different from direct $n\bar n$ mixing, $n\rightarrow\bar n$ transitions induced via $n\rightarrow n'$ and $n\rightarrow n'$ mixings has a tiny effect on the stability 
of nuclei \cite{Berezhiani:2020nzn}.

\subsection{Antineutron-nucleon annihilation}\label{sec:annsig}

The distribution of final states following $\bar{n}$ annhilation in target nuclei Of critical importance for understanding the needs of the annihilation detector system for the $n\rightarrow \bar{n}$ (and $n\rightarrow n' \rightarrow \bar{n}$) searches. To date, the target material has been $^{12}$C, with $4$-$5$ pions in the final state, but their intranuclear origin and dynamics can be significant. Studies of these effects have been made \cite{Golubeva:1997fs,Golubeva:2018mrz,Barrow:2019viz}. %The simulations agree well with antiproton interactions data sets by accounting for $\sim100$ annihilation channels (shown in tables in Ref.~\cite{Golubeva:2018mrz}) across $\bar{p}p$ and $\bar{p}n$ annihilation processes, including heavy resonances. It is assumed that annihilation channels for $\bar{n}n$ are identical to $\bar{p}p$, and annihilation channels for $\bar{n}p$ are charge conjugated to $\bar{p}n$. These processes can then be considered for $\bar{n}p$ and $\bar{n}n$ annihilations; further computations model the particle transport through the nuclear medium (final state interactions) \cite{Golubeva:1997fs,Golubeva:2018mrz,Barrow:2019viz}. 
In this approach, a model of elementary $\bar{p}N$ annihilation is used, described in detail in \cite{Golubeva:1997fs,Barrow:2019viz}, taking into account  $\sim100$ annihilation channels for $\bar{p}p$ and $\sim80$ channels for $\bar{p}n$, including heavy resonances. The simulations of the elementary annihilation agree well with $\bar{p}$-$p$ interaction data sets (for instance, see Table II of \cite{Barrow:2019viz}). For $\bar{n}N$ annihilation, it is assumed that annihilation channels for $\bar{n}n$ are identical to $\bar{p}p$, and annihilation channels for $\bar{n}p$ are charge conjugated to $\bar{p}n$. Thus, annihilation processes can indeed be considered for $\bar{n}p$ and $\bar{n}n$. Further computations model the particle transport through the nuclear medium (final state interactions). The proposed model for $\bar{N}{}^{12}C$ was tested on available experimental data sets from $\bar{p}{}^{12}C$ annihilations at rest, showing good agreement.

Tab.~\ref{tab:multiplicities_exp_v_sim} shows the simulated and measured particle multiplicities following $\bar{p} {}^{12}C$ interactions based on 10,000 Monte Carlo events.  Pionic states dominate after the decay of heavy resonances, and are in good agreement with experimental data. The total energy of the final state particles is also shown, for which the measurement is well reproduced by the calculations.

\begin{center}
\begin{table}[H]  %table font is tiny, could this be vertical and side-by-side with figure, which is very large?
% \vskip .1cm
\tiny{
    \begin{tabular}{c|ccccccc}
     & $M(\pi)$ & $M(\pi^+)$ & $M(\pi^-)$ & $M(\pi^0)$ & $E_{tot}$\,(MeV) & $M(p)$ & $M(n)$\\ \hline
    $\bar{p}{\rm C}$ Experiment & $4.57 \pm 0.15$ & $1.25 \pm 0.06$ & $1.59 \pm 0.09$ & $1.73 \pm 0.10$ & $1758 \pm 59$ & ----- & -----\\
    $\bar{p}{\rm C}$ Simulation & $4.60$ & $1.22$ & $1.65$ & $1.73$ & $1762$ & $0.96$ & $1.03$\\
    \end{tabular}}
    \caption{A list of multiplicities $M$ from experimental data and simulations taking into account $\bar{p}$ annihilation branching ratios \cite{Golubeva:2018mrz,Barrow:2019viz} while also considering intranuclear (anti)nucleon potentials with associated nucleon mass defects and nuclear medium response. Based on simulations of 10,000 events. Measurements of proton and neutron multiplicities were not made.}
\label{tab:multiplicities_exp_v_sim}
\end{table}
\end{center}

\begin{figure}[H]
    \begin{center}
    \includegraphics[width=0.69\linewidth]{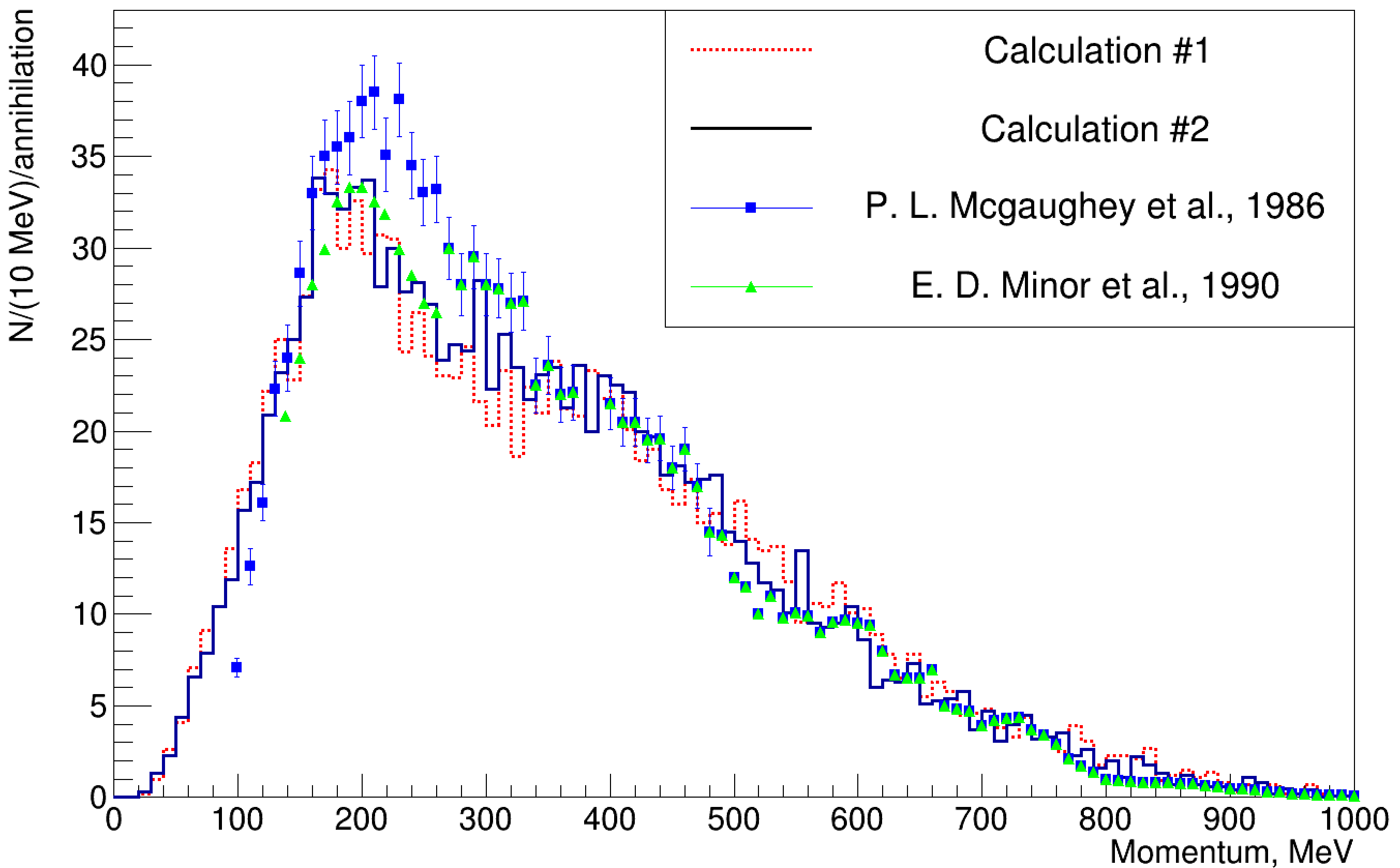} \\
    \includegraphics[width=0.69\linewidth]{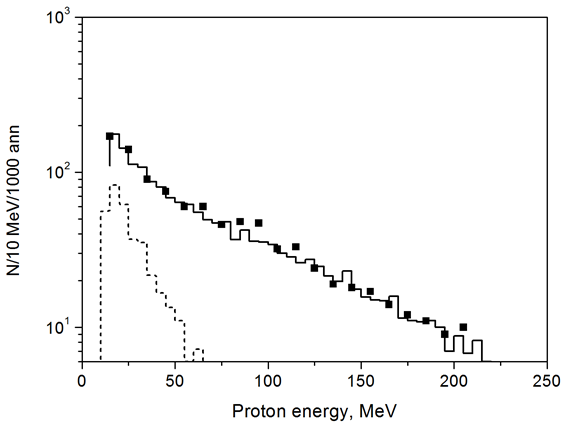}
    \caption{Final state kinematic distributions for 10,000 $\bar{p}{\rm C}$ simulated annihilation events at rest. Top: momentum distribution of positively charged pions \cite{Barrow:2019viz}. The solid histogram shows the distribution generated from the simulation mentioned in Table~\ref{tab:multiplicities_exp_v_sim}, while the red histogram shows the behavior when (anti)nucleon potentials and mass defects are not considered. Available pertinent data are shown. Bottom: energy distribution of final state protons \cite{Golubeva:2018mrz}. The solid line shows the full spectrum and the dotted line represents the contribution from evaporative processes. All points are taken from experimental data in \cite{Minor1990,Mcgaughey:1986kz}. See \cite{Barrow:2019viz} for detailed discussions.}
    \label{fig:pbarCarbon-PiPlusMomentum}
    \end{center}
\end{figure}

The momentum distribution of positively charged pions is shown in Fig.~\ref{fig:pbarCarbon-PiPlusMomentum}. The momentum peaks around $\sim 250$~MeV, albeit within a broad distribution which extends up to around $1000$~MeV. The data are reasonably well described by the simulation. The figure also shows the distribution of kinetic energies of protons. As before the data are well described. The contribution from evaporative protons is also shown and is seen to correspond to low values of kinetic energy ($\sim 20$~MeV). 
Given the agreement with data observed in Fig.~\ref{fig:pbarCarbon-PiPlusMomentum}, there is some measure of confidence in the simulations of extranuclear $\bar{n} {}^{12}C$ final states. 

\begin{figure}[H]
%    \begin{center}
    \includegraphics[width=0.49\linewidth]{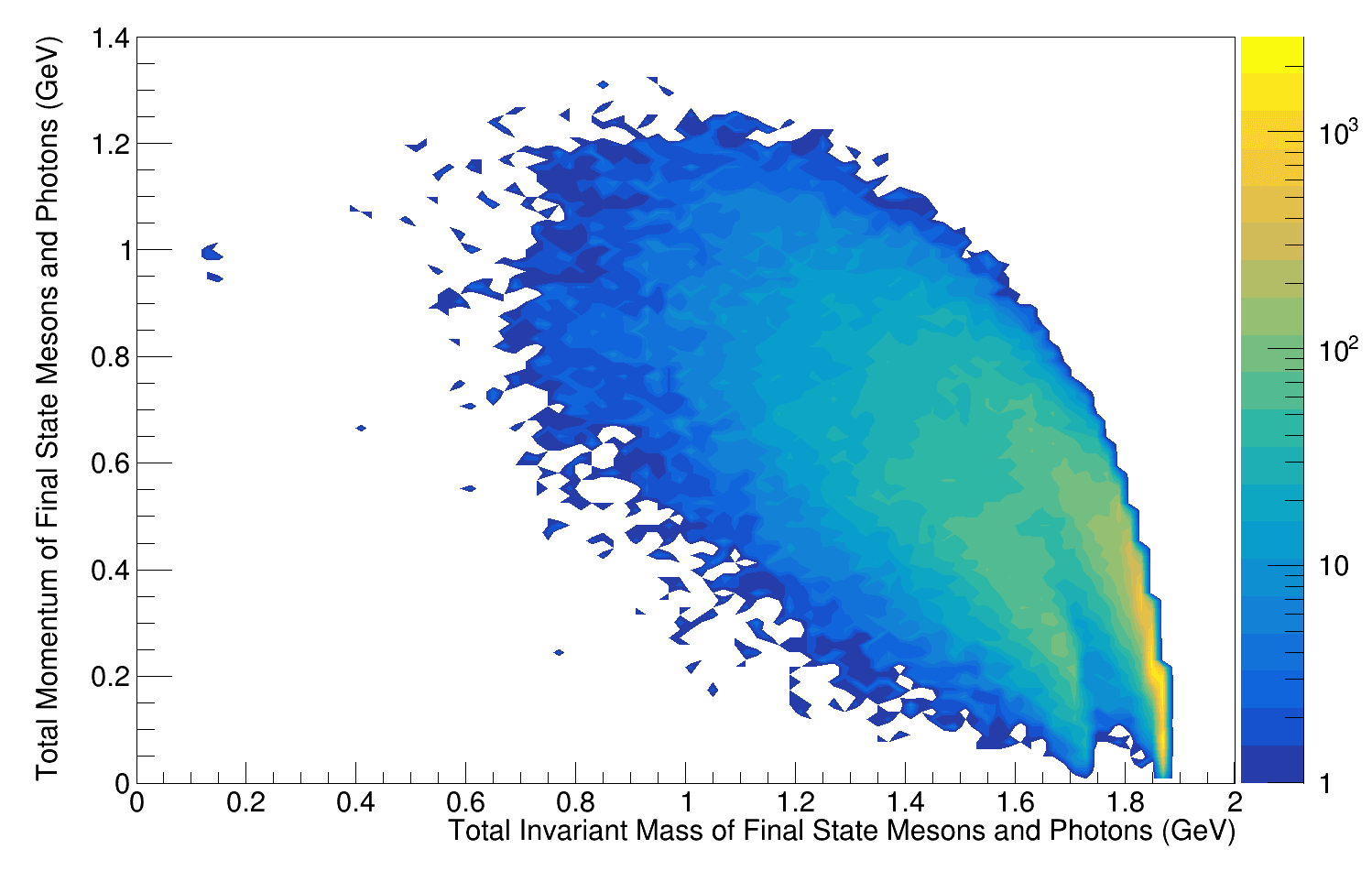}
    \includegraphics[width=0.49\linewidth]{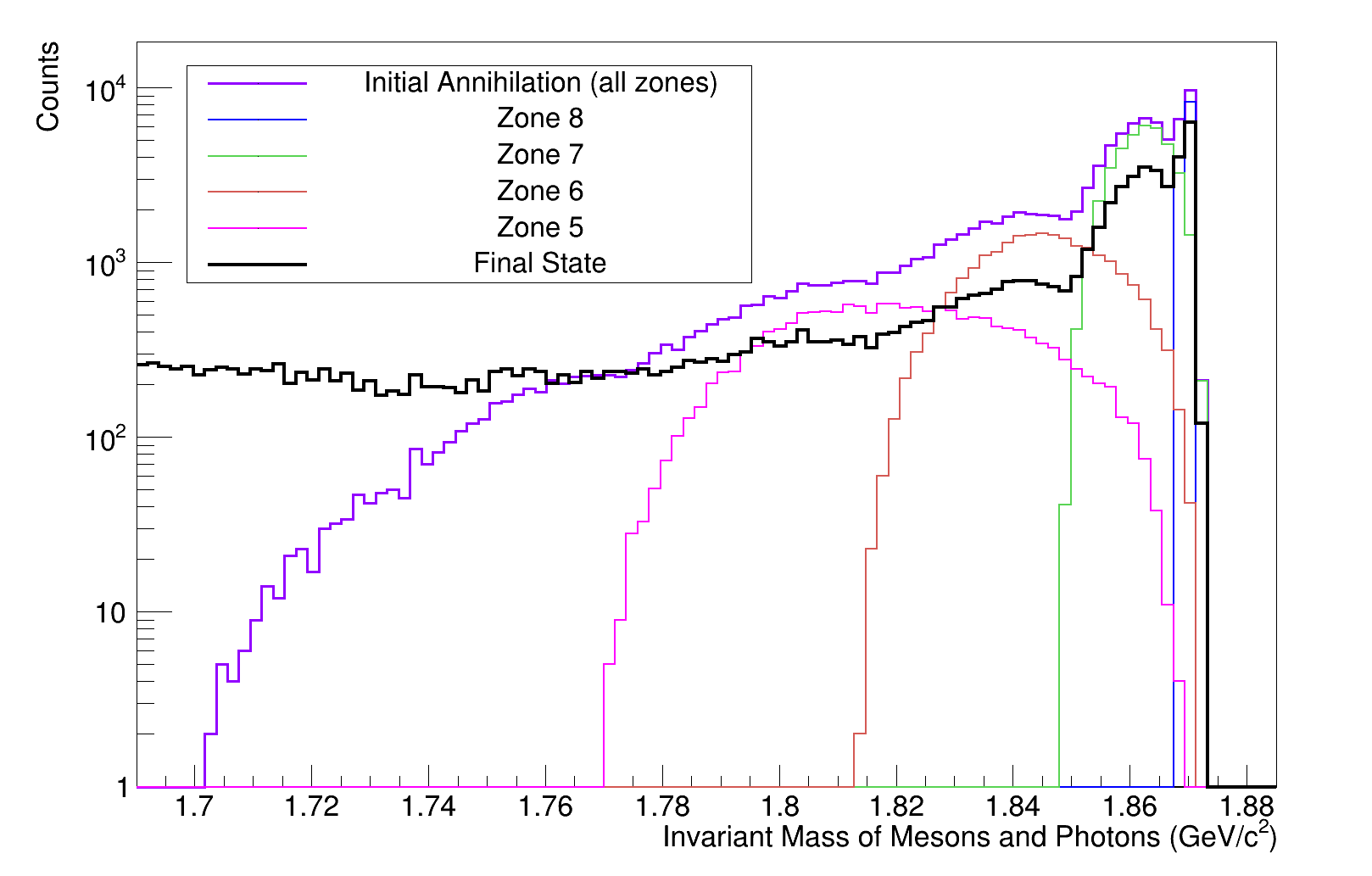}
    \caption{Final state kinematic distributions following an extranuclear $\bar{n} C$ annihilation for 100,000 events using an antineutron potential as described in \cite{Barrow:2019viz} (calculation number 2 therein).
    Left: the final states' sum total momentum of emitted \textit{pions and photons} versus the total invariant mass of the \textit{pion and photon} system. or similar figures and discussions, see \cite{Abe:2011ky}.
    Right: one dimensional distributions of the same invariant mass as at right. The purple histogram characterizes the initial state's invariant mass of all annihilations over the whole of the nucleus. The black histogram shows the final state pions and photons' after  undergoing transport. Other colors show the origin of the spectra's structure, with the leftward march of the distributions arising from the annihilation taking place further and further into the interior of the nucleus where isotropically distributed Fermi motion can become large within zones of smaller radii (only some zones are shown). For further details, see \cite{Golubeva:2018mrz,Barrow:2019viz}.}
%    \end{center}
  \label{fig:InitMesMomvsInvMass} \end{figure}

Fig.~\ref{fig:InitMesMomvsInvMass}, which shows the total momentum of the final state system of emitted \textit{mesons and photons} versus the system's total invariant mass. Owing to nuclear effects (final state interactions, rescattering, absorption), the final state invariant mass distribution for mesons and photons falls to less than $1$~GeV, lower than would be expected for a naive $\bar{n}N$ annihilation at around $1.9$~GeV. The figure also shows the distribution of invariant mass arising only from original annihilation mesons before and after transport. The kinematic distributions shown in Figs.~\ref{fig:pbarCarbon-PiPlusMomentum} and \ref{fig:InitMesMomvsInvMass} have implications for the detection strategy of annihilation events.

\section{Searches for neutron conversions}\label{Sec:previous_searches}

\subsection{Previous searches for $n\rightarrow \bar{n}$}\label{sec:search-nnbar}
\begin{figure}[tb] 
  \setlength{\unitlength}{1mm}
  \begin{center}
  \includegraphics[width=0.75\linewidth, angle=0]{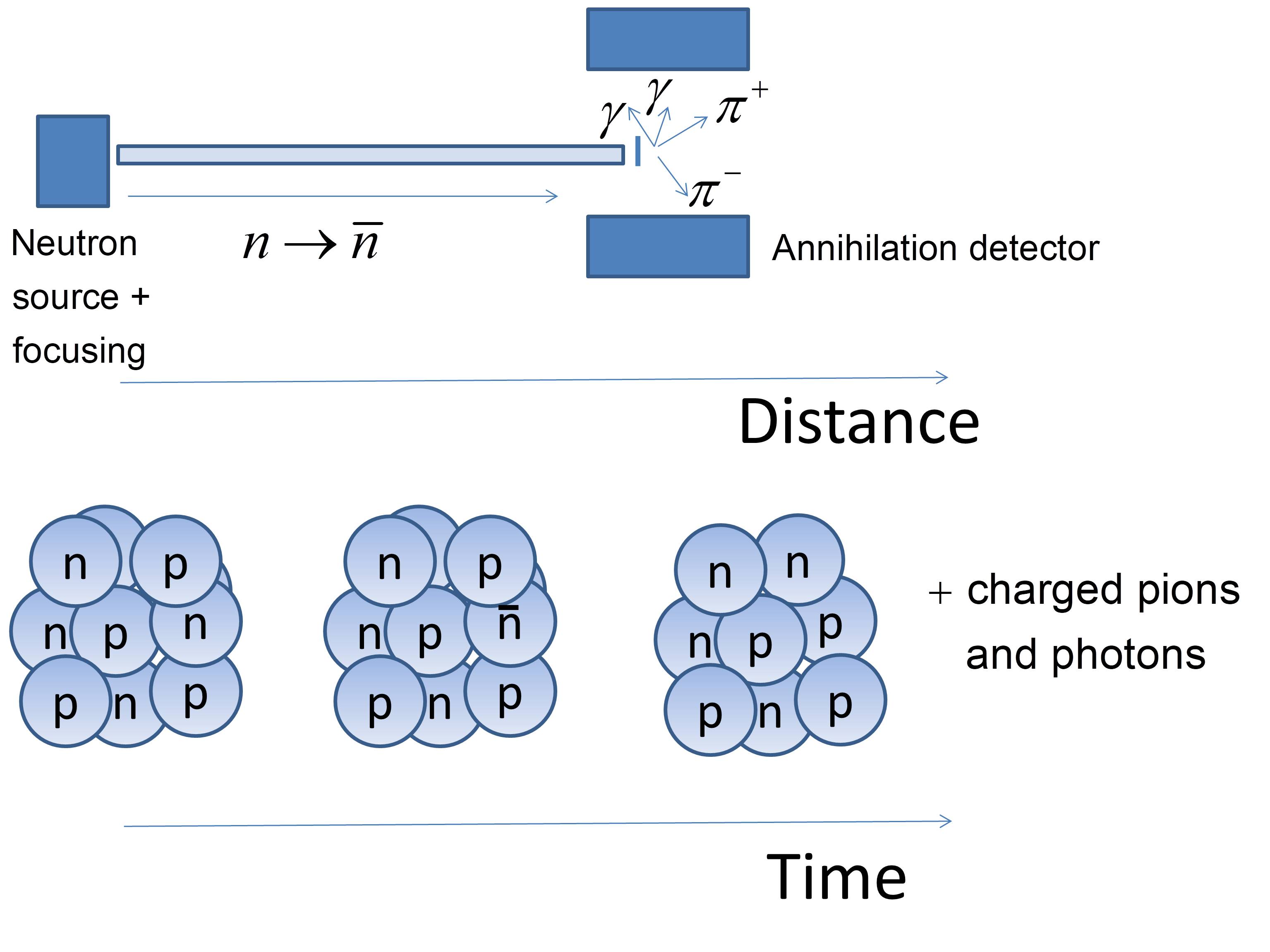}
  \end{center}
  \caption{Illustration of the principles of free (top) and bound (bottom) searches for $n\rightarrow \bar{n}$ .}
  \label{fig:nnbar-cartoon}
\end{figure}

As illustrated in Fig.~\ref{fig:nnbar-cartoon}, free (or extranuclear) searches consist of a beam of focused free neutrons propagating in field-free (or quasi-free) regions to an annihilation detector at which any antineutrons would annihilate with a thin target, giving rise to a final state of charged pions and photons. Searches for free $n\rightarrow \bar{n}$ oscillation have taken place at the Pavia Triga Mark II reactor\cite{Bressi:1989zd,Bressi:1990zx} and at the ILL~\cite{Fidecaro:1985cm,BaldoCeolin:1994jz}. The latter ILL search~\cite{BaldoCeolin:1994jz} provides the most competitive limit for the free neutron oscillation time: $\sim 8.6 \times 10^7$s. 

The figure of merit ($FOM$) of sensitivity for a free $n\rightarrow\bar{n}$ search is best estimated not by the oscillation time sensitivity but by the quantity below:
\begin{equation}
    FOM=\sum_i N_{n_{i}} \cdot t_{n_{i}}^2 \sim <N_n\cdot t_n^2>, 
\end{equation}\label{eq:fomnnbar}
where $N_{n_{i}}$ is the number of neutrons per unit time reaching the annihilation detector after $t_{n_{i}}$ seconds of flight through a magnetically protected, quasi-free conditioned vacuum region. As Eq.~\ref{eq:freennbarosc} shows, the probability of a conversion is proportional to the (transit time)$^2$. Thus $FOM = <N_n \cdot t_n^2>$ is proportional to the approximate number of the conversions per unit time in a neutron beam which impinge on a target. 

A high precision search therefore requires a large flux of slow neutrons produced at a low emission temperature which are allowed to propagate over a long time prior to allow conversions to antineutrons. As shown in the subsequent Sections, these conditions are satisfied in searches at the ESS.

Searches for $n\rightarrow \bar{n}$ in bound neutrons in large volume detectors look for a signature of pions and photons consistent with a $\bar{n}N$ annihilation event inside a nucleus, as illustrated in Fig.~\ref{fig:nnbar-cartoon}. Searches have taken place at Homestake~\cite{Homestake}, KGF~\cite{KGF}, NUSEX~\cite{NUSEX}, IMB~\cite{IMB}, Kamiokande~\cite{Kamiokande}, Frejus~\cite{Frejus}, Soudan-2~\cite{Soudan-2}, the Sudbury Neutrino Observatory~\cite{Aharmim:2017jna}, and Super-Kamiokande~\cite{Abe:2011ky,Gustafson:2015qyo}. A signature of pions and photons consistent with a $\bar{n}N$ annihilation event was sought, with Super-Kamiokande providing the most competitive search, for which an inferred free neutron oscillation time lower limit of $\sim 2.7 \times 10^8$s was obtained. Super-Kamiokande has also searched for dinucleon decays to specific hadronic final states, such as $nn \to 2\pi^0$ and $np \to \pi^+\pi^0$, as well as dinucleon decays into purely leptonic and lepton+photon final states~\cite{Bryman:2014tta,Takhistov:2015fao,Sussman:2018ylo}. Further limits on BNV decays have been obtained by relating these types of decays~\cite{Girmohanta:2019xya,Nussinov:2020wri,Girmohanta:2019cjm,Girmohanta:2020qfd}.

\subsection{Previous searches for $n\rightarrow n'$}\label{sec:search-nnprime}

\begin{figure}[tb]
  \setlength{\unitlength}{1mm}
  \begin{center}
  \includegraphics[width=0.9\linewidth, angle=0]{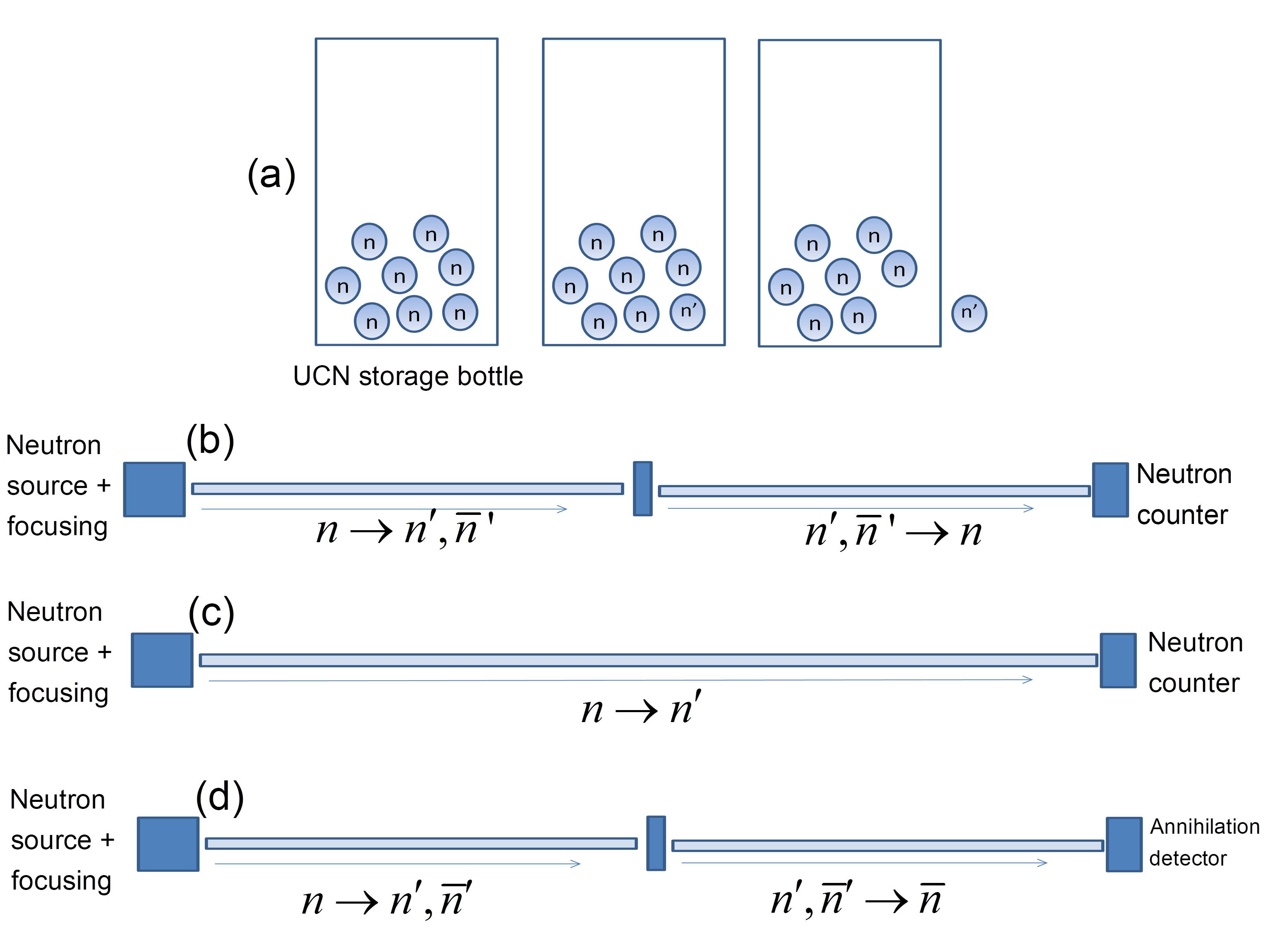}
  \end{center}
  \caption{Illustration of the principles of searches for sterile neutrons in a UCN trap (a) and for beam neutrons.   Regeneration, disappearance and $n\rightarrow\{\bar{n},n'\}\rightarrow \bar{n}$ modes for neutrons along a beamline are shown in (b),(c), and (d) respectively.}
  \label{fig:nnprime-cartoon}
\end{figure}

Two main experimental approaches are used to search for sterile neutrons: measurements of neutrons trapped in a UCN bottle and measurements of beam neutrons\footnote{In principle, although large volume experiments could have a sensitivity to sterile neutrons, searches for sterile neutron-induced destabilised neutrons are problematic. Any interpretation of results would depend strongly on the composition and properties of the dark sector to which a sterile neutron would belong~\cite{Berezhiani:2005hv}.}. The principles behind these approaches are 
illustrated in Fig.~\ref{fig:nnprime-cartoon}. There would be anomalous loss of neutrons from the UCN trap via their conversion to sterile neutrons (Fig~\ref{fig:nnprime-cartoon} (a)). With beam neutrons, experiments can look for the regeneration of neutrons following a beam stop (Fig~\ref{fig:nnprime-cartoon} (b)), an unexplained disappearance of neutron flux (Fig~\ref{fig:nnprime-cartoon} (c)) and $n\rightarrow  \bar{n}$ via a sterile neutron state. For a comprehensive set of searches with both UCN and beam neutrons, the experiments should scan as wide a range of magnetic fields as possible to induce a neutron-sterile neutron transitions.  

%LJB  trimming the history exposition and moving motivation to previous section
Early searches for sterile neutrons were performed using UCN gravitational storage traps to correlate the possible disappearance of neutrons with the variation of the laboratory magnetic fields \cite{Tanabashi:2018oca,Ban:2007tp,Serebrov:2007gw,Serebrov:2008hw,Bodek:2009zz,Serebrov:2009zz}, assuming the Earth's magnetic field should be compensated to near zero (to satisfy the quasi-free condition) to permit the $n \rightarrow n'$ process to occur. With this assumption, the best limit for a free oscillation time $\tau_{n \rightarrow n'}$was obtained by \cite{Serebrov:2008hw}, where $\tau_{n \rightarrow n'} \geq 448$ s (90 \% CL). More recent measurements and analyses~\cite{Berezhiani:2012rq,Altarev:2009tg} have accounted for the possibility of a modest sterile sector magnetic field by including a wider 
%In further experiments, following the paper \cite{Berezhiani:2008bc}, it was assumed that the sterile sector magnetic field $\bf{B'}$ may exist with unknown magnitude and direction (though probably smaller than the Earth's magnetic field). Such a sterile sector field could arise within the gravitational well of the Earth due to the accumulation of ionized gas clouds of sterile hydrogen and sterile helium through the dynamo effect \cite{Berezhiani:2008bc}. 
%Thus, more measurements were performed with 
variation of the laboratory magnetic field $\bf{B}$ in the UCN traps.
%, assuming that $\bf{B}$ could partially compensate for the sterile magnetic field $\bf{B'}$ and energetically permit transitions. 
From the analysis of all existing UCN experimental data, the lower limits on $\tau_{n \rightarrow n'}$ as a function of $|\bf{B'}|$ were obtained \cite{Berezhiani:2017jkn} in the range of tens of seconds for the sterile sector magnetic fields less than $\sim 0.3$ G. However, one UCN experiment \cite{Serebrov:2007gw,Serebrov:2008hw} reanalyzed in \cite{Berezhiani:2012rq} has reported a non-zero asymmetry with a significance of $5\sigma$ in the storage time of unpolarized neutrons in a Be-coated trap when a laboratory magnetic field was regularly changed from +$0.2$ G to $-0.2$ G. This anomalous result was interpreted \cite{Berezhiani:2012rq} as a $n \rightarrow n'$ oscillation signal with the asymmetry caused by the variation of the angle $\beta$ between vectors of magnetic fields of sterile and laboratory fields, $\bf{B'}$ and $\pm \bf{B}$. Thus, $n \rightarrow n'$ transitions with e.g. $\tau_{n \rightarrow n'}\sim 30$ s are not excluded for a region of $|\bf{B'}|\sim 0.25$ G. 
%move details to motivation/anni section
%Besides that the $N-N'$ Collaboration at ORNL plans to explore other mechanisms of neutron to sterile neutron transformations described in the papers: \cite{Berezhiani:2018eds} with slightly non-degenerate masses of sterile and ordinary neutrons, \cite{Berezhiani:2018qqw} utilizing a new concept of neutron Transition Magnetic Moment, \cite{Berezhiani:2018zvs}. For some of these processes \cite{Berezhiani:2018eds,Berezhiani:2018qqw,Berezhiani:2018zvs} 

%%searches
\section{Overview of ESS}\label{sec:ess}

The European Spallation Source, ESS, currently under construction in Lund~\cite{Peggs:2013sgv}, will be the world's most powerful facility for research using neutrons. It will have a higher useful flux of neutrons than any research reactor, and its neutron beams will have a brightness that is up to two orders of magnitude higher than at any existing neutron source.

ESS is organised as a European Research Infrastructure Consortium (ERIC) and currently has 13 member states: Czech Republic, Denmark, Estonia, France, Germany, Hungary, Italy, Norway, Poland, Spain, Sweden, Switzerland and the United Kingdom. Sweden and Denmark are the host countries, providing nearly half of the budget for the construction phase. More than half of the budget from the non-host countries is in the form of in-kind contributions, meaning that the countries are delivering components to the facility (accelerator, target, integrated control system and neutron scattering systems) rather than cash.

The project has been driven by the neutron-scattering community, and the construction budget includes 15 instruments covering a wide range of topics in neutron science. ESS will also offer opportunities for fundamental physics with neutrons, for instance as described in this paper.

Most of the existing spallation neutron sources use a linear accelerator to accelerate particles to high energy. The particles are stored in an accumulator ring and are then extracted in a short pulse to the spallation target. A notable exception is SINQ at PSI, which uses a cyclotron that produces a DC beam on the spallation target. ESS will use a linear accelerator but no accumulator ring, and it will thus have longer neutron pulses. This will allow more neutrons to be produced for a given budget, and for most studies in neutron scattering the long pulses will not be at any disadvantage, often rather the opposite. For experiments in fundamental physics where total integrated flux is a main figure of merit, the ESS concept is clearly of major benefit.

The high neutron flux at ESS is also due to the fact that it will have the world's most powerful particle accelerator, in terms of MW of beam on target. It will have a proton beam of 62.5\,mA accelerated to 2\,GeV, with most of the energy gain coming from superconducting RF cavities cooled to 2\,K. Together with a 14 Hz pulse structure, each pulse being 2.86\,ms long, this gives 5\,MW average power and 125\,MW of peak power. For proton energies around a few GeV, the neutron production is nearly proportional to beam power, so the ratio between beam current and beam energy is to a large extent the result of a cost optimisation, while the pulse structure is set by requirements from neutron science.

\begin{figure}[p]
  \setlength{\unitlength}{1mm}
  \begin{center}
  \includegraphics[width=1.0\linewidth, angle=0]{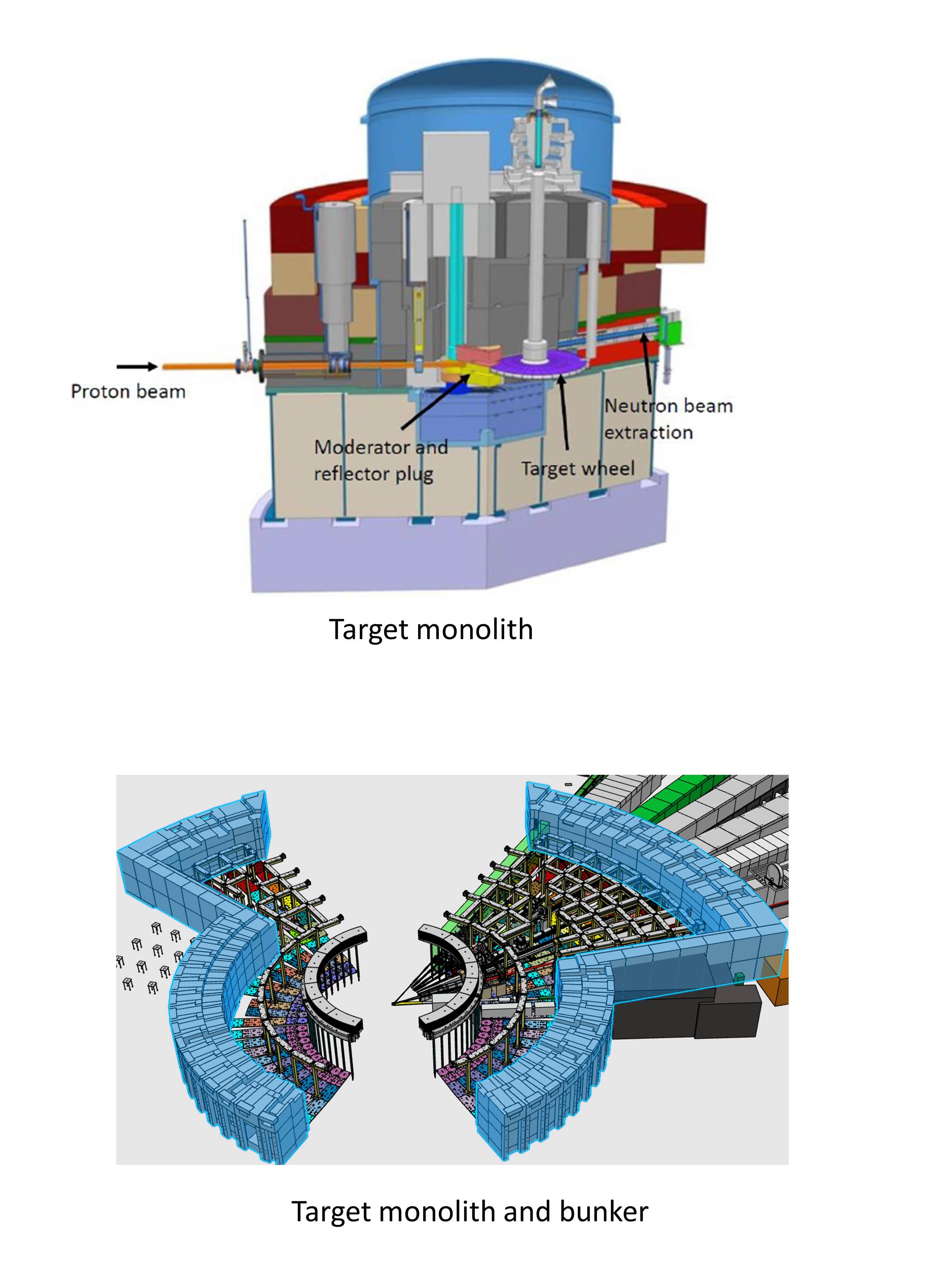}
  \end{center}
  \caption{Top: the ESS target monolith.Bottom: the ESS target monolith and bunker, view from above.}
  \label{fig:ess-monobunker}
\end{figure}

The neutrons are produced when the protons hit a rotating tungsten target. The target wheel consists of sectors of tungsten blocks inside a stainless-steel disk. It is cooled by helium gas, and it rotates at approx. 0.4\,Hz, such that successive beam pulses hit adjacent sectors, allowing adequate heat dissipation and limiting radiation damage. Fig.~\ref{fig:ess-monobunker} shows a cut-out of the target monolith, having the tungsten wheel in the centre. High-energy spallation neutrons are slowed down in an adjacent cold neutron moderator surrounded by a beryllium reflector, exiting the moderator-reflector system to be fed to beam extraction points placed within the monolith wall. The monolith extends to a radius of 5.5\,m and contains 3.5\,m of steel shielding extending from the beamline opening that are located 2\,m after the moderator
center.

The neutron radiation dose coming out of the monolith is substantial, and further shielding is needed in the structure referred to as ``the bunker". The bunker, comprised mainly of concrete, ensures that dose levels at the outer bunker surface are less than $3\mu$Sv/h. Within the bunker, neutron beams are delivered to multiple instruments, which are distributed in two wide angle regions on both sides of the target area. 
%The radii of these angular segments individually extend to $15\,$m and $28\,$m, and each region %corresponds to an angular coverage of $120\degree$, giving the facility a capacity of up to %$\sim20$ neutron instruments with an angular separation of $\sim6\degree$. 
Neutrons from the monolith are fed into neutron guides in the bunker, pass through the bunker wall, and, ultimately, on to ESS instruments. In addition to the shielding, the bunker contains components related to the instruments such as guides, choppers, shutters and collimators \cite{ess-bunker}.

\begin{figure}[tb]
  \setlength{\unitlength}{1mm}
  \begin{center}
  \includegraphics[width=0.75\linewidth, angle=0]{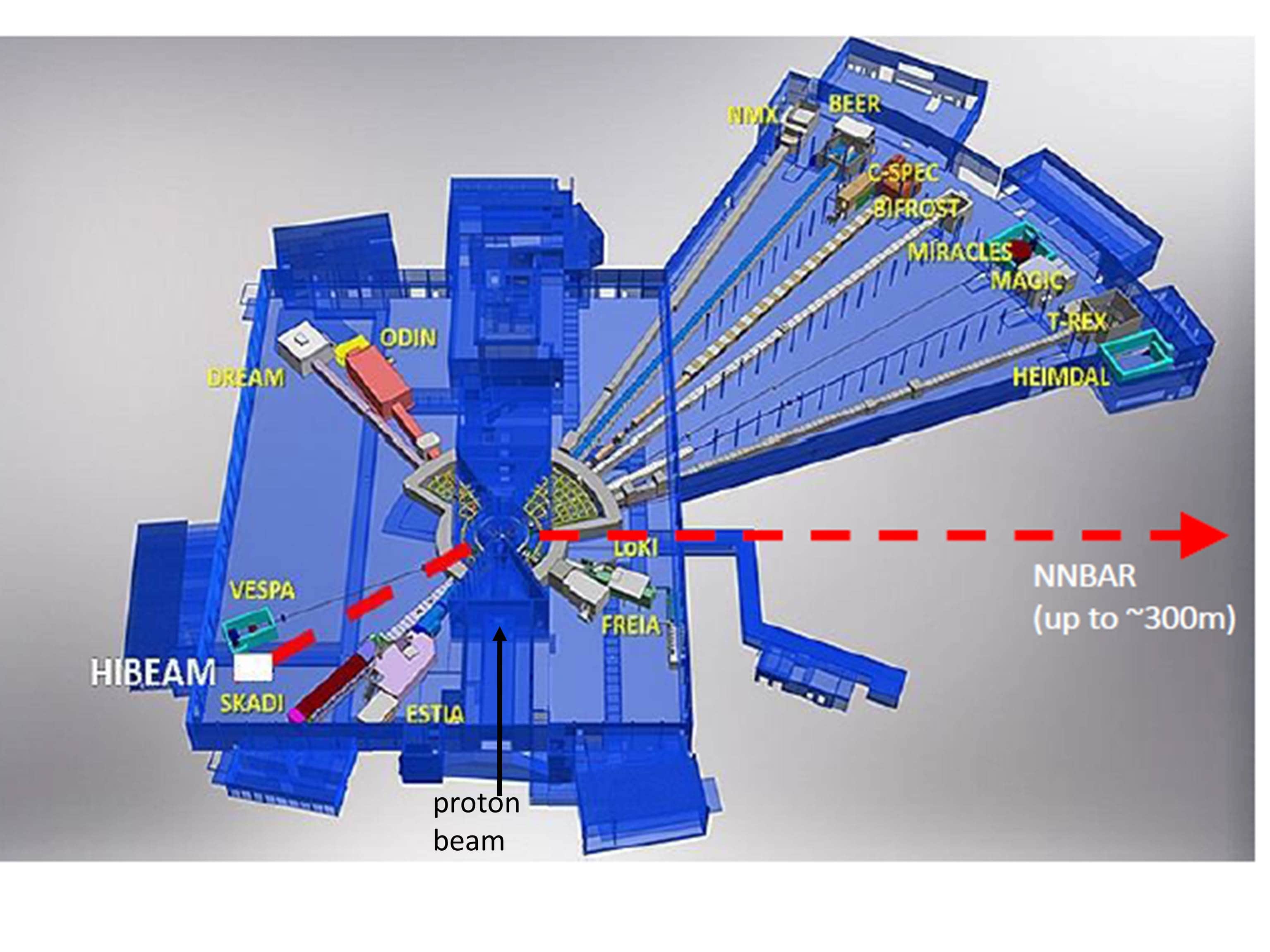}
  \end{center}
  \caption{Overview of the ESS, beamlines and instruments. The locations for the proposed HIBEAM  and NNBAR experiments are also shown. }
  \label{fig:ESSinstruments}
\end{figure}

In Fig. \ref{fig:ESSinstruments}, an overview of the ESS beamlines and instruments is shown. 
There are $15$ instruments currently under construction at ESS, representing only a subset of the full $22$-instrument suite required for the facility to fully realize its scientific objectives as defined in ESS statutes.
In addition to the 15 instruments, a test beamline will be installed among the very first instruments which serves the primary purpose to characterise the target-reflector-moderator system, verifying the performance of the neutron source at the start of operations. It also allows to test and develop relevant neutron technologies.
Regarding instruments $16$-$22$, an ESS analysis of the facility's scientific diversity has identified that the addition of a fundamental physics beamline is of the highest priority~\cite{ess-gap}. The location of the foreseen ESS fundamental physics beamline, HIBEAM/ANNI at beamport E5, is shown. The prospective beamline from the LBP leading to NNBAR is also shown, though would extend far beyond the radii of other instruments.

\subsection{ESS timescales and power usage projections}
\begin{figure}[tb]
  \setlength{\unitlength}{1mm}
  \begin{center}
  \includegraphics[width=1.\linewidth, angle=0]{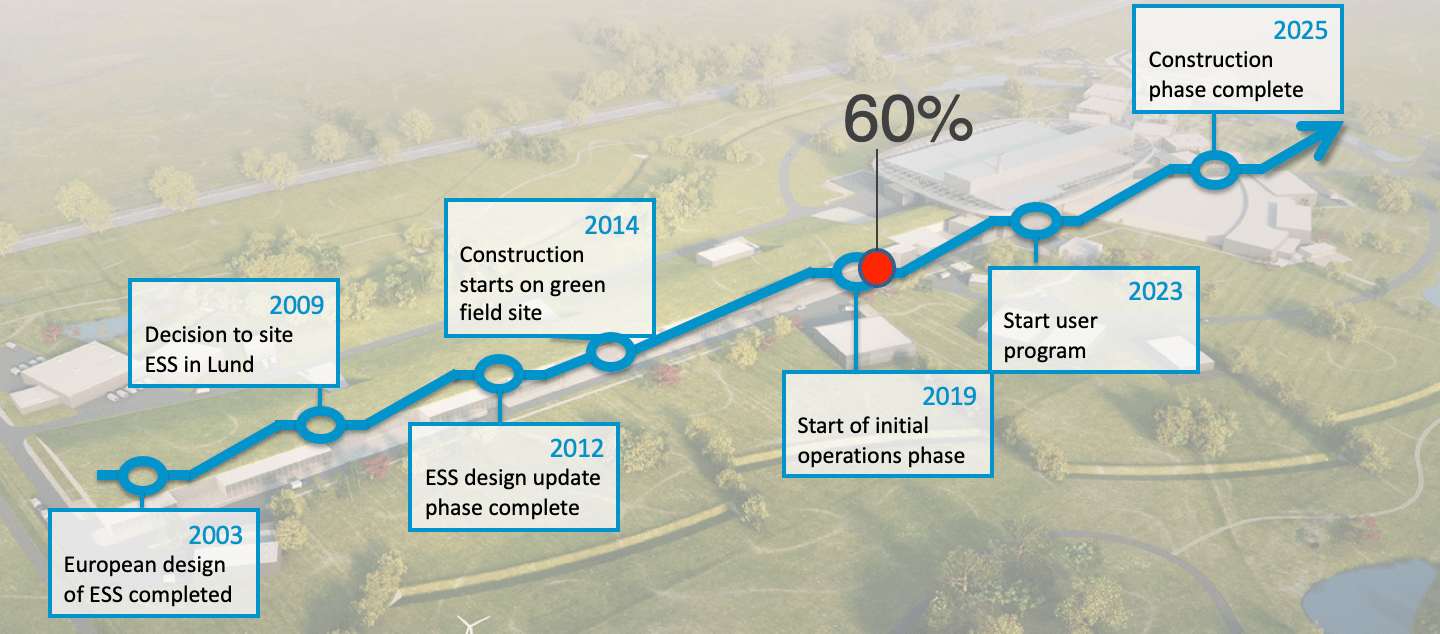}
  \end{center}
  \caption{The ESS overall schedule after re-baselining. Milestones are shown, as is the current stage of development.}
  \label{fig:ESSschedule}
  \end{figure}

In 2018, a re-optimisation of the ESS schedule took place, due mainly to the fact that the original ESS timeline was established before construction began in 2014, and so was impacted by building delays resulting from ESS's implementation of updated seismic and security standards adopted in recent years within Sweden (and globally). This has primarily impacted the design and construction of the ESS Target Building, as the strengthened standards were formulated in the aftermath of the Fukushima accident, and amid a general increase in concern over global security threats. Following this new baseline, the new plan has the goal to start instrument commissioning in 2022, with early scientific experiments to be carried out on the first three instruments as soon as possible thereafter. The Start Of the User Programme (SOUP) at ESS is expected to begin in 2023, as seen in the ESS overall Schedule after re-baselining shown in Fig. \ref{fig:ESSschedule}. In the ESS initial stages (July 2022-December 2023), the first neutrons will be produced from the target at very low beam power, where commissioning will include ramp up and testing from Accelerator, Target and Integrated Control Systems (ICS). As also shown in Fig.~ \ref{fig:ESSschedule}, the ESS is approximately 60\% along its development and commissioning path. The planned ESS ramp up time frame is shown in Tab. \ref{tab:ESSrampup}.

\begin{center}
\begin{table}
\begin{tabular}{ |c|c|c|c|c| }
 \hline
  & Jan. $2024$ & Jan. $2025$ & Jan. $2026$ & Jan. $2027$ \\
 \hline
 Source operator power (MW) & $>0.57$ & 1.25 & 2 & 2 \\
 \hline
 Source availability & $80\%$ & $85\%$ & $90\%$ & $95\%$ \\
 \hline
 Source installed capacity (MW) & $1$ & $2$ & $2$ & $2$ \\
 \hline
 Instruments in operation & $3$ & $8$ & $12$ & $15$ \\
 \hline
 Days of neutron production & $200$ minus long & $200$ minus long & $200$ & $200$ \\
 & shutdown days & shutdown days & & \\
 \hline
 \end{tabular}
\caption{Target values for the ramp up of the ESS accelerator and instrument availability.}
\label{tab:ESSrampup}
\end{table}
\end{center}

\subsection{Moderator}\label{sec:moderatorsec}
The configuration of the ESS neutron source and moderator systems present excellent opportunities for the fundamental physics research of HIBEAM and NNBAR. This is achieved thanks to the high brightness of the ESS source, the configuration of the beam extraction system, and the upgradeability options available for the source which are attractive for proposed fundamental physics applications. The upper and lower moderators have been designed for the initial suite of $16$ instruments ($15$ neutron scattering instruments, plus the test beamline, located at W11, the same position as NNBAR). The design of the moderators is fully described in \cite{zanini2019design} and \cite{andersen2018optimization}. The features of interest for HIBEAM and NNBAR include the following:
\begin{enumerate}
  \item The retaining of a monolith configuration for shielding openings and beam extraction ports such that moderators can be placed above and below the tungsten target. However, a design optimization for the initial instrument suite led to the choice of a single (upper) moderator system, leaving open an option for future upgrades \textit{below} the target useful for NNBAR.
  
  \begin{figure}[tb]
  \setlength{\unitlength}{1mm}
  \begin{center}
  \includegraphics[width=0.70\linewidth]{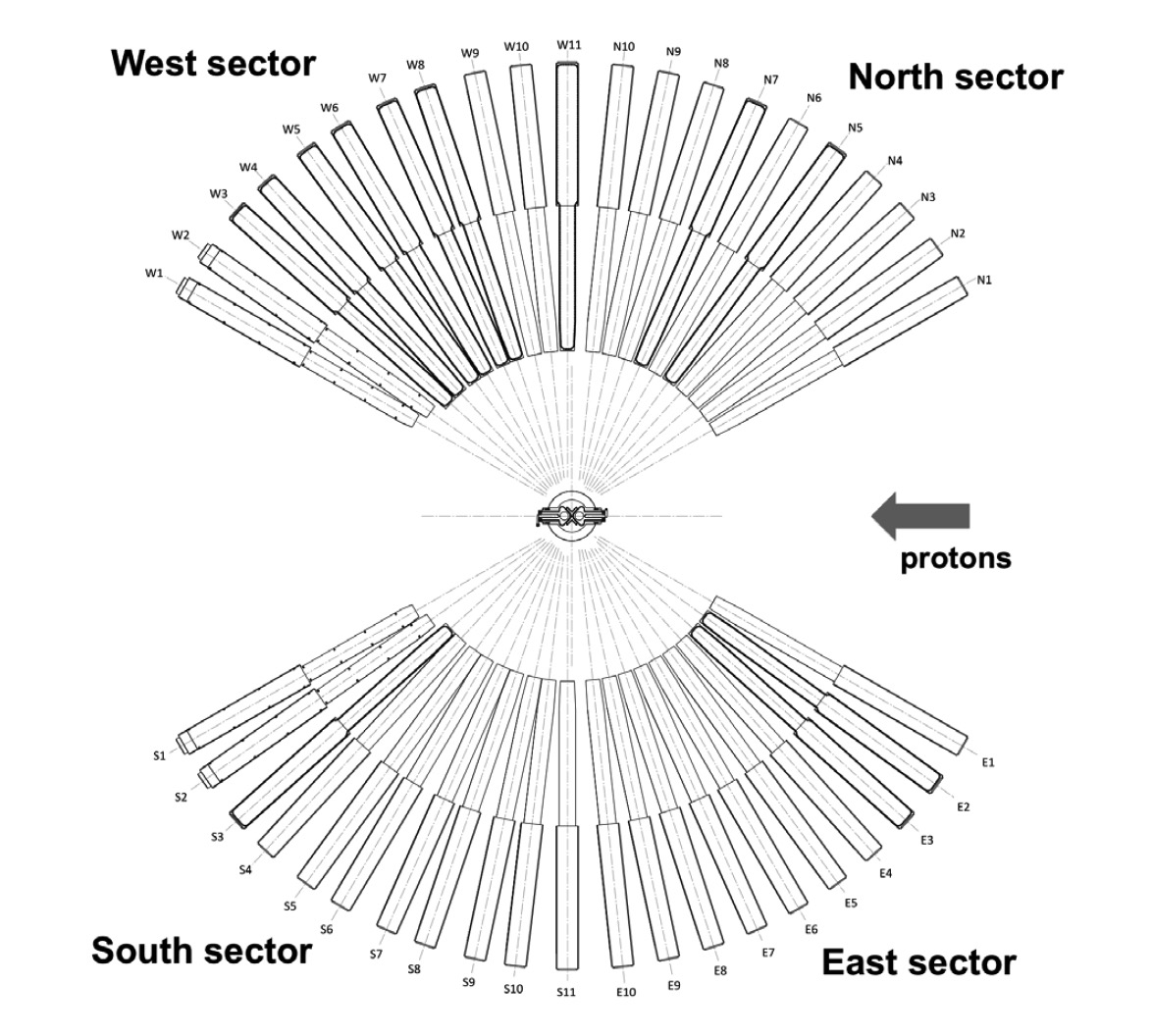}
  \end{center}
  \caption{The beamport systems of ESS arranged around the moderators. Note that neutron beam extraction is possible \textit{above} and \textit{below} the target. The location for the proposed HIBEAM experiment is at beamport E5. The location for the proposed NNBAR experiment is at beamports N10, W11 and W10.}
  \label{fig:luca-1}
\end{figure}
  
  \item The upper cold parahydrogen moderator is $3\,$cm thick, with a shape optimized for beam extraction in the $42$ beamports arranged in two 120\degree sectors, as seen in Fig. \ref{fig:luca-1}. For HIBEAM/NNBAR, the total available width of the moderator for beam extraction is about $17\,$cm.
  \item Moderators are placed in plugs that are replaced frequently (the average lifetime of a moderator system at full power is presently assessed to be $\sim1$ year)
  \item In consideration of possible future upgrades and fundamental physics experiments like NNBAR (which may need a larger moderator system), the inner shielding openings have been designed to be taller at the bottom than at the top of the tungsten target. A cross sectional view of the region of the target showing the moderators, inner shielding, and beam extraction openings is shown in Fig. \ref{fig:luca-2}.
  \end{enumerate}

\begin{figure}[tb]
  \setlength{\unitlength}{1mm}
  \begin{center}
\includegraphics[width=1.00\linewidth]{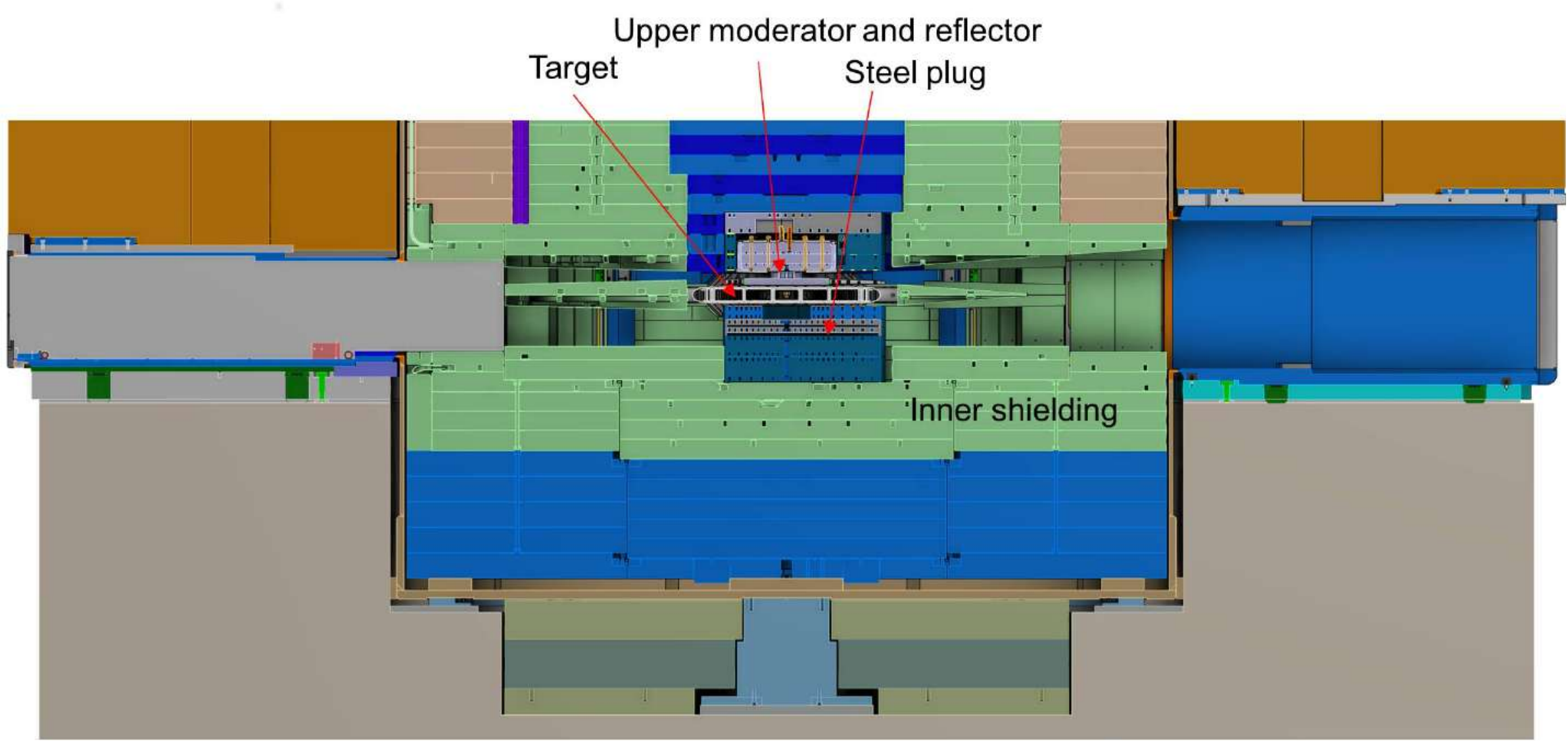}
    \end{center}
\caption{A cross sectional view of the target/moderator area, including the inner shielding.}
  \label{fig:luca-2}
%  \vspace{2.75cm}
\end{figure}

The HIBEAM experiment will use the upper moderator. By the time the experiment starts, it is expected that the Mark II moderator (“butterfly-1”), optimized for maximum cold brightness to all the instruments, will be in place. This moderator's main characteristic is its reduced height ($3\,$cm) compared to its lateral dimensions ($\sim24\,$cm), which was found to deliver maximum brightness to the sample areas across the instrument suite \cite{zanini2019design,andersen2018optimization}.

 A design study, termed HighNESS, is set to begin for a new cold neutron source and associated instruments at the ESS, including NNBAR. This is funded as a Research and Innovation Action within the EU Horizon 2020 program~\cite{euh2020,Santoro:2020nke}. A liquid deuterium moderator is envisaged which provides a high-flux, slow-neutron source required by NNBAR. This would be of benefit both to NNBAR and neutron scattering experiments for the following reasons: 1) NNBAR is an experiment limited to a few years of operation, and the rather frequent change of the moderator/reflector plugs is also in favor of the use of a dedicated moderator for NNBAR; 2) A large moderator for NNBAR may have other applications of interest at ESS, which could include neutron scattering (e.g., for neutron spin-echo imaging) or for ultracold neutron production. 

The lower moderator to be used by the NNBAR experiment should be tailored specifically for high-intensity neutron extraction, unlike other ESS moderators which are generally developed for brightness. This high intensity can be achieved by increasing the dimensions of the moderator. However, it has been shown \cite{zanini2019design} that a $3\,$cm parahydrogen moderator already delivers about $80\%$ of the maximum intensity achievable by increasing the moderator height. The only way to have a worthwhile increase in intensity is by using {a different type of moderator}. The choice under study is a liquid \isotope[2]{H} moderator, similar to what is used at a reactor source like the ILL (also in dimensions), or at the SINQ facility.  Preliminary studies of this options were performed in \cite{Klinkby:2014cma}, indicating an increase in intensity of a factor of $\sim3$ compared to a $3\,$cm flat moderator. This increase in intensity is related to the overall larger dimensions, as well as the absence of neutron absorption in deuterium. However, it should be noted that the response time of the deuterium is considerably longer than for hydrogen, making the use of such a moderator not optimal for some applications at a pulsed source (even a long pulse), but this should not be a problem for NNBAR.

Some possible parameters of the neutron source for HIBEAM and NNBAR are listed in Tab.~\ref{tab:mod}.
\begin{table}
\begin{center}
\begin{tabular}{ |c|c|c| }
 \hline
  & Upper & Lower  \\
 \hline
 Moderator temperature & $20$K & $20$K   \\
 \hline
 Total beam extraction & $17 \times 3\,$cm$^2$ & $25 \times 20$ cm$^2$  \\
 window at moderator face &   & \\
 \hline
Moderator time average & $5\times 10^{13}$ ncm$^{-2}$s$^{-1}$sr$^{-1}$ & $1.5\times 10^{13}$ ncm$^{-2}$s$^{-1}$sr$^{-1}$ \\
brightness ($E<20$ meV) &   & \\
\hline
Moderator time average & $2.5\times 10^{15}$ ns$^{-1}$sr$^{-1}$ & $7.5\times 10^{15}$ ns$^{-1}$sr$^{-1}$ \\
intensity ($E<20$ meV) &   & \\
 \hline
\end{tabular}
\end{center}
\caption{Parameters of the upper moderator, and possible parameters for the lower moderator, of relevance for HIBEAM/NNBAR. Note that the lower moderator's \isotope[2]{H} brightness and intensity values are highly approximate estimates at this stage of research and design, while expected values for the upper moderator are more precise~\cite{zanini2019design}.}
\label{tab:mod}
\end{table}

\subsubsection{The Large Beam Port for NNBAR}
\begin{figure}[tb]
  \setlength{\unitlength}{1mm}
  \begin{center}
\includegraphics[width=1.00\linewidth]{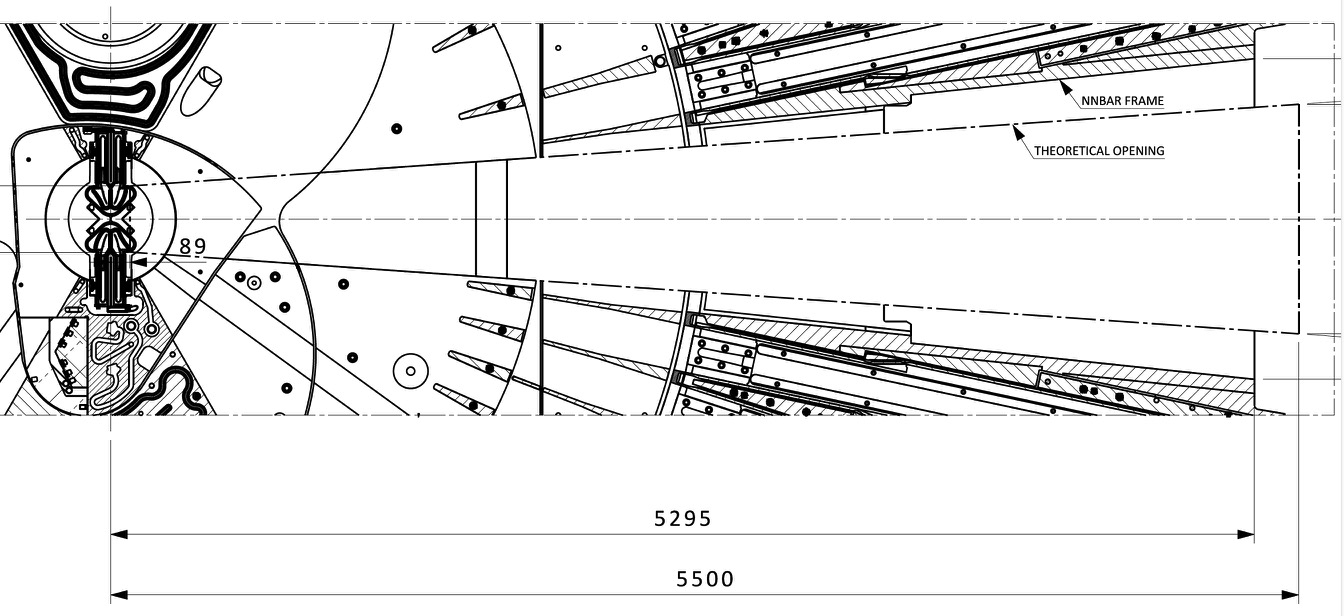}
    \end{center}
\caption{Top view of the NNBAR beamport.}
  \label{fig:valentina1-2}
%  \vspace{2.75cm}
\end{figure}

In the current baseline design of the ESS monolith, a critical provision has been made for the NNBAR experiment. A normal ESS beamport would be too small for NNBAR to reach its ambitious sensitivity goals. Therefore, part of the beam extraction system in the ESS monolith has been engineered so that a large frame covering the size of three beamports will be constructed. Initially, the frame will be filled by three regular-size beamports plus additional shielding for other experiments, including the ESS test beamline. The three beamports can be removed to provide a Large Beam Port to NNBAR for the duration of the experiment, and eventually replaced at the end of the experiment. Two views of the beam extraction region at the NNBAR beamport are shown in Figs. \ref{fig:luca-2} and \ref{fig:valentina1-2}, both from the moderator to the monolith exit. At the time of this writing, no other existing or planned neutron facility will have a large beam port of similar dimensions, making the ESS the ideal and site for a full scale NNBAR experiment. 

%Fig.~\ref{fig:moderatorandreflector} gives an overview of the moderator and internal reflectors. As seen, losses will occur due to the presence of the Fe shield and Be reflector system. Fig.~\ref{fig:moderatorandreflector} also shows a possible adjustment to the design of the beam port for the proposed work.
%Here, parts of the shield and reflector system would be removed to allow a greater conical penetration. 

%\begin{figure}[tb]
%  \setlength{\unitlength}{1mm}
%  \begin{center}
%\includegraphics[width=0.80\linewidth, angle=0]{reflectormoderator.pdf}
%    \end{center}
%    \vspace{-.75cm}
%  \caption{ Nominal (white region to the left of the neutron source) and enlarged (region enclosed %by dashed lines) conical penetration through the Be reflector and Fe shield).
%  }
%  \label{fig:moderatorandreflector}
%%  \vspace{2.75cm}
%\end{figure}

%%overview
\section{ANNI beamline}
\label{sec:HIBEAMProgram}
The search for $n \rightarrow \bar{n}$ with a sensitivity $\sim$ 1000 times higher than in the previous ILL-based experiments \cite{BaldoCeolin:1994jz} remains an ultimate goal of the NNBAR collaboration. Due to the commissioning schedule of ESS, the floor for the start of construction of $n \rightarrow \bar{n}$ experiment might be 
available not earlier than by year 2026 and the designed power of 5 MW could be obtainable after year 2030. The NNBAR Collaboration would exploit the opportunity of low-power operation and commissioning time of ESS during the intervening years to exploit opportunities at the ESS to search for sterile neutrons, at the HIBEAM stage of the experiment using the ESS beamline developed by ANNI Collaboration \cite{Soldner:2018ycf}.  As shown in Section~\ref{Sec:motivation}, the physics of $n \rightarrow n{'}$ is close to and possibly generically related to the $n \rightarrow \bar{n}$ process. Several smaller scale and relatively inexpensive experiments can be made in this area. A further goal of HIBEAM is the development of a search for $n\rightarrow {\bar{n}}$, albeit at a likely lower sensitivity than that achieved by the ILL experiment. The pilot experiment will enable detector research and development together with background mitigation techniques, necessary for the full NNBAR experiment.

This Section is organised as follows. First, the ANNI beamline is described, followed by the searches for neutron conversion processes involving sterile neutrons. 

\subsection{ANNI beamline and beam properties}\label{sec:anni}
\begin{center}
\begin{figure}[H]
  \setlength{\unitlength}{1mm}
  %\begin{center}
  \includegraphics[width=1.0\linewidth, angle=0]{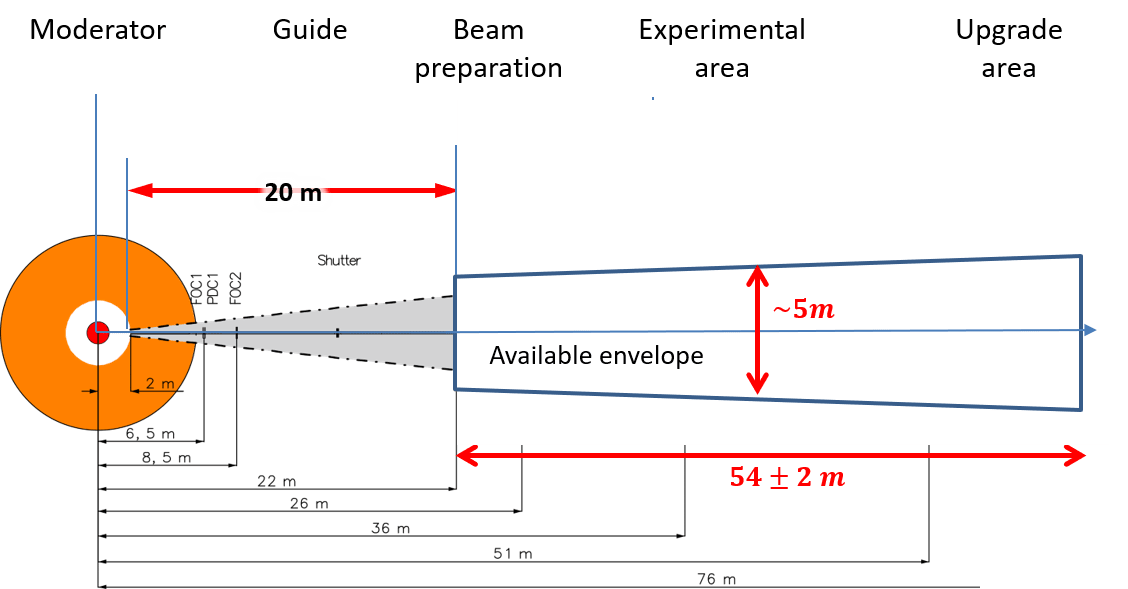}
  %\end{center}
  \caption{A schematic overview of the ANNI fundamental physics beamline floor plan which would be used in HIBEAM. The figure is adapted from~\cite{Soldner:2018ycf}.}
  \label{fig:hibeam@anni}
\end{figure}
\end{center}

\begin{center}
\begin{figure}[H]
  %\begin{center}
  \includegraphics[width=1.0\linewidth, angle=0]{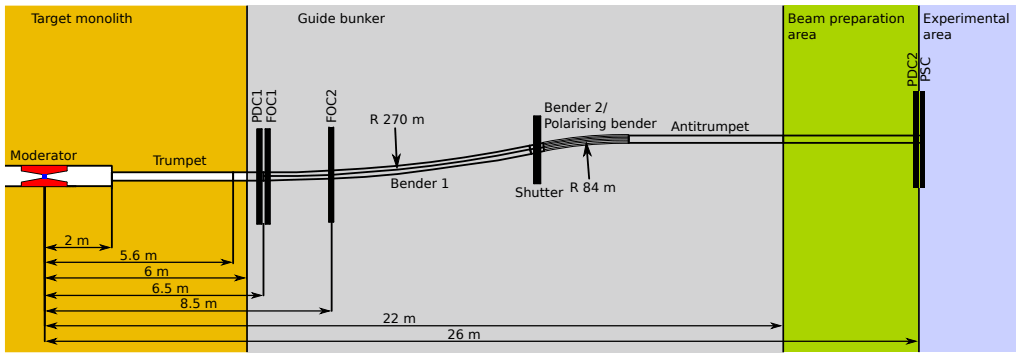}
  %\end{center}
  \caption{A basic schematic overview of the optimized vertically-curved $n$ guide system used in the ANNI design preventing direct sight of the cold moderator, thus reducing backgrounds, as discussed in \cite{Soldner:2018ycf}.}
  \label{fig:ANNIGuideDesign}
\end{figure}
\end{center}

Ample discussion on the properties and usage of the ANNI beamline for fundamental physics searches has been considered in \cite{Soldner:2018ycf}. Fig.~\ref{fig:hibeam@anni} gives a schematic outline of the ESS/ANNI fundamental physics beamport. Neutrons emerging from the moderator pass through a guide system (Fig.~\ref{fig:ANNIGuideDesign}) and then into an experimental area (Fig.~\ref{fig:hibeam@anni} ) with a length of around $54$ m and a width of around $5$ m (visualized here without all other experimental apparatuses in the hall). Due to the beam hall size constraints, there would be very limited room to use focusing reflectors to   increase sensitivity for $n$ oscillation  searches. HIBEAM considerations are thus minimally based on the full-length beamline with possible aperture collimation. 

\begin{figure}[H]
  \begin{center}
  \includegraphics[width=0.90\linewidth, angle=0]{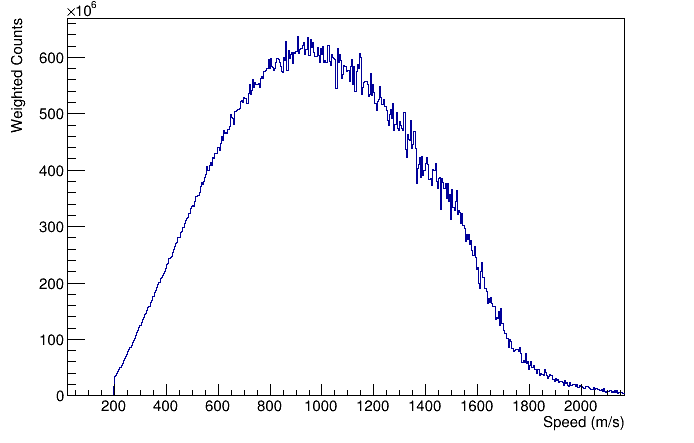}
  \end{center}
  \caption{The \textit{incident} beam velocity spectrum coming from the ANNI/HIBEAM beamport. The results use a simulation event file provided by the authors of \cite{Soldner:2018ycf}.} 
  \label{fig:ANNIVelocity}
\end{figure}

Simulations using the ESS butterfly moderator design in {\sc McStas} 2.4 \cite{Nielsen:2016fce} have been performed by ANNI Collaboration, fully modeling the S-curved $n$ guide in a background-reduced, cold-spectrum selected, beam-shape-optimized way \cite{Soldner:2018ycf}. {\sc McStas} output of $n$ source file with coordinates, momenta, and weights of the neutrons normalized to one initial proton with energy 2 GeV on the ESS tungsten target transported through the ANNI beam optics to the collimator exit at $z = 22$ m was produced. Fig. \ref{fig:ANNIVelocity} shows the spectrum of velocities of neutrons from this simulation flying through ANNI beamline.

\begin{figure}[H]
  \begin{center}
  \includegraphics[width=1.00\linewidth, angle=0]{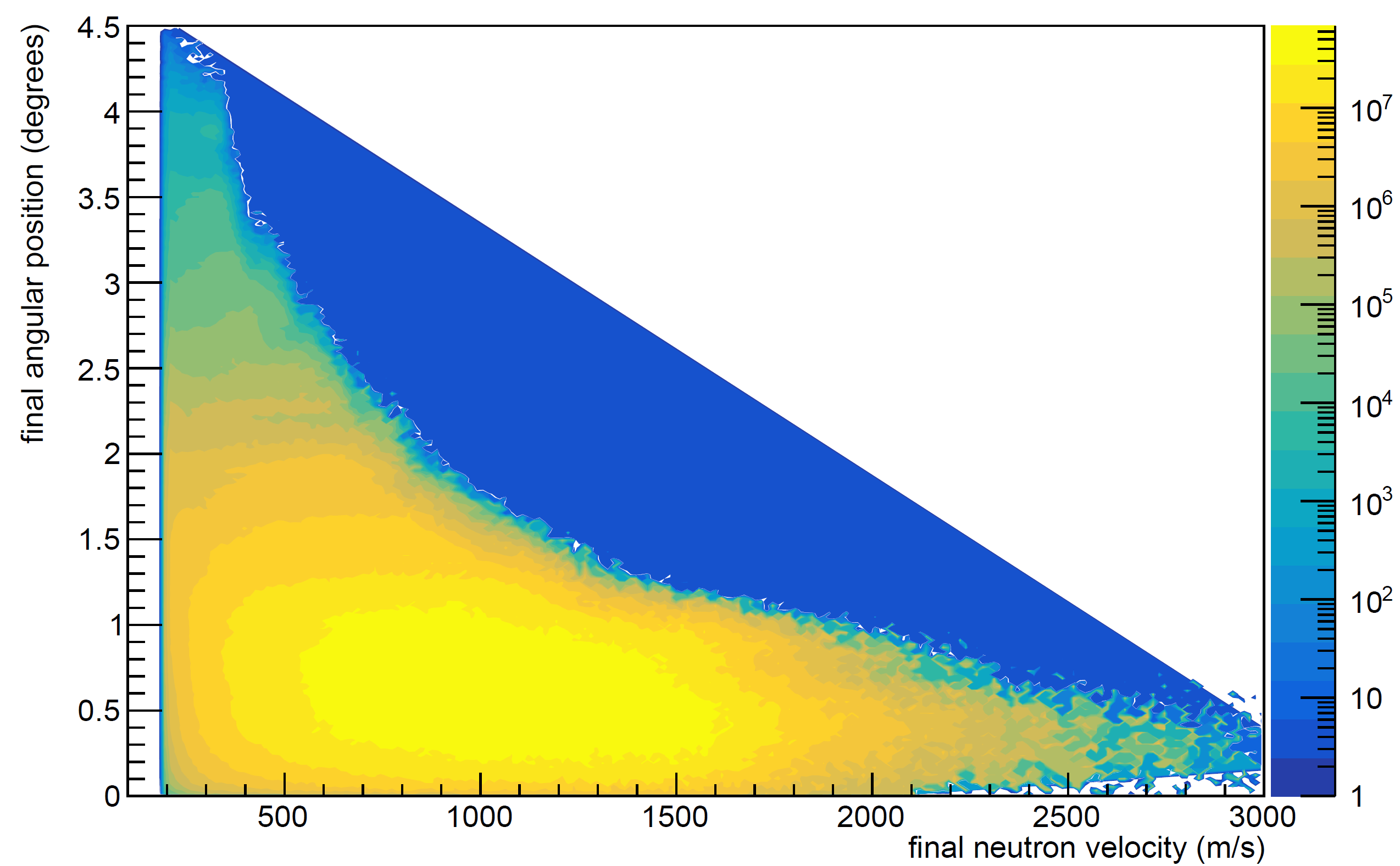}
  \end{center}
  \caption{The ANNI beam divergence as a function of velocity at a distance of $50$m from the beamport is shown; gravity is taken into account, and the \textit{entire flux} (irrespective of any virtual detector's $\sim\infty$ size) is considered. The results use a simulation event file provided by the authors of \cite{Soldner:2018ycf}, and includes the effects of gravity.} 
  \label{fig:ANNIDivergence}
\end{figure}

 As seen in Figs. \ref{fig:ANNIDivergence}, the initial beam characterization shows a large swath of slow neutrons over the beam length of $50$ m having larger divergence at smaller velocities. Most neutrons do not fall outside a $1.5$m radius (angle $\leq$ 1.72 degree), though even that would constitute an enormous and financially untenable detector. Thus there is a sacrifice of a fraction of the valuable slowest neutrons by choosing a detector with a practical radius of 0.25 m - 0.50 m. Assuming $1$ MW of operating power, simulations show an absolute beam normalization of $1.5\times 10^{11} n$/s (at the beamport exit). For a conservatively designed $1$-m \textit{diameter} detector downstream of the $50$-m propagation length, this flux becomes $6.4\times 10^{10} n$/s. This is without the installation of any further beam optimization or neutron reflectors, no lowering of the detector due to gravitational drop or optimisation of beampipe-shape, and assuming an inherently perfect detector efficiency.

\begin{figure}[H]
  \setlength{\unitlength}{1mm}
  \begin{center}
\includegraphics[width=0.89\linewidth, angle=0]{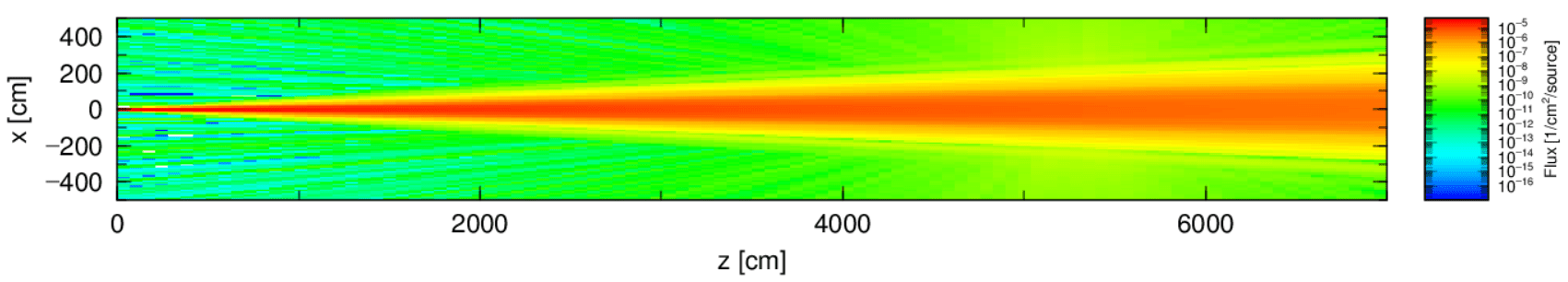}
    \end{center}
%    \vspace{-1.75cm}
  \caption{\footnotesize A top view of the the ANNI neutron beam tracks obtained by {\sc Phits}. The origin of the coordinate system is in the experimental area, after ANNI's curved guide extraction. Gravitational effects are \textit{not} taken into account, but do little to effect this view.}
  \label{fig:neutron_xz}
%  \vspace{2.75cm}
\end{figure}

Fig. \ref{fig:neutron_xz} shows a top-view of the neutron tracks estimated by {\sc Phits} due to the interaction of the ANNI neutron beam with the carbon-12 target. Most of the neutrons pass directly through the target, or are scattered at it and a smaller fraction, given its relatively lower cross section, are absorbed, a process which induce the emission of MeV photons. The origin of the coordinate system is in the experimental area, after ANNI's curved guide extraction, i.e., it is located in the so-called "available envelope" shown in Fig. \ref{fig:hibeam@anni}. 

\begin{figure}[H]
  \setlength{\unitlength}{1mm}
  \begin{center}
%  \captionsetup{width=.0.69\linewidth}
\includegraphics[width=0.59\linewidth,angle=0]{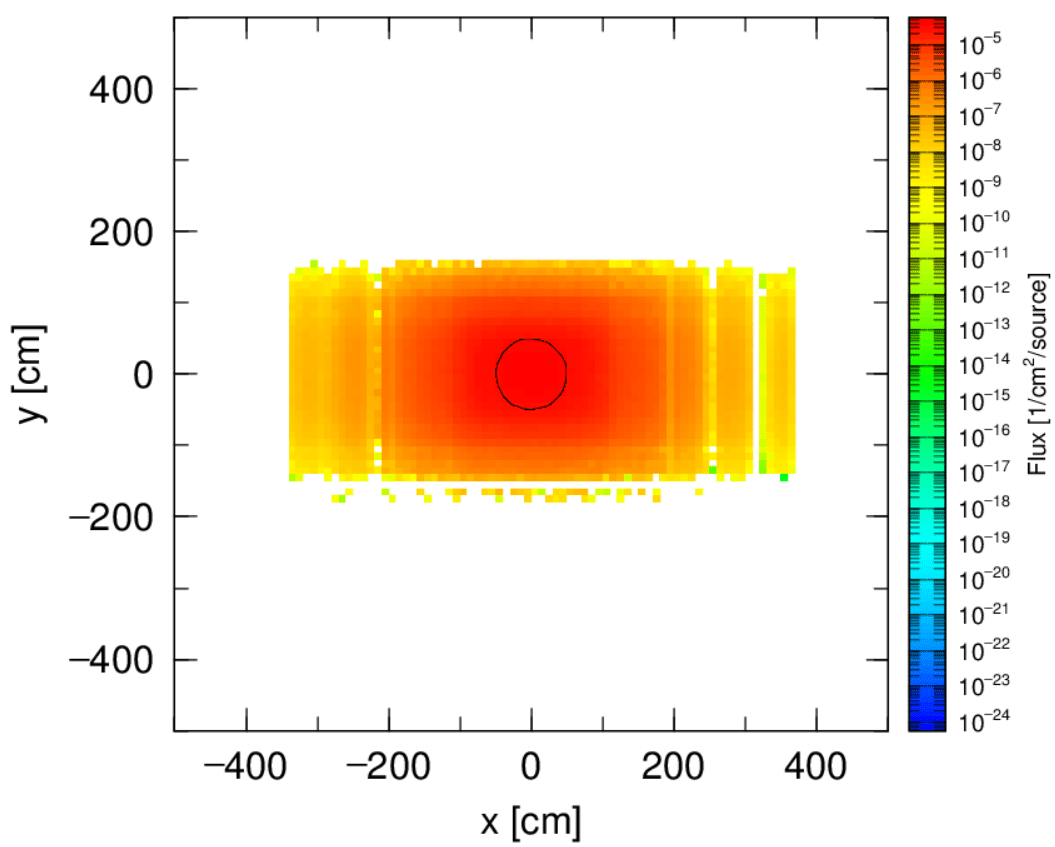}
    \end{center}
%    \vspace{-1.75cm}
  \caption{\footnotesize A cross section view of the target region showing the ANNI neutron beam tracks obtained by {\sc Phits}. The observed interference-like pattern is due to bounce-to-detector distances along ANNI's $S$-shaped curved guide. The black circle at the origin represents the prospective 1m diameter ${}^{12}C$ target. Gravitational changes to this distribution are not included for simplicity in {\sc Phits}, but are marginal within the detector region.  }
  \label{fig:neutron_xy}
%  \vspace{2.75cm}
\end{figure}

Fig. \ref{fig:neutron_xy} shows a cross-sectional view of the ANNI neutrons at the annihilation target. The observed neutron interference pattern is caused by the different bounce distances the ANNI neutrons take when being transported through the S-curved guide.

%The photons are produced in virtue of the neutron capture process.

\begin{figure}[H]
  \setlength{\unitlength}{1mm}
  \begin{center}
  \includegraphics[width=0.90\linewidth, angle=0]{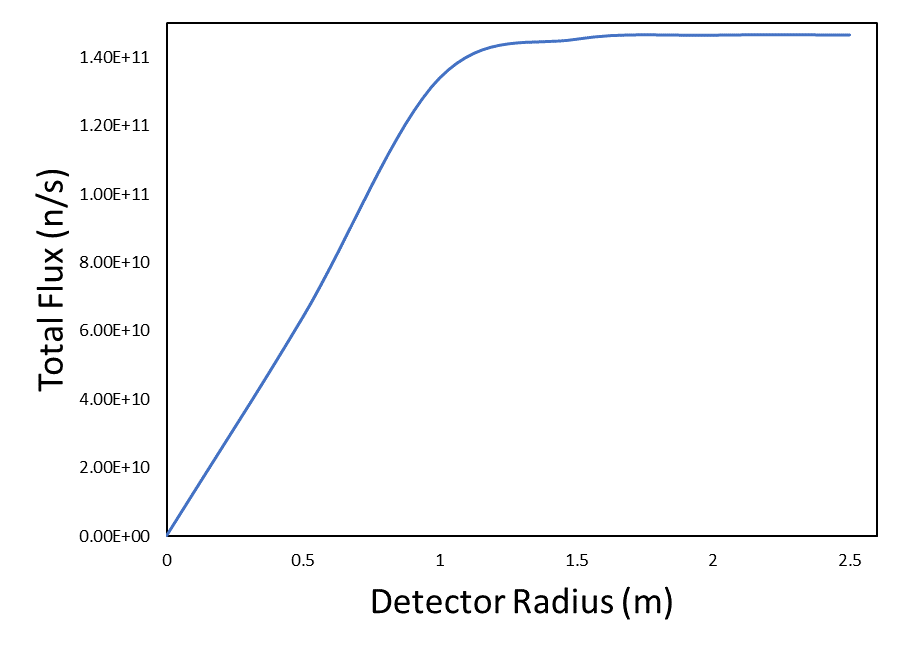}
  \end{center}
  \caption{The smoothed total unreflected flux per $1$MW of spallation power for the ANNI beamline as a function of final detector radius assuming $50$m of flight.}
  \label{fig:ANNITotalFlux}
\end{figure}

The capabilities of the HIBEAM beamline can be further contextualized when the full final flux is considered as a function of a detector radius, as seen in Fig. \ref{fig:ANNITotalFlux}. This hints at the need for greater beam control via $n$ reflectors. However, space constraints will limit this prospect. 

\section{Searches for sterile neutrons at  HIBEAM}\label{sec:nnprimesearches}

The envisaged program is based on the theoretical possibilities for $n\rightarrow n'$ described in Section~\ref{sec:nnprimeform}. The program will include, but will not be limited to the following experiments with each being able to be performed for relatively short times at the ANNI beamline. Sections~\ref{sec:nnprime} to ~\ref{sec:nnprimebarn} (Sections~\ref{sec:nprimetmmgrad} and ~\ref{sec:nprimetmmcom}) cover searches for evidence of sterile neutrons generated by mass mixing (a non-zero TMM). As shown in this Section, by employing complementary configurations for sterile neutron searches, a characterisation of the sterile neutron mixing sector can be made in the event of a discovery. Section~\ref{sec:nnprimebarn} describes a search for neutrons transforming to antineutrons via a sterile neutron state in a regeneration-style experiment. This provides a well-motivated opportunity to refine the technical approach to the high efficiency detection of antineutron annihilation events with small backgrounds. Taken together, the range of experiments envisaged enables a discovery made with one set-up to be supported by an observation of a signal with a different experimental configuration.  

\subsection{ Search for $n \rightarrow n{'}$ via disappearance.}\label{sec:nnprime} This search looks for $n \rightarrow n{'},\bar{n}'$ and is sensitive to a scenario in which at least one of the mass mixing parameters $\alpha_{nn'}$ and $\beta_{n\bar{n}'}$ (see Eq.~\ref{eq:nbarmatrix}) is non-zero. The search assumes the presence of an unknown sterile magnetic field $\mathbf{B{'}}$ which would be matched by a magnetic field in the visible sector.
  
\begin{figure}[H]
  \begin{center}
  \includegraphics[width=1.0\linewidth, angle=0]{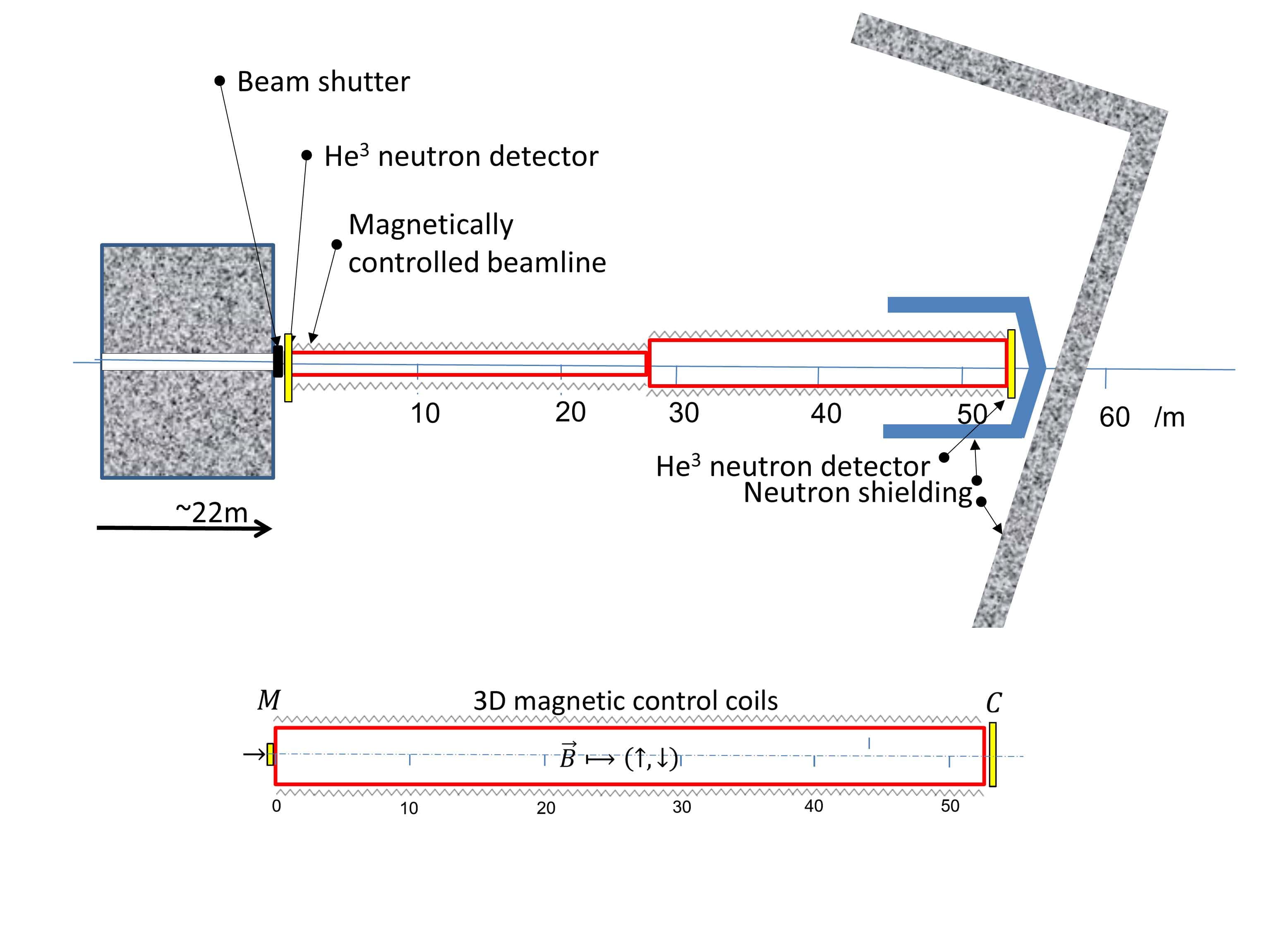}
  \end{center}
  \caption{Schematic overviews of the $n \rightarrow n{'}$ search by disappearance at HIBEAM. Top: diagram showing apparatus components and flux values. Bottom:  diagram of a simplified schematic illustrating the basic principles of the search. The symbol $M$ represents a current-integrating beam monitor with efficiency 20 – 30 \%. The symbol $C$ represents a current-integrating beam absorption counter with an efficiency $\sim$100 \%. An axial magnetic field is applied in different directions in the two tubes (shown by the up and down arrows within the parentheses). }
  \label{fig:DisappearanceScheme}
\end{figure}

A schematic overview of the experiment is shown in the Fig. \ref{fig:DisappearanceScheme}. A more detailed diagram showing all relevant apparatus is shown, together with a simpler schematic picture illustrating the basic principles of the search. Neutrons propagate along an Al vacuum tube of length around 50m and varying diameter and the neutron rates at the start and the end of the propagation zone are measured. The symbol $M$ represents a current-integrating beam monitor with efficiency 20 – 30 \%. The symbol $C$ represents a current-integrating beam absorption counter with an efficiency $\sim$100 \%. The assumed beam intensity used here and for subsequent HIBEAM projections is $6.4 \times 10^{10} n$/s.
The sterile magnetic field is assumed to be constant, uniform, and not exceeding the 
magnitude of Earth magnetic field \cite{Berezhiani:2008bc}. The measurements of the change of neutron flux will be made for a range of axial laboratory magnetic field values in different directions for the range from $-0.5$G to $+0.5$G with a step of few mG, a few times less than the resonance width. Thus, the counting rate (determined by charge integration) of the counter $C$ in Fig. \ref{fig:DisappearanceScheme} will be controlled by the magnitude of magnetic field. The charge integrating counter $M$ will monitor variations of the beam intensity independent of variations of magnetic field. 

The detection of a resonance would appear as the reduction of the total counting rate in the $C/M$ ratio vs $|\mathbf{B}-\mathbf{B{'}}|$. This experimental signal is sensitive to multiple parameters of the sterile sector. From this measurement, a limit on the mass mixing parameter $\epsilon$, or the $n \rightarrow n{'}$ oscillation time $\tau_{n\rightarrow n'}^{dis}$, can be extracted. A positive signal would indicate not only the existence of the sterile state $n{'}$, but also the existence of sterile photons $\gamma{'}$, required for the transformation to occur at non-zero $\mathbf{B{'}}$. %, i.e. that mirror magnetic  exist that together with $n{'}$ can be at least two components of the Mirror Dark Matter. 
With more detailed scans, the 3-dimensional direction of the sterile magnetic field $\mathbf{B{'}}$ can be established.
%In case of no resonance detected in the counting rate ratio $C/M$, a limit will be set on the $n \rightarrow n{'}$ oscillation time $\tau_{n\rightarrow n'}^{dis}$. 
As also shown in Refs.~\cite{Berezhiani:2017azg,Broussard:2017yev}, the disappearance method is the most statistically sensitive approach for setting a limit on $\epsilon$ or $\tau_{n\rightarrow n'}^{dis}$.

\begin{figure}[H]
  \setlength{\unitlength}{1mm}
  \begin{center}
  \includegraphics[width=0.90\linewidth, angle=0]{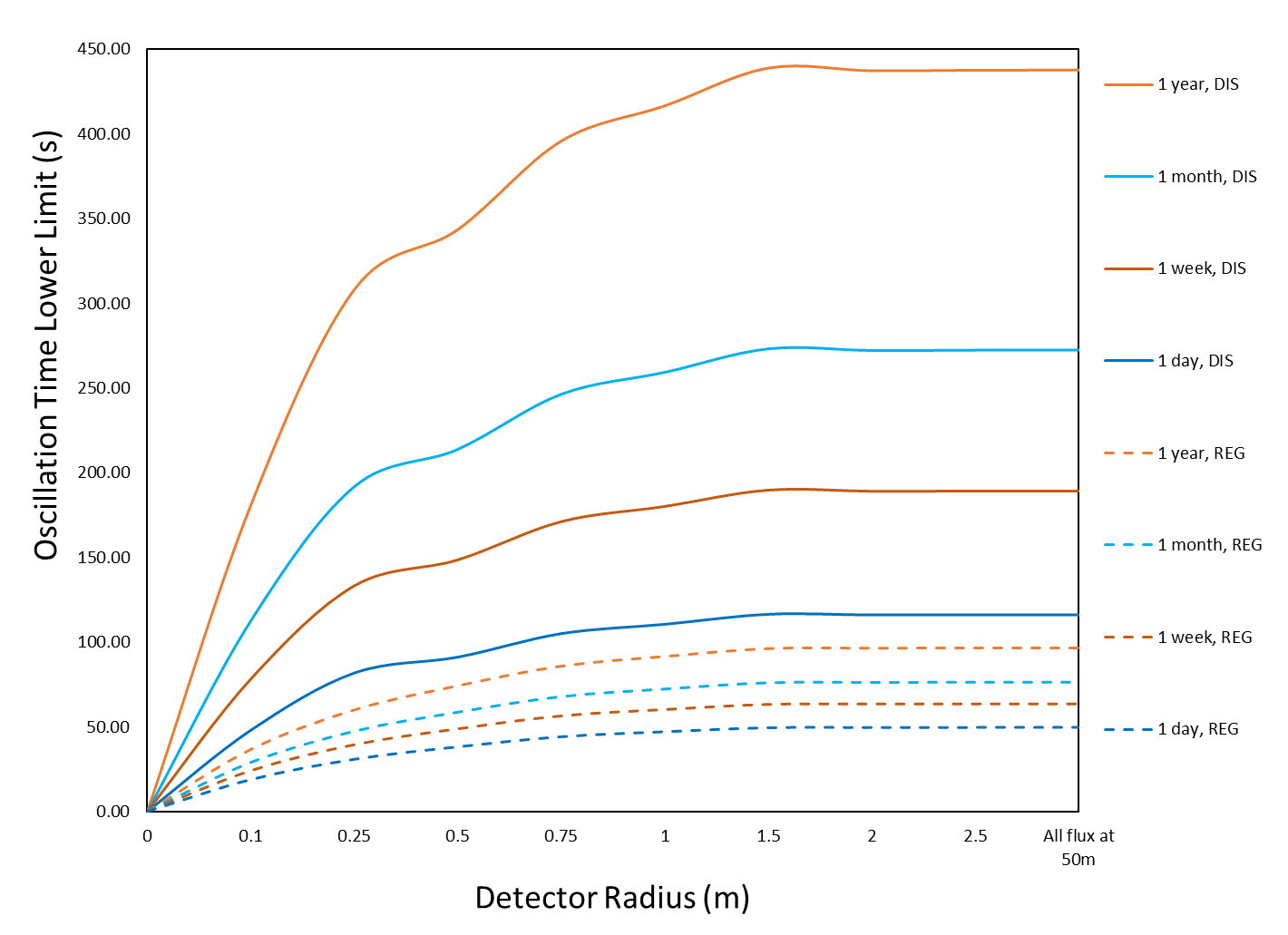}
  \end{center}
  \caption{Sensitivity at $95\%$ CL for the discovery of $\tau_{n\rightarrow n'}^{dis}$ (disappearance, ``DIS") and $\tau_{n\rightarrow n'}^{reg}$ (regeneration, ``REG") for various detector radii for the nominal 1MW HIBEAM/ANNI flux at $50$m. A background rate of $1$n/s is assumed for the regeneration search. Plots have been smoothed.}
  \label{fig:DisappearanceRegenerationTaus}
\end{figure}
Measurements of $n \rightarrow n{'}$ disappearance with a cold $n$ beam will require full magnetic control in the flight volume of the vacuum tube shown in Fig. \ref{fig:DisappearanceScheme}. The 3D 
magnetic field should be uniform and be preset to the desired 3D value in any direction with accuracy better than 2 mG in the range from 0 mG 
to $\sim 500$mG, presenting a technical challenge for this experiment. Another challenge will be the construction of the charge-integrating counters which can achieve a measured charge proportional to the $n$ flux with high accuracy, typically $10^{-7}$. %this was true (but marginal) at HFIR but we want much higher accuracy at HIBEAM...
It has been recently shown~\cite{doi:10.1063/1.4919412} that such stability and accuracy can be achieved with a $^3He$ detector in charge-integration mode. Measurements for positive and negative $\mathbf{B}$-field magnitudes would allow the determination of the oscillation time $\tau_{n\rightarrow n'}^{dis}$ independently of the value of the unknown angle $\beta$ between the vectors $\mathbf{B}$ and $\mathbf{B{'}}$, as well as an estimate of the angle $\beta$ itself.

The dependence of the sensitivity of $n\rightarrow n'$ searches on the properties of the beam and apparatus can be rather complex given the regenerative nature of the oscillation under certain magnetic field conditions, $n$ monitor efficiencies, and environmental background rates.
The sensitivity for low magnetic field disappearance in the absence of visible resonance signal was best (if briefly) discussed in \cite{Berezhiani:2017azg}, but are reiterated here. For disappearance, the main dependencies concern the bare and square-normalized integrals of the neutron velocity spectrum, $S(v)$:
\begin{equation}
    \label{eq:J0J2}
    J_0 = \int S(v) dv, ~ J_2 = \int \frac{S(v)}{v^2} dv
\end{equation}
which are then used to calculate the lower limit for the $n\rightarrow n'$ oscillation time, $\tau_{n\rightarrow n'}^{dis}$:
\begin{equation}
    \label{eq:disappearancesensitivity}
    \tau_{n\rightarrow n{'}}^{dis} > {\left(\frac{J_2 \sqrt{J_0 T \epsilon}}{J_0} \cdot \frac{L^2}{2} \cdot \frac{1-1.7\epsilon+0.76\epsilon^2}{g\sqrt{1-\epsilon}} \cdot \frac{\sqrt{2K}}{\sqrt{2+K}}\right)}^{\frac{1}{2}},
\end{equation}
where $T$ (s) is the accumulated time for each individual magnetic field measurement point, $\epsilon$ is the $n$-monitor efficiency (taken to be e.g. $30\%$), $L$ (m) the length of magnetically controlled flight, the $g$-factor parameterizes the confidence level (for instance, $g_{95\%}=3.283$) obtained from statistical simulations, and $K$ is the number of ''zero-effect" measurements (e.g. at magnetic field $B = 0$)  exceeding the time $T$ for the effect measurement by factor $K$. This calculation is based on the single maximum deviation of one of the $+B$ and $-B$ folded together measured points from the mean value of 200 individual measurements of the ratio $C/M$. With appropriate calculation of $J_0$ and $J_2$ (see Eq.~\ref{eq:J0J2}), Eq.~\ref{eq:disappearancesensitivity} can be used for 
scaling different measurements and configurations at the same beamline. However, better limits can be obtained with
more detailed analysis based on the line shape fit to 
experimental magnetic field scan data.

The structure of Eqs. \ref{eq:disappearancesensitivity} is  mainly analytical in origin, though their dependence upon factors of $g$ was ascertained by thousands of independent Monte Carlo experiments. To obtain signal sensitivity to a $95\%$ CL with $200$ separate magnetic field point measurements and an additional $25$ background runs with equidistributed run folding over a finite magnetic field range (eg $[-200,200]$mG), a (conservative) background of $1$ n/s, a $30\%$ $n$ monitor efficiency, and two $25$m magnetically controlled sections of beamline, the sensitivity in oscillation time $\tau_{n\rightarrow n'}$ can be calculated for various detector radii over different periods of running without any exploitation of the pulsed beam time-structure. One ESS operating year is considered to be approximately $200$ days (see Tab. \ref{tab:ESSrampup}) when discounting for routine maintenance and seasonal shutdowns. The sensitivity of the disappearance method as a function of detector radius for $\tau_{n\rightarrow n'}^{dis}$  is shown in Fig. \ref{fig:DisappearanceRegenerationTaus} together with sensitivities for regeneration modes (discussed in Sections~\ref{sec:nnprimen} and \ref{sec:regen}).

\begin{figure}[H]
  \setlength{\unitlength}{1mm}
  \begin{center}
  \includegraphics[width=1.15\linewidth, angle=0]{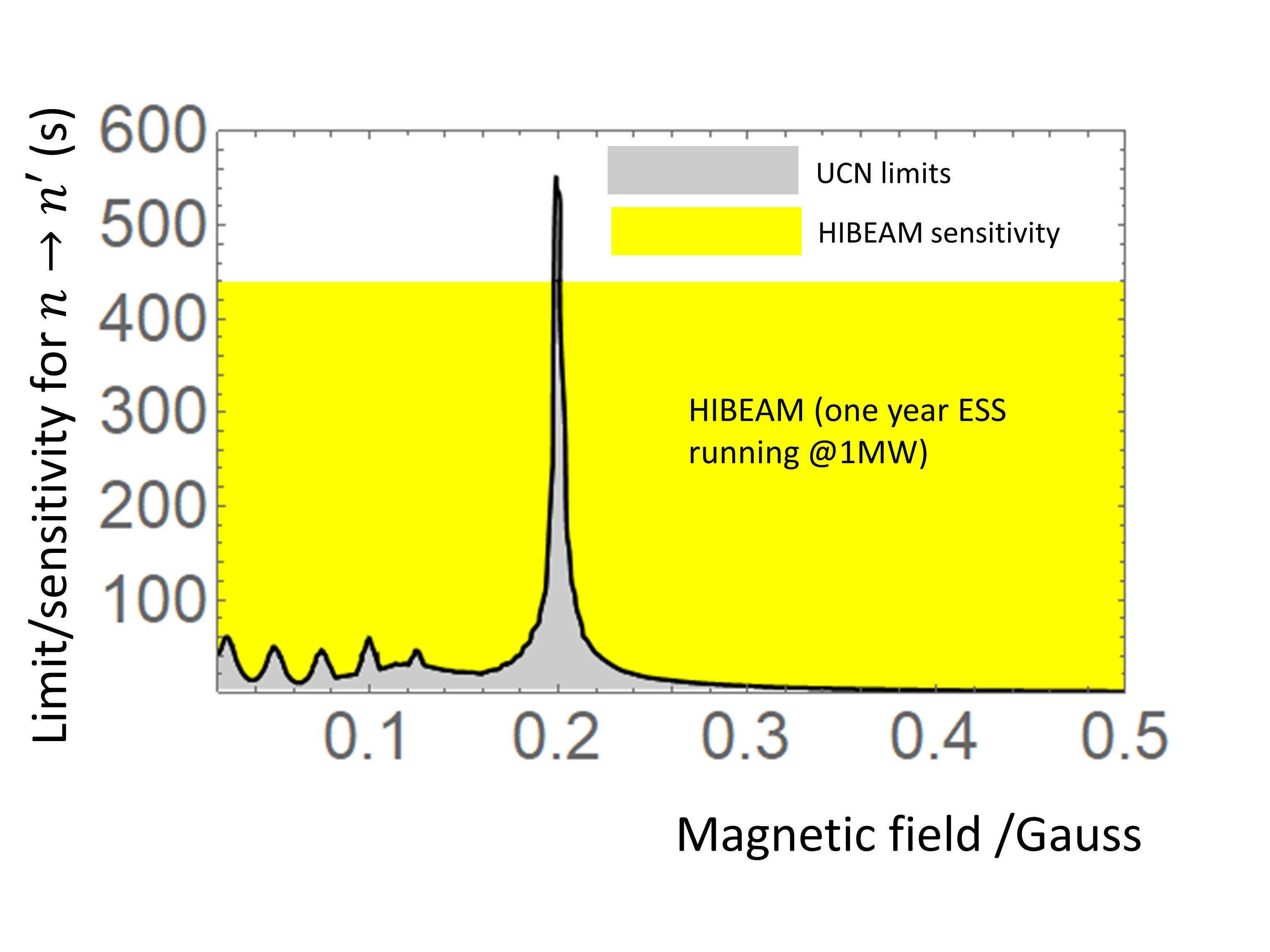}
  \end{center}
  \vspace{-1.0cm}
  \caption{Excluded neutron oscillation times in grey for $n\rightarrow n'$ from UCN experiments~\cite{Ban:2007tp,Serebrov:2009zz,Altarev:2009tg,Berezhiani:2012rq,Berezhiani:2017jkn} as a function of the magnetic field $\mathbf{B{'}}$. The projected sensitivity in yellow for HIBEAM (disappearance mode) is also shown for one year's running at the ESS assuming a power of 1MW. The HIBEAM sensitivity region should also be taken correspond to the area representing UCN limits, where there is overlap.  }
  \label{fig:nnprime-global}
\end{figure}

Fig.~\ref{fig:nnprime-global} shows the current limits from bound neutrons together with the expected sensitivity of the HIBEAM experiment (in the disappearance mode) after one year's ESS running for a power usage of 1MW. Increases in sensitivity of greater than an order of magnitude are possible depending on the value of the magnetic field used.  It can be seen that HIBEAM covers a wide range of oscillation times for a given magnetic field value (up to and beyond an order of magnitude) which are unexplored by UCN-based experiments and free of the model assumptions of those searches.  In the limit of a vanishing TMM it would reduce to the form given in Eq.~\ref{eq:fomnnbar}, implying a sensitivity for an observation which can increase quadratically with the observation time. However, the possible contribution of a TMM complicates this picture (Section~\ref{sec:nnprimeform}). For simplicity, the figure of merit of sensitivity when comparing experiments is therefore taken here to be the oscillation time.

%add yuri sns paper
It should be noted that the interpretations and limits rely on the experimental assumptions, which may be poorly understood, for neutron collisions on UCN material trap walls. This source of systematic uncertainty can be removed by performing dedicated searches with propagating cold neutrons in a magnetic field, as planned for the HIBEAM experiment. 

%Whilst it is non-trivial to quantify the reach of other planned searches, a basic description is given here. %standard practice is to include self-reported expected reach, but not needed here
High precision searches for $n \rightarrow n'$ are also being pursued using UCN at PSI by the $n$EDM Collaboration \cite{Abel:2018tib} albeit for the magnetic field range $B<0.2$~G so far considered. A series of searches for $n\rightarrow n'$ conversions due to various processes along a beamline (e.g. Fig.~\ref{fig:nnprime-cartoon}) are planned at the High Flux Isotope Reactor (HFIR) at Oak Ridge National Laboratory~\cite{Broussard:2017yev,Broussard:2019tgw}. The higher beam intensity of ESS and the longer available flight paths will allow exploring these mechanisms with higher sensitivity at ESS than at the HFIR reactor.

\subsection{Search for the regenerative $n \rightarrow n{'} \rightarrow n$ process}\label{sec:nnprimen}
The regeneration search derives from a similar theoretical basis as the disappearance search 
\cite{Berezhiani:2017azg,Broussard:2017yev} but corresponds to a two-stage process with a consequently quadratically smaller probability. In the first stage the $n \rightarrow n{'}$ transformation takes place in an intense cold $n$ beam at the quasi-free environment limit corresponding to $|\mathbf{B}-\mathbf{B{'}}|\sim 0$. The $n$ beam will be blocked by a high suppression beam absorber, but the sterile $n'$ will continue unabated through the absorber. In a second volume behind the absorber (stage two) under the same condition of $|\mathbf{B}-\mathbf{B{'}}|\sim 0$, the $n{'} \rightarrow n$ transformation produces detectable $n'$s with momentum conserved, as though the totally-absorbing wall were not present. The resonance-behaviour depends primarily on the magnitude of the laboratory $\mathbf{B}$; if the vectors of $\mathbf{B}$ and $\mathbf{B{'}}$ are not well aligned, i.e. the angle $\beta \neq 0$, the oscillation can still occur with somewhat reduced amplitude \cite{Berezhiani:2008bc}. This feature provides a robust systematic check for the experiment: oscillations can be turned off simply by changing the magnitude of $\mathbf{B}$ out-of-resonance in the volume before and/or after absorber. Taking measurements at the positive and negative magnitudes of the field $\mathbf{B}$ in both volumes (four combinations) allows for a determination of the oscillation time independent of the angle $\beta$.

A schematic overview of the principle of the regeneration experiment is shown in Fig.~\ref{fig:RegenerationScheme1}. The lowest possible $n$-background rate in the counter $R$ will be important for a high sensitivity of regeneration search in $\mathbf{B}$-scan, and sufficiently shielding $R$ represents an important challenge for this measurement.

\begin{figure}[H]
  \setlength{\unitlength}{1mm}
  \begin{center}
  \includegraphics[width=0.90\linewidth, angle=0]{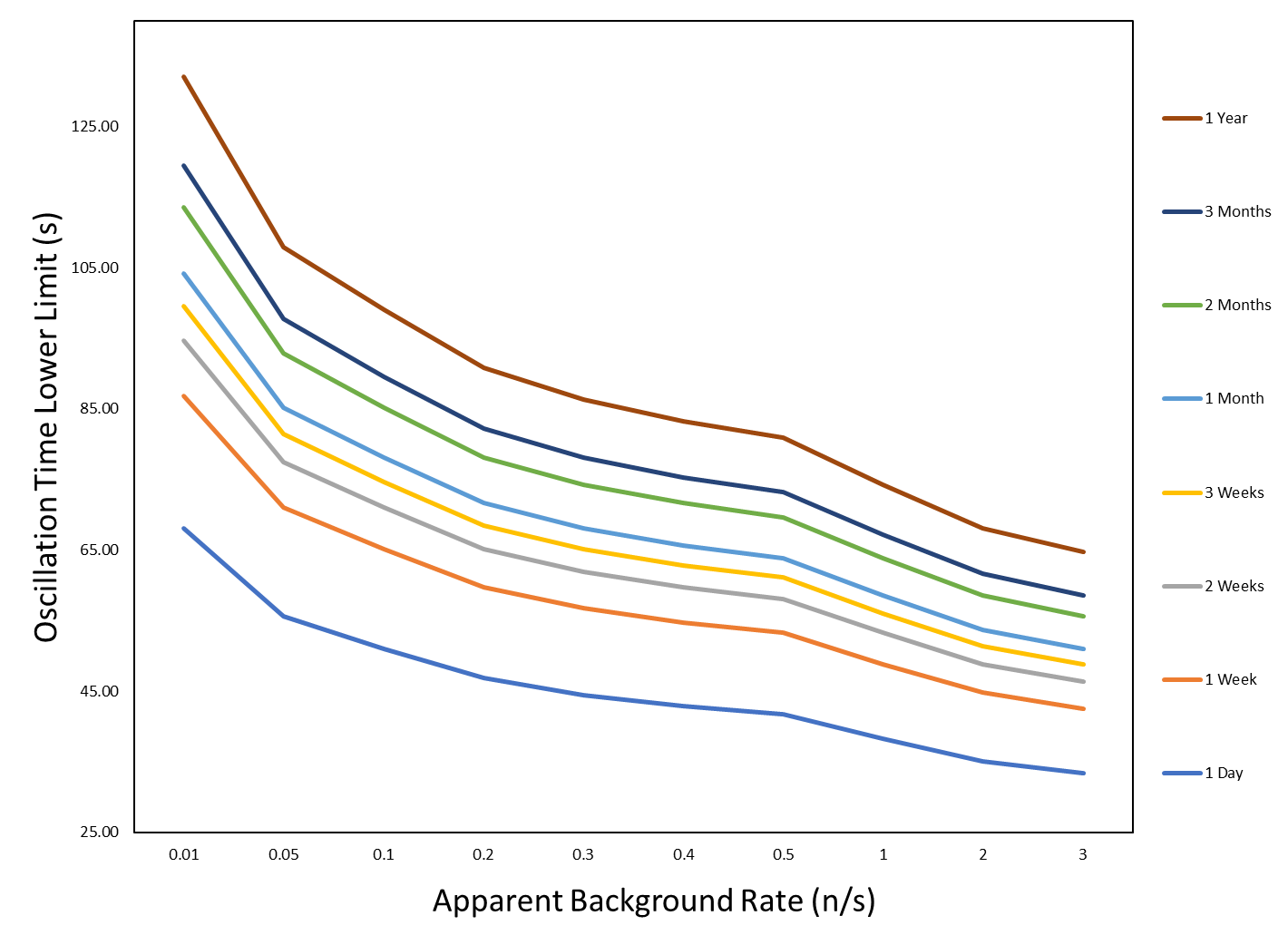}
  \end{center}
  \caption{Sensitivity at $95\%$ CL on the oscillation time as a function of the apparent background count for regeneration searches, $\tau_{n\rightarrow n'\rightarrow n}$ and $\tau_{n\rightarrow \bar{n}'\rightarrow n}$, in a low magnetic field configuration after $50$m of flight shown for a $0.5$m radius detector for the nominal 1MW HIBEAM/ANNI flux.}
  \label{fig:RegenerationBackgrounds}
\end{figure}

The observation of the resonance in the $\mathbf{B}$-scan would be defined by a sudden appearance of regenerated $n$'s when the $\mathbf{B}-\mathbf{B{'}}=0$ condition in both volumes is met. Like with disappearance, a positive signal would be a demonstration of the $n{'}\rightarrow n$ transformation as well as the existence of the sterile $n'$ and $\gamma{'}$. The requirement of matching conditions in both volumes ensures this type of measurement is significantly more robust to systematic uncertainties that could cause a false signal to be observed, and provides an unambiguous test of that hypothesis.

\begin{figure}[H]
  \setlength{\unitlength}{1mm}
  \begin{center}
  \includegraphics[width=1.00\linewidth, angle=0]{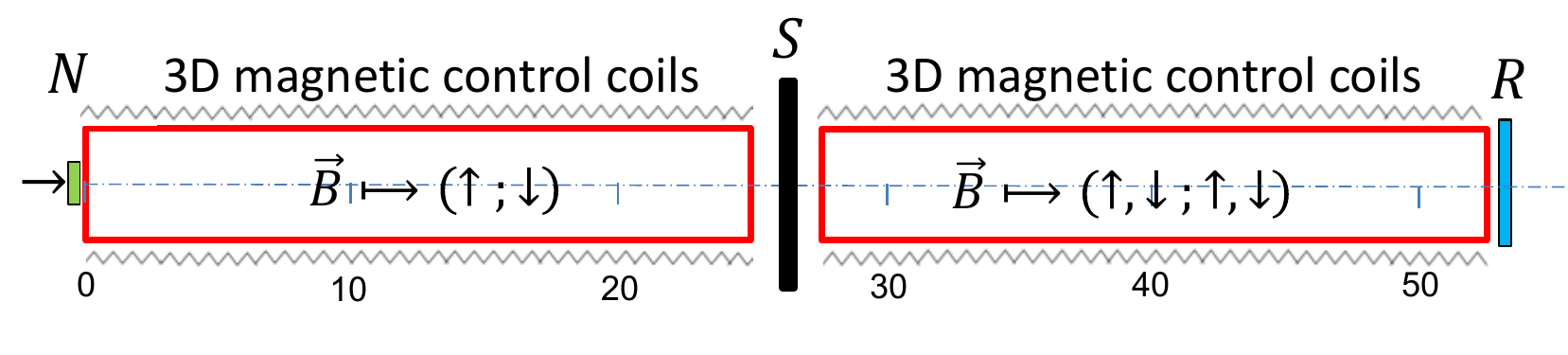}
  \end{center}
  \caption{A simplified schematic of the $n\rightarrow n{'} \rightarrow n$ and $n\rightarrow \bar{n}{'} \rightarrow n$  regeneration searches. Two vacuum Al tubes with lengths 25 m are shown. The symbol $N$ represents a  low efficiency beam monitor, $R$ shows a $^3$He high-efficiency low-background counter and $S$ is a beam absorber. An axial magnetic field is applied in different directions (shown by up and down arrows within parentheses) in the two vacuum tubes, and the configurations can be alternated to choose between hypothetically identical (opposite) magnetic moments of $n$ and $n{'}$ ($\bar{n}{'}$).}
  \label{fig:RegenerationScheme1}
\end{figure}

For the regeneration experiments, in the absence of an observed signal above background level, an upper limit on the oscillation time
$\tau_{n\rightarrow n'}^{reg}$ can be established from a statistical analysis. This limit was parameterized in \cite{Berezhiani:2017azg} through the quartic-normalized integral of the velocity spectrum
\begin{equation}
    J_4 = \int \frac{S(v)}{v^4} dv
\end{equation}
providing the following estimate for the oscillation $n\rightarrow n{'}$ time:

\begin{equation}
    \label{eq:regenerationsensitivity}
    \tau_{n\rightarrow n'}^{reg} > {\left(\sqrt{4T} \cdot \frac{L^4}{4 g \sqrt{\bar{n}_b}} \cdot J_4\right)}^{\frac{1}{4}}
\end{equation}
where $\bar{n}_b$ is the average background rate in the detector $R$ (see Fig. \ref{fig:RegenerationScheme1}) in n/s. This calculation is again based on the single maximum deviation of one of the folded magnetic scan 
of 200 measurements and $T$ is time for one 
individual measurement.
Running over possible values of this background rate, the behavior of the upper oscillation limit can be constructed and is shown in Fig. \ref{fig:RegenerationBackgrounds}. It should be noted that the regeneration mode searches are susceptible to environmental background rates only due to full beam absorption at the halfway-point of the beamline. From Eq. (\ref{eq:regenerationsensitivity}), it can be seen that the length $L$ of each of two vacuum tubes in the regeneration scheme is the only parameter that can essentially increase the limit for $\tau_{n\rightarrow n'}^{reg}$. The regeneration oscillation time sensitivity is lower for the same running time than for the disappearance mode as the former (latter) is a two-transitions (one-transitions) process. However, both processes are complementary with different experimental configurations and neither sharing the same sets of experimental uncertainties. Furthermore, any observation in the disappearance mode could be verified by a regeneration experiment running for a longer time.

\subsection{Search for $n\rightarrow \bar{n}' \rightarrow n$}\label{sec:regen}
As discussed in Section~\ref{sec:nnprimeform}, the symmetry between ordinary matter and mirror matter in general allows a range of  transformations between the visible and sterile neutron sectors, beyond the simple $n\rightarrow n'$ process which is tackled in Sections~\ref{sec:nnprime} and~\ref{sec:nnprimen}.  A neutron can be transformed into a mirror antineutron which then 
regenerates back to a detectable $n$ state: 
$n \rightarrow \bar{n}{'} \rightarrow n$. Since the angular momentum  of the neutron is conserved, the magnetic moment of the  mirror antineutron will be oppositely aligned to the magnetic moment of the sterile neutron due to the mirror $CPT$ theorem. Therefore, a resonance should be observed when magnetic fields in the first and second flight tubes are opposite in direction. This field configuration is included in the anticipated set of measurements shown in Fig.~\ref{fig:RegenerationScheme1}. This search will therefore be be made as a complement to $n \rightarrow n{'} \rightarrow n$  discussed in Section~\ref{sec:nnprimen} and with a similar sensitivity.
\subsection{Search for $n\rightarrow \bar{n}$  by regeneration through mirror states.} \label{sec:nnprimebarn}

Searches for $n\rightarrow \bar{n}$ assume that the transformation occurs via mixing with a non-zero mass amplitude $\varepsilon_{n\bar{n}}$ term. This necessitates the need for magnetic shielding in a search. 
However, $n\rightarrow \bar{n}$ can also arise due to the second order oscillation processes: $n\rightarrow n{'}\rightarrow\bar{n}$ and $n\rightarrow \bar{n}{'}\rightarrow \bar{n}$, with an amplitude comprising a $(\beta_{n\bar{n}'} \alpha_{nn'})$ and interference terms. The earlier body of  searches~\cite{Bressi:1989zd,Bressi:1990zx,Fidecaro:1985cm,BaldoCeolin:1994jz} for free $n\rightarrow \bar{n}$ would be insensitive to this scenario. 

\begin{figure}[H]
  \setlength{\unitlength}{1mm}
  \begin{center}
  \includegraphics[width=1.00\linewidth, angle=0]{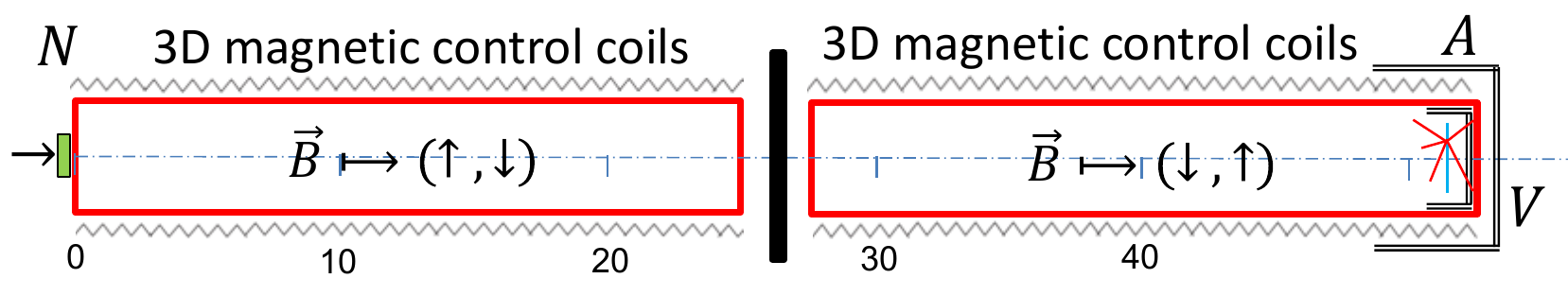}
  \end{center}
  \caption{$n\rightarrow [n{'},\bar{n}'] \rightarrow \bar{ n}$ regeneration search schematic. Two vacuum Al tubes with diameter 1m are used. The symbol $N$ represents a low efficiency beam monitor, $A$ shows an annihilation tracking detector enclosing a carbon foil target to capture the annihilation $\bar{n}C$ event, all of which is surrounded by $V$, a cosmic veto system. An axial magnetic field is applied in different directions in the two tubes.}
  \label{fig:RegenerationScheme2}
\end{figure}

A schematic layout of the search for $n\rightarrow \bar{n}$ through regeneration is shown in Fig.\ref{fig:RegenerationScheme2}. Construction of the $\bar{n}$ annihilation detector at the end of second vacuum volume will be required for this experiment. 

A search for $n\rightarrow \bar{n}$ through regeneration is a complement to the classic $n\rightarrow \bar{n}$ search assuming no sterile neutron mixing. 
Discussion of the details of the annihilation detector is therefore deferred to Section~\ref{sec:nnbardet} where it is described in the context of the classic $n\rightarrow \bar{n}$ search. 
%However, it can be noted here that an efficiency substantially in excess of that achieved at the ILL~\cite{BaldoCeolin:1994jz}) (50\%) for the detection of the annihilation of antineutrons can be expected owing to the use of more modern detector techniques. 

\subsection{Complementarity of searches for sterile neutrons generated via mass mixing}
The program of searches has the potential to both make a fundamental discovery of a dark sector and to quantify the processes underpinning the observations. For example, if the search discussed in Section~\ref{sec:nnprime} would detect a signal in the disappearance mode, this would imply a disappearance for all possible final states of the oscillating neutrons. Since the magnetic moment of a neutron is oppositely aligned to that of an antineutron due to $CPT$ theorem (and as for sterile neutron and antineutrons), it will be advantageous to use non-polarized beams. Here, the compensation $\mathbf{B=B{'}}$ would imply transformations to $\bar{n},n{'},\bar{n}{'}$ for different initial polarization states of neutrons. In regeneration searches, all four magnetic field combinations in two flight volumes would be possible, with $\mathbf{\pm B_1}$ versus $\mathbf{\pm B_2}$ used to detect all possible channels of $n\rightarrow$ regeneration and $n\rightarrow \bar{n}$ due to mixing mass constants $\alpha_{nn'}$ and $\beta_{n\bar{n}'}$.

%\vspace{2 pt}
\subsection{Search for regeneration through a neutron transition magnetic moment (gradients method)}\label{sec:nprimetmmgrad}

As discussed in Section~\ref{sec:nnprimeform} and shown in Eq.~\ref{eq:nnprimetimm}, the probability of the process $n\leftrightarrow n'$ due to a TMM with magnitude $\kappa$ for sufficiently large magnetic fields is approximately  constant,  $P_{n'} = {\kappa}^2$. Due to the independence of $P_{n'}$ on the magnitude of magnetic fields the oscillating $(n,n')$ system can travel through strong magnetic fields with large gradients while retaining the same probability of transformation. The gradients of the magnetic field potential $\boldsymbol{\mu} \cdot \mathbf{B(r)}$ in a classical sense causes a force acting only upon the neutron part of the $(n,n')$ system, but not on the sterile neutron part. Thus the components of $(n,n')$ system are separated, like the two spin components in the Stern-Gerlach experiment. When passing through a difference of magnetic potential, the difference in kinetic energies of the $(n,n')$ components can become larger than the energy-width of the wave packet of the system: this ``measures" the system by collapsing it into either a pure $n$ state or pure $n'$ state. The required magnetic field gradient corresponding to the ``measurement'' event can be found according to 
\cite{Berezhiani:2018qqw} from the following equation:
\begin{equation}
    \label{eq:gradient}
 \frac{\Delta B}{\Delta x}>\frac{1}{\mu v (\Delta t)^2}=\frac{v}{\mu (\Delta x)^2},
\end{equation}
where $\Delta x$ is the distance traveled in the magnetic field for time $\Delta t$, $\mu$ is neutron magnetic moment, and $v$ is the neutron velocity.

The presence of strong magnetic field gradients  destroys the entanglement of the oscillating $(n,n')$ system. A surprising consequence is that as the system continues through the gradient and has its initial state repeatedly reset, this creates additional opportunities for the transformations $n\rightarrow n'$ or $n'\rightarrow n$, effectively increasing the transformation rate. In Ref.~\cite{Berezhiani:2018qqw} this mechanism was suggested as an explanation for the neutron lifetime anomaly~\cite{nlife:2018wfe}.
The disappearance of neutrons due to the magnetic gradients present in UCN trap experiments could explain the $\sim 1\%$ lower value of the measured $n$ lifetime in bottle experiments than is seen in measurements using the beam method. This explanation of the $n$ lifetime anomaly together with existing limits from the direct experimental $n\rightarrow n'$ searches implies~\cite{Berezhiani:2018qqw} that $\kappa$ is in the range  of $\sim 10^{-4}-10^{-5}$.

To test this hypothesis, solenoidal coils with alternating currents in each coil can be implemented around the two vacuum tubes to create a magnetic field along the beam axis with a ``zig-zag" shape, providing an almost constant gradient along the  tube length. These coils can be applied in a regeneration experiment scheme as shown in Fig. \ref{fig:RegenerationScheme1} to search for $n$TMM-induced $n\rightarrow n'\rightarrow n$ regeneration effect with $\kappa < 10^{-5}$. 

%\vspace{5 pt}
\subsection{Search for regeneration through a neutron TMM (compensation method.)}
\label{sec:nprimetmmcom}

As described in \cite{Berezhiani:2018qqw} and Section~\ref{sec:nnprimeform}, an enhanced $n\rightarrow n'$ transformation rate can be produced in a gas atmosphere due to the nTMM. A constant magnetic field $\mathbf{B}$ will be applied in the flight volume to give rise to a negative magnetic potential which will compensate the positive Fermi potential of the gas. The gas density should be sufficiently low to avoid incoherent scattering or absorption of the neutrons. This results in a pure oscillation with probability (from Eq.~\ref{eq:Probc1}) $P_{nn'}=(\epsilon_{nn'}+\kappa \mu B)^2 t^2$. The magnetic field is then scanned in order to search for a resonance condition resulting in a regeneration signal. The magnitude of the laboratory magnetic field at which the resonance might occur depends on the magnitude of the neglected hypothetical sterile magnetic field. The magnitude of $\mathbf{B{'}}$ can be determined by setting the laboratory magnetic field to zero and instead scanning the pressure in the flight tube. In this scenario, the probability is described by the corresponding equation:
\begin{equation}
P_{nn'}=(\epsilon_{nn'} \pm \kappa \mu B{'})^2 t^2
\end{equation}\label{eq:Probc2} where $\pm$ is due to the different possible parities of the sterile magnetic field. Thus, the magnitude of $\mathbf{B{'}}$ can be also independently determined.

\subsection{Neutron detection for sterile neutron searches}
All forms of the sterile neutron searches rely on measurement of the visible state of the neutrons.  Detection of cold and thermal neutrons is a major technical competence for the scattering experiments ESS. A standard solution for neutron detection uses gas detectors based on \isotope[3]{He} in a single wire proportional chamber. The detectors can be operated at low gain since the $n+\isotope[3]{He}\rightarrow t+p$ reaction produces a very large ionization signal. While this is the baseline technology assumed for HIBEAM, it is also possible that modern readout solutions from high energy physics can augment the performance of such a neutron detection scheme further. 
As an example, the most challenging readout scenario is considered here, in which each neutron in the flux would be individually detected.  Assuming a neutron flux of $10^{11}\,$n/s evenly spread out over a circular surface of diameter 2m, and assuming each detector element to be about 1 square cm in transverse area, one would have 30000 readout channels with a singles counting rate in each channel of about 3MHz. This rate is indeed not trivial to deal with but it can be accommodate by the ATLAS TRT detector, where each detector element is a single wire proportional chamber (operated at high avalanche gain) read out on the wire. Typical singles counting rates in the ATLAS TRT are in the range of 6--20\,MHz, so 3\,MHz seems quite feasible in comparison.  The ATLAS TRT electronics will be taken out from the ATLAS setup in 2024, presenting a timely opportunity for HIBEAM, and are therefore an interesting possibility to investigate for use in HIBEAM. The energy of the reaction products do not provide any useful information about the originating neutron, therefore only the number of neutrons detected would be recorded in the data stream. The use of coincidence criteria with neighbouring detector cells will also be investigated, to avoid double counting when nuclear fragments leak into neighboring cells. 

If the individual counting of neutrons is not necessary, an integration of the released charge in the detector material gives a measurement proportional to the number of incoming neutrons. After proper calibration, such current integration is less sensitive to the actual rate of neutrons. It should also be mentioned that in view in the shortage of \isotope[3]{He} much R\&D is done for scattering experiments on other methods for cold neutron detection. Synergies with readout systems developed for high energy experiments are expected and neutron detection in HIBEAM is a good example of an application with a specific scientific use as motivation for exploring such synergies.
\section{Search for $n\rightarrow \bar{n}$ at HIBEAM}
\label{sec:small_nnbar_exp}

In addition to the suite of sterile neutron searches described in Section~\ref{sec:nnprimesearches}, a major aim of HIBEAM is to act as a pilot experimental program during the early, developmental stages of the ESS, with the aim of performing a new search for $n\rightarrow \bar{n}$ transitions, at first without exploiting the planned full beam power of the facility. While this search will likely not surpass the sensitivity in Ref.~\cite{BaldoCeolin:1994jz}, HIBEAM will be used to develop the design and prototyping of technologies necessary for the second stage NNBAR program, which is dedicated to world-leading, complementary, high precision searches for $n\rightarrow \bar{n}$ at the Large Beam Port. As described in Section~\ref{sec:nnprimebarn}, the HIBEAM experiment will also perform searches for $n\rightarrow \bar{n}$ via regeneration from a mirror sector for which the target and annihilation detector described in this Section would also be used.  

This Section describes the apparatus needed to perform an experimental search: magnetic shielding, the vacuum vessel, target and the annihilation detector. Here, the detector requirements and possible technology choices are outlined. The sensitivity of the HIBEAM experiment for a HIBEAM search for $n\rightarrow \bar{n}$ is then estimated. Since HIBEAM is a pilot experiment ahead of the NNBAR stage, it would be expected that the experience of designing and operating HIBEAM would inform the final design of the NNBAR annihilation detector.

\subsection{Magnetic shielding}\label{sec:magnetics}
As explained in Section~\ref{sec:nnbar}, the $n$'s must be transported in a magnetically shielded vacuum. For quasi-free $n$'s, this corresponds to a vacuum of $10^{-5}$ mbar and a magnetic field of less than around 10~nT along the $n$ flight path~\cite{Davis:2016uyk}.

\begin{figure}[tb]
  \setlength{\unitlength}{1mm}
  \begin{center}
\includegraphics[width=0.80\linewidth, angle=0]{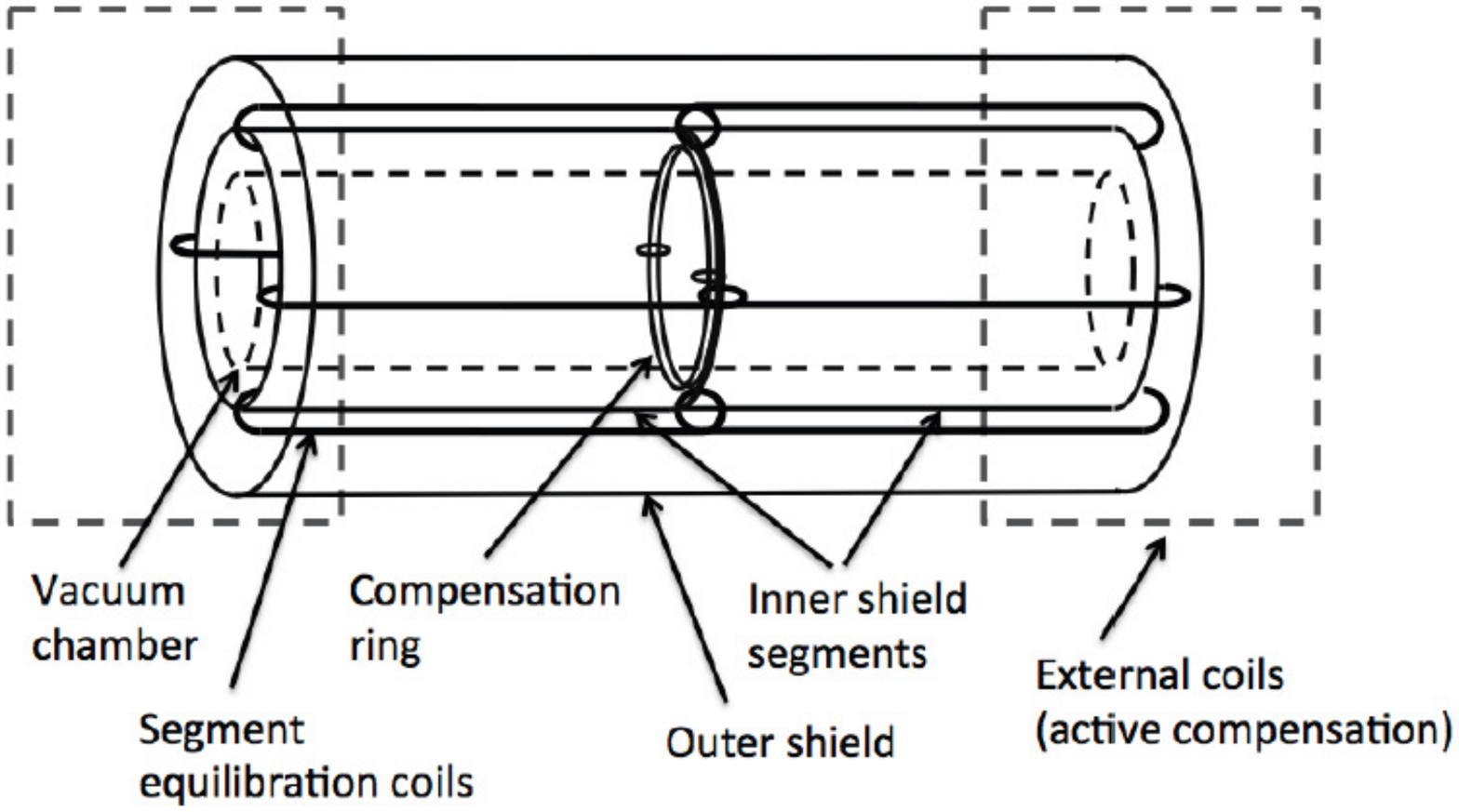}
    \end{center}
  \caption{ Schematic overview of the planned shielding.
  }
  \label{fig:shield}
%  \vspace{2.75cm}
\end{figure}

The target vacuum can be achieved with a vacuum chamber comprising highly non-magnetic materials, e.g. Al, with turbo molecular pumps mounted outside of the magnetically shielded area. Magnetic fields of  less than 10~nT have been achieved over large volumes (see, for example, Ref.~\cite{Altarev:2015fra}). For the planned experiment, a shielding concept will be used based on an aluminium vacuum chamber, a two layer passive shield made from magnetizable alloy for transverse shielding, and end sections made from passive and active components for longitudinal shielding, as shown in Fig.~\ref{fig:shield}.

%\vspace{-2.0cm}

%section is duplicated in detectorsystem
%\subsection{The vacuum vessel}
%The vacuum vessel will be formed with connected cylinders with varying diameter $0.5-1$~m. The vessel is currently envisioned to have $2$~cm thick aluminum walls, but the final design and material choice will be informed by simulations of neutron transport and background calculations. The pressure should be less than around $10^{-5}$~mbar. 

\subsection{Detector components for $n\rightarrow\bar{n}$ searches}\label{sec:nnbardet}
A key experimental task of any $n\rightarrow\bar{n}$ search is to isolate and detect the annihilation of $\bar{n}$'s from a beam of free $n$'s. The transformation has an extremely low probability, and although an ESS experiment could have by far the longest experimental observation time of a free $n$ beam,
%the rate of produced antineutrons is expected to be extremely low (on the level one per month or less).
it may be probable that any experiment would measure only $\mathcal{O}(1)$ candidates.

The overarching goal for the detector system is to provide the highest possible sensitivity for detecting an $\bar{n}$ annihilation. These ambitions go beyond a statistical significance analysis, allowing for a claim of discovery from even only a few observed annihilation events; in the case of non-observation, a robust upper limit can be imposed, and multiple compelling theories of baryogenesis eliminated or severely constrained.

As much as possible, the detector system must provide a reliable and complete reconstruction of each annihilation event. Statistical correction of experimental shortcomings cannot be performed on the individual event level; thus, the design goal must be to record as many observable parameters as possible, taking in all available information about the subsequent annihilation products, and, if possible, compensate directly for detector effects that are statistical in nature via over-sampling. From this, one understands immediately that special attention must also be paid to $\sim4\pi$ detector coverage, and similarly must avoid permanently and temporarily dead detection areas, due either to support structures or failing detector components, respectively. Serviceability, too, must then be a key design feature.

\begin{figure}[tb]
  \setlength{\unitlength}{1mm}
  \begin{center}
\includegraphics[width=0.80\linewidth, angle=90]{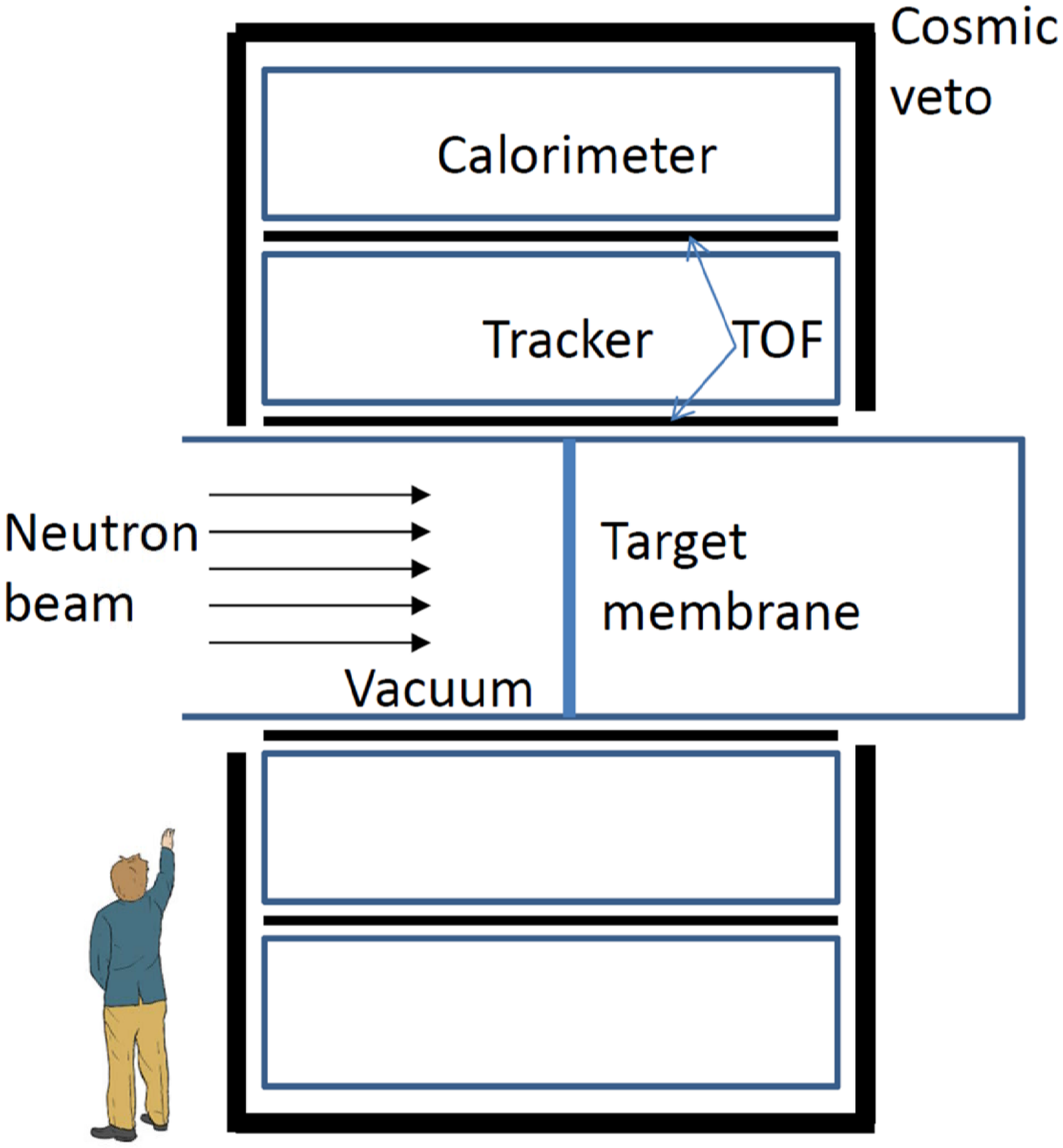}
    \end{center}
  \caption{Schematic overview of the NNBAR detector.
  }
  \label{fig:detector}
%  \vspace{2.75cm}
\end{figure}

An important detector constraint is that a magnetic field cannot be used. Thus, momentum cannot be directly measured, only the kinetic energy deposited in the detector and the direction of the particles. A sensitive $n\rightarrow\bar{n}$ experiment should therefore:
\begin{enumerate}
  \item Identify all charged and neutral pions, properly reconstructing their energy and direction
  \item Reconstruct the energy and direction of most higher energy, charged nuclear fragments (mostly protons; fast $n$'s will likely escape undetected)
\end{enumerate}
The positive identification of an annihilation event would ideally comprise the identity of all pions, a total invariant mass which amounts to two nucleons ($\sim1.9\,$GeV), and a reconstructed common point of origin in space and time of all emitted particles including nuclear fragments from the plane of the annihilation foil. The directionality of an event should be verified by checking that particles move outwards, acting as an important discriminant against backgrounds from cosmic rays and atmospheric neutrinos. A generic detector must thus include tracking, energy loss, calorimetry, timing, and cosmic ray veto systems, as illustrated in Fig.~\ref{fig:detector}.

\subsection{Overall geometry}
%At this time, the diameter of the annihilation foil has not been optimized. 
The optimization of the diameter of the annihilation foil is a balance between a large diameter which maximizes the visible $n$ flux and overall manufacturing cost, which grows dramatically with diameter. The baseline design assumes a $200\,$cm diameter for NNBAR and substantially lower for HIBEAM. The discussion below is guided by NNBAR while HIBEAM features are presented in Section \ref{sec:HIBEAMProgram}. During the long time of flight, the $n$ beam will be affected by gravitation such that the flux incident upon the annihilation foil will not be regular and uniform but rather elongated vertically and containing a vertical gradient in observation time such that the slowest $n$'s in the velocity spectrum will pass through the bottom of the foil. Since the oscillation probability (and so too the annihilation probability) grows as $t^2$, the largest number of annihilations will hypothetically take place at the bottom of the foil.

The geometry of the detection system will also be optimized balancing practical considerations for event reconstruction. A consequence of the broad spread of annihilations in the foil disc is that there is no strong argument for having a cylindrical detector layout. A rectangular layout is likely to be cheaper and will serve the purpose equally well, if not even better than a circular cross-section setup. On the one hand a large area for annihilation presents a difficulty since particle detection normally assumes some average angle of incidence and a central point of emission. On the other hand, a common vertex for a number of secondaries in a large area becomes a strong constraint for an annihilation event: they contain at least two tracks in the majority of cases.

The momentum for each particle can be reconstructed if particles are identified and their direction and kinetic energies measured. If the missing momentum (higher than the Fermi momentum) points into any uninstrumented areas of the detector, one can still have a good understanding of the event, though the topology would not be as well characterized as in the case of all secondaries being observed. While full efficiency and $4\pi$ coverage for tracking and reconstruction of annihilation products is impossible due to the $n$ beam path, very high coverage should be achievable. A $\sim10$m long detector with a $1$m inner radius with full coverage in azimuth would result in a geometrical acceptance of $90$\%, such that about half of all annihilation events will be fully reconstructed. Even with excellent geometric efficiency, loss of information about the total energy is unavoidable. The energy carried away by nuclear fragments may in most cases go unrecorded, though this should be a small fraction of the kinetic energy imparted to other heavy charged fragments like protons. On the other hand, up to $10$\% of the annihilation energy is lost to fast neutrons emitted from the nucleus. For momentum balance, the loss of nuclear fragments can be more significant due to their higher mass.

\subsection{The annihilation foil and the vacuum vessel}
\subsubsection{The ${}^{12} C$ annihilation foil}
The annihilation cross-section is very large (kilobarns compared to millibarns) for cold $n$ capture in e.g. carbon. A very thin carbon foil is thus sufficient to provide high probability for the $\bar{n}$ to annihilate.
However, the very large cold $n$ flux will produce gamma emitting nuclear reactions which will be a source of background that contributes to the singles counting rate in detector channels, though it will not be a severe background to the annihilation signal itself. Although the gamma energies are low at the MeV scale, gamma production occurs at a large rate. The pileup of many gammas is therefore a potential problem to consider in the detector design, in particular the granularity.

A claim of discovery of $n\rightarrow \bar{n}$ should be supported by non-observation under conditions where no $\bar{n}$ should occur. One strategy is to switch off the magnetic shielding in the beamline, effectively prohibiting oscillation. However, this approach requires excessive additional running time to verify that the signal vanishes. An attractive alternative is to install two identical foils, separated by a distance of less than roughly a meter\cite{Bressi:1990zx}. True $\bar{n}$ annihilation would then occur only in the first foil while false annihilation signals should occur with nearly equal abundance in the second foil, as the neutron beam is not significantly attenuated in the foil. A downside of this approach is that the background is also doubled from nuclear physics processes in the foils, requiring an increase in the background rejection capability of the setup, as well as tracking of particles in three dimensions, which is generally advantageous.

\subsubsection{The vacuum vessel}
The cylindrical vacuum chamber is envisioned as $2\,$m in diameter with $2\,$cm thick aluminum walls, and must achieve a vacuum pressure less than $10^{-5}\,$mbar. The final engineering design and material choice will be informed by simulations of particle transport and background calculations. Some material effects have advantages to the experimental approach, but other effects may motivate locating the tracking detectors inside the vacuum vessel.

\textit{Advantages of thick vacuum vessel walls}
\begin{itemize}
\item Stops electrons from $n$ beta decay in flight
\item Shields to some extent nuclear physics gamma background
\item Pair production of $\gamma$s from $pi^0$ decay well measured by tracking and calorimetry
\end{itemize}

\textit{Drawbacks of thick vacuum vessel walls}

\begin{itemize}
\item Energy loss in the material makes higher threshold for detection outside the wall
\item Multiple scattering leads to worse resolution when pointing back to the annihilation vertex
\end{itemize}

With good tracking in the vacuum one could even consider to benefit better from the material in the vessel walls by using a high-$Z$ material instead of Aluminum.

\subsubsection{Background considerations for higher sensitivity searches}
From the ILL experiment one knows that the annihilation signal was background free at the sensitivity of this experiment. The NNBAR experiment aims at a factor of around $1000$X higher sensitivity. In the ILL experiment, cosmic rays produced the dominant contributions to background, with the total number of these events naively scaling with the exposure time and the detector volume.  These backgrounds will be present at the ESS as well, and will ultimately place more stringent demands on the efficiency for background rejection than the experiment at ILL.  Because the planned sensitivity increase at the ESS relies on an increase of several orders of magnitude in the cold neutron beam, the probability for false annihilation signals due to cold neutron beam-induced background will increase accordingly as well.

A major new difficulty compared to ILL is the presence of high energy background induced by the proton beam at ESS. The contribution from this high energy background to the annihilation signal is very hard to estimate. However the fact that it should occur only in beam-related, rather narrow time windows, makes it possible to inhibit events when this background reaches the detector. 
The large spread of $n$ velocities around $800$~m/s means that the arrival of cold $n$'s to the detector area at $200$~m distance from the moderator will be quite uncorrelated with the cycles of the ESS linac.
Based on the experience at the ILL, background to the annihilation signal is expected to be very low, but since the aim is for substantially higher sensitivity, the background discrimination should be improved as much as one can to allow discovery with minimum number of events.
Thus having the best possible resolution in total energy and vertex definition shall be the the design goal. A very strong constraint is to reconstruct the vertex in three dimensions, not only two dimensions as in previous experiments. For charged particles $3$D tracking is fairly straightforward to introduce. For neutral pions, pointing back to a vertex is an interesting development task for the calorimetry.

\subsection{Tracking and energy loss measurements}
\subsubsection{Tracking outside the vacuum}
Around $99$\% of the annihilation events will have two or more charged pions, many of which will be visible above any detector thresholds (consider Fig. \ref{fig:InitMesMomvsInvMass}). The full reconstruction is based on these tracks, which must be reliably extrapolated through space, inwards to the annihilation vertex, and outwards to the calorimetry. The tracking of these charged particles shall thus be of very high reliability. A Time Projection Chamber (TPC) is ideal for the purpose having:

\begin{itemize}
\item   3D tracking
\item	Similar position response in all dimensions.
\item	Independent of track direction.
\item	Each ionization event recorded in all space points.
\item	One space point per cm gas thickness – continuous track image.
\item	Tracks reconstructed with very few combinatorial mistakes.
\item	Excellent for $\frac{dE}{dx}$ reconstruction with an eliminated Landau tail.
\item	Excellent granularity by the time dimension.
\end{itemize}

A TPC tracking detector can record the most information of each track in the most reliable way. By using both the space and time dimension it has a very high granularity. The TPC effectively divides up the sensitive volume in independent cells of about $1\,$cm${}^{3}$ volume each. The drawbacks are that it has poor time resolution and requires a trigger, and is thus limited to a few kHz trigger rate. 

Particle identification is important since a large fraction (typically $30$\%) of the annihilation energy is bound in the rest masses of 	the pions. Identifying the pions would constrain the requirement on the energy balance considerably. Neutral pion identification is discussed in the calorimeter section. Measurement of $\frac{dE}{dx}$ together with a kinetic energy of charged particles can provide particle identification. At the high end of the proton spectrum, $400\,$MeV, a proton has velocity $0.7c$. A pion with that velocity (i.e. same energy loss) has $60\,$MeV kinetic energy due to the large mass difference. Setting this as a limit will cut away all protons and only lead to a loss of pions below $60\,$MeV, a small fraction of the pions. If energy loss is measured outside the vacuum chamber wall the difference between pions and protons will be even larger.

Since many $\frac{dE}{dx}$ measurements are performed on each track one can eliminate the exceptional $\frac{dE}{dx}$ samples due to delta electrons. Since track direction and $\frac{dE}{dx}$ is obtained from the same information, combinatorial mistakes when combining track and energy loss information become negligible. For a robust $\frac{dE}{dx}$ measurement the track length should be reasonably long ($>50\,$cm). Since the track in the TPC reliably points in 3D, inwards and outwards, information about the track from different detectors are assembled with minimum risk of combinatorial mistakes. This is essential since statistical correction for combinatorial mistakes cannot be performed on individual events.

\iffalse
\begin{figure}[tb!]
  \setlength{\unitlength}{1mm}
  \begin{center}
    \includegraphics[width=0.65\linewidth, angle=0]{etplot.jpg}
  \end{center}
%    \vspace{-1.75cm}
  \caption{Kinetic energy distributions of hadrons following an annihilation event.}
  \label{fig:figet}
%  \vspace{2.75cm}
\end{figure}
\fi

\subsubsection{Tracking inside the vacuum chamber}
The thick wall of the vacuum vessel will cause multiple scattering such that the track direction measured outside may lose some pointing resolution. Since the charged pions have fairly low mass, and since low kinetic energies are of interest, this is a considerable effect and the vertex resolution is an important discriminating parameter for claiming discovery. Including track coordinates on the vacuum side of the wall resolves this situation. Two space points on each track should be sufficient since the pointing from outside is good enough for the track definition.

The detector inside the vacuum will face heavy background from both $n$ beta decaying in flight and gammas from $n$ capture reactions. High granularity is thus necessary. Two Si strip stations with stereo angle between strips to allow for a space point from each station would serve the purpose. Arranged as a cylinder with 1m radius and 10m length this would be a very costly detector and one may have to consider cheaper options with gas detectors or scintillating fibers. Possibly, one could compromise on the $z$-coordinate for the inner detectors if multiple scattering does not prohibit safe determination of the foil responsible for the vertex. In that case, more options for inner detectors may be considered, as long as the issue of high singles rate due to background can be avoided, which calls for a high granularity.
An energy loss measurement in the vacuum also presents an advantage. With Si detectors this could be realized but is a matter of cost, while this appears to be more difficult to accomplish with other detector types. The benefit of energy loss measurement on the inside of the vacuum envelope has to carefully evaluated.

Cases where no charged pion goes through the wall cannot be handled. At least one charged particle must give a track in the tracking outside to enable a search for stopping charged particles from a potential annihilation, hypothesizing an emission point at the intercept of the extrapolated track and the annihilation foil. Rather than pions it will be more likely that charged nuclear fragments will stop in the chamber wall. For stopping particles a maximum energy corresponding to the range in the wall material can be set.

\subsection{Calorimetry}
The energy measurement by the calorimeters will be a crucial part of the evidences that an annihilation event has been observed in two ways. It will identify the neutral pions which occur in $90$\% of the annihilations, and ensure that the sum of absorbed energy (kinetic and tied up in pion rest masses) shall sum to two nucleon masses. In addition the calorimetry shall provide the kinetic energy of the charged pions and charged nuclear fragments.
Energy measurements at these energies are notoriously difficult. Several processes are involved.

\subsubsection{Charged hadronic particles} 
As long as the incoming particle is stopped by ionization energy loss only, its kinetic energy can be measured with good resolution. However already at pion energies around 100MeV this means traversing much material and the probability for nuclear reactions becomes sizeable. The energy can still be correctly measured as long as secondaries are charged and all energy is absorbed in sensitive detector material. Energy carried by fast neutrons will not be absorbed and remain unmeasured. For this reason low-$Z$ materials are preferred. For the actual annihilation products (fig 1.), protons will be mostly stopped by ionization energy loss while the bulk part of the charged pions cause nuclear interactions resulting in an energy deficit. If the calorimeter only samples a fraction of the energy of the charged particles as in a sampling calorimeter, the actual energy signal in case of a nuclear reactions can instead become larger due to the larger energy loss by slow nuclear fragments. On the other hand, some nuclear fragments will stop in insensitive absorber materials. Other effects that obscure the energy measurement of charged pions are additional 4MeV energy from weak decay of positive pions where an undetected neutrino carries away most of the energy of the pion mass. This mass could always be inferred from the charged pion identification so this is a small effect. A larger effect is that negative pions stopping in the material will be captured by nuclei. Energy corresponding to the pion mass shall be carried away but some will be by fast neutrons and be undetected. This gives an uncertainty in the energy by one pion mass.

All these unavoidable effects can normally be corrected for by averages based on simulations. Here, where we want the energy to be measured as accurately as possible for each individual particle one can not correct and the calorimeter design must be optimized differently. In some sense these fundamental unavoidable problems are arguments against expensive materials with very good energy resolution. Possibly, measuring $\frac{dE}{dx}$ and the range of particles can give can the most reliable information.

\subsubsection{Photons from neutral pion decay}
The two photons from neutral pion decay have at least $67.5\,$MeV energy. This is far above the energy of any natural sources of particles except cosmic origin. $90$\% of all annihilations have at least one neutral pion. A single photon energy threshold should thus be a simple and reliable trigger on annihilation events. For electromagnetic calorimetry the energies are quite low leading to poor shower statistics. As long as the detector medium is sensitive over the whole volume and large enough to absorb all energy, the shower fluctuations do not influence the energy resolution. In a sampling calorimeter (mixed absorber and sensitive materials) shower fluctuations lead to lower energy resolution. For both calorimeter types, the position resolution of the incoming photon is worsened by the shower fluctuations.

The calorimetry of photons serves three purposes: triggering on annihilation events, identifying neutral pions, and determination of the pion kinetic energy. A pointing ability towards the annihilation vertex would verify the neutral pion as having the same origin as the charged particles and add to the constraints on the annihilation event. Since $98$\% of the annihilation events have at least two charged pions, a vertex based on these should be identified for all events one would analyze.
With this vertex known (which is also the decay point of the neutral pion) and the impact positions on the calorimeter measured, one can reconstruct the invariant mass of any pair of photons from the opening angle between photons and the measured photon energies. The invariant mass resolution is key to a firm, particle by particle statement about the potential neutral pion since:

\begin{itemize}
\item A narrow cut on invariant mass minimizes combinatorial background which is key to a statement that these photons come from a neutral pion.
\item The more accurately the invariant mass is measured, the better confirmed is the assumption of the photon origin at the vertex of the charged particles.
\end{itemize}

Crystals of high $Z$, sensitive over the whole volume, are superior in terms of energy resolution since the limited shower statistics is irrelevant for the resolution. The position resolution will be rather poor both for crystals and sampling calorimetry since it is deteriorated by shower fluctuations. The way to improve position resolution is to choose materials with small Moliere radius. Moving the calorimeter to a larger radial distance from the cold neutron beam axis improves the opening angle resolution as well.

Uniformly sensitive crystals can be either based on scintillation light (several high Z materials exist) or Cerenkov light (lead glass being mostly used). Both types share the ability to measure the total energy as deposited by electrons and positrons, and from a fundamental point of view they could have equally good resolution for photons. In the readout stage, one can expect to have more light from a scintillator and thus somewhat better energy resolution. However, scintillators will give a signal corresponding to all deposited energy while Lead glass is basically blind to nuclear fragments due to the Cerenkov threshold. Charged pions of at least $30$~MeV produce Cherenkov light. The energy calibration of lead glass for charged pions in the actual energy range desired for NNBAR is not trivial. On the other hand, the different nuclear effects for stopping pions will not give arbitrary additions to the measured energy.

While fully sensitive crystals are highly desirable, they are also the most expensive solution. One will not benefit fully from the expensive materials because of the large spread in angle of incidence of the photons. Placing the calorimeter at larger radial distance gives a more perpendicular angle of incidence but at dramatically increased cost. Crystals will also have substantial sensitivity to high intensity energy gamma background with nuclear origin.

A sampling calorimeter may be a more cost effective option. It can be made so that the response does not depend on angle of incidence and it is less sensitive to gamma backgrounds since the conversions happen mostly in the high Z absorber material. If photoelectrons escape into a readout scintillator they will give a signal only in one of the readout planes i.e. much lower signal than if created by showers of many electrons propagating through many layers of scintillator. A useful feature of the measurement situation is that the rate of ionizing particles going through more than one sensitive layer of the sampling calorimeter is very low while the rate of ionizing events in single layers is large, calling for a high granularity in the readout. High granularity can however just as well be achieved by segmentation in depth instead of laterally as one would be normally, which can be useful for the charged particle energy/range measurement. The readout of individual range segments can then be performed over a large transverse area as long as the readout is position sensitive. Then the tower structure of calorimeter cells can be avoided while still obtaining the same calorimetric response, irrespective of angle of incidence. By a 2-D coordinate for each range segment one has a pointing vector for the photon. Longer air-gaps in the calorimeter stack can even be included to improve the pointing resolution. The requirements on the calorimetry are quite different from state of the art calorimetry in high energy physics. This calls for an interesting R\&D program to find the best solution for NNBAR.

Motivated by the arguments discussed in this Section, a {\sc Geant} study of the response of a calorimeter module to charged hadrons and photons has been performed, ahead of detailed study of a physical prototype at an {\it in-situ} ESS neutron test beam in 2023 and test beams at other facilities. The module is based on lead-glass and scintillators and exploits the Cerenkov signature for electromagnetic energy, caused by the interaction of photons, charged hadrons (mainly via $\frac{dE}{dx}$), and hadronic energy. A charged particle range telescope comprising ten layers of plastic scintillator lies in front of the lead-glass. More details are given in the appendix.

\subsection{Cosmic veto, timing and triggering}
\subsubsection{Cosmic veto}
The sum of two $n$ masses represents a high energy which cannot be produced by any background source in nature other than cosmic rays. Therefore an active veto detector against charged cosmic ray particles must surround the detector package. It should consist of two layers of active material such as plastic scintillators. Two close-by detector layers in hardware coincidence reject induced background in order to avoid false vetoes, which, if too frequent, reduce efficiency by vetoing good events. The detector material shall be several cm thick to allow discrimination of $n$ induced background such as Compton scattered gammas, with signal independent of detector thickness, from charged minimum ionizing particles, with signal increasing linearly with detector thickness, by a simple threshold. The cosmic veto is expected to be a part of the hardware trigger logic, and vetoed events will not be stored. However, it may prove possible to postpone the rejection from the cosmic veto to the offline analysis. Thus the cosmic veto should be designed with sufficient timing resolution to determine the direction, inwards or outwards, of the particles associated with the signal.

Charged cosmic rays producing high energy deposits and tracks in the NNBAR detection system are rather straightforward to discriminate. Much more problematic is if the deposit is induced by energetic neutral particles ($\gamma$s or $n$s). A geometrically long sampling calorimeter divided in depth opens the possibility to measure the direction of the showers by timing measurements within the calorimeter. Also a Cherenkov based calorimeter is sensitive to the direction of the shower, and could be made essentially blind to showers directed inwards.  Finally, dE/dx measurements may be helpful for an additional layer of veto for fast neutrons.

\subsubsection{Timing}
For charged particles in the tracking system it is desirable to verify that particles of interest travel outwards. Over a $1\,$m distance there will be a $6\,$ns timing difference between relativistic particles for the two cases. Such timing resolution is not very demanding and detectors for this purpose shall be placed at the entrance and exit of the tracking system. More demanding than time resolution is the singles counting rate which is large in these regions of the setup. Plastic scintillators are adequate for the timing resolution, and background signals from $n$ induced nuclear physics processes can be discriminated similarly as was described for the cosmic veto, with a double detector layer for each station and several cm of scintillator thickness.

Other detector solutions with good timing resolution and 2-dimensional readout (to give the space point together with the chamber plane) would be resistive plate chambers (RPC). One could achieve a high granularity by the 2 dimensional readout, and since the signal is formed by the particle passing multiple gaps of avalanche gain one could discriminate Compton gammas since they will only give signal in at most one avalanche gap. Making this a specific design goal one could probably arrive at a good solution in the respect of preventing background to destroy the time measurement. RPC could therefore also be a viable option for the cosmic veto.

\subsubsection{Triggering}
A trigger to catch energy deposits of more than $67.5$~MeV for one of the gammas from a neutral pion decay is straightforward to implement in hardware as a signal threshold that will catch $90$\% of the annihilations. If taken in anticoincidence with the cosmic ray shield, the trigger should be easy to handle by modern DAQ systems. With modern computational approaches and powerful signal processing on the detector, online data-reduction can be powerful enough to take data without hardware triggering, as with the upgrade experiment at LHC. In addition to a trigger on electromagnetic energy in the calorimeter, a track trigger can be implemented with the timing detectors (plastic scintillators or RPC) as described above. The lack of a point defined by a known collision vertex makes it necessary to allow large directional freedom in the track matching of the track trigger. Of course, straight line tracking (no magnetic field) helps motivate three tracking trigger stations. Powerful background rejection on the signal level is mandatory.

\subsection{Search for $n\rightarrow \bar{n}$}
For $n\rightarrow \bar{n}$ via mass mixing, the quasi-free condition is needed, implying magnetic field-free transmission of neutrons. Any antineutrons which are produced would then annihilate with a target surrounded by a detector. The detector would reconstruct the characteristic multi-pion signal to infer the existence of $n\rightarrow \bar{n}$. 

As shown in Section~\ref{sec:search-nnbar}, the FOM for a free $n\rightarrow\bar{n}$ search is given by  $\langle N_n\, t_n^2\rangle$. The FOM is proportional to the rate of converted neutrons impinging on a target To achieve a high FOM, the following criteria must be met:
\begin{enumerate}
  \item The $n$ source must deliver a beam of slow, cold $n$'s (energy $<5$meV) at high intensity, maximising both $t_n$ and $N_n$, respectively,  for a given beamline length.
  \item The beamport must correspond to a large opening angle for $n$ emission.
  \item A long beamline is needed to maximize $t_n$.
  \item A long overall running time is needed due to the rareness of the process.
\end{enumerate}
Note that neutron beams with lower average energies have higher transport efficiencies when supermirror reflectors are utilized, as in the second stage NNBAR experiment described in Section~\ref{sec:nnbarsec}. 

\subsection{Sensitivity of HIBEAM for $n\rightarrow\bar{n}$}\label{sec:hibeam-nnbar}

\begin{figure}[tb]
  \setlength{\unitlength}{1mm}
  \begin{center}
  \includegraphics[width=1.0\linewidth, angle=0]{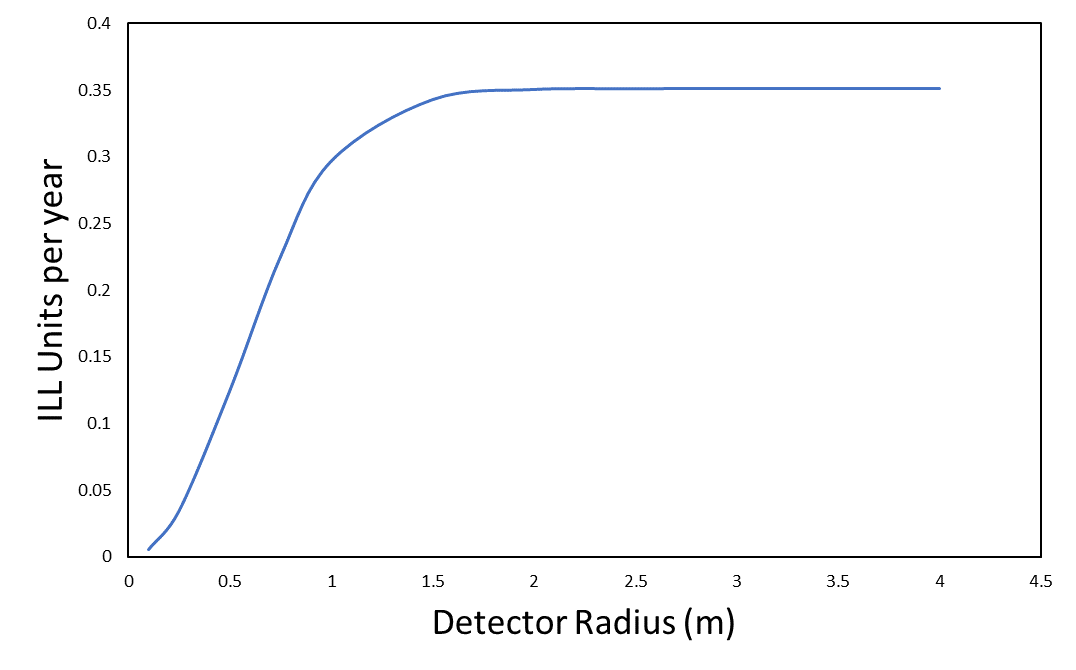}
  \end{center}
  \caption{Sensitivity (in ILL units) of the $n\rightarrow \bar{n}$ search at HIBEAM/ANNI as a function of the radius of the annihilation target, assuming $1$MW of operating power, perfect reconstruction, and zero background.}
  \label{fig:nnbaranni}
\end{figure}

Considering Fig. \ref{fig:RegenerationScheme2}, if the central $n$ absorber were to be removed, and two vacuum tubes were be combined to one with the common magnetic compensating/shielding system, one would recover the essential elements of a $n\rightarrow \bar{n}$ experiment at the HIBEAM/ANNI beamline, although with shorter neutron flight path. Fig.~\ref{fig:nnbaranni} shows the sensitivity in ILL units per year normalized to the ESS running year, i.e. 
\begin{equation}
  \textrm{ILL units per year} = \frac{ \langle N_n\, t_n^2\rangle_{ESS}}{\langle N_n\, t_n^2\rangle_{ILL} \cdot~ \textrm{(Operational Factor)}}  =\frac{\langle N_n\, t_n^2\rangle_{ESS}}{(1.5\times10^9) \cdot (1.2)}
\end{equation}\label{eq:fom-ill} 
of a $n\rightarrow \bar{n}$ search as a function of the radius of the detector, assuming a $1$MW operating power. One ILL unit is defined using the FOM and the flux and running time used in Ref.~\cite{BaldoCeolin:1994jz} as an observable for the number of converting neutrons for a given mass mixing term. The operational factor is a correction factor for the different annual running times expected at the ESS compared to the ILL for the latter's total running period. The sensitivity estimate given by this approach conservatively assumes that the detection efficiency at HIBEAM would be the same as at the ILL experiment ($\sim 50\%$). 

The sensitivity reaches a plateau for a detector radius of $\sim 2$~m. It can be seen that an ILL-level sensitivity can be achieved after running for several ($\sim3$) years with an appropriately sized detector, but this should be considered a generous possibility, as cost considerations may lead to a smaller detector. These can be only linearly offset by a longer running period and higher operating power. $\bar{n} {}^{12} C$ annihilation and outgoing product tracking efficiencies, along with their associated cosmic, atmospheric and fast $n$ backgrounds, have not yet been considered entirely, though state of the art simulations of the underlying microscopic processes have been completed \cite{Golubeva:2018mrz,Barrow:2019viz}.

Nonlinear, though fractional, increases in sensitivity can be achieved with the design and construction of focusing (pseudo-)ellipsoidal super-mirrors starting near the beamport to increase (anti)neutron flux on the ${}^{12} C$ annihilation target. Highly preliminary computations using $0.25$m minor-axes and major-axis lengths of $27$-$50$m for half-ellipsoidal reflector geometries assuming perfect $n$ reflectivity have shown some $\sim40\%$ increase in overall sensitivity. More realistic configurations and reflectivity modeling must be completed and geometrically optimized.

\section{Neutronics and the NNBAR experiment}\label{sec:nnbarsec}
\begin{figure}[H]
  \setlength{\unitlength}{1mm}
  \begin{center}
\includegraphics[width=0.80\linewidth, angle=0]{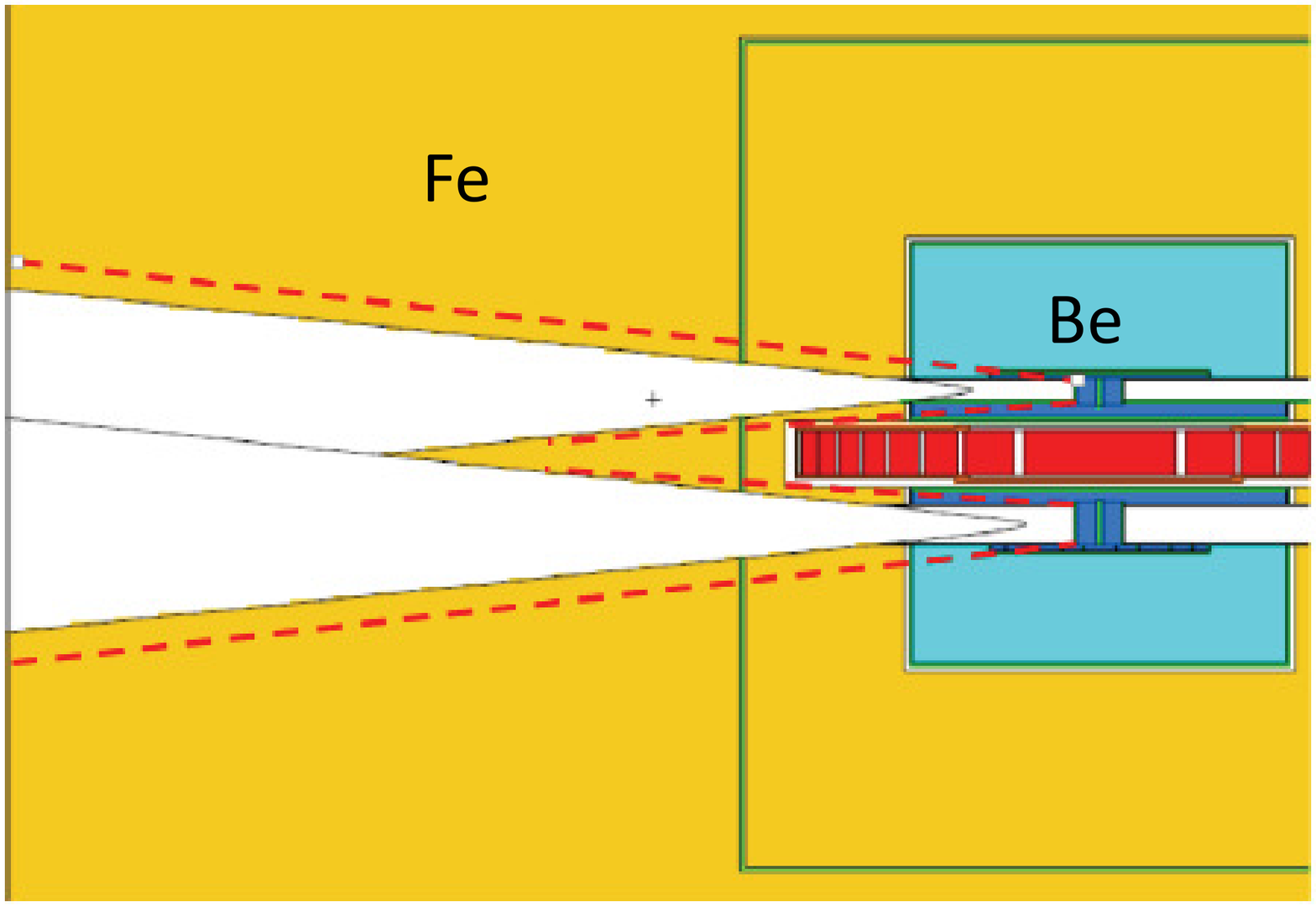}
    \end{center}
    \vspace{-.75cm}
  \caption{ Nominal (white region to the left of the neutron source) and enlarged (region enclosed by dashed lines) conical penetration through the Be reflector and Fe shield. }
  \label{fig:moderatorandreflector}
%  \vspace{2.75cm}
\end{figure}

In order to realize a modern NNBAR experiment that would provide a substantial improvement in sensitivity ($\sim$1000X) than that achieved at the ILL experiment~\cite{BaldoCeolin:1994jz}, a higher overall cold neutron intensity must be utilized. As stated in the 2013 ESS Technical Design Report~\cite{Peggs:2013sgv}, there is a requirement that the ESS provide a level of time averaged cold brilliance comparable to the current cold source at ILL. This, along with novel neutron optical design concepts, can facilitate an experiment with orders of magnitude improvement over previous experiments. Furthermore, as discussed in Section~\ref{sec:ess}, a lower liquid deuterium moderator can be installed. Fig.~\ref{fig:moderatorandreflector} illustrates how parts of the shield and reflector system removed to allow a greater conical penetration and to minimise losses due to the presence of the Fe shield and Be reflector system.

In this Section, the sensitivity of the NNBAR experiment is quantified. A baseline outline of the NNBAR experiment is given followed by a description of a Monte Carlo-based estimation~\cite{Matt-thesis} of the performance of a neutron reflector coupled to a large volume, lower liquid deuterium moderator~\cite{Klinkby:2014cma} using simulation geometry implemented within {\sc MCNP%X
}~\cite{Waters:2007zza}%,x5}
. The aim is for a zero background search, as achieved at the ILL. The background reduction strategy is described in Section~\ref{Sec:backgrounds}.

\subsection{Baseline NNBAR Experiment}

\begin{figure}[H]
  \setlength{\unitlength}{1mm}
  \begin{center}
  \includegraphics[width=0.95\linewidth, angle=0]{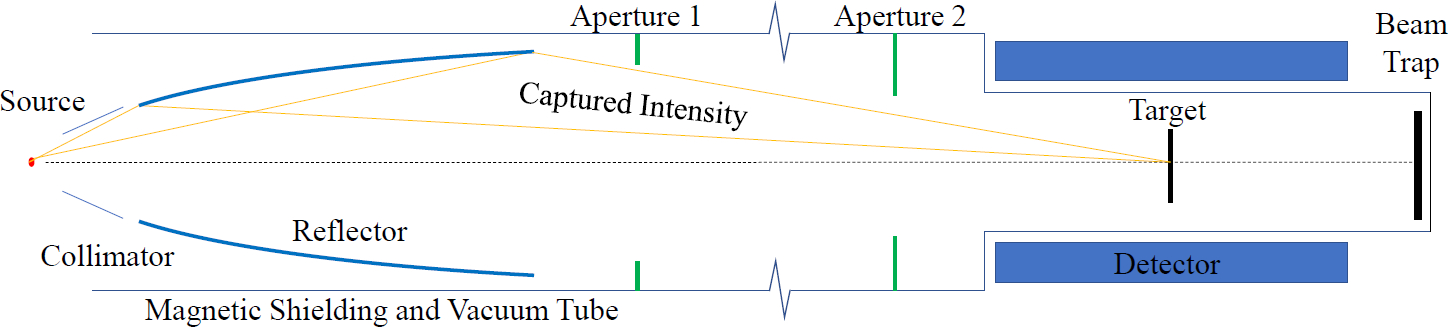}
  \end{center}
  %\vspace{-0.75cm}
  \caption{Baseline NNBAR experiment. Neutrons from the moderator are focused on a distant target foil surrounded by an annihilation detector.}
  \label{fig:nnbar-baseline}
\end{figure}

A simple baseline NNBAR experiment is shown in Fig.~\ref{fig:nnbar-baseline}. A longitudinal distance of 200m separates the source and the target foil. Neutrons passing through the foil are absorbed in a beam trap.

In order for NNBAR to make the most of the impressive neutron intensity at the ESS, there must a be a ``gathering'' reflector to ensure that a large neutron flux  is directed and focused via a magnetically shielded region on the annihilation target. To first order, the most obvious reflector geometry is that of an ellipsoid \cite{YK_icans_1995}, which would facilitate efficient transport of neutrons that otherwise would miss the annihilation target completely. The start and end points of the reflector on the longitudinal axis are 10m and 50m, respectively. The semi-minor axis of the ellipsoid is 2m. The radius of the annihilation foil is 1m. 

An effective reflector must be fully illuminated by the source. This means that a substantial amount of the overall intensity will have trajectories that deviate significantly from the nominal beam trajectory axis.
To achieve this, reflecting angles for even fairly cold neutrons ($\sim$1000 m/s) will exceed that of the limit of the best traditional reflectors. NNBAR will thus use neutron supermirror technology~\cite{mezei1976novel} which has been used with success for many years as a means to guide thermal and cold neutrons to many scattering instruments at both pulsed and continuous neutron sources. 
NNBAR will utilize the same multi-layered surface treatment on its reflector to gather and focus the wide range of neutron trajectories at the source. To do so, surfaces with  surface reflectivity up to $m=6$, i.e. a reflection capability as high as six times better than the reflectivity limit for polished nickel. Neutrons from a liquid deuterium source are collimated in the structure housing the moderator. Neutrons emerging are reflected via an ellipsoid supermirror along a magnetically shielded region towards a target foil, surrounded by an annihilation detector. A beam trap absorbs the beam. The detection efficiency of an annihilation event in the foil is 50\%.

%\begin{figure}[tb]
%  \setlength{\unitlength}{1mm}
%  \begin{center}
%\includegraphics[width=0.80\linewidth, angle=0]{overview-matt.jpg}
%    \end{center}
%%    \vspace{-.75cm}
%  \caption{Schematic of the NNBAR experiment. Neutrons from the source are %reflected by a supermirror towards the target foil. 
%  }
%  \label{fig:overview-matt}
%%  \vspace{2.75cm}
%\end{figure}

\subsection{Differential reflectors}

%In Fig.~\ref{overview-matt} the source lies along the longitudinal axis.
At the LBP, the liquid deuterium source would be offset from symmetry axis of the LBP which is aligned with the target. To deal with this, a segmented differential reflector was designed. This has a distorted ellpsoid-like shape, albeit one which can be optimised via the solution of coupled differential equations for neutron reflections to allow specific reflection angles at certain distances from the foci, providing maximum intensity to the target foil~\cite{Matt-thesis}. Segmentation also allows optimisation of the $m$ value for different panels in the supermirror complex, reducing costs and allowing for easier large scale manufacturing.

\begin{figure}[H]
  \setlength{\unitlength}{1mm}
  \begin{center}
\includegraphics[width=0.80\linewidth, angle=0]{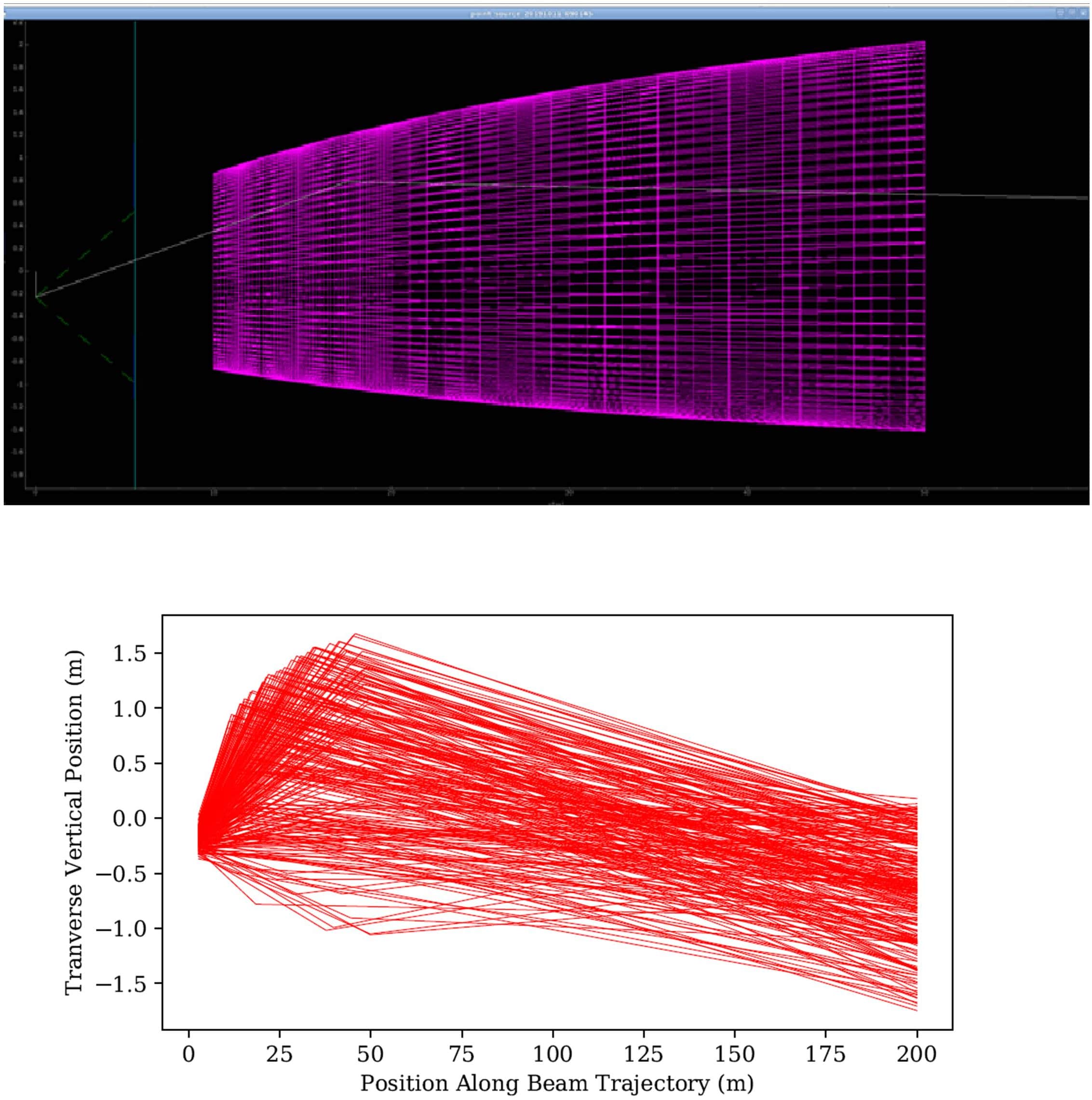}
    \end{center}
%    \vspace{-.75cm}
  \caption{Top: Neutron reflection and focusing from a segmented differential reflector. Bottom: sampling of traced rays showing the transverse vertical displacement of the neutrons as a function of longitudinal distance.  
  }
  \label{fig:diff-ref}
%  \vspace{2.75cm}
\end{figure}

Fig.~\ref{fig:diff-ref} (top) shows a simulation of a neutron reflection and focusing towards a target at 200m distance along the longitudinal axis (referred to as beam trajectory position) from the position of the reflector in a differential reflector. It is shown how the neutron emerges from a source which is offset from the central-axis of the ellipsoid-like reflector. Fig.~\ref{fig:diff-ref} (bottom) shows a sampling of traced rays representing neutron trajectories, also including gravity. The neutrons can be restricted to a range in the vertical direction transverse to the longitudinal axis of around $2$m.  

\subsection{Sensitivity of the NNBAR Experiment}\label{sec:optim}

\begin{figure}[H]
  \setlength{\unitlength}{1mm}
  \begin{center}
\includegraphics[width=0.80\linewidth, angle=0]{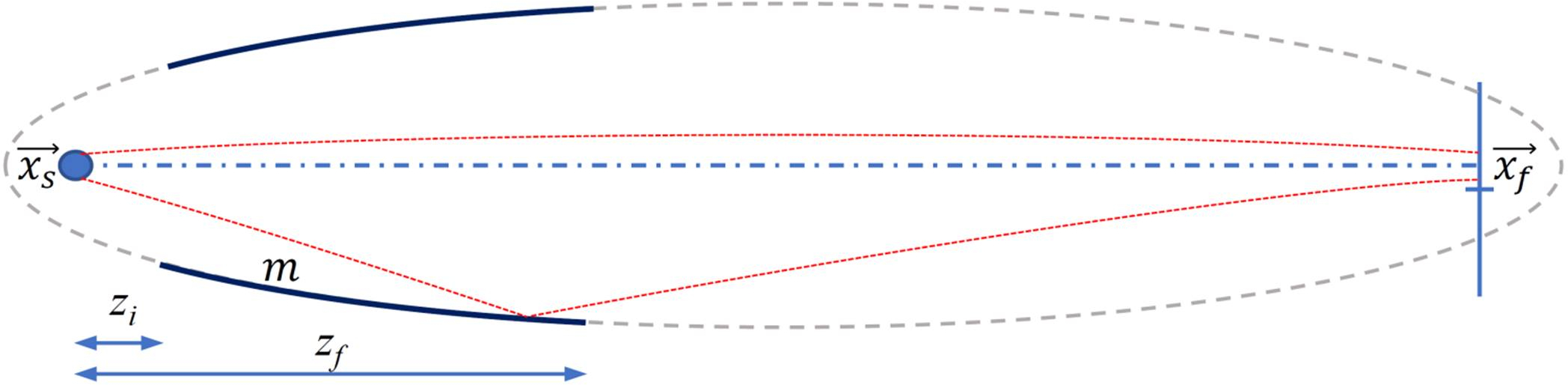}
    \end{center}
%    \vspace{-.75cm}
  \caption{Parameters for a differential reflector relevant to the design and sensitivity of the NNBAR experiment.
  }
  \label{fig:diff-ref-ov}
%  \vspace{2.75cm}
\end{figure}

Fig.~\ref{fig:diff-ref-ov} shows a sketch of a differential reflector configuration with parameters that can be optimised to maximise the sensitivity of the NNBAR experiment. Shown are the source focal point position, $\overrightarrow{x_s}$, the target focal point position , $\overrightarrow{x_f}$, the reflector start position, $z_i$, and the reflector end position, $z_f$. 

\begin{figure}[H]
  \setlength{\unitlength}{1mm}
  \begin{center}
  \includegraphics[width=1.0\linewidth, angle=0]{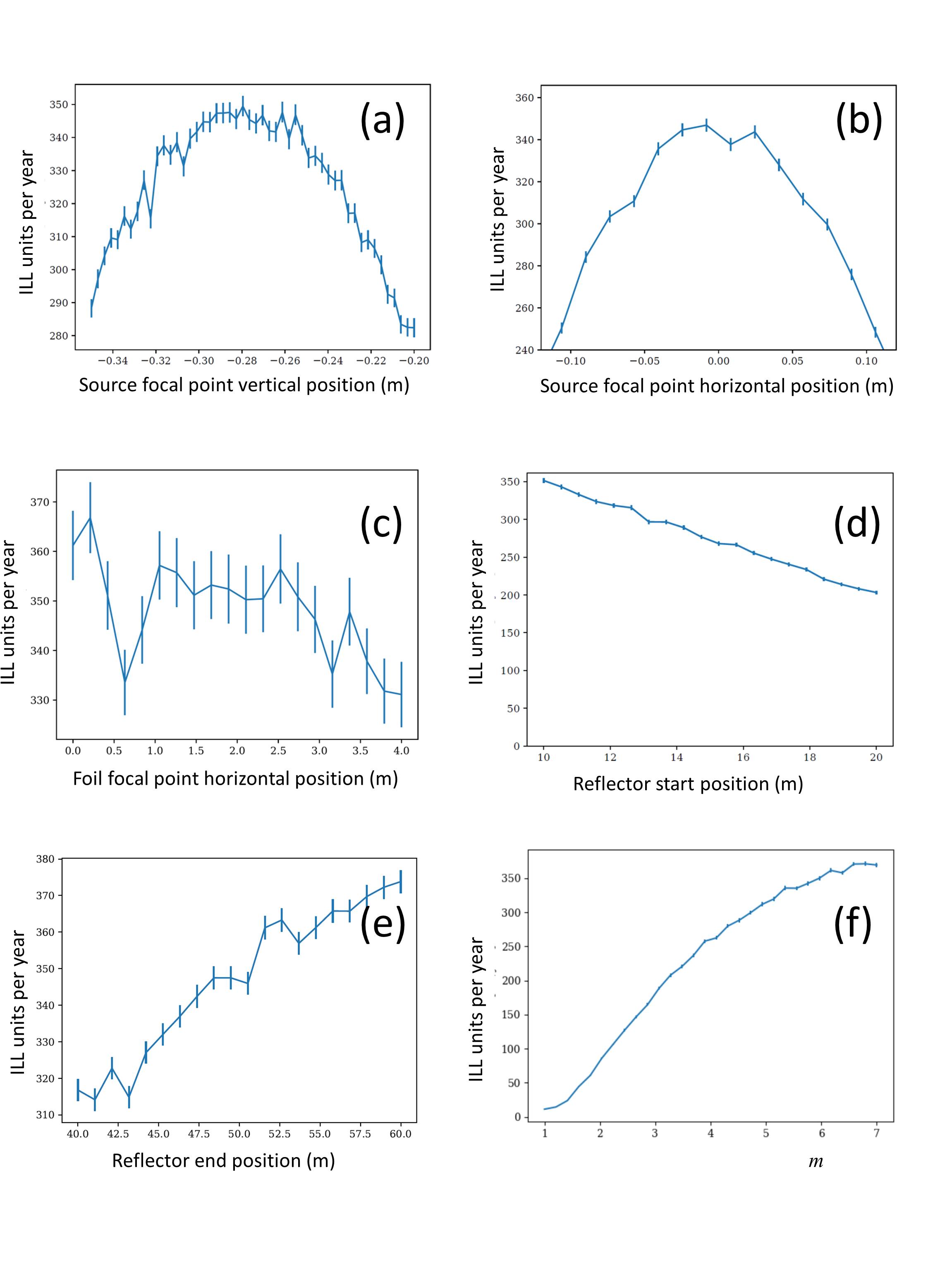}
  \end{center}
  \vspace{-.75cm}
  \caption{Sensitivity of the NNBAR experiment for one year of running in ILL units as a function of (a) the source focal point vertical position,
  (b) the source focal point horizontal position, (c) the target foil focal point horizontal position (d) the reflector start position (e) the reflector end position and (f) the surface reflectivity $m$.}
  \label{fig:optimisation-diff-rev}
%  \vspace{2.75cm}
\end{figure}

 An optimisation of the shape parameters (keeping the others constant) was made. Figs.~\ref{fig:optimisation-diff-rev} show how the sensitivity for NNBAR varies as a function of the source focal point vertical position, the source focal point horizontal position, the target foil focal point horizontal position, the reflector start position, the reflector end position and the surface reflectivity $m$. The sensitivity is expressed as ILL units per year, as defined in Section~\ref{sec:hibeam-nnbar}.  

As can be seen, a sensitivity of around 350 ILL units per year is obtained.
The sensitivity is strongly dependent on the surface reflectivity $m$. Indeed, developments in the super-mirror technology are one of the driving factors in allowing NNBAR a large sensitivity increase since the ILL experiment. The peak sensitivity of $\sim 350$ ILL units per year is achieved for $m \sim 6,7$ and falls to $\sim 10$ ILL units per year for $m\sim 1$. A doubling of start position from $10$m to $20m$ leads to an approximate halving of the sensitivity. Extending the end position from $40$m to $60$m increases the sensitivity by around $20$\%. The foil focal point horizontal position is less sensitive, changing by about $10$\% for shifts of up to $4$m.  Optimising the source focal points (horizontal and vertical) changes the sensitivity by up to around $30$\% in the considered range.   

\subsection{Gain with respect to the ILL experiment}
As was shown in Section~\ref{sec:optim}, simulations predict a sensitivity of around 350 ILL unit per year. For an experiment running for three years this provides an improvement of three orders of magnitude. The increase in sensitivity can be broadly decomposed into the gain factors given in Table~\ref{tab:gain}. Sensitivity increases are due to the greater source intensity, propagation length, and run time. The largest gain is from now widely available high $m$ reflectors.

To further investigate differences between the ILL and ESS for a search for $n\rightarrow \bar{n}$, the performance of only the moderators was studied. In this case, the calculated performance of a liquid deuterium lower moderator of ESS was compared to the ILL Horizontal Cold Source (HCS), used for the original ILL search. For this comparison, updated brightness data was considered, indicating that the cold brightness of the ILL moderators is about three times larger than the official data~\cite{ILL-yellow}. Even with this correction, it was found that an optimized ESS lower moderator together with the upper “butterfly” moderator would deliver a higher intensity, larger than the HCS used for the ILL search. The expected gain is due to several factors, including an optimal beam extraction (aiming at viewing the full surface of the moderator), and an optimization of the moderator specifically for the NNBAR experiment.

\begin{table}
\begin{center}
\begin{tabular}{| c | c |}
 \hline
 Factor & Gain wrt ILL  \\
 \hline
 Source Intensity & $\geq 2$   \\
 \hline
 Neutron Reflector & $40$  \\
\hline
 Length & $5$  \\
 \hline
 Run time & $3$ \\
 \hline
 {\bf Total gain} & $\geq 1000$ \\
 \hline
\end{tabular}
\end{center}
\caption{Breakdown of gain factors for NNBAR with respect to the last search for free neutron-antineutron conversions at the ILL. }
\label{tab:gain}
\end{table}

\subsection{Discussion of Sensitivity}
The results shown above exploit a simulation of a liquid deuterium lower moderator. One of the goals of the HighNESS project~\cite{euh2020,Santoro:2020nke} will be to deliver an engineering design of such a moderator. The performance will be different to that previously simulated and a new NNBAR sensitivity will be quantitatively determined. Based on experience and results from previous moderator designs, it is possible to list and give quantitative estimates of factors influencing the performance of such a moderator. This includes several contributions, which either increase or decrease the performance of the moderator for NNBAR's ultimate capability.

The first group of contributions enhance NNBAR's sensitivity, and include optimization of the design, including moderator size, positioning with respect of the target, and the use of reentrant holes, a proven technology to increase neutron intensity by a factor of 1.5X for a specific direction~\cite{bergmann2018simulation}. Furthermore, if pure orthodeuterium is assumed instead of the mixture of ortho-and-paradeuterium used in Ref.~\cite{Klinkby:2014cma}, the expected spectrum should be colder than previously calculated in the simulation~\cite{Klinkby:2014cma} used for the NNBAR sensitivity estimates above. Another option might be to employ a single-crystal reflector filter in front of the moderator to enhance the thermal and cold flux while reducing the epithermal and fast flux
~\cite{Muhrer2016}.

The second group comprises contributions which can potentially degrade the sensitivity of NNBAR. This can occur with the refinement of engineering details: the fact that other beamlines, other than NNBAR, might view the moderator, may consequently require the removal of some reflector material surrounding the moderator, resulting in decreased performance~\cite{zanini2019design}. Also, a possible shadowing effect of the inner shielding in the monolith could reduce the effective (viewed) surface area of the moderator. Naive estimations suggest a cancellation of these competing effects.

A full quantification of the NNBAR sensitivity is part of the HighNESS program. However, it should be noted that, in principle, running times can be extended to mitigate against any loss of sensitivity. Furthermore, estimates provided in this paper are rather conservative with an assumed selection efficiency for an annihilation event of around $50$\%, as obtained at the ILL; indeed, detector technology and data analysis methods in experimental particle physics are substantially more advanced compared to the early 1990's, and so a far higher efficiency would be expected for a modern-day experiment. Finally, only a lower liquid deuterium moderator was considered for the sensitivity calculations given in this Section. The upper butterfly moderator would also provide an additional flux of cold neutrons. Mitigation by longer running and the conservative nature of the current estimates could thus also protect against an unexpected lowering of the planned full power of the ESS from $5$~MW to $2-3$~MW.

\subsection{Sensitivity of NNBAR and other experiments}
\begin{figure}[H]
  \setlength{\unitlength}{1mm}
  \begin{center}
  \includegraphics[width=0.95\linewidth, angle=0]{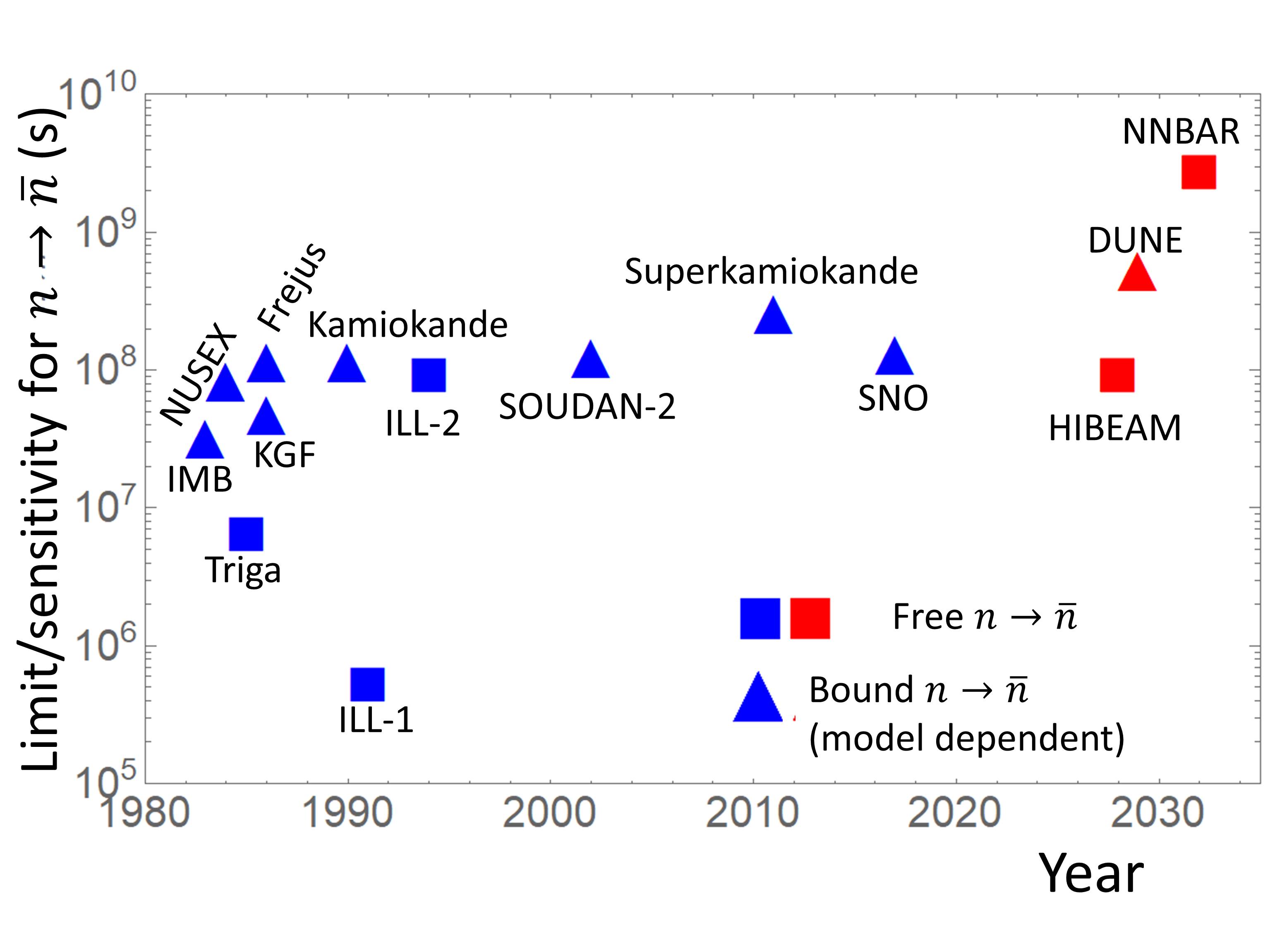}
  \end{center}
  %\vspace{-0.75cm}
  \caption{Lower limits on the free neutron oscillation time from past (blue) experiments on free and bound neutrons. Projected future (red) sensitivities from HIBEAM and NNBAR are shown together with the expected sensitivity for DUNE. Searches with free neutrons with made with the Pavia Triga Mark II reactor\cite{Bressi:1989zd,Bressi:1990zx} and at the ILL~\cite{Fidecaro:1985cm,BaldoCeolin:1994jz}, denoted ILL-1~\cite{Fidecaro:1985cm} and ILL-2~\cite{BaldoCeolin:1994jz}. Limits from bound neutron searches 
   are given from KGF~\cite{KGF}, NUSEX~\cite{NUSEX}, IMB~\cite{IMB}, Kamiokande~\cite{Kamiokande}, Frejus~\cite{Frejus}, Soudan-2~\cite{Soudan-2}, the Sudbury Neutrino Observatory~\cite{Aharmim:2017jna}, and Super-Kamiokande~\cite{Abe:2011ky,Gustafson:2015qyo}. For all bound neutron experiments, model-dependent intranuclear suppression factors are used to estimate a free neutron oscillation time lower limit.}
  \label{fig:nnbar-limits}
\end{figure}

Limits on the free $n\rightarrow\bar{n}$ oscillation time, together with the potential sensitivities of HIBEAM (Section~\ref{sec:small_nnbar_exp}) (assuming three year running at 1MW) and NNBAR (assuming three years running at 5MW, and a three orders of magnitude improvement in ILL units) are shown in Fig.~\ref{fig:nnbar-limits}. 

Also shown in Fig.~\ref{fig:nnbar-limits} is a projected \textit{converted} free oscillation time lower limit for bound neutron conversions within ${}^{40}Ar$ nuclei within the future DUNE experiment, where $\tau_{n\rightarrow\bar{n}}\geq5.53 \times 10^8$s~\cite{Abi:2020evt} for an assumed exposure of 400 kt$\cdot$years. Note that this limit does not take account of new ${}^{40}Ar$ intranuclear suppression factor calculations completed in \cite{Barrow:2019viz}; systematics simulation studies continue within the DUNE collaboration, and further automated analysis improvements are underway~\cite{Abi:2020evt} in hopes of eliminating atmospheric neutrino backgrounds. There is as yet no estimate for the expected $n\rightarrow \bar{n}$ sensitivity for Hyper-Kamiokande.

%} experiments \cite{Abe:2018uyc} also have enhanced sensitivity to $n\rightarrow \bar{n}$ for bound neutrons. 
%\cite{Abi:2020evt} have been completed, and others are underway \cite{Barrow:2019viz,Barrow:2020tba}; see also \cite{Grojean:2018fus}.

As discussed in Section~\ref{sec:nnbar}, consideration of limits or sensitivities on the free neutron oscillation time for free and bound neutron searches is, to a certain extent, an \textit{apples and pears} comparison. There is overlap in physics potential, but neither renders the other redundant. Indeed, both would play essential and complementary roles in both the definitive establishment of and cross reference for any future discovery.

\section{Backgrounds}\label{Sec:backgrounds}
In this Section, the background sources for HIBEAM and NNBAR are described. Common backgrounds for HIBEAM and NNBAR in the context of $n\rightarrow\bar{n}$ transformations 
%highlighting \textit{common} 
are discussed first, followed by a description of backgrounds which are exclusive to NNBAR. Backgrounds for HIBEAM in the context of $n-n'$ oscillations, for both disappearance and regeneration modes, are discussed separately.

\subsection{HIBEAM-NNBAR backgrounds}
As discussed in Section~\ref{sec:annsig} the experimental signature for $n\rightarrow \bar{n}$ is striking: the annihilation of the antineutron, releasing roughly $1.9$ GeV of total energy, typically in the form of pions (4 to 5 on average). 

It is instructive to consider the background mitigation strategy of the most sensitive cold neutron experiment performed to date at the ILL~\cite{BaldoCeolin:1989qd}. Using a steady reactor neutron source, this experiment confirmed an absolutely zero background while expecting a signal efficiency of $\sim50\%$ by using multiple track and kinematic cuts. Such a unique, state of the art feature is impressive, especially given the large integrated flux of slow neutrons which passed through the annihilation target every second ($~10^{11}$ n/s). Developing a detector scheme with zero background is extremely important for the potential detection of an antineutron appearance where even a single event can represent a fundamental discovery.

Both HIBEAM and NNBAR must contend with cold neutron beam-generated and cosmic ray backgrounds. Beam-generated backgrounds comprise fast neutrons and high energy particles from the spallation source, as well as charged and neutral cosmic rays generated in the upper atmosphere. NNBAR will be especially likely to be affected by high energy products from the spallation process given the large opening angle of the beamport. These all are now discussed. 
%\textit{(add free neutron decay backgrounds?)}

\subsubsection{Cold neutron beam backgrounds}
Cold neutron beam backgrounds include an irreducible component in the form of MeV gammas from neutron capture in the annihilation target and beamtube, as well as fast neutrons from the source. To achieve backgroundless operation (below one stray event per year of operation), the HIBEAM and NNBAR detectors should use a similar configuration to the ILL experiment, where only $5.2\%$ of the cold neutron beam was lost to the beam optics and in the journey to the target. At the ILL, the beam halo was absorbed by boron-loaded glass collimators, and the beam dump was constructed of ${}^{6}$Li-loaded tiles. Similar technology is being considered.

Compared to ILL, the (NNBAR) detectors will see a significant increase of $\geq100$ times) cold neutron current through the annihilation target. With a neutron absorption probability $\sim$ 5 $\times$ $10^{-6}$, this rate will produce $\sim$ $10^{8}$ reactions in the target film per second or 5 $\times$ $10^{5}$ GeV per second of isotropically emitted low energy ($\sim$~MeV) photons inside the detector. This energy deposition rate will not be too challenging for a modern trigger system, but it can provide high counting rates in the sub-detector elements and can be a potential source of background in random combination with cosmic events or (for NNBAR) with fast neutrons.
%Never-the-less, this is most likely not a problem considering the past ILL search and their associated cuts.
%Also, walls of the vacuum chamber in the central part of the detector should be recessed not to provide additional (n, g) interactions with the wall material of the vacuum tube. Improvement of the tracking detector spatial resolution might be limited by the multiple scattering in the material of the vacuum tube walls. This suggests the selection of the tracking technology that can operate inside the vacuum tube in the recessed space.
Given that the ESS will have such an enormous increase in incident flux compared to the ILL, and an even larger increase in the portion of the beam which does not reach the target, more stringent requirements on the beam line shielding, detector granularity, tracking resolution, and trigger cuts will be required. In the ILL experiment, spurious events above threshold produced by multiple gamma-ray hits during the 150 ns trigger timing window account for about $32\%$ ($1$Hz) of the total trigger rate of $\sim 4$Hz.

\subsubsection{Cosmic Rays}
Cosmic rays (CRs) were the dominant backgrounds for all previous free neutron transformation experiments. For the ILL experiment, these alone accounted for a remaining $3$Hz of trigger rate, with $2.7$Hz coming from CR muons which evaded the veto (an efficiency of $\sim99.5\%$), and $0.3$Hz due to neutral CRs. The neutral CRs were of particular concern, as they evade the CR veto and can appear to produce events which originate from the target; these were assessed to be the leading contributors to possible backgrounds in the signal window. Given the larger annihilation target area and detector volume for HIBEAM, and especially for NNBAR, these events are expected to potentially contribute to backgrounds and a corresponding improvement in vertex reconstruction and event identification will be required. 

\subsubsection{High energy products from the spallation source - (NNBAR only)}
Previously at the ILL, moderated neutrons from the reactor source were directed into a curved neutron reflecting guide, such that slow neutrons with low transverse momenta were transported along while fast neutrons and photons were filtered out. This will occur for HIBEAM by virtue of using the $S$-curved ANNI guide system (see Fig. \ref{fig:ANNIGuideDesign}).

For the NNBAR experiment there are also other sources of backgrounds specific to the spallation source that were not present for the ILL experiment due to the usage of the large beamport. When protons with an energy of 1-2 GeV interact with a heavy nuclear target, spallation can produce high-energy particles (such as protons, pions, muons, gammas), most essentially among them neutrons with an energy range of MeV to GeV; once moderated, these become the slower neutrons of interest (reaching the detector after $\sim0.1$s). In the NNBAR layout, in order to achieve higher cold neutron currents, the detector will directly view the source through the large beamport, where all fast charged and neutral particle components will contribute to additional backgrounds. Fortunately, due to the pulsed operation of the ESS, these fast particles can be vetoed by time of flight (during a period of $\sim10\mu$s after the beginning of the pulse) by excluding the beginning of a proton beam spill on the spallation target. It is estimated that this would lead to a $<5\%$ loss of the total cold beam intensity for a neutron-antineutron transformation search.

\subsection{HIBEAM backgrounds: neutron--sterile-neutron oscillations}
% From https://arxiv.org/abs/1703.06735:
Both disappearance and regeneration of sterile neutrons would rely on the observation of a magnetic field dependent resonant oscillation signal. These two methods have different systematic difficulties: \textit{i)} for disappearance, one must detect a small reduction in the counting rate of total incident neutrons to the order of $\sim 10^{-6}$--$10^{-8}$ of the total flux, a challenge due to the need for precisely characterized, ultrahigh efficiency neutron monitors able to handle the large beam intensity; \textit{ii)} for regeneration, one must have a very low background count rate in the final neutron detector following total incident beam absorption, requiring an essential detector shielding effort.

From knowledge of other currently operating cold neutron sources such as the High-Flux Isotope Reactor and the Spallation Neutron Source at Oak Ridge Natinal Laboratory, it is known that the ambient background rates in ${}^{3}$He tube-style cold neutron detectors are rather low (few n/s), even when large in overall area and closer to their respective sources than HIBEAM plans to be; this is also the case despite higher duty cycles compared to the ESS. This generalized rate usually includes no vetoes or purpose-built particle tracking equipment around the detectors and only modest shielding, and so can be still improved. Assuming an effective background rate of $\sim1$n/s should thus be achievable for a regeneration experiment at HIBEAM. 

%\\
%(\textit{further discuss backgrounds here, if possible)}

\subsection{Background mitigation strategies}

The main methods to suppress the expected backgrounds for HIBEAM and NNBAR in neutron-antineutron searches are:

\begin{itemize}
  \item (1) Prepare a control background sample by ``switching off" all active magnetic shielding elements of the experiment, allowing the Earth's field to suppress the neutron transformation effect, but leaving the entire experiment unperturbed in the process. This could be done during beam-off and beam-on periods, as well as partially characterized with a similar target-detector configuration during a HIBEAM $n\rightarrow n' \rightarrow \bar{n}$ experiment
  \item (2) Add one or more targets downstream of the annihilation target but within the sensitive volume of the detector to produce additional ``sources" for background events without an annihilation signal. Any antineutron produced in the original cold neutron beam would be removed by the primary annihilation target \cite{Bressi:1990zx}.
  %This method to characterize backgrounds was explored in the $n - \bar n$ search at  Pavia University's Triga Mark II reactor, which employed a second downstream target for this purpose~\cite{Bressi90}.
  \item (3) Suppress the generation of gamma backgrounds produced by neutron capture on target via demanding multiple ``track-like" cuts on the detector, since these events do not create tracks.
    \subitem (i) Tracking these particles back to a common vertex ($\pm$ several mm${}^3$) to resolution smaller than the total beam spot on target will be important
  \item (4) Design a CR veto or, with a modern fast-timing calorimeter, use the entire calorimeter as a CR veto, coupled with vertex reconstruction capability, in order to reduce muon events.
    \subitem (i) Rejection of neutral CR events is to be accomplished by background subtraction via directly measured rates within the detector setup, along with energy deposition and multiple track cuts
\end{itemize}

\subsubsection{Gamma backgrounds from target} 
% have to add proper comments for each figure, for now just adding them with minimal description

A preliminary estimation of gamma emission from the annihilation target due to neutron capture was done for HIBEAM. {\sc McStas} simulated events \cite{Soldner:2018ycf} for the ANNI beam were used for the neutron source assuming ESS operating at 1 MW and the ANNI neutron current 6.4 $\times$ $10^{10}$ neutrons/s . The simulation was implemented in {\sc Phits}~\cite{sato2013particle} using a carbon-12 target of 1 m diameter. The distance from the neutron source to the target was assumed to be of 53 m, with a target thickness of 100 $\mu$m. Gravitational effects acting on the neutrons were not taken into account in the simulation. 

\begin{figure}[H]
  \setlength{\unitlength}{1mm}
  \begin{center}
\includegraphics[width=0.89\linewidth, angle=0]{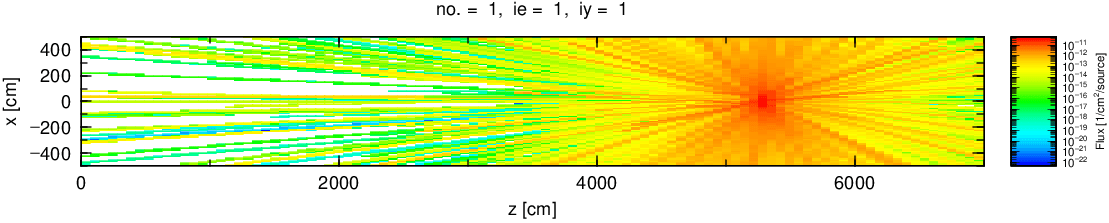}
    \end{center}
%    \vspace{-1.75cm}
  \caption{\footnotesize Photon tracks produced from the interaction of the ANNI neutrons with the carbon-12 target. The origin of the coordinate system is in the experimental area, after ANNI's curved guide extraction.
  }
  \label{fig:gamma_xz}
%  \vspace{2.75cm}
\end{figure}

Fig. \ref{fig:gamma_xz} shows a top-view of the photon tracks obtained by {\sc Phits} due to the interaction of the ANNI neutron beam with the carbon-12 target. As the HIBEAM detector will completely surround the target, these photon tracks represent an important background source that will be further studied in dedicated detector simulations in the near future.

 The photon current from the target is obtained by \textsc{Phits}. The current is calculated such that any photon crossing the target surface adds 1 to the current. By multiplying the photon current with the incident neutron current and target area, the photon rate emission from the target over the full energy spectrum can be estimated. The photon rate calculated by \textsc{Phits} is 3.15 $\times$ $10^{5}$ photons/s. 
Furthermore, since the collaboration is also exploring the use of Beryllium-9 as the material for the target, the photon rate from neutron capture for it was also estimated for it in \textsc{Phits} and found to be 6.68 $\times$ $10^{5}$ photons/s.

%\begin{figure}[tb!]
%\centering
%  \setlength{\unitlength}{1mm}
%  \begin{center}
%\includegraphics[width=0.69\linewidth, %angle=0]{gamma_fluence.PNG}
%    \end{center}
%%    \vspace{-1.75cm}
%  \caption{\footnotesize Photon current  obtained by %\textsc{Phits} through a carbon-12 target area of 1 m %diameter. The photons are generated through capture of the %ANNI beam neutrons.
%  }
%  \label{fig:gamma_fluence}
%%  \vspace{2.75cm}
%\end{figure}

\section{Future directions}\label{Sec:future}
In addition to the work outlined in the previous Sections, the collaboration is pursuing  possibilities to further enhance the sensitivity to neutron conversions and quantify and suppress backgrounds. A fundamental aim is to design a background-free $n\rightarrow n$ search, as achieved at the ILL. Similarly, backgrounds to mirror neutron searches need also to be minimised.

To achieve the above, a full Monte Carlo-based model of different designs of the annihilation detector is being constructed. Simulated background and signal processes will be used to optimise the experiment's sensitivity. These simulations will be informed and optimised with the experience of early running and tests at the ESS. 

The work will dovetail with the Horizon2020 HighNESS project to design a liquid deuterium moderator and associated instruments including NNBAR. 

Hardware R\&D is also planned to support the simulation effort. Work has started on a prototype calorimeter module which would take part in a neutron test beam at the ESS in 2023 for {\it in situ} background measurements. The calorimeter module is described in the Appendix. 

Cost estimation and reduction will also be done. A potentially important way to reduce the cost is to exploit the possibility of allowing neutrons to bounce without ''resetting the clock" ~\ref{sec:bounce-idea}. There are  laboratory experiments which can shed light on the validity of these ideas. The formula used for the antineutron-nucleus reflectivity deviates strongly from the usual Fresnel shape familiar from light optics due to the very large effect of the imaginary part of the antineutron optical potential. The accuracy of the neutron reflectivity formula in this extreme limit has been verified by neutron reflectometry measurements on gadolinium, which is an element with two isotopes that possess a very high neutron absorption cross section comparable in size to that possessed by antineutrons~\cite{Korneev1992,Nikova2019}. It would also be very interesting to test theoretical calculations of nbar-A scattering amplitudes with data from slow antiprotons (there is essentially no hope to get slow antineutrons for scattering experiments) as long as theory can handle the Coulomb corrections to extract the nuclear component of the scattering from the data. Experiments with very cold antiprotons available at CERN could also be envisaged.

Other strategies which may impact cost are optimization of the distance from the entrance aperture of the neutron reflector system to the moderator, potentially reducing the size of the optical system dramatically, and relaxing magnetic shielding requirements while still sufficiently satisfying the quasi-free condition\cite{Davis:2016uyk}.

\section{Summary}\label{sec:summary}
Strong theoretical motivations addressing open questions in modern physics such as the matter-antimatter asymmetry, the nature of dark matter, and the possible Majorana nature of the neutrino imply the existence of neutron conversions into anti-neutrons and/or sterile neutrons. A remarkable opportunity has arisen to conduct a  two-stage experiment (HIBEAM leading to NNBAR) to be hosted by the European Spallation Source  which will perform a series of high precision, world-leading neutron conversion searches. The final goal is to achieve a sensitivity in searching for $n\rightarrow \bar{n}$ with free neutrons which is three orders of magnitude higher than the last such  experiment. A collaboration to carry out this program has been formed with the aim of performing the experiment.

\section{Acknowledgements}\label{sec:acknowledgements}
D. Milstead gratefully acknowledges support from the Swedish Research Council. The work of Z. Berezhiani is supported in part by CETEMPs at the University di L’Aquila.
The work of Y. Kamyshkov was supported in part by US DOE Grant DE-SC0014558. The work of S. Girmohanta and R. Shrock is supported in part by US NSF grant NSF-PHY-1915093.
The work of L. Varriano is supported by a National Science Foundation Graduate Research Fellowship under Grant No. DGE-1746045.
B. Kerbikov is supported by RSF grant 16-12-10414.B. Kopeliovich and I. Potashnikova acknowledge grants ANID - Chile FONDECYT 1170319,
and ANID PIA/APOYO AFB180002. A. Young is supported by the NSF grant PHY-1914133 and the DOE grant DE-FG02-ER41042.
 J. L. Barrow’s work was supported by the U.S. Department of Energy, Office of Science, Office of Workforce Development for Teachers and Scientists, Office of Science Graduate Student Research (SCGSR) program. The SCGSR program is administered by the Oak Ridge Institute for Science and Education for the DOE under Contract No. DESC0014664. J. L. Barrow is also supported in part by the Visiting Scholars Award Program of the Universities Research Association.

\section{Bibliography}\label{sec:Bibliography}

%\begin{figure}
% \includegraphics{example.eps}
%\label{fig:1}     % Give a unique label
%\end{figsure}
%

  %do it manually for the arxiv
%\bibliographystyle{utphys}       % APS-like style for physics
%\bibliography{nnbar}%%pap}   % name your BibTeX data base

\begin{thebibliography}{10}
\providecommand{\url}[1]{{#1}}
\providecommand{\urlprefix}{URL }
\expandafter\ifx\csname urlstyle\endcsname\relax
  \providecommand{\doi}[1]{DOI \discretionary{}{}{}#1}\else
  \providecommand{\doi}{DOI \discretionary{}{}{}\begingroup
  \urlstyle{rm}\Url}\fi

\bibitem{Sakharov:1967dj}
A.D. Sakharov, Pisma Zh. Eksp. Teor. Fiz. \textbf{5}, 32 (1967).
\newblock \doi{10.1070/PU1991v034n05ABEH002497}.
\newblock [Usp. Fiz. Nauk161,no.5,61(1991)]

\bibitem{Mohapatra:1980qe}
R.N. Mohapatra, R.E. Marshak, Phys. Rev. Lett. \textbf{44}, 1316 (1980).
\newblock \doi{10.1103/PhysRevLett.44.1644.2, 10.1103/PhysRevLett.44.1316}.
\newblock [Erratum: Phys. Rev. Lett.44,1643(1980)]

\bibitem{Adler:1969gk}
S.L. Adler, Phys. Rev. \textbf{177}, 2426 (1969).
\newblock \doi{10.1103/PhysRev.177.2426}.
\newblock [,241(1969)]

\bibitem{tHooft:1976snw}
G.~'t~Hooft, Phys. Rev. \textbf{D14}, 3432 (1976).
\newblock \doi{10.1103/PhysRevD.18.2199.3, 10.1103/PhysRevD.14.3432}.
\newblock [,70(1976)]

\bibitem{Smith:1999cr}
G.L. Smith, C.D. Hoyle, J.H. Gundlach, E.G. Adelberger, B.R. Heckel, H.E.
  Swanson, Phys. Rev. \textbf{D61}, 022001 (2000).
\newblock \doi{10.1103/PhysRevD.61.022001}

\bibitem{Schlamminger:2007ht}
S.~Schlamminger, K.Y. Choi, T.A. Wagner, J.H. Gundlach, E.G. Adelberger, Phys.
  Rev. Lett. \textbf{100}, 041101 (2008).
\newblock \doi{10.1103/PhysRevLett.100.041101}

\bibitem{Barbier:2004ez}
R.~Barbier, et~al., Phys. Rept. \textbf{420}, 1 (2005).
\newblock \doi{10.1016/j.physrep.2005.08.006}

\bibitem{EOInnbar}
G.~Brooijmans, et~al.,   (2015)

\bibitem{Riotto:1999yt}
A.~Riotto, M.~Trodden, Ann. Rev. Nucl. Part. Sci. \textbf{49}, 35 (1999).
\newblock \doi{10.1146/annurev.nucl.49.1.35}

\bibitem{Dine:2003ax}
M.~Dine, A.~Kusenko, Rev. Mod. Phys. \textbf{76}, 1 (2003).
\newblock \doi{10.1103/RevModPhys.76.1}

\bibitem{Cline:2006ts}
J.M. Cline, in \emph{{Les Houches Summer School - Session 86: Particle Physics
  and Cosmology: The Fabric of Spacetime Les Houches, France, July 31-August
  25, 2006}} (2006)

\bibitem{Weinberg:1979bt}
S.~Weinberg, Phys. Rev. Lett. \textbf{42}, 850 (1979).
\newblock \doi{10.1103/PhysRevLett.42.850}

\bibitem{Fry:1980bd}
J.N. Fry, K.A. Olive, M.S. Turner, Phys. Rev. Lett. \textbf{45}, 2074 (1980).
\newblock \doi{10.1103/PhysRevLett.45.2074}

\bibitem{Yoshimura:1978ex}
M.~Yoshimura, Phys. Rev. Lett. \textbf{41}, 281 (1978).
\newblock \doi{10.1103/PhysRevLett.42.746, 10.1103/PhysRevLett.41.281}.
\newblock [Erratum: Phys. Rev. Lett.42,746(1979)]

\bibitem{Ellis:1978xg}
J.R. Ellis, M.K. Gaillard, D.V. Nanopoulos, Phys. Lett. \textbf{80B}, 360
  (1979).
\newblock \doi{10.1016/0370-2693(79)91190-0}.
\newblock [Erratum: Phys. Lett.82B,464(1979)]

\bibitem{Kuzmin:1985mm}
V.A. Kuzmin, V.A. Rubakov, M.E. Shaposhnikov, Phys. Lett. \textbf{155B}, 36
  (1985).
\newblock \doi{10.1016/0370-2693(85)91028-7}

\bibitem{Shaposhnikov:1986jp}
M.E. Shaposhnikov, JETP Lett. \textbf{44}, 465 (1986).
\newblock [Pisma Zh. Eksp. Teor. Fiz.44,364(1986)]

\bibitem{Shaposhnikov:1987tw}
M.E. Shaposhnikov, Nucl. Phys. \textbf{B287}, 757 (1987).
\newblock \doi{10.1016/0550-3213(87)90127-1}

\bibitem{Babu:2006xc}
K.S. Babu, R.N. Mohapatra, S.~Nasri, Phys. Rev. Lett. \textbf{97}, 131301
  (2006).
\newblock \doi{10.1103/PhysRevLett.97.131301}

\bibitem{Babu:2013yca}
K.S. Babu, P.S. Bhupal~Dev, E.C.F.S. Fortes, R.N. Mohapatra, Phys. Rev.
  \textbf{D87}(11), 115019 (2013).
\newblock \doi{10.1103/PhysRevD.87.115019}

\bibitem{Bringmann:2018sbs}
T.~Bringmann, J.M. Cline, J.M. Cornell, Phys. Rev. \textbf{D99}(3), 035024
  (2019).
\newblock \doi{10.1103/PhysRevD.99.035024}

\bibitem{Berezhiani:2018zvs}
Z.~Berezhiani, Int. J. Mod. Phys. \textbf{A33}(31), 1844034 (2018).
\newblock \doi{10.1142/S0217751X18440347}

\bibitem{Essig:2013lka}
R.~Essig, et~al., in \emph{{Proceedings, 2013 Community Summer Study on the
  Future of U.S. Particle Physics: Snowmass on the Mississippi (CSS2013):
  Minneapolis, MN, USA, July 29-August 6, 2013}} (2013).
\newblock
  \urlprefix\url{http://www.slac.stanford.edu/econf/C1307292/docs/IntensityFrontier/NewLight-17.pdf}

\bibitem{Holdom:1985ag}
B.~Holdom, Phys. Lett. \textbf{166B}, 196 (1986).
\newblock \doi{10.1016/0370-2693(86)91377-8}

\bibitem{Berezhiani:2018eds}
Z.~Berezhiani,   (2018)

\bibitem{Mohapatra:1980de}
R.N. Mohapatra, R.E. Marshak, Phys. Lett. \textbf{94B}, 183 (1980).
\newblock \doi{10.1016/0370-2693(80)90853-9, 10.1016/0370-2693(80)90805-9}.
\newblock [Erratum: Phys. Lett.96B,444(1980)]

\bibitem{Calibbi:2016ukt}
L.~Calibbi, G.~Ferretti, D.~Milstead, C.~Petersson, R.~Pöttgen, JHEP
  \textbf{05}, 144 (2016).
\newblock \doi{10.1007/JHEP05(2016)144, 10.1007/JHEP10(2017)195}.
\newblock [Erratum: JHEP10,195(2017)]

\bibitem{Winslow:2010wf}
P.T. Winslow, J.N. Ng, Phys. Rev. \textbf{D81}, 106010 (2010).
\newblock \doi{10.1103/PhysRevD.81.106010}

\bibitem{Mohapatra:2009wp}
R.N. Mohapatra, J. Phys. \textbf{G36}, 104006 (2009).
\newblock \doi{10.1088/0954-3899/36/10/104006}

\bibitem{Dover:1982wv}
C.B. Dover, A.~Gal, J.M. Richard, Phys. Rev. \textbf{D27}, 1090 (1983).
\newblock \doi{10.1103/PhysRevD.27.1090}

\bibitem{Alberico:1982nu}
W.M. Alberico, A.~Bottino, A.~Molinari, Phys. Lett. \textbf{114B}, 266 (1982).
\newblock \doi{10.1016/0370-2693(82)90493-2}

\bibitem{Alberico:1984wk}
W.M. Alberico, J.~Bernabeu, A.~Bottino, A.~Molinari, Nucl. Phys. \textbf{A429},
  445 (1984).
\newblock \doi{10.1016/0375-9474(84)90691-2}

\bibitem{Dover:1989zz}
C.B. Dover, A.~Gal, J.M. Richard, Nucl. Instrum. Meth. \textbf{A284}, 13
  (1989).
\newblock \doi{10.1016/0168-9002(89)90239-8}

\bibitem{Alberico:1990ij}
W.M. Alberico, A.~De~Pace, M.~Pignone, Nucl. Phys. \textbf{A523}, 488 (1991).
\newblock \doi{10.1016/0375-9474(91)90032-2}

\bibitem{Hufner:1998gu}
J.~Hufner, B.Z. Kopeliovich, Mod. Phys. Lett. \textbf{A13}, 2385 (1998).
\newblock \doi{10.1142/S0217732398002540}.
\newblock [,171(1998)]

\bibitem{Friedman:2008es}
E.~Friedman, A.~Gal, Phys. Rev. \textbf{D78}, 016002 (2008).
\newblock \doi{10.1103/PhysRevD.78.016002}

\bibitem{Abe:2011ky}
K.~Abe, et~al., Phys. Rev. \textbf{D91}, 072006 (2015).
\newblock \doi{10.1103/PhysRevD.91.072006}

\bibitem{Bressi:1990zx}
G.~Bressi, et~al., Nuovo Cim. \textbf{A103}, 731 (1990).
\newblock \doi{10.1007/BF02789025}

\bibitem{Sharma:2010vv}
N.~Sharma, H.~Dahiya, P.K. Chatley, M.~Gupta, Phys. Rev. \textbf{D81}, 073001
  (2010).
\newblock \doi{10.1103/PhysRevD.81.073001}

\bibitem{Broggini:2012df}
C.~Broggini, C.~Giunti, A.~Studenikin, Adv. High Energy Phys. \textbf{2012},
  459526 (2012).
\newblock \doi{10.1155/2012/459526}

\bibitem{Berezhiani:2006je}
Z.~Berezhiani, L.~Bento, Phys. Lett. \textbf{B635}, 253 (2006).
\newblock \doi{10.1016/j.physletb.2006.03.008}

\bibitem{Fidecaro:1985cm}
G.~Fidecaro, et~al., Phys. Lett. \textbf{156B}, 122 (1985).
\newblock \doi{10.1016/0370-2693(85)91367-X}

\bibitem{BaldoCeolin:1994jz}
M.~Baldo-Ceolin, et~al., Z. Phys. \textbf{C63}, 409 (1994).
\newblock \doi{10.1007/BF01580321}

\bibitem{Chung:2002fx}
J.~Chung, et~al., Phys. Rev. \textbf{D66}, 032004 (2002).
\newblock \doi{10.1103/PhysRevD.66.032004}

\bibitem{Takita:1986zm}
M.~Takita, et~al., Phys. Rev. \textbf{D34}, 902 (1986).
\newblock \doi{10.1103/PhysRevD.34.902}

\bibitem{Jones:1983ij}
T.W. Jones, et~al., Phys. Rev. Lett. \textbf{52}, 720 (1984).
\newblock \doi{10.1103/PhysRevLett.52.720}

\bibitem{Aharmim:2017jna}
B.~Aharmim, et~al., Phys. Rev. \textbf{D96}(9), 092005 (2017).
\newblock \doi{10.1103/PhysRevD.96.092005}

\bibitem{Ban:2007tp}
G.~Ban, et~al., Phys. Rev. Lett. \textbf{99}, 161603 (2007).
\newblock \doi{10.1103/PhysRevLett.99.161603}

\bibitem{Serebrov:2007gw}
A.P. Serebrov, et~al., Phys. Lett. \textbf{B663}, 181 (2008).
\newblock \doi{10.1016/j.physletb.2008.04.014}

\bibitem{Serebrov:2008hw}
A.P. Serebrov, et~al., Nucl. Instrum. Meth. \textbf{A611}, 137 (2009).
\newblock \doi{10.1016/j.nima.2009.07.041}

\bibitem{Bodek:2009zz}
K.~Bodek, et~al., Nucl. Instrum. Meth. \textbf{A611}, 141 (2009).
\newblock \doi{10.1016/j.nima.2009.07.047}

\bibitem{Altarev:2009tg}
I.~Altarev, et~al., Phys. Rev. \textbf{D80}, 032003 (2009).
\newblock \doi{10.1103/PhysRevD.80.032003}

\bibitem{Berezhiani:2017jkn}
Z.~Berezhiani, R.~Biondi, P.~Geltenbort, I.A. Krasnoshchekova, V.E. Varlamov,
  A.V. Vassiljev, O.M. Zherebtsov, Eur. Phys. J. \textbf{C78}(9), 717 (2018).
\newblock \doi{10.1140/epjc/s10052-018-6189-y}

\bibitem{Peggs:2013sgv}
S.~Peggs,   (2013)

\bibitem{Davis:2016uyk}
E.D. Davis, A.R. Young, Phys. Rev. \textbf{D95}(3), 036004 (2017).
\newblock \doi{10.1103/PhysRevD.95.036004}

\bibitem{Altarev:2015fra}
I.~Altarev, et~al., J. Appl. Phys. \textbf{117}, 183903 (2015).
\newblock \doi{10.1063/1.4919366}

\bibitem{Golubeva:2018mrz}
E.S. Golubeva, J.L. Barrow, C.G. Ladd, Phys. Rev. \textbf{D99}(3), 035002
  (2019).
\newblock \doi{10.1103/PhysRevD.99.035002}

\bibitem{Bargholtz:2008aa}
C.~Bargholtz, et~al., Nucl. Instrum. Meth. \textbf{A594}, 339 (2008).
\newblock \doi{10.1016/j.nima.2008.06.011}

\end{thebibliography}


\begin{thebibliography}{100}

\bibitem{tHooft:1976snw}
G.~'t~Hooft, ``{Computation of the Quantum Effects Due to a Four-Dimensional
  Pseudoparticle},'' \href{http://dx.doi.org/10.1103/PhysRevD.14.3432}{{\em
  Phys. Rev. D} {\bfseries 14} (1976) 3432--3450}. [Erratum: Phys.Rev.D 18,
  2199 (1978)].

\bibitem{Kuzmin:1987wn}
V.~A. Kuzmin, V.~A. Rubakov, and M.~E. Shaposhnikov, ``{Anomalous Electroweak
  Baryon Number Nonconservation and GUT Mechanism for Baryogenesis},''
\href{http://dx.doi.org/10.1016/0370-2693(87)91340-2}{{\em Phys. Lett.}
  {\bfseries B191} (1987) 171--173}.
%%CITATION = PHLTA,B191,171;%%.

\bibitem{Dolgov:1991fr}
A.~Dolgov, ``{Non-GUT baryogenesis},''
  \href{http://dx.doi.org/10.1016/0370-1573(92)90107-B}{{\em Phys. Rept.}
  {\bfseries 222} (1992) 309--386}.

\bibitem{Smith:1999cr}
G.~L. Smith, C.~D. Hoyle, J.~H. Gundlach, E.~G. Adelberger, B.~R. Heckel, and
  H.~E. Swanson, ``{Short range tests of the equivalence principle},''
\href{http://dx.doi.org/10.1103/PhysRevD.61.022001}{{\em Phys. Rev.} {\bfseries
  D61} (2000) 022001}.
%%CITATION = PHRVA,D61,022001;%%.

\bibitem{Schlamminger:2007ht}
S.~Schlamminger, K.~Y. Choi, T.~A. Wagner, J.~H. Gundlach, and E.~G.
  Adelberger, ``{Test of the equivalence principle using a rotating torsion
  balance},'' \href{http://dx.doi.org/10.1103/PhysRevLett.100.041101}{{\em
  Phys. Rev. Lett.} {\bfseries 100} (2008) 041101},
\href{http://arxiv.org/abs/0712.0607}{{\ttfamily arXiv:0712.0607 [gr-qc]}}.
%%CITATION = ARXIV:0712.0607;%%.

\bibitem{Cowsik:2018jbq}
R.~Cowsik {\em et~al.}, ``{Test of Einstein's Equivalence Principle with a
  Long-Period Torsion Balance},''
\href{http://arxiv.org/abs/1808.09925}{{\ttfamily arXiv:1808.09925 [gr-qc]}}.
%%CITATION = ARXIV:1808.09925;%%.

\bibitem{Sakharov:1967dj}
A.~D. Sakharov, ``{Violation of CP Invariance, C asymmetry, and baryon
  asymmetry of the universe},''
  \href{http://dx.doi.org/10.1070/PU1991v034n05ABEH002497}{{\em Pisma Zh. Eksp.
  Teor. Fiz.} {\bfseries 5} (1967) 32--35}.
[Usp. Fiz. Nauk161,no.5,61(1991)].
%%CITATION = ZFPRA,5,32;%%.

\bibitem{Kuzmin:1970nx}
V.~A. Kuzmin, ``{Cp violation and baryon asymmetry of the universe},''
{\em Pisma Zh. Eksp. Teor. Fiz.} {\bfseries 12} (1970) 335--337.
%%CITATION = ZFPRA,12,335;%%.

\bibitem{Mohapatra:1980de}
R.~N. Mohapatra and R.~E. Marshak, ``{Phenomenology of neutron oscillations},''
  \href{http://dx.doi.org/10.1016/0370-2693(80)90853-9,
  10.1016/0370-2693(80)90805-9}{{\em Phys. Lett.} {\bfseries 94B} (1980) 183}.
[Erratum: Phys. Lett.96B,444(1980)].
%%CITATION = PHLTA,94B,183;%%.

\bibitem{Phillips:2014fgb}
D.~G. Phillips, II {\em et~al.}, ``{Neutron-Antineutron Oscillations:
  Theoretical Status and Experimental Prospects},''
  \href{http://dx.doi.org/10.1016/j.physrep.2015.11.001}{{\em Phys. Rept.}
  {\bfseries 612} (2016) 1--45},
\href{http://arxiv.org/abs/1410.1100}{{\ttfamily arXiv:1410.1100 [hep-ex]}}.
%%CITATION = ARXIV:1410.1100;%%.

\bibitem{Chang:1980ey}
L.~Chang and N.~Chang, ``{$B-L$ Nonconservation And Neutron Oscillation},''
  \href{http://dx.doi.org/10.1016/0370-2693(80)90314-7}{{\em Phys. Lett. B}
  {\bfseries 92} (1980) 103--106}.

\bibitem{Kuo:1980ew}
T.-K. Kuo and S.~T. Love, ``{Neutron Oscillations and the Existence of Massive
  Neutral Leptons},''
\href{http://dx.doi.org/10.1103/PhysRevLett.45.93}{{\em Phys. Rev. Lett.}
  {\bfseries 45} (1980) 93}.
%%CITATION = PRLTA,45,93;%%.

\bibitem{Cowsik:1980np}
R.~Cowsik and S.~Nussinov, ``{Some Constraints On $\Delta B =2$
  Neutron-antineutron Oscillations},''
  \href{http://dx.doi.org/10.1016/0370-2693(81)90302-6}{{\em Phys. Lett. B}
  {\bfseries 101} (1981) 237--240}.

\bibitem{Rao:1982gt}
S.~Rao and R.~Shrock, ``{$n \leftrightarrow \bar{n}$ Transition Operators and
  Their Matrix Elements in the {MIT} Bag Model},''
  \href{http://dx.doi.org/10.1016/0370-2693(82)90333-1}{{\em Phys. Lett. B}
  {\bfseries 116} (1982) 238--242}.

\bibitem{Rao:1983sd}
S.~Rao and R.~E. Shrock, ``{Six Fermion ($B-L$) Violating Operators of
  Arbitrary Generational Structure},''
\href{http://dx.doi.org/10.1016/0550-3213(84)90365-1}{{\em Nucl. Phys.}
  {\bfseries B232} (1984) 143--179}.
%%CITATION = NUPHA,B232,143;%%.

\bibitem{Caswell:1982qs}
W.~E. Caswell, J.~Milutinovic, and G.~Senjanovic, ``{Matter-Antimatter
  Transition Operators: A Manual for Modeling},''
  \href{http://dx.doi.org/10.1016/0370-2693(83)91585-X}{{\em Phys. Lett. B}
  {\bfseries 122} (1983) 373--377}.

\bibitem{Berezhiani:2005hv}
Z.~Berezhiani and L.~Bento, ``{Neutron - mirror neutron oscillations: How fast
  might they be?},''
  \href{http://dx.doi.org/10.1103/PhysRevLett.96.081801}{{\em Phys. Rev. Lett.}
  {\bfseries 96} (2006) 081801},
\href{http://arxiv.org/abs/hep-ph/0507031}{{\ttfamily arXiv:hep-ph/0507031
  [hep-ph]}}.
%%CITATION = HEP-PH/0507031;%%.

\bibitem{Berezhiani:2008bc}
Z.~Berezhiani, ``{More about neutron - mirror neutron oscillation},''
  \href{http://dx.doi.org/10.1140/epjc/s10052-009-1165-1}{{\em Eur. Phys. J.}
  {\bfseries C64} (2009) 421--431},
\href{http://arxiv.org/abs/0804.2088}{{\ttfamily arXiv:0804.2088 [hep-ph]}}.
%%CITATION = ARXIV:0804.2088;%%.

\bibitem{Grojean:2018fus}
C.~Grojean, B.~Shakya, J.~D. Wells, and Z.~Zhang, ``{Implications of an
  Improved Neutron-Antineutron Oscillation Search for Baryogenesis: A Minimal
  Effective Theory Analysis},''
  \href{http://dx.doi.org/10.1103/PhysRevLett.121.171801}{{\em Phys.\ Rev.\
  Lett.} {\bfseries 121} no.~17, (2018) 171801},
  \href{http://arxiv.org/abs/1806.00011}{{\ttfamily arXiv:1806.00011
  [hep-ph]}}.

\bibitem{Bringmann:2018sbs}
T.~Bringmann, J.~M. Cline, and J.~M. Cornell, ``{Baryogenesis from neutron-dark
  matter oscillations},''
  \href{http://dx.doi.org/10.1103/PhysRevD.99.035024}{{\em Phys. Rev.}
  {\bfseries D99} no.~3, (2019) 035024},
\href{http://arxiv.org/abs/1810.08215}{{\ttfamily arXiv:1810.08215 [hep-ph]}}.
%%CITATION = ARXIV:1810.08215;%%.

\bibitem{Mohapatra:1980qe}
R.~N. Mohapatra and R.~Marshak, ``{Local $B-L$ Symmetry of Electroweak
  Interactions, Majorana Neutrinos and Neutron Oscillations},''
  \href{http://dx.doi.org/10.1103/PhysRevLett.44.1316}{{\em Phys. Rev. Lett.}
  {\bfseries 44} (1980) 1316--1319}. [Erratum: Phys.Rev.Lett. 44, 1643 (1980)].

\bibitem{Babu:2006xc}
K.~S. Babu, R.~N. Mohapatra, and S.~Nasri, ``{Post-Sphaleron Baryogenesis},''
  \href{http://dx.doi.org/10.1103/PhysRevLett.97.131301}{{\em Phys. Rev. Lett.}
  {\bfseries 97} (2006) 131301},
\href{http://arxiv.org/abs/hep-ph/0606144}{{\ttfamily arXiv:hep-ph/0606144
  [hep-ph]}}.
%%CITATION = HEP-PH/0606144;%%.

\bibitem{Dev:2015uca}
P.~S.~B. Dev and R.~N. Mohapatra, ``{TeV scale model for baryon and lepton
  number violation and resonant baryogenesis},''
  \href{http://dx.doi.org/10.1103/PhysRevD.92.016007}{{\em Phys. Rev.}
  {\bfseries D92} no.~1, (2015) 016007},
\href{http://arxiv.org/abs/1504.07196}{{\ttfamily arXiv:1504.07196 [hep-ph]}}.
%%CITATION = ARXIV:1504.07196;%%.

\bibitem{Allahverdi:2017edd}
R.~Allahverdi, P.~S.~B. Dev, and B.~Dutta, ``{A simple testable model of baryon
  number violation: Baryogenesis, dark matter, neutron–antineutron
  oscillation and collider signals},''
  \href{http://dx.doi.org/10.1016/j.physletb.2018.02.019}{{\em Phys. Lett.}
  {\bfseries B779} (2018) 262--268},
\href{http://arxiv.org/abs/1712.02713}{{\ttfamily arXiv:1712.02713 [hep-ph]}}.
%%CITATION = ARXIV:1712.02713;%%.

\bibitem{Barbier:2004ez}
R.~Barbier {\em et~al.}, ``{R-parity violating supersymmetry},''
  \href{http://dx.doi.org/10.1016/j.physrep.2005.08.006}{{\em Phys. Rept.}
  {\bfseries 420} (2005) 1--202},
\href{http://arxiv.org/abs/hep-ph/0406039}{{\ttfamily arXiv:hep-ph/0406039
  [hep-ph]}}.
%%CITATION = HEP-PH/0406039;%%.

\bibitem{Calibbi:2016ukt}
L.~Calibbi, G.~Ferretti, D.~Milstead, C.~Petersson, and R.~Pöttgen, ``{Baryon
  number violation in supersymmetry: n-nbar oscillations as a probe beyond the
  LHC},'' \href{http://dx.doi.org/10.1007/JHEP05(2016)144,
  10.1007/JHEP10(2017)195}{{\em JHEP} {\bfseries 05} (2016) 144},
  \href{http://arxiv.org/abs/1602.04821}{{\ttfamily arXiv:1602.04821
  [hep-ph]}}.
[Erratum: JHEP10,195(2017)].
%%CITATION = ARXIV:1602.04821;%%.

\bibitem{Nussinov:2001rb}
S.~Nussinov and R.~Shrock, ``{Neutron-antineutron oscillations in models with
  large extra dimensions},''
  \href{http://dx.doi.org/10.1103/PhysRevLett.88.171601}{{\em Phys. Rev. Lett.}
  {\bfseries 88} (2002) 171601},
\href{http://arxiv.org/abs/hep-ph/0112337}{{\ttfamily arXiv:hep-ph/0112337
  [hep-ph]}}.
%%CITATION= HEP-PH/0112337;%%.

\bibitem{Girmohanta:2019fsx}
S.~Girmohanta and R.~Shrock, ``{Baryon-Number-Violating Nucleon and Dinucleon
  Decays in a Model with Large Extra Dimensions},''
  \href{http://dx.doi.org/10.1103/PhysRevD.101.015017}{{\em Phys. Rev. D}
  {\bfseries 101} no.~1, (2020) 015017},
  \href{http://arxiv.org/abs/1911.05102}{{\ttfamily arXiv:1911.05102
  [hep-ph]}}.

\bibitem{Girmohanta:2020qfd}
S.~Girmohanta and R.~Shrock, ``{Nucleon decay and $n$-$\bar n$ oscillations in
  a left-right symmetric model with large extra dimensions},''
  \href{http://dx.doi.org/10.1103/PhysRevD.101.095012}{{\em Phys. Rev. D}
  {\bfseries 101} no.~9, (2020) 095012},
  \href{http://arxiv.org/abs/2003.14185}{{\ttfamily arXiv:2003.14185
  [hep-ph]}}.

\bibitem{Berezhiani:2006je}
Z.~Berezhiani and L.~Bento, ``{Fast neutron: Mirror neutron oscillation and
  ultra high energy cosmic rays},''
  \href{http://dx.doi.org/10.1016/j.physletb.2006.03.008}{{\em Phys. Lett.}
  {\bfseries B635} (2006) 253--259},
\href{http://arxiv.org/abs/hep-ph/0602227}{{\ttfamily arXiv:hep-ph/0602227
  [hep-ph]}}.
%%CITATION = HEP-PH/0602227;%%.

\bibitem{Berezhiani:2011da}
Z.~Berezhiani and A.~Gazizov, ``{Neutron Oscillations to Parallel World:
  Earlier End to the Cosmic Ray Spectrum?},''
  \href{http://dx.doi.org/10.1140/epjc/s10052-012-2111-1}{{\em Eur. Phys. J.}
  {\bfseries C72} (2012) 2111},
\href{http://arxiv.org/abs/1109.3725}{{\ttfamily arXiv:1109.3725
  [astro-ph.HE]}}.
%%CITATION = ARXIV:1109.3725;%%.

\bibitem{Mohapatra:2009wp}
R.~N. Mohapatra, ``{Neutron-antineutron Oscillation: Theory and
  Phenomenology},''
  \href{http://dx.doi.org/10.1088/0954-3899/36/10/104006}{{\em J. Phys.}
  {\bfseries G36} (2009) 104006},
\href{http://arxiv.org/abs/0902.0834}{{\ttfamily arXiv:0902.0834 [hep-ph]}}.
%%CITATION = ARXIV:0902.0834;%%.

\bibitem{Berezhiani:2015afa}
Z.~Berezhiani, ``{Neutron–antineutron oscillation and baryonic majoron: low
  scale spontaneous baryon violation},''
  \href{http://dx.doi.org/10.1140/epjc/s10052-016-4564-0}{{\em Eur. Phys. J.}
  {\bfseries C76} no.~12, (2016) 705},
\href{http://arxiv.org/abs/1507.05478}{{\ttfamily arXiv:1507.05478 [hep-ph]}}.
%%CITATION = ARXIV:1507.05478;%%.

\bibitem{Arnold:2012sd}
J.~M. Arnold, B.~Fornal, and M.~B. Wise, ``{Simplified models with baryon
  number violation but no proton decay},''
  \href{http://dx.doi.org/10.1103/PhysRevD.87.075004}{{\em Phys. Rev. D}
  {\bfseries 87} (2013) 075004},
  \href{http://arxiv.org/abs/1212.4556}{{\ttfamily arXiv:1212.4556 [hep-ph]}}.

\bibitem{Mohapatra:1982xz}
R.~N. Mohapatra and G.~Senjanovic, ``{Spontaneous Breaking of Global $B-L$
  Symmetry and Matter-Antimatter Oscillations in Grand Unified Theories},''
  \href{http://dx.doi.org/10.1103/PhysRevD.27.254}{{\em Phys. Rev. D}
  {\bfseries 27} (1983) 254}.

\bibitem{Senjanovic:1982np}
G.~Senjanovic, ``{Higgs Mass Scales And Matter-Antimatter Oscillations In Grand
  Unified Theories},'' in {\em {Workshop on Neutrino-Antineutrino
  Oscillations}}.
\newblock 4, 1982.

\bibitem{BaldoCeolin:1994jz}
M.~Baldo-Ceolin {\em et~al.}, ``{A New experimental limit on neutron -
  antineutron oscillations},''
\href{http://dx.doi.org/10.1007/BF01580321}{{\em Z. Phys.} {\bfseries C63}
  (1994) 409--416}.
%%CITATION = ZEPYA,C63,409;%%.

\bibitem{EOInnbar}
G.~Brooijmans {\em et~al.} {\em {Expression of interest for a new search for
  neutron-antineutron oscillations at the ESS}} (2015) .

\bibitem{Soldner:2018ycf}
T.~Soldner, H.~Abele, G.~Konrad, B.~Märkisch, F.~M. Piegsa, U.~Schmidt,
  C.~Theroine, and P.~T. Sánchez, ``{ANNI - A pulsed cold neutron beam
  facility for particle physics at the ESS},'' in {\em {International Workshop
  on Particle Physics at Neutron Sources 2018 (PPNS 2018) Grenoble, France, May
  24-26, 2018}}.
\newblock 2018.
\newblock
\href{http://arxiv.org/abs/1811.11692}{{\ttfamily arXiv:1811.11692
  [physics.ins-det]}}.
\newblock
%%CITATION = ARXIV:1811.11692;%%.

\bibitem{Ban:2007tp}
G.~Ban {\em et~al.}, ``{A Direct experimental limit on neutron: Mirror neutron
  oscillations},'' \href{http://dx.doi.org/10.1103/PhysRevLett.99.161603}{{\em
  Phys. Rev. Lett.} {\bfseries 99} (2007) 161603},
\href{http://arxiv.org/abs/0705.2336}{{\ttfamily arXiv:0705.2336 [nucl-ex]}}.
%%CITATION = ARXIV:0705.2336;%%.

\bibitem{Serebrov:2007gw}
A.~P. Serebrov {\em et~al.}, ``{Experimental search for neutron: Mirror neutron
  oscillations using storage of ultracold neutrons},''
  \href{http://dx.doi.org/10.1016/j.physletb.2008.04.014}{{\em Phys. Lett.}
  {\bfseries B663} (2008) 181--185},
\href{http://arxiv.org/abs/0706.3600}{{\ttfamily arXiv:0706.3600 [nucl-ex]}}.
%%CITATION = ARXIV:0706.3600;%%.

\bibitem{Altarev:2009tg}
I.~Altarev {\em et~al.}, ``{Neutron to Mirror-Neutron Oscillations in the
  Presence of Mirror Magnetic Fields},''
  \href{http://dx.doi.org/10.1103/PhysRevD.80.032003}{{\em Phys. Rev.}
  {\bfseries D80} (2009) 032003},
\href{http://arxiv.org/abs/0905.4208}{{\ttfamily arXiv:0905.4208 [nucl-ex]}}.
%%CITATION = ARXIV:0905.4208;%%.

\bibitem{Bodek:2009zz}
K.~Bodek {\em et~al.}, ``{Additional results from the first dedicated search
  for neutron-mirror neutron oscillations},''
\href{http://dx.doi.org/10.1016/j.nima.2009.07.047}{{\em Nucl. Instrum. Meth.}
  {\bfseries A611} (2009) 141--143}.
%%CITATION = NUIMA,A611,141;%%.

\bibitem{Serebrov:2008hw}
A.~P. Serebrov {\em et~al.}, ``{Search for neutronmirror neutron oscillations
  in a laboratory experiment with ultracold neutrons},''
  \href{http://dx.doi.org/10.1016/j.nima.2009.07.041}{{\em Nucl. Instrum.
  Meth.} {\bfseries A611} (2009) 137--140},
\href{http://arxiv.org/abs/0809.4902}{{\ttfamily arXiv:0809.4902 [nucl-ex]}}.
%%CITATION = ARXIV:0809.4902;%%.

\bibitem{Berezhiani:2012rq}
Z.~Berezhiani and F.~Nesti, ``{Magnetic anomaly in UCN trapping: signal for
  neutron oscillations to parallel world?},''
  \href{http://dx.doi.org/10.1140/epjc/s10052-012-1974-5}{{\em Eur. Phys. J.}
  {\bfseries C72} (2012) 1974},
\href{http://arxiv.org/abs/1203.1035}{{\ttfamily arXiv:1203.1035 [hep-ph]}}.
%%CITATION = ARXIV:1203.1035;%%.

\bibitem{Berezhiani:2017jkn}
Z.~Berezhiani, R.~Biondi, P.~Geltenbort, I.~A. Krasnoshchekova, V.~E. Varlamov,
  A.~V. Vassiljev, and O.~M. Zherebtsov, ``{New experimental limits on neutron
  - mirror neutron oscillations in the presence of mirror magnetic field},''
  \href{http://dx.doi.org/10.1140/epjc/s10052-018-6189-y}{{\em Eur. Phys. J.}
  {\bfseries C78} no.~9, (2018) 717},
\href{http://arxiv.org/abs/1712.05761}{{\ttfamily arXiv:1712.05761 [hep-ex]}}.
%%CITATION = ARXIV:1712.05761;%%.

\bibitem{Schmidt2007}
U.~Schmidt, ``An experimental limit on neutron mirror-neutron oscillation.''
  presented at Search for Baryon and Lepton number Violations International
  Workshop, 2007.

\bibitem{Berezhiani:2015uya}
Z.~Berezhiani and A.~Vainshtein, ``{Neutron-Antineutron Oscillation as a Signal
  of CP Violation},''
\href{http://arxiv.org/abs/1506.05096}{{\ttfamily arXiv:1506.05096 [hep-ph]}}.
%%CITATION = ARXIV:1506.05096;%%.

\bibitem{Dvali:1999gf}
G.~Dvali and G.~Gabadadze, ``{Nonconservation of global charges in the brane
  universe and baryogenesis},''
  \href{http://dx.doi.org/10.1016/S0370-2693(99)00766-2}{{\em Phys. Lett. B}
  {\bfseries 460} (1999) 47--57},
  \href{http://arxiv.org/abs/hep-ph/9904221}{{\ttfamily arXiv:hep-ph/9904221}}.

\bibitem{Dutta:2005af}
B.~Dutta, Y.~Mimura, and R.~Mohapatra, ``{Observable neutron-antineutron
  oscillation in high scale seesaw models},''
  \href{http://dx.doi.org/10.1103/PhysRevLett.96.061801}{{\em Phys. Rev. Lett.}
  {\bfseries 96} (2006) 061801},
  \href{http://arxiv.org/abs/hep-ph/0510291}{{\ttfamily arXiv:hep-ph/0510291}}.

\bibitem{Dvali:2009ne}
G.~Dvali and M.~Redi, ``{Phenomenology of $10^{32}$ Dark Sectors},''
  \href{http://dx.doi.org/10.1103/PhysRevD.80.055001}{{\em Phys. Rev. D}
  {\bfseries 80} (2009) 055001},
  \href{http://arxiv.org/abs/0905.1709}{{\ttfamily arXiv:0905.1709 [hep-ph]}}.

\bibitem{Abe:2011ky}
{\bfseries Super-Kamiokande} Collaboration, K.~Abe {\em et~al.}, ``{The Search
  for $n-\bar{n}$ oscillation in Super-Kamiokande I},''
  \href{http://dx.doi.org/10.1103/PhysRevD.91.072006}{{\em Phys. Rev.}
  {\bfseries D91} (2015) 072006},
\href{http://arxiv.org/abs/1109.4227}{{\ttfamily arXiv:1109.4227 [hep-ex]}}.
%%CITATION = ARXIV:1109.4227;%%.

\bibitem{Gustafson:2015qyo}
{\bfseries Super-Kamiokande} Collaboration, J.~Gustafson {\em et~al.},
  ``{Search for dinucleon decay into pions at Super-Kamiokande},''
  \href{http://dx.doi.org/10.1103/PhysRevD.91.072009}{{\em Phys. Rev.}
  {\bfseries D91} no.~7, (2015) 072009},
\href{http://arxiv.org/abs/1504.01041}{{\ttfamily arXiv:1504.01041 [hep-ex]}}.
%%CITATION = ARXIV:1504.01041;%%.

\bibitem{Sussman:2018ylo}
{\bfseries Super-Kamiokande} Collaboration, S.~Sussman {\em et~al.},
  ``{Dinucleon and Nucleon Decay to Two-Body Final States with no Hadrons in
  Super-Kamiokande},''
\href{http://arxiv.org/abs/1811.12430}{{\ttfamily arXiv:1811.12430 [hep-ex]}}.
%%CITATION = ARXIV:1811.12430;%%.

\bibitem{Girmohanta:2019cjm}
S.~Girmohanta and R.~Shrock, ``{Improved Upper Limits on Baryon-Number
  Violating Dinucleon Decays to Dileptons},''
  \href{http://dx.doi.org/10.1016/j.physletb.2020.135296}{{\em Phys. Lett. B}
  {\bfseries 803} (2020) 135296},
  \href{http://arxiv.org/abs/1910.08356}{{\ttfamily arXiv:1910.08356
  [hep-ph]}}.

\bibitem{Weinberg:1979bt}
S.~Weinberg, ``{Cosmological Production of Baryons},''
\href{http://dx.doi.org/10.1103/PhysRevLett.42.850}{{\em Phys. Rev. Lett.}
  {\bfseries 42} (1979) 850--853}.
%%CITATION = PRLTA,42,850;%%.

\bibitem{Fry:1980bd}
J.~N. Fry, K.~A. Olive, and M.~S. Turner, ``{Hierarchy of Cosmological Baryon
  Generation},''
\href{http://dx.doi.org/10.1103/PhysRevLett.45.2074}{{\em Phys. Rev. Lett.}
  {\bfseries 45} (1980) 2074}.
%%CITATION = PRLTA,45,2074;%%.

\bibitem{Yoshimura:1978ex}
M.~Yoshimura, ``{Unified Gauge Theories and the Baryon Number of the
  Universe},'' \href{http://dx.doi.org/10.1103/PhysRevLett.42.746,
  10.1103/PhysRevLett.41.281}{{\em Phys. Rev. Lett.} {\bfseries 41} (1978)
  281--284}.
[Erratum: Phys. Rev. Lett.42,746(1979)].
%%CITATION = PRLTA,41,281;%%.

\bibitem{Ellis:1978xg}
J.~R. Ellis, M.~K. Gaillard, and D.~V. Nanopoulos, ``{Baryon Number Generation
  in Grand Unified Theories},''
  \href{http://dx.doi.org/10.1016/0370-2693(79)91190-0}{{\em Phys. Lett.}
  {\bfseries 80B} (1979) 360}.
[Erratum: Phys. Lett.82B,464(1979)].
%%CITATION = PHLTA,80B,360;%%.

\bibitem{Morrissey:2012db}
D.~E. Morrissey and M.~J. Ramsey-Musolf, ``{Electroweak baryogenesis},''
  \href{http://dx.doi.org/10.1088/1367-2630/14/12/125003}{{\em New J. Phys.}
  {\bfseries 14} (2012) 125003},
\href{http://arxiv.org/abs/1206.2942}{{\ttfamily arXiv:1206.2942 [hep-ph]}}.
%%CITATION = ARXIV:1206.2942;%%.

\bibitem{Fukugita:1986hr}
M.~Fukugita and T.~Yanagida, ``{Baryogenesis Without Grand Unification},''
\href{http://dx.doi.org/10.1016/0370-2693(86)91126-3}{{\em Phys. Lett.}
  {\bfseries B174} (1986) 45--47}.
%%CITATION = PHLTA,B174,45;%%.

\bibitem{Weinberg:1979sa}
S.~Weinberg, ``{Baryon and Lepton Nonconserving Processes},''
\href{http://dx.doi.org/10.1103/PhysRevLett.43.1566}{{\em Phys. Rev. Lett.}
  {\bfseries 43} (1979) 1566--1570}.
%%CITATION = PRLTA,43,1566;%%.

\bibitem{Mohapatra:1979ia}
R.~N. Mohapatra and G.~Senjanovic, ``{Neutrino Mass and Spontaneous Parity
  Nonconservation},'' \href{http://dx.doi.org/10.1103/PhysRevLett.44.912}{{\em
  Phys. Rev. Lett.} {\bfseries 44} (1980) 912}.

\bibitem{Kuzmin:1985mm}
V.~A. Kuzmin, V.~A. Rubakov, and M.~E. Shaposhnikov, ``{On the Anomalous
  Electroweak Baryon Number Nonconservation in the Early Universe},''
\href{http://dx.doi.org/10.1016/0370-2693(85)91028-7}{{\em Phys. Lett.}
  {\bfseries 155B} (1985) 36}.
%%CITATION = PHLTA,155B,36;%%.

\bibitem{Asaka:2005pn}
T.~Asaka and M.~Shaposhnikov, ``{The $\nu$MSM, dark matter and baryon asymmetry
  of the universe},''
  \href{http://dx.doi.org/10.1016/j.physletb.2005.06.020}{{\em Phys. Lett. B}
  {\bfseries 620} (2005) 17--26},
  \href{http://arxiv.org/abs/hep-ph/0505013}{{\ttfamily arXiv:hep-ph/0505013}}.

\bibitem{Asaka:2005an}
T.~Asaka, S.~Blanchet, and M.~Shaposhnikov, ``{The nuMSM, dark matter and
  neutrino masses},''
  \href{http://dx.doi.org/10.1016/j.physletb.2005.09.070}{{\em Phys. Lett. B}
  {\bfseries 631} (2005) 151--156},
  \href{http://arxiv.org/abs/hep-ph/0503065}{{\ttfamily arXiv:hep-ph/0503065}}.

\bibitem{Bento:2001rc}
L.~Bento and Z.~Berezhiani, ``{Leptogenesis via collisions: The Lepton number
  leaking to the hidden sector},''
  \href{http://dx.doi.org/10.1103/PhysRevLett.87.231304}{{\em Phys. Rev. Lett.}
  {\bfseries 87} (2001) 231304},
\href{http://arxiv.org/abs/hep-ph/0107281}{{\ttfamily arXiv:hep-ph/0107281
  [hep-ph]}}.
%%CITATION = HEP-PH/0107281;%%.

\bibitem{Bento:2002sj}
L.~Bento and Z.~Berezhiani, ``{Baryon asymmetry, dark matter and the hidden
  sector},''
\href{http://dx.doi.org/10.1002/9783527610853.ch8}{{\em Fortsch. Phys.}
  {\bfseries 50} (2002) 489--495}.
%%CITATION = FPYKA,50,489;%%.

\bibitem{Berezhiani:2008zza}
Z.~Berezhiani, ``{Unified picture of ordinary and dark matter genesis},''
\href{http://dx.doi.org/10.1140/epjst/e2008-00824-6}{{\em Eur. Phys. J. ST}
  {\bfseries 163} (2008) 271--289}.
%%CITATION = 00619,163,271;%%.

\bibitem{Babu:2008rq}
K.~Babu, P.~Bhupal~Dev, and R.~Mohapatra, ``{Neutrino mass hierarchy,
  neutron-antineutron oscillation from baryogenesis},''
  \href{http://dx.doi.org/10.1103/PhysRevD.79.015017}{{\em Phys. Rev. D}
  {\bfseries 79} (2009) 015017},
  \href{http://arxiv.org/abs/0811.3411}{{\ttfamily arXiv:0811.3411 [hep-ph]}}.

\bibitem{Babu:2013yca}
K.~S. Babu, P.~S. Bhupal~Dev, E.~C. F.~S. Fortes, and R.~N. Mohapatra,
  ``{Post-Sphaleron Baryogenesis and an Upper Limit on the Neutron-Antineutron
  Oscillation Time},'' \href{http://dx.doi.org/10.1103/PhysRevD.87.115019}{{\em
  Phys. Rev.} {\bfseries D87} no.~11, (2013) 115019},
\href{http://arxiv.org/abs/1303.6918}{{\ttfamily arXiv:1303.6918 [hep-ph]}}.
%%CITATION = ARXIV:1303.6918;%%.

\bibitem{Berezhiani:2018zvs}
Z.~Berezhiani, ``{Matter, dark matter, and antimatter in our Universe},''
\href{http://dx.doi.org/10.1142/S0217751X18440347}{{\em Int. J. Mod. Phys.}
  {\bfseries A33} no.~31, (2018) 1844034}.
%%CITATION = IMPAE,A33,1844034;%%.

\bibitem{Feng:2010gw}
J.~L. Feng, ``{Dark Matter Candidates from Particle Physics and Methods of
  Detection},''
  \href{http://dx.doi.org/10.1146/annurev-astro-082708-101659}{{\em Ann. Rev.
  Astron. Astrophys.} {\bfseries 48} (2010) 495--545},
  \href{http://arxiv.org/abs/1003.0904}{{\ttfamily arXiv:1003.0904
  [astro-ph.CO]}}.

\bibitem{Battaglieri:2017aum}
M.~Battaglieri {\em et~al.}, ``{US Cosmic Visions: New Ideas in Dark Matter
  2017: Community Report},'' in {\em {U.S. Cosmic Visions: New Ideas in Dark
  Matter}}.
\newblock 7, 2017.
\newblock \href{http://arxiv.org/abs/1707.04591}{{\ttfamily arXiv:1707.04591
  [hep-ph]}}.

\bibitem{Berezhiani:1995am}
Z.~G. Berezhiani, A.~D. Dolgov, and R.~N. Mohapatra, ``{Asymmetric inflationary
  reheating and the nature of mirror universe},''
  \href{http://dx.doi.org/10.1016/0370-2693(96)00219-5}{{\em Phys. Lett.}
  {\bfseries B375} (1996) 26--36},
\href{http://arxiv.org/abs/hep-ph/9511221}{{\ttfamily arXiv:hep-ph/9511221
  [hep-ph]}}.
%%CITATION = HEP-PH/9511221;%%.

\bibitem{Mohapatra:2000qx}
R.~N. Mohapatra and V.~L. Teplitz, ``{Mirror dark matter and galaxy core
  densities of galaxies},''
  \href{http://dx.doi.org/10.1103/PhysRevD.62.063506}{{\em Phys. Rev.}
  {\bfseries D62} (2000) 063506},
\href{http://arxiv.org/abs/astro-ph/0001362}{{\ttfamily arXiv:astro-ph/0001362
  [astro-ph]}}.
%%CITATION = ASTRO-PH/0001362;%%.

\bibitem{Berezhiani:2000gw}
Z.~Berezhiani, D.~Comelli, and F.~L. Villante, ``{The Early mirror universe:
  Inflation, baryogenesis, nucleosynthesis and dark matter},''
  \href{http://dx.doi.org/10.1016/S0370-2693(01)00217-9}{{\em Phys. Lett.}
  {\bfseries B503} (2001) 362--375},
\href{http://arxiv.org/abs/hep-ph/0008105}{{\ttfamily arXiv:hep-ph/0008105
  [hep-ph]}}.
%%CITATION = HEP-PH/0008105;%%.

\bibitem{Foot:2014mia}
R.~Foot, ``{Mirror dark matter: Cosmology, galaxy structure and direct
  detection},'' \href{http://dx.doi.org/10.1142/S0217751X14300130}{{\em Int. J.
  Mod. Phys.} {\bfseries A29} (2014) 1430013},
\href{http://arxiv.org/abs/1401.3965}{{\ttfamily arXiv:1401.3965
  [astro-ph.CO]}}.
%%CITATION = ARXIV:1401.3965;%%.

\bibitem{Essig:2013lka}
R.~Essig {\em et~al.}, ``{Working Group Report: New Light Weakly Coupled
  Particles},'' in {\em {Proceedings, 2013 Community Summer Study on the Future
  of U.S. Particle Physics: Snowmass on the Mississippi (CSS2013): Minneapolis,
  MN, USA, July 29-August 6, 2013}}.
\newblock 2013.
\newblock
\href{http://arxiv.org/abs/1311.0029}{{\ttfamily arXiv:1311.0029 [hep-ph]}}.
\newblock
%%CITATION = ARXIV:1311.0029;%%.

\bibitem{Holdom:1985ag}
B.~Holdom, ``{Two U(1)'s and Epsilon Charge Shifts},''
\href{http://dx.doi.org/10.1016/0370-2693(86)91377-8}{{\em Phys. Lett.}
  {\bfseries 166B} (1986) 196--198}.
%%CITATION = PHLTA,166B,196;%%.

\bibitem{Berezhiani:1995yi}
Z.~G. Berezhiani and R.~N. Mohapatra, ``{Reconciling present neutrino puzzles:
  Sterile neutrinos as mirror neutrinos},''
  \href{http://dx.doi.org/10.1103/PhysRevD.52.6607}{{\em Phys. Rev.} {\bfseries
  D52} (1995) 6607--6611},
\href{http://arxiv.org/abs/hep-ph/9505385}{{\ttfamily arXiv:hep-ph/9505385
  [hep-ph]}}.
%%CITATION = HEP-PH/9505385;%%.

\bibitem{Berezhiani:1996sz}
Z.~G. Berezhiani, ``{Astrophysical implications of the mirror world with broken
  mirror parity},'' {\em Acta Phys. Polon.} {\bfseries B27} (1996) 1503--1516,
\href{http://arxiv.org/abs/hep-ph/9602326}{{\ttfamily arXiv:hep-ph/9602326
  [hep-ph]}}.
%%CITATION = HEP-PH/9602326;%%.

\bibitem{Berezhiani:2003xm}
Z.~Berezhiani, ``{Mirror world and its cosmological consequences},''
  \href{http://dx.doi.org/10.1142/S0217751X04020075}{{\em Int. J. Mod. Phys.}
  {\bfseries A19} (2004) 3775--3806},
\href{http://arxiv.org/abs/hep-ph/0312335}{{\ttfamily arXiv:hep-ph/0312335
  [hep-ph]}}.
%%CITATION = HEP-PH/0312335;%%.

\bibitem{Berezhiani:2005ek}
Z.~Berezhiani, ``{Through the looking-glass: Alice's adventures in mirror
  world},''
\href{http://arxiv.org/abs/hep-ph/0508233}{{\ttfamily arXiv:hep-ph/0508233
  [hep-ph]}}.
%%CITATION = HEP-PH/0508233;%%.

\bibitem{Okun:2006eb}
L.~B. Okun, ``{Mirror particles and mirror matter: 50 years of speculations and
  search},'' \href{http://dx.doi.org/10.1070/PU2007v050n04ABEH006227}{{\em
  Phys. Usp.} {\bfseries 50} (2007) 380--389},
\href{http://arxiv.org/abs/hep-ph/0606202}{{\ttfamily arXiv:hep-ph/0606202
  [hep-ph]}}.
%%CITATION = HEP-PH/0606202;%%.

\bibitem{Berezhiani:2018eds}
Z.~Berezhiani, ``{Neutron lifetime puzzle and neutron–mirror neutron
  oscillation},'' \href{http://dx.doi.org/10.1140/epjc/s10052-019-6995-x}{{\em
  Eur. Phys. J.} {\bfseries C79} no.~6, (2019) 484},
\href{http://arxiv.org/abs/1807.07906}{{\ttfamily arXiv:1807.07906 [hep-ph]}}.
%%CITATION = ARXIV:1807.07906;%%.

\bibitem{Berezhiani:2018qqw}
Z.~Berezhiani, R.~Biondi, Y.~Kamyshkov, and L.~Varriano, ``{On the Neutron
  Transition Magnetic Moment},''
  \href{http://dx.doi.org/10.3390/physics1020021}{{\em MDPI Physics} {\bfseries
  1} no.~2, (2019) 271--289},
\href{http://arxiv.org/abs/1812.11141}{{\ttfamily arXiv:1812.11141 [nucl-th]}}.
%%CITATION = ARXIV:1812.11141;%%.

\bibitem{nlife:2018wfe}
F.~E. Wietfeldt, ``{Measurements of the Neutron Lifetime},''
  \href{http://dx.doi.org/10.3390/atoms6040070}{{\em Atoms} {\bfseries 6(4)}
  (2018) 70}. \url{https://www.mdpi.com/2218-2004/6/4/70}.

\bibitem{Georgi:1974sy}
H.~Georgi and S.~L. Glashow, ``{Unity of All Elementary Particle Forces},''
\href{http://dx.doi.org/10.1103/PhysRevLett.32.438}{{\em Phys. Rev. Lett.}
  {\bfseries 32} (1974) 438--441}.
%%CITATION = PRLTA,32,438;%%.

\bibitem{Glashow:1979nm}
S.~Glashow, ``{The Future of Elementary Particle Physics},''
  \href{http://dx.doi.org/10.1007/978-1-4684-7197-7\_15}{{\em NATO Sci. Ser. B}
  {\bfseries 61} (1980) 687}.

\bibitem{Dolgov:2006ay}
A.~Dolgov and F.~Urban, ``{Baryogenesis by $R$-parity violating top quark
  decays and neutron-antineutron oscillations},''
  \href{http://dx.doi.org/10.1016/j.nuclphysb.2006.06.035}{{\em Nucl. Phys. B}
  {\bfseries 752} (2006) 297--315},
  \href{http://arxiv.org/abs/hep-ph/0605263}{{\ttfamily arXiv:hep-ph/0605263}}.

\bibitem{Bambi:2006mi}
C.~Bambi, A.~Dolgov, and K.~Freese, ``{A Black Hole Conjecture and Rare Decays
  in Theories with Low Scale Gravity},''
  \href{http://dx.doi.org/10.1016/j.nuclphysb.2006.11.010}{{\em Nucl. Phys. B}
  {\bfseries 763} (2007) 91--114},
  \href{http://arxiv.org/abs/hep-ph/0606321}{{\ttfamily arXiv:hep-ph/0606321}}.

\bibitem{Schwarz:1982jn}
J.~H. Schwarz, ``{Superstring Theory},''
\href{http://dx.doi.org/10.1016/0370-1573(82)90087-4}{{\em Phys. Rept.}
  {\bfseries 89} (1982) 223--322}.
%%CITATION = PRPLC,89,223;%%.

\bibitem{Senjanovic:1975rk}
G.~Senjanovic and R.~N. Mohapatra, ``{Exact Left-Right Symmetry and Spontaneous
  Violation of Parity},''
  \href{http://dx.doi.org/10.1103/PhysRevD.12.1502}{{\em Phys. Rev. D}
  {\bfseries 12} (1975) 1502}.

\bibitem{Senjanovic:1978ev}
G.~Senjanovic,
  \href{http://dx.doi.org/10.1016/0550-3213(79)90604-7}{``{Spontaneous
  Breakdown of Parity in a Class of Gauge Theories},''} other thesis, 1979.

\bibitem{Mohapatra:1980yp}
R.~N. Mohapatra and G.~Senjanovic, ``{Neutrino Masses and Mixings in Gauge
  Models with Spontaneous Parity Violation},''
  \href{http://dx.doi.org/10.1103/PhysRevD.23.165}{{\em Phys. Rev. D}
  {\bfseries 23} (1981) 165}.

\bibitem{deGouvea:2014lva}
A.~de~Gouvea, J.~Herrero-Garcia, and A.~Kobach, ``{Neutrino Masses, Grand
  Unification, and Baryon Number Violation},''
  \href{http://dx.doi.org/10.1103/PhysRevD.90.016011}{{\em Phys. Rev. D}
  {\bfseries 90} no.~1, (2014) 016011},
  \href{http://arxiv.org/abs/1404.4057}{{\ttfamily arXiv:1404.4057 [hep-ph]}}.

\bibitem{Keung:1983uu}
W.-Y. Keung and G.~Senjanovic, ``{Majorana Neutrinos and the Production of the
  Right-handed Charged Gauge Boson},''
  \href{http://dx.doi.org/10.1103/PhysRevLett.50.1427}{{\em Phys. Rev. Lett.}
  {\bfseries 50} (1983) 1427}.

\bibitem{Tello:2010am}
V.~Tello, M.~Nemevsek, F.~Nesti, G.~Senjanovic, and F.~Vissani, ``{Left-Right
  Symmetry: from LHC to Neutrinoless Double Beta Decay},''
  \href{http://dx.doi.org/10.1103/PhysRevLett.106.151801}{{\em Phys. Rev.
  Lett.} {\bfseries 106} (2011) 151801},
  \href{http://arxiv.org/abs/1011.3522}{{\ttfamily arXiv:1011.3522 [hep-ph]}}.

\bibitem{Babu:2014tra}
K.~S. Babu and R.~N. Mohapatra, ``{Determining Majorana Nature of Neutrino from
  Nucleon Decays and $n-\bar{n}$ oscillations},''
  \href{http://dx.doi.org/10.1103/PhysRevD.91.013008}{{\em Phys. Rev.}
  {\bfseries D91} no.~1, (2015) 013008},
\href{http://arxiv.org/abs/1408.0803}{{\ttfamily arXiv:1408.0803 [hep-ph]}}.
%%CITATION = ARXIV:1408.0803;%%.

\bibitem{Berezhiani:2018xsx}
Z.~Berezhiani and A.~Vainshtein, ``{Neutron--Antineutron Oscillations: Discrete
  Symmetries and Quark Operators},''
  \href{http://dx.doi.org/10.1016/j.physletb.2018.11.014}{{\em Phys. Lett.}
  {\bfseries B788} (2019) 58--64},
\href{http://arxiv.org/abs/1809.00997}{{\ttfamily arXiv:1809.00997 [hep-ph]}}.
%%CITATION = ARXIV:1809.00997;%%.

\bibitem{Berezhiani:2018pcp}
Z.~Berezhiani and A.~Vainshtein, ``{Neutron–antineutron oscillation and
  discrete symmetries},''
\href{http://dx.doi.org/10.1142/S0217751X18440165}{{\em Int. J. Mod. Phys.}
  {\bfseries A33} no.~31, (2018) 1844016}.
%%CITATION = IMPAE,A33,1844016;%%.

\bibitem{Addazi:2016rgo}
A.~Addazi, Z.~Berezhiani, and Y.~Kamyshkov, ``{Gauged $B-L$ number and
  neutron–antineutron oscillation: long-range forces mediated by
  baryophotons},'' \href{http://dx.doi.org/10.1140/epjc/s10052-017-4870-1}{{\em
  Eur. Phys. J.} {\bfseries C77} no.~5, (2017) 301},
\href{http://arxiv.org/abs/1607.00348}{{\ttfamily arXiv:1607.00348 [hep-ph]}}.
%%CITATION = ARXIV:1607.00348;%%.

\bibitem{Babu:2016rwa}
K.~Babu and R.~N. Mohapatra, ``{Limiting Equivalence Principle Violation and
  Long-Range Baryonic Force from Neutron-Antineutron Oscillation},''
  \href{http://dx.doi.org/10.1103/PhysRevD.94.054034}{{\em Phys. Rev. D}
  {\bfseries 94} no.~5, (2016) 054034},
  \href{http://arxiv.org/abs/1606.08374}{{\ttfamily arXiv:1606.08374
  [hep-ph]}}.

\bibitem{BITTER1985461}
T.~Bitter and D.~Dubbers, ``Test of the quasifree condition in neutron
  oscillation experiments,''
  \href{http://dx.doi.org/https://doi.org/10.1016/0168-9002(85)90024-5}{{\em
  Nuclear Instruments and Methods in Physics Research Section A: Accelerators,
  Spectrometers, Detectors and Associated Equipment} {\bfseries 239} no.~3,
  (1985) 461 -- 466}.
  \url{http://www.sciencedirect.com/science/article/pii/0168900285900245}.

\bibitem{SCHMIDT1992569}
U.~Schmidt, T.~Bitter, P.~El-Muzeini, D.~Dubbers, and O.~Scharpf, ``Long
  distance propagation of a polarized neutron beam in zero magnetic field,''
  \href{http://dx.doi.org/https://doi.org/10.1016/0168-9002(92)90952-Z}{{\em
  Nuclear Instruments and Methods in Physics Research Section A: Accelerators,
  Spectrometers, Detectors and Associated Equipment} {\bfseries 320} no.~3,
  (1992) 569 -- 573}.
  \url{http://www.sciencedirect.com/science/article/pii/016890029290952Z}.

\bibitem{Davis:2016uyk}
E.~D. Davis and A.~R. Young, ``{Neutron-antineutron oscillations beyond the
  quasifree limit},'' \href{http://dx.doi.org/10.1103/PhysRevD.95.036004}{{\em
  Phys. Rev.} {\bfseries D95} no.~3, (2017) 036004},
\href{http://arxiv.org/abs/1611.04205}{{\ttfamily arXiv:1611.04205 [nucl-ex]}}.
%%CITATION = ARXIV:1611.04205;%%.

\bibitem{Bressi:1989zd}
G.~Bressi {\em et~al.}, ``{Search for Free Neutron antineutron Oscillations},''
\href{http://dx.doi.org/10.1007/BF01588203}{{\em Z. Phys.} {\bfseries C43}
  (1989) 175--179}.
%%CITATION = ZEPYA,C43,175;%%.

\bibitem{Bressi:1990zx}
G.~Bressi {\em et~al.}, ``{Final results of a search for free neutron
  antineutron oscillations},''
\href{http://dx.doi.org/10.1007/BF02789025}{{\em Nuovo Cim.} {\bfseries A103}
  (1990) 731--750}.
%%CITATION = NUCIA,A103,731;%%.

\bibitem{Fidecaro:1985cm}
{\bfseries CERN-Grenoble-Padua-Rutherford-Sussex} Collaboration, G.~Fidecaro
  {\em et~al.}, ``{Experimental search for neutron antineutron transitions with
  free neutrons},''
\href{http://dx.doi.org/10.1016/0370-2693(85)91367-X}{{\em Phys. Lett.}
  {\bfseries 156B} (1985) 122--128}.
%%CITATION = PHLTA,156B,122;%%.

\bibitem{Costa:1983wc}
G.~Costa and P.~Kabir, ``{Environmental Effects On Possible Neutron-Antineutron
  Transitions},'' \href{http://dx.doi.org/10.1103/PhysRevD.28.667}{{\em Phys.
  Rev. D} {\bfseries 28} (1983) 667--668}.

\bibitem{Kerbikov:2017spv}
B.~Kerbikov, ``{Lindblad and Bloch equations for conversion of a neutron into
  an antineutron},''
  \href{http://dx.doi.org/10.1016/j.nuclphysa.2018.04.006}{{\em Nucl. Phys. A}
  {\bfseries 975} (2018) 59--72},
  \href{http://arxiv.org/abs/1704.07117}{{\ttfamily arXiv:1704.07117
  [hep-ph]}}.

\bibitem{Gudkov:2019gro}
V.~Gudkov, V.~Nesvizhevsky, K.~Protasov, W.~Snow, and A.~Voronin, ``{A new
  approach to search for free neutron-antineutron oscillations using coherent
  neutron propagation in gas},''
  \href{http://arxiv.org/abs/1912.06730}{{\ttfamily arXiv:1912.06730
  [hep-ph]}}.

\bibitem{Homestake}
M.~l. Cherry, K.~Lande, C.~k. Lee, R.~i. Steinberg, and B.~T. Cleveland,
  ``{Experimental test of baryon conservation: a new limit on neutron
  antineutron oscillations in oxygen},''
\href{http://dx.doi.org/10.1103/PhysRevLett.50.1354}{{\em Phys. Rev. Lett.}
  {\bfseries 50} (1983) 1354--1356}.
%%CITATION = PRLTA,50,1354;%%.

\bibitem{KGF}
M.~R. Krishnaswamy, M.~G.~K. Menon, N.~K. Mondal, V.~S. Narasimham, B.~V.
  Sreekantan, Y.~Hayashi, N.~Ito, S.~Kawakami, and S.~Miyake, ``{Results From
  the Kgf Proton Decay Experiment},''
  \href{http://dx.doi.org/10.1007/BF02514839}{{\em Nuovo Cim.} {\bfseries C9}
  (1986) 167--181}.
[Conf. Proc.C850418,97(1985)].
%%CITATION = NUCIA,C9,167;%%.

\bibitem{NUSEX}
G.~Battistoni {\em et~al.}, ``{Nucleon Stability, Magnetic Monopoles and
  Atmospheric Neutrinos in the Mont Blanc Experiment},''
\href{http://dx.doi.org/10.1016/0370-2693(83)90827-4}{{\em Phys. Lett.}
  {\bfseries 133B} (1983) 454--460}.
%%CITATION = PHLTA,133B,454;%%.

\bibitem{IMB}
{\bfseries Irvine-Michigan-Brookhaven} Collaboration, T.~W. Jones {\em et~al.},
  ``{A Search for $N \bar{N}$ Oscillation in Oxygen},''
\href{http://dx.doi.org/10.1103/PhysRevLett.52.720}{{\em Phys. Rev. Lett.}
  {\bfseries 52} (1984) 720}.
%%CITATION = PRLTA,52,720;%%.

\bibitem{Kamiokande}
{\bfseries Kamiokande} Collaboration, M.~Takita {\em et~al.}, ``{A Search for
  Neutron - antineutron Oscillation in a $^{16}$O Nucleus},''
\href{http://dx.doi.org/10.1103/PhysRevD.34.902}{{\em Phys. Rev.} {\bfseries
  D34} (1986) 902}.
%%CITATION = PHRVA,D34,902;%%.

\bibitem{Frejus}
{\bfseries Frejus} Collaboration, C.~Berger {\em et~al.}, ``{Search for Neutron
  - antineutron Oscillations in the Frejus Detector},''
\href{http://dx.doi.org/10.1016/0370-2693(90)90441-8}{{\em Phys. Lett.}
  {\bfseries B240} (1990) 237--242}.
%%CITATION = PHLTA,B240,237;%%.

\bibitem{Soudan-2}
J.~Chung {\em et~al.}, ``{Search for neutron antineutron oscillations using
  multiprong events in Soudan 2},''
  \href{http://dx.doi.org/10.1103/PhysRevD.66.032004}{{\em Phys. Rev.}
  {\bfseries D66} (2002) 032004},
\href{http://arxiv.org/abs/hep-ex/0205093}{{\ttfamily arXiv:hep-ex/0205093
  [hep-ex]}}.
%%CITATION = HEP-EX/0205093;%%.

\bibitem{Aharmim:2017jna}
{\bfseries SNO} Collaboration, B.~Aharmim {\em et~al.}, ``{Search for
  neutron-antineutron oscillations at the Sudbury Neutrino Observatory},''
  \href{http://dx.doi.org/10.1103/PhysRevD.96.092005}{{\em Phys. Rev.}
  {\bfseries D96} no.~9, (2017) 092005},
\href{http://arxiv.org/abs/1705.00696}{{\ttfamily arXiv:1705.00696 [hep-ex]}}.
%%CITATION = ARXIV:1705.00696;%%.

\bibitem{Abi:2020evt}
{\bfseries DUNE} Collaboration, B.~Abi {\em et~al.}, ``{Deep Underground
  Neutrino Experiment (DUNE), Far Detector Technical Design Report, Volume II
  DUNE Physics},'' \href{http://arxiv.org/abs/2002.03005}{{\ttfamily
  arXiv:2002.03005 [hep-ex]}}.

\bibitem{Barrow:2019viz}
J.~L. Barrow, E.~S. Golubeva, E.~Paryev, and J.-M. Richard, ``{Progress and
  simulations for intranuclear neutron-antineutron transformations in
  ${}^{40}_{18} Ar$},''
  \href{http://dx.doi.org/10.1103/PhysRevD.101.036008}{{\em Phys. Rev. D}
  {\bfseries 101} no.~3, (2020) 036008},
  \href{http://arxiv.org/abs/1906.02833}{{\ttfamily arXiv:1906.02833
  [hep-ex]}}.

\bibitem{Hewes:2017xtr}
J.~E.~T. Hewes, \href{http://dx.doi.org/10.2172/1426674}{{\em {Searches for
  Bound Neutron-Antineutron Oscillation in Liquid Argon Time Projection
  Chambers}}}.
\newblock PhD thesis, Manchester U., 2017.
\newblock
\url{http://lss.fnal.gov/archive/thesis/2000/fermilab-thesis-2017-27.pdf}.
\newblock
%%CITATION = FERMILAB-THESIS-2017-27;%%.

\bibitem{Labarga:2018owv}
{\bfseries Hyper-Kamiokande-proto} Collaboration, L.~Labarga,
  \href{http://dx.doi.org/10.22323/1.314.0117}{``Potential of hyper-kamiokande
  at some non-accelerator physics and nucleon decay searches,''} in {\em
  Potential of Hyper-Kamiokande at some non-Accelerator Physics and Nucleon
  Decay Searches}, vol.~EPS-HEP2017, p.~117.
\newblock SISSA, 2018.

\bibitem{Dover:1982wv}
C.~B. Dover, A.~Gal, and J.~M. Richard, ``{Neutron antineutron oscillations in
  nuclei},''
\href{http://dx.doi.org/10.1103/PhysRevD.27.1090}{{\em Phys. Rev.} {\bfseries
  D27} (1983) 1090--1100}.
%%CITATION = PHRVA,D27,1090;%%.

\bibitem{Alberico:1982nu}
W.~Alberico, A.~Bottino, and A.~Molinari, ``{A New Evaluation Of The
  Neutron-Antineutron Oscillation Time},''
  \href{http://dx.doi.org/10.1016/0370-2693(82)90493-2}{{\em Phys. Lett. B}
  {\bfseries 114} (1982) 266--270}.

\bibitem{Alberico:1984wk}
W.~Alberico, J.~Bernabeu, A.~Bottino, and A.~Molinari, ``{Neutron-Antineutron
  Mixing Inside Nuclei},''
  \href{http://dx.doi.org/10.1016/0375-9474(84)90691-2}{{\em Nucl. Phys. A}
  {\bfseries 429} (1984) 445--461}.

\bibitem{Dover:1989zz}
C.~B. Dover, A.~Gal, and J.~M. Richard, ``{Neutron antineutron Oscillations in
  Nuclei},''
\href{http://dx.doi.org/10.1016/0168-9002(89)90239-8}{{\em Nucl. Instrum.
  Meth.} {\bfseries A284} (1989) 13}.
%%CITATION = NUIMA,A284,13;%%.

\bibitem{Alberico:1990ij}
W.~M. Alberico, A.~De~Pace, and M.~Pignone, ``{Neutron - antineutron
  oscillations in nuclei},''
\href{http://dx.doi.org/10.1016/0375-9474(91)90032-2}{{\em Nucl. Phys.}
  {\bfseries A523} (1991) 488--498}.
%%CITATION = NUPHA,A523,488;%%.

\bibitem{Hufner:1998gu}
J.~Hufner and B.~Z. Kopeliovich, ``{Neutron - antineutron oscillations in
  nuclei revisited},'' \href{http://dx.doi.org/10.1142/S0217732398002540}{{\em
  Mod. Phys. Lett. A} {\bfseries 13} (1998) 2385--2392},
  \href{http://arxiv.org/abs/hep-ph/9807210}{{\ttfamily arXiv:hep-ph/9807210}}.

\bibitem{Friedman:2008es}
E.~Friedman and A.~Gal, ``{Realistic calculations of nuclear disappearance
  lifetimes induced by neutron-antineutron oscillations},''
  \href{http://dx.doi.org/10.1103/PhysRevD.78.016002}{{\em Phys. Rev. D}
  {\bfseries 78} (2008) 016002},
  \href{http://arxiv.org/abs/0803.3696}{{\ttfamily arXiv:0803.3696 [hep-ph]}}.

\bibitem{Nussinov:2020wri}
S.~Nussinov and R.~Shrock, ``{Using $\bar p p$ and $e^+e^-$ Annihilation Data
  to Refine Bounds on the Baryon-Number-Violating Dinucleon Decays $nn \to
  e^+e^-$ and $nn \to \mu^+\mu^-$},''
  \href{http://arxiv.org/abs/2005.12493}{{\ttfamily arXiv:2005.12493
  [hep-ph]}}.

\bibitem{Oosterhof:2019dlo}
F.~Oosterhof, B.~Long, J.~de~Vries, R.~Timmermans, and U.~van Kolck,
  ``{Baryon-number violation by two units and the deuteron lifetime},''
  \href{http://dx.doi.org/10.1103/PhysRevLett.122.172501}{{\em Phys. Rev.
  Lett.} {\bfseries 122} no.~17, (2019) 172501},
  \href{http://arxiv.org/abs/1902.05342}{{\ttfamily arXiv:1902.05342
  [hep-ph]}}.

\bibitem{Haidenbauer:2019fyd}
J.~Haidenbauer and U.-G. Meißner, ``{Neutron-antineutron oscillations in the
  deuteron studied with $NN$ and $\bar NN$ interactions based on chiral
  effective field theory},''
  \href{http://dx.doi.org/10.1088/1674-1137/44/3/033101}{{\em Chin. Phys. C}
  {\bfseries 44} no.~3, (2020) 033101},
  \href{http://arxiv.org/abs/1910.14423}{{\ttfamily arXiv:1910.14423
  [hep-ph]}}.

\bibitem{Wagner:2012ui}
T.~A. Wagner, S.~Schlamminger, J.~H. Gundlach, and E.~G. Adelberger,
  ``{Torsion-balance tests of the weak equivalence principle},''
  \href{http://dx.doi.org/10.1088/0264-9381/29/18/184002}{{\em Class. Quant.
  Grav.} {\bfseries 29} (2012) 184002},
\href{http://arxiv.org/abs/1207.2442}{{\ttfamily arXiv:1207.2442 [gr-qc]}}.
%%CITATION = ARXIV:1207.2442;%%.

\bibitem{WagmanPrivComm}
M.~Wagman, ``{Private communication},''  (2020) .

\bibitem{Rinaldi:2018osy}
E.~Rinaldi, S.~Syritsyn, M.~L. Wagman, M.~I. Buchoff, C.~Schroeder, and
  J.~Wasem, ``{Neutron-antineutron oscillations from lattice QCD},''
  \href{http://dx.doi.org/10.1103/PhysRevLett.122.162001}{{\em Phys. Rev.
  Lett.} {\bfseries 122} no.~16, (2019) 162001},
  \href{http://arxiv.org/abs/1809.00246}{{\ttfamily arXiv:1809.00246
  [hep-lat]}}.

\bibitem{Rinaldi:2019thf}
E.~Rinaldi, S.~Syritsyn, M.~L. Wagman, M.~I. Buchoff, C.~Schroeder, and
  J.~Wasem, ``{Lattice QCD determination of neutron-antineutron matrix elements
  with physical quark masses},''
  \href{http://dx.doi.org/10.1103/PhysRevD.99.074510}{{\em Phys. Rev. D}
  {\bfseries 99} no.~7, (2019) 074510},
  \href{http://arxiv.org/abs/1901.07519}{{\ttfamily arXiv:1901.07519
  [hep-lat]}}.

\bibitem{Buchoff:2015qwa}
M.~I. Buchoff and M.~Wagman, ``{Perturbative Renormalization of
  Neutron-Antineutron Operators},''
  \href{http://dx.doi.org/10.1103/PhysRevD.93.016005}{{\em Phys. Rev. D}
  {\bfseries 93} no.~1, (2016) 016005},
  \href{http://arxiv.org/abs/1506.00647}{{\ttfamily arXiv:1506.00647
  [hep-ph]}}. [Erratum: Phys.Rev.D 98, 079901 (2018)].

\bibitem{Buchoff:2015wwa}
M.~I. Buchoff and M.~Wagman, ``{Neutron-Antineutron Operator
  Renormalization},'' \href{http://dx.doi.org/10.22323/1.214.0290}{{\em PoS}
  {\bfseries LATTICE2014} (2015) 290},
  \href{http://arxiv.org/abs/1502.00044}{{\ttfamily arXiv:1502.00044
  [hep-lat]}}.

\bibitem{BhupalPrivComm}
B.~Dev, ``{Private communication},''  (2020) .

\bibitem{Nesvizhevsky2019}
V.~Nesvizhevsky, V.~Gudkov, K.~Protasov, W.~Snow, and A.~Y. Voronin,
  ``{Experimental Approach to Search for Free Neutron-Antineutron Oscillations
  Based on Coherent Neutron and Antineutron Mirror Reflection},''
  \href{http://dx.doi.org/10.1103/PhysRevLett.122.221802}{{\em Phys. Rev.
  Lett.} {\bfseries 122} no.~22, (2019) 221802},
  \href{http://arxiv.org/abs/1810.04988}{{\ttfamily arXiv:1810.04988
  [hep-ex]}}.

\bibitem{Kerbikov:2018mct}
B.~O. Kerbikov, ``{The effect of collisions with the wall on
  neutron-antineutron transitions},''
  \href{http://dx.doi.org/10.1016/j.physletb.2019.06.041}{{\em Phys. Lett. B}
  {\bfseries 795} (2019) 362--365},
  \href{http://arxiv.org/abs/1810.02153}{{\ttfamily arXiv:1810.02153
  [hep-ph]}}.

\bibitem{Kazarnovskii80}
K.~Chetyrkin, M.~Kazarnovsky, V.~Kuzmin, and M.~Shaposhnikov, ``{Neutron -
  antineutron oscillations},'' {\em Pisma Zh. Eksp. Teor. Fiz.} {\bfseries 32}
  (1980) 88--91.

\bibitem{Chetyrkin81}
K.~Chetyrkin, M.~Kazarnovsky, V.~Kuzmin, and M.~Shaposhnikov,
  ``{Neutron-Antineutron Oscillations: How Fast Could They Be?},''
  \href{http://dx.doi.org/10.1016/0370-2693(81)90117-9}{{\em Phys. Lett. B}
  {\bfseries 99} (1981) 358--360}.

\bibitem{Yoshiki89}
R.~Golub and H.~Yoshiki, ``{Ultracold antineutrons (UCN-bar). 1: The approach
  to the semiclassical limit},''
\href{http://dx.doi.org/10.1016/0375-9474(89)90166-8}{{\em Nucl. Phys.}
  {\bfseries A501} (1989) 869--876}.
%%CITATION = NUPHA,A501,869;%%.

\bibitem{Yoshiki92}
H.~Yoshiki and R.~Golub, ``{Ultracold antineutrons (UC antineutrons). 2:
  Production probability under magnetic and gravitational fields},''
  \href{http://dx.doi.org/10.1016/0375-9474(92)90117-3}{{\em Nucl. Phys. A}
  {\bfseries 536} (1992) 648--668}.

\bibitem{Glashow:1985ud}
S.~L. Glashow, ``{Positronium Versus the Mirror Universe},''
\href{http://dx.doi.org/10.1016/0370-2693(86)90540-X}{{\em Phys. Lett.}
  {\bfseries 167B} (1986) 35--36}.
%%CITATION = PHLTA,167B,35;%%.

\bibitem{Carlson:1987si}
E.~D. Carlson and S.~L. Glashow, ``{Nucleosynthesis Versus the Mirror
  Universe},''
\href{http://dx.doi.org/10.1016/0370-2693(87)91216-0}{{\em Phys. Lett.}
  {\bfseries B193} (1987) 168--170}.
%%CITATION = PHLTA,B193,168;%%.

\bibitem{Gninenko:1994dr}
S.~N. Gninenko, ``{Limit on 'disappearance' of orthopositronium in vacuum},''
\href{http://dx.doi.org/10.1016/0370-2693(94)91329-3}{{\em Phys. Lett.}
  {\bfseries B326} (1994) 317--319}.
%%CITATION = PHLTA,B326,317;%%.

\bibitem{Berezhiani:1996ii}
Z.~Berezhiani, ``{Unified picture of the particle and sparticle masses in SUSY
  GUT},'' \href{http://dx.doi.org/10.1016/S0370-2693(97)01359-2}{{\em Phys.
  Lett.} {\bfseries B417} (1998) 287--296},
\href{http://arxiv.org/abs/hep-ph/9609342}{{\ttfamily arXiv:hep-ph/9609342
  [hep-ph]}}.
%%CITATION = HEP-PH/9609342;%%.

\bibitem{Berezhiani:2018ill}
B.~Belfatto and Z.~Berezhiani, ``{How light the lepton flavor changing gauge
  bosons can be},''
  \href{http://dx.doi.org/10.1140/epjc/s10052-019-6724-5}{{\em Eur. Phys. J.}
  {\bfseries C79} no.~3, (2019) 202},
\href{http://arxiv.org/abs/1812.05414}{{\ttfamily arXiv:1812.05414 [hep-ph]}}.
%%CITATION = ARXIV:1812.05414;%%.

\bibitem{Belfatto:2019swo}
B.~Belfatto, R.~Beradze, and Z.~Berezhiani, ``{The CKM unitarity problem: A
  trace of new physics at the TeV scale?},''
  \href{http://dx.doi.org/10.1140/epjc/s10052-020-7691-6}{{\em Eur. Phys. J.}
  {\bfseries C80} no.~2, (2020) 149},
\href{http://arxiv.org/abs/1906.02714}{{\ttfamily arXiv:1906.02714 [hep-ph]}}.
%%CITATION = ARXIV:1906.02714;%%.

\bibitem{Foot:2004pa}
R.~Foot, ``{Mirror matter-type dark matter},''
  \href{http://dx.doi.org/10.1142/S0218271804006449}{{\em Int. J. Mod. Phys.}
  {\bfseries D13} (2004) 2161--2192},
\href{http://arxiv.org/abs/astro-ph/0407623}{{\ttfamily arXiv:astro-ph/0407623
  [astro-ph]}}.
%%CITATION = ASTRO-PH/0407623;%%.

\bibitem{Addazi:2015cua}
A.~Addazi, Z.~Berezhiani, R.~Bernabei, P.~Belli, F.~Cappella, R.~Cerulli, and
  A.~Incicchitti, ``{DAMA annual modulation effect and asymmetric mirror
  matter},'' \href{http://dx.doi.org/10.1140/epjc/s10052-015-3634-z}{{\em Eur.
  Phys. J.} {\bfseries C75} no.~8, (2015) 400},
\href{http://arxiv.org/abs/1507.04317}{{\ttfamily arXiv:1507.04317 [hep-ex]}}.
%%CITATION = ARXIV:1507.04317;%%.

\bibitem{Cerulli:2017jzz}
R.~Cerulli, P.~Villar, F.~Cappella, R.~Bernabei, P.~Belli, A.~Incicchitti,
  A.~Addazi, and Z.~Berezhiani, ``{DAMA annual modulation and mirror Dark
  Matter},'' \href{http://dx.doi.org/10.1140/epjc/s10052-017-4658-3}{{\em Eur.
  Phys. J.} {\bfseries C77} no.~2, (2017) 83},
\href{http://arxiv.org/abs/1701.08590}{{\ttfamily arXiv:1701.08590 [hep-ex]}}.
%%CITATION = ARXIV:1701.08590;%%.

\bibitem{Mohapatra:2005ng}
R.~N. Mohapatra, S.~Nasri, and S.~Nussinov, ``{Some implications of neutron
  mirror neutron oscillation},''
  \href{http://dx.doi.org/10.1016/j.physletb.2005.08.101}{{\em Phys. Lett.}
  {\bfseries B627} (2005) 124--130},
\href{http://arxiv.org/abs/hep-ph/0508109}{{\ttfamily arXiv:hep-ph/0508109
  [hep-ph]}}.
%%CITATION = HEP-PH/0508109;%%.

\bibitem{Pokotilovski:2006gq}
{\relax Yu}.~N. Pokotilovski, ``{On the experimental search for neutron --->
  mirror neutron oscillations},''
  \href{http://dx.doi.org/10.1016/j.physletb.2006.06.005}{{\em Phys. Lett.}
  {\bfseries B639} (2006) 214--217},
\href{http://arxiv.org/abs/nucl-ex/0601017}{{\ttfamily arXiv:nucl-ex/0601017
  [nucl-ex]}}.
%%CITATION = NUCL-EX/0601017;%%.

\bibitem{Berezhiani:2017azg}
Z.~Berezhiani, M.~Frost, Y.~Kamyshkov, B.~Rybolt, and L.~Varriano, ``{Neutron
  Disappearance and Regeneration from Mirror State},''
  \href{http://dx.doi.org/10.1103/PhysRevD.96.035039}{{\em Phys. Rev.}
  {\bfseries D96} no.~3, (2017) 035039},
\href{http://arxiv.org/abs/1703.06735}{{\ttfamily arXiv:1703.06735 [hep-ex]}}.
%%CITATION = ARXIV:1703.06735;%%.

\bibitem{Berezhiani:2008gi}
Z.~Berezhiani and A.~Lepidi, ``{Cosmological bounds on the 'millicharges' of
  mirror particles},''
  \href{http://dx.doi.org/10.1016/j.physletb.2009.10.023}{{\em Phys. Lett.}
  {\bfseries B681} (2009) 276--281},
\href{http://arxiv.org/abs/0810.1317}{{\ttfamily arXiv:0810.1317 [hep-ph]}}.
%%CITATION = ARXIV:0810.1317;%%.

\bibitem{Raaijmakers:2019hqj}
C.~Vigo, L.~Gerchow, B.~Radics, M.~Raaijmakers, A.~Rubbia, and P.~Crivelli,
  ``{New bounds from positronium decays on massless mirror dark photons},''
  \href{http://dx.doi.org/10.1103/PhysRevLett.124.101803}{{\em Phys. Rev.
  Lett.} {\bfseries 124} no.~10, (2020) 101803},
\href{http://arxiv.org/abs/1905.09128}{{\ttfamily arXiv:1905.09128
  [physics.atom-ph]}}.
%%CITATION = ARXIV:1905.09128;%%.

\bibitem{Ignatiev:2000yw}
A.~{\relax Yu}. Ignatiev and R.~R. Volkas, ``{Geophysical constraints on mirror
  matter within the earth},''
  \href{http://dx.doi.org/10.1103/PhysRevD.62.023508}{{\em Phys. Rev.}
  {\bfseries D62} (2000) 023508},
\href{http://arxiv.org/abs/hep-ph/0005125}{{\ttfamily arXiv:hep-ph/0005125
  [hep-ph]}}.
%%CITATION = HEP-PH/0005125;%%.

\bibitem{Berezhiani:2016ong}
Z.~Berezhiani, ``{Anti-dark matter: a hidden face of mirror world},''
  \href{http://arxiv.org/abs/1602.08599}{{\ttfamily arXiv:1602.08599
  [astro-ph.CO]}}.

\bibitem{Berezhiani:2013dea}
Z.~Berezhiani, A.~D. Dolgov, and I.~I. Tkachev, ``{Dark matter and generation
  of galactic magnetic fields},''
  \href{http://dx.doi.org/10.1140/epjc/s10052-013-2620-6}{{\em Eur. Phys. J.}
  {\bfseries C73} (2013) 2620},
\href{http://arxiv.org/abs/1307.6953}{{\ttfamily arXiv:1307.6953
  [astro-ph.CO]}}.
%%CITATION = ARXIV:1307.6953;%%.

\bibitem{Sharma:2010vv}
N.~Sharma, H.~Dahiya, P.~K. Chatley, and M.~Gupta, ``{Spin $\frac{1}{2}^+$,
  spin $\frac{3}{2}^+$ and transition magnetic moments of low lying and charmed
  baryons},'' \href{http://dx.doi.org/10.1103/PhysRevD.81.073001}{{\em Phys.
  Rev.} {\bfseries D81} (2010) 073001},
\href{http://arxiv.org/abs/1003.4338}{{\ttfamily arXiv:1003.4338 [hep-ph]}}.
%%CITATION = ARXIV:1003.4338;%%.

\bibitem{Broggini:2012df}
C.~Broggini, C.~Giunti, and A.~Studenikin, ``{Electromagnetic Properties of
  Neutrinos},'' \href{http://dx.doi.org/10.1155/2012/459526}{{\em Adv. High
  Energy Phys.} {\bfseries 2012} (2012) 459526},
\href{http://arxiv.org/abs/1207.3980}{{\ttfamily arXiv:1207.3980 [hep-ph]}}.
%%CITATION = ARXIV:1207.3980;%%.

\bibitem{Berezhiani:2020nzn}
Z.~Berezhiani, ``{A possible shortcut for neutron--antineutron oscillation},''
\href{http://arxiv.org/abs/2002.05609}{{\ttfamily arXiv:2002.05609 [hep-ph]}}.
%%CITATION = ARXIV:2002.05609;%%.

\bibitem{Golubeva:1997fs}
E.~Golubeva and L.~Kondratyuk, ``{Annihilation of low energy antineutrons on
  nuclei},'' \href{http://dx.doi.org/10.1016/S0920-5632(97)00260-0}{{\em Nucl.
  Phys. B Proc. Suppl.} {\bfseries 56} (1997) 103--107}.

\bibitem{Golubeva:2018mrz}
E.~S. Golubeva, J.~L. Barrow, and C.~G. Ladd, ``{Model of $\bar n$ annihilation
  in experimental searches for $\bar n$ transformations},''
  \href{http://dx.doi.org/10.1103/PhysRevD.99.035002}{{\em Phys. Rev.}
  {\bfseries D99} no.~3, (2019) 035002},
\href{http://arxiv.org/abs/1804.10270}{{\ttfamily arXiv:1804.10270 [hep-ex]}}.
%%CITATION = ARXIV:1804.10270;%%.

\bibitem{Minor1990}
E.~D. Minor, T.~A. Armstrong, R.~Bishop, V.~Harris, R.~A. Lewis, and G.~A.
  Smith, ``Charged pion spectra and energy transfer following antiproton
  annihilation at rest in carbon and uranium,''
  \href{http://dx.doi.org/10.1007/BF01294119}{{\em Zeitschrift f{\"u}r Physik A
  Atomic Nuclei} {\bfseries 336} no.~4, (Dec, 1990) 461--468}.
  \url{https://doi.org/10.1007/BF01294119}.

\bibitem{Mcgaughey:1986kz}
P.~Mcgaughey {\em et~al.}, ``{Dynamics of Low-energy Antiproton Annihilation in
  Nuclei as Inferred From Inclusive Proton and Pion Measurements},''
  \href{http://dx.doi.org/10.1103/PhysRevLett.56.2156}{{\em Phys. Rev. Lett.}
  {\bfseries 56} (1986) 2156--2159}.

\bibitem{Bryman:2014tta}
D.~Bryman, ``{Two nucleon (B - L)-conserving reactions involving tau leptons
  },'' \href{http://dx.doi.org/10.1016/j.physletb.2014.04.042}{{\em Phys.
  Lett.} {\bfseries B733} (2014) 190--192},
\href{http://arxiv.org/abs/1404.7776}{{\ttfamily arXiv:1404.7776 [hep-ex]}}.
%%CITATION = ARXIV:1404.7776;%%.

\bibitem{Takhistov:2015fao}
{\bfseries Super-Kamiokande} Collaboration, V.~Takhistov {\em et~al.},
  ``{Search for Nucleon and Dinucleon Decays with an Invisible Particle and a
  Charged Lepton in the Final State at the Super-Kamiokande Experiment},''
  \href{http://dx.doi.org/10.1103/PhysRevLett.115.121803}{{\em Phys. Rev.
  Lett.} {\bfseries 115} no.~12, (2015) 121803},
\href{http://arxiv.org/abs/1508.05530}{{\ttfamily arXiv:1508.05530 [hep-ex]}}.
%%CITATION = ARXIV:1508.05530;%%.

\bibitem{Girmohanta:2019xya}
S.~Girmohanta and R.~Shrock, ``{Improved Lower Bounds on Partial Lifetimes for
  Nucleon Decay Modes},''
  \href{http://dx.doi.org/10.1103/PhysRevD.100.115025}{{\em Phys. Rev.}
  {\bfseries D100} no.~11, (2019) 115025},
\href{http://arxiv.org/abs/1910.08106}{{\ttfamily arXiv:1910.08106 [hep-ph]}}.
%%CITATION = ARXIV:1910.08106;%%.

\bibitem{Tanabashi:2018oca}
{\bfseries Particle Data Group} Collaboration, M.~Tanabashi {\em et~al.},
  ``{Review of Particle Physics},''
\href{http://dx.doi.org/10.1103/PhysRevD.98.030001}{{\em Phys. Rev.} {\bfseries
  D98} no.~3, (2018) 030001}.
%%CITATION = PHRVA,D98,030001;%%.

\bibitem{Serebrov:2009zz}
A.~Serebrov {\em et~al.}, ``{Search for neutronmirror neutron oscillations in a
  laboratory experiment with ultracold neutrons},''
  \href{http://dx.doi.org/10.1016/j.nima.2009.07.041}{{\em Nucl. Instrum. Meth.
  A} {\bfseries 611} (2009) 137--140},
  \href{http://arxiv.org/abs/0809.4902}{{\ttfamily arXiv:0809.4902 [nucl-ex]}}.

\bibitem{Peggs:2013sgv}
S.~Peggs, ``{ESS Technical Design Report},''.
\url{https://europeanspallationsource.se/documentation/tdr.pdf}.
%%CITATION = ESS-DOC-274;%%.

\bibitem{ess-bunker}
V.~Santoro {\em et~al.}, ``{Neutronic design of the bunker},''
{\em ESS General Document, ESS-0052649} (2016) .
%%CITATION = ZFPRA,5,32;%%.

\bibitem{ess-gap}
``"the ess instrument suite, a capability gap analysis",'' 2018.
\newblock
  \url{https://europeanspallationsource.se/instruments/capability-gap-analysis}.

\bibitem{zanini2019design}
L.~Zanini, K.~Andersen, K.~Batkov, E.~Klinkby, F.~Mezei, T.~Sch{\"o}nfeldt, and
  A.~Takibayev, ``Design of the cold and thermal neutron moderators for the
  european spallation source,'' {\em Nuclear Instruments and Methods in Physics
  Research Section A: Accelerators, Spectrometers, Detectors and Associated
  Equipment} {\bfseries 925} (2019) 33--52.

\bibitem{andersen2018optimization}
K.~H. Andersen, M.~Bertelsen, L.~Zanini, E.~B. Klinkby, T.~Sch{\"o}nfeldt,
  P.~M. Bentley, and J.~Saroun, ``Optimization of moderators and beam
  extraction at the ess,'' {\em Journal of applied crystallography} {\bfseries
  51} no.~2, (2018) 264--281.

\bibitem{euh2020}
 \url{https://ec.europa.eu/research/participants/data/ref/h2020/other/wp/2018-2020/annexes/h2020-wp1820-annex-d-ria_en.pdf}.

\bibitem{Santoro:2020nke}
V.~Santoro {\em et~al.}, ``{Development of High Intensity Neutron Source at the
  European Spallation Source},'' in {\em {Accepted by Journal of Neutron
  Research}}.
\newblock 2, 2020.
\newblock \href{http://arxiv.org/abs/2002.03883}{{\ttfamily arXiv:2002.03883
  [physics.ins-det]}}.

\bibitem{Klinkby:2014cma}
E.~Klinkby, K.~Batkov, F.~Mezei, T.~Schønfeldt, A.~Takibayev, and L.~Zanini,
  ``{Voluminous D2 source for intense cold neutron beam production at the
  ESS},''
\href{http://arxiv.org/abs/1401.6003}{{\ttfamily arXiv:1401.6003
  [physics.ins-det]}}.
%%CITATION = ARXIV:1401.6003;%%.

\bibitem{Nielsen:2016fce}
T.~R. Nielsen, A.~J. Markvardsen, and P.~K. Willendrup, ``{McStas and Mantid
  integration},''
\href{http://arxiv.org/abs/1607.02498}{{\ttfamily arXiv:1607.02498
  [physics.ins-det]}}.
%%CITATION = ARXIV:1607.02498;%%.

\bibitem{Broussard:2017yev}
L.~J. Broussard {\em et~al.}, ``{New Search for Mirror Neutrons at HFIR},'' in
  {\em {Proceedings, Meeting of the APS Division of Particles and Fields (DPF
  2017): Fermilab, Batavia, Illinois, USA, July 31 - August 4, 2017}}.
\newblock 2017.
\newblock
\href{http://arxiv.org/abs/1710.00767}{{\ttfamily arXiv:1710.00767 [hep-ex]}}.
\newblock
%%CITATION = ARXIV:1710.00767;%%.

\bibitem{doi:10.1063/1.4919412}
W.~M. Snow {\em et~al.}, ``A slow neutron polarimeter for the measurement of
  parity-odd neutron rotary power,''
  \href{http://dx.doi.org/10.1063/1.4919412}{{\em Review of Scientific
  Instruments} {\bfseries 86} no.~5, (2015) 055101}.

\bibitem{Abel:2018tib}
C.~Abel {\em et~al.}, ``{Statistical sensitivity of the nEDM apparatus at PSI
  to neutron mirror-neutron oscillations},'' in {\em {International Workshop on
  Particle Physics at Neutron Sources 2018 (PPNS 2018) Grenoble, France, May
  24-26, 2018}}.
\newblock 2018.
\newblock
\href{http://arxiv.org/abs/1811.01906}{{\ttfamily arXiv:1811.01906 [nucl-ex]}}.
\newblock
%%CITATION = ARXIV:1811.01906;%%.

\bibitem{Broussard:2019tgw}
L.~Broussard {\em et~al.}, ``{New search for mirror neutron regeneration},''
  \href{http://dx.doi.org/10.1051/epjconf/201921907002}{{\em EPJ Web Conf.}
  {\bfseries 219} (2019) 07002},
  \href{http://arxiv.org/abs/1912.08264}{{\ttfamily arXiv:1912.08264
  [physics.ins-det]}}.

\bibitem{Altarev:2015fra}
I.~Altarev {\em et~al.}, ``{A large-scale magnetic shield with $10^6$ damping
  at mHz frequencies},'' \href{http://dx.doi.org/10.1063/1.4919366}{{\em J.
  Appl. Phys.} {\bfseries 117} (2015) 183903},
\href{http://arxiv.org/abs/1501.07861}{{\ttfamily arXiv:1501.07861
  [physics.ins-det]}}.
%%CITATION = ARXIV:1501.07861;%%.

\bibitem{Matt-thesis}
M.~J. Frost, {\em {Searching for Baryon Number Violation at Cold Neutron
  Sources}}.
\newblock PhD thesis, University of Tennessee, 2019.

\bibitem{Waters:2007zza}
L.~S. Waters, G.~W. McKinney, J.~W. Durkee, M.~L. Fensin, J.~S. Hendricks,
  M.~R. James, R.~C. Johns, and D.~B. Pelowitz, ``{The MCNPX Monte Carlo
  radiation transport code},'' \href{http://dx.doi.org/10.1063/1.2720459}{{\em
  AIP Conf. Proc.} {\bfseries 896} no.~1, (2007) 81--90}.

\bibitem{YK_icans_1995}
Y.~Kamyshkov, ``Use of a cold source and large reflector mirror guide for a
  neutron-antineutron oscillation search,'' in {\em {Proceedings of the
  ICANS-XIII meeting, Villigen, PSI.}}
\newblock 1995.

\bibitem{mezei1976novel}
F.~Mezei, ``Novel polarized neutron devices: supermirror and spin component
  amplifier,'' {\em Communications on Physics (London)} {\bfseries 1} no.~3,
  (1976) 81--85.

\bibitem{ILL-yellow}
{Institut Laue-Langevin}, ``{ILL Yellow Book 2008},''.

\bibitem{bergmann2018simulation}
R.~Bergmann, U.~Filges, D.~Kiselev, C.~Klauser, E.~Rantsiou, V.~Talanov,
  M.~Wohlmuther, and M.~Yamada, ``Simulation methods and results of the sinq
  cold neutron source upgrade study,'' in {\em Journal of Physics: Conference
  Series}, vol.~1021, p.~012081, IOP Publishing.
\newblock 2018.

\bibitem{Muhrer2016}
G.~Muhrer, T.~Schonfeldt, E.~Iverson, M.~Mocko, D.~Baxter, T.~H{\"u}gle,
  F.~Gallmeier, and E.~Klinkby, ``Demonstration of a single-crystal
  reflector-filter for enhancing slow neutron beams,''
  \href{http://dx.doi.org/10.1016/j.nima.2016.06.047}{{\em Nuclear Instruments
  and Methods in Physics Research Section A: Accelerators, Spectrometers,
  Detectors and Associated Equipment} {\bfseries 830} (2016) 454--460}.

\bibitem{BaldoCeolin:1989qd}
M.~Baldo-Ceolin {\em et~al.}, ``{A new experimental limit on neutron
  antineutron transitions},''
\href{http://dx.doi.org/10.1016/0370-2693(90)90601-2}{{\em Phys. Lett.}
  {\bfseries B236} (1990) 95--101}.
%%CITATION = PHLTA,B236,95;%%.

\bibitem{sato2013particle}
T.~Sato, K.~Niita, N.~Matsuda, S.~Hashimoto, Y.~Iwamoto, S.~Noda, T.~Ogawa,
  H.~Iwase, H.~Nakashima, T.~Fukahori, {\em et~al.}, ``Particle and heavy ion
  transport code system, phits, version 2.52,'' {\em Journal of Nuclear Science
  and Technology} {\bfseries 50} no.~9, (2013) 913--923.

\bibitem{Korneev1992}
V.~V.~P. D.~A.~Korneev and A.~V. Petrenko, ``{Absorbing sublayers and their
  influence on the polarizing efficiency of magnetic neutron mirrors},'' {\em
  Nucl. Inst. Meth. B} {\bfseries 63} (1992) 328.

\bibitem{Nikova2019}
E.~Nikova {\em et~al.}, ``{Experimental determination of gadolinium scattering
  characteristics in neutron reflectometry with reference layer},''
  \href{http://dx.doi.org/https://doi.org/10.1016/j.physb.2018.09.033}{{\em
  Physica B} {\bfseries 552} (2019) 58}.

\bibitem{Agostinelli:2002hh}
{\bfseries GEANT4} Collaboration, S.~Agostinelli {\em et~al.}, ``{GEANT4: A
  Simulation toolkit},''
  \href{http://dx.doi.org/10.1016/S0168-9002(03)01368-8}{{\em Nucl. Instrum.
  Meth. A} {\bfseries 506} (2003) 250--303}.

\end{thebibliography}
\newpage
\appendix
\section{Prototype development and the ESS test beam line (TBL)}\label{sec:app}

In this appendix, a description is given of the prototype detector development which would be deployed at an ESS test beam. The test beam is also described.

\subsection{Prototype calorimeter module}
As described in Section~\ref{sec:small_nnbar_exp} an annihilation detector must provide as reliable and selective information as possible in each individual event. The crucial information is event topology and conservation of energy and momentum. 
For free $\bar{n}N$ annihilation events, identifying and measuring total energy and momentum of all outgoing particles give unique annihilation signature and one should measure these parameters with the best possible resolution since the discriminating cuts will be limited by the resolution by which the kinetic energy has been measured. However $\bar{n}N$ annihilation in nuclei (carbon in this case) introduces  additional means to dissipate energy and momentum by nuclear fragments. Some of the energy cannot be captured as  it is carried by neutrons. Energy carried with protons (and to some minor extent composite fragments) can be measured but, particle identification is important since the proton rest mass shall not be included in the total invariant mass. Taken together, the total invariant mass of pions and photons will only be equal to two neutron masses in about 30\% of the cases when the annihilation takes place in  a carbon nucleus (see Fig.~\ref{fig:InitMesMomvsInvMass}).  

If the kinetic energy of charged nuclear fragments is also measured, this will narrow the distribution but it is clear that a selection on invariant mass  has to be quite generous in order to avoid cutting away good annihilation events. As a consequence, it makes no sense to strive at highest possible energy resolution. This is actually quite satisfying since, in particular the charged hadrons are in a  a very difficult energy regime where energies are often too high to be absorbed by electromagnetic processes only, and the statistical significance on energy deposit by strong interactions is extremely poor. Thus optimisation of the calorimetry is important. Fig.~\ref{fig:pbarCarbon-PiPlusMomentum} illustrates the simulated kinetic energy/momentum distributions of the charged hadrons expected in an annihilation event. Given the distributions shown, an 
energy resolution for charged hadrons of $\pm 5$ MeV appears to be more than sufficient.      

\subsection*{The principles}
The topological aspects of annihilation events to be handled by the tracking must be considered together with the need that the tracking will provide a $\frac{dE}{dx}$ measurement. Since tracks are resolved in 3 dimensional space, the track direction for charged particles into the calorimeter is accurately determined. Thus safe particle identification by combining $\frac{dE}{dx}$ and energy is achieved. 
Background of gamma radiation in the MeV range will be very high. To cope with this background a high granularity, in space and time, using a large number of detector elements and electronic readout channels is needed. This will be one guideline for
a calorimeter design.
As pointed out, the kinetic energy resolution can be rather relaxed. For charged hadrons this may pay off in a simple calorimeter design. For photons however one should still strive at good energy resolution in order to obtain a narrow invariant mass peak for the identification of the neutral pions.

\subsubsection{The tentative calorimeter design}
\begin{figure}[H]
  \setlength{\unitlength}{1mm}
  \begin{center}
  \includegraphics[width=0.7\linewidth, angle=0]{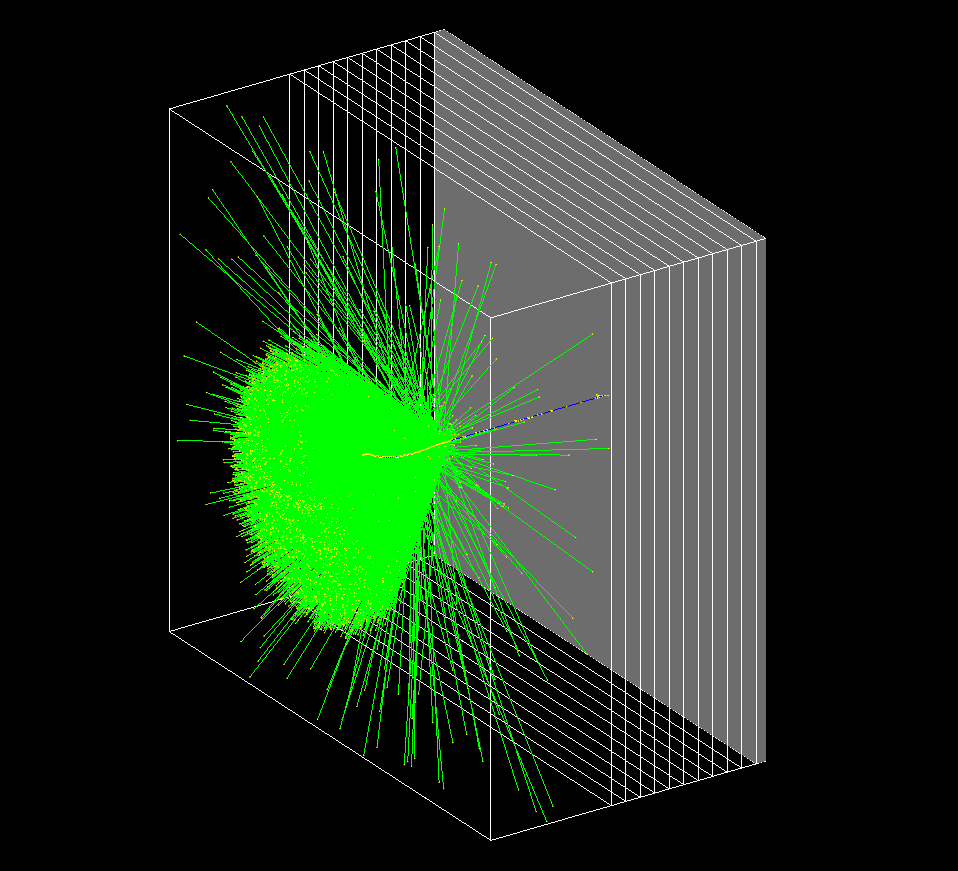}
  \end{center}
  \caption{A {\sc Geant}-4 visualisation of a calorimeter module with ten layers of plastic scintillator and a lead-glass block. The Cerenkov photons (green) are shown as is the pion track.}
  \label{fig:cal-ev-display}
\end{figure}

A hybrid approach is adopted for the calorimeter with ten, 3cm thick layers of plastic scintillators reaching a total thickness of 30cm followed by a 25cm (ca 20 radiation lengths) of lead-glass for the electromagnetic calorimetry. A {\sc Geant-4}~\cite{Agostinelli:2002hh} visualisation of the response the set-up to a charged pion with 240 MeV kineatic energy is shown in Fig.~\ref{fig:cal-ev-display}. The pion punches through the scintillators leaving a cone a Cerenkov light.  

\begin{figure}[h]
  \setlength{\unitlength}{1mm}
  %\begin{left}
  \includegraphics[width=0.8\linewidth]{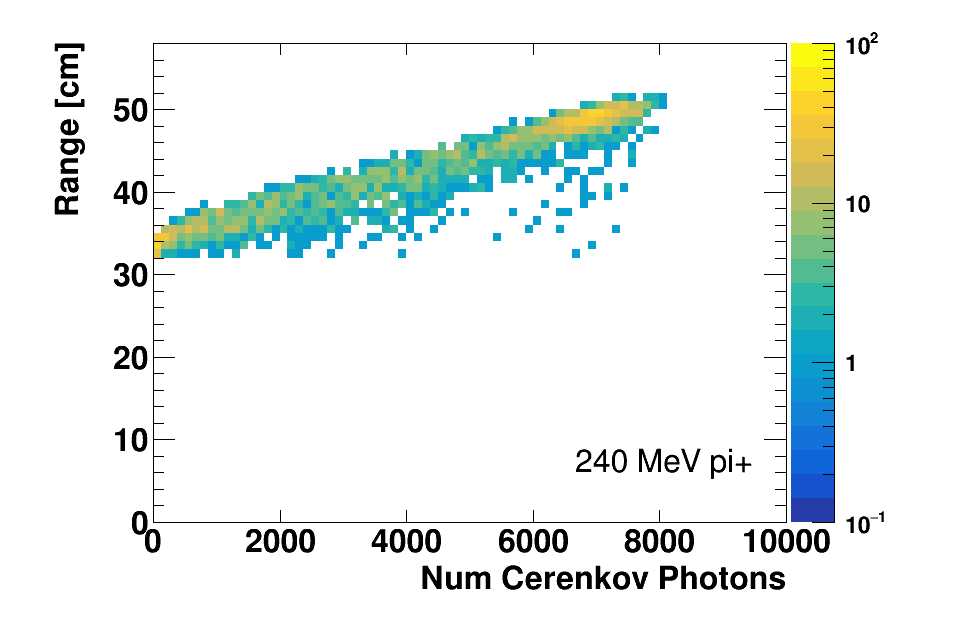}
  %\end{left}
  \caption{The relationship between range and amount of Cherenkov light for charged pions with 240~MeV kinetic energy, as predicted by {\sc Geant}-4. Pions with this energy reach into around 20cm into the lead glass.}
  \label{fig:2d-calo}
\end{figure}

The hadron calorimeter with 30 cm thickness is sufficient to stop protons with more than 200 MeV kinetic energy. About 80\% of protons in the high energy end of the spectrum will come to rest without a nuclear interaction. This will cover effectively all protons that happen to be emitted from the annihilation point (Fig.~\ref{fig:pbarCarbon-PiPlusMomentum}, bottom). However 30cm of plastic will only stop pions up to about 80~MeV kinetic energy (260~MeV in momentum, Fig.~\ref{fig:pbarCarbon-PiPlusMomentum}, top) which is only about 30\% of the expected pion spectrum. Charged pions passing through 
additional 25cm of lead-glass will have about 250MeV kinetic energy i.e, momentum about 400MeV which would account for another 30\% of the pion spectrum. Even if one can calibrate the energy response by Cherenkov radiation one cannot expect to make a proper energy measurement since most charged hadrons passing this amount of matter will suffer one or a few hadron nucleus collisions with a rather unpredictable energy signal as result. The very low number of collisions is the reason that hadron calorimetry by showering is not a viable technique at these energies. Actually, the nuclear fragments from these hadronic collisions will not produce Cherenkov light. Thus calibrated Cherenkov light for charged pions will mostly reflect the range of the pion until its makes a collision. Fig.~\ref{fig:2d-calo} shows the clear almost linear relation between range and the amount of Cherenkov light, as predicted using {\sc Geant}-4. Note that only pions with 240 MeV have been injected. Ideally one should have only a sharp peak at about 7000 photons, but anything lower than that is due to collisions on the way. 

Considering all of the above, the strategy is to use the plastic scintillator stack for a range measurement, where each scintillator defines a kinetic energy bin in the proton and pion spectra and the Cherenkov light in the lead-glass as a measure of the charged pion range in the lead-glass. For cases where a particle comes to rest by electromagnetic energy loss only, the range will give the kinetic energy (since the particle mass is known by the particle identification) with an adequate resolution. 
In case of a hadronic interaction in the detector materials one must differ between protons and pions.  For protons one should  be able to know if a nuclear reaction has taken place since there will be a clear correlation between $\frac{dE}{dx}$ and range which is violated if a reaction takes place. The $\frac{dE}{dx}$ value itself can, if good enough energy loss resolution is achieved, provide a complementary energy information, unaffected by the hadronic collision.  For pions however,  which almost all will be near minimum ionizing, there is no such correlation between $\frac{dE}{dx}$ and range and one will only be able to state the range of the pion for as long as it was a pion. The charged pions produce Cherenkov light down to quite low kinetic energies in lead-glass.  

By limiting the information from the plastic layers to a hit/no hit information the light readout and the readout electronics can be made simple (a discriminator threshold only) and cheap thus allowing a very high segmentation to the benefit of handling the gamma background. The scintillator thicknesses will be chosen such that minimum ionizing particles will give substantially larger signals than Compton electrons from gammas. With 3 cm thickness of a MIP will deposit about 6 MeV, much above gammas with nuclear physics origin.

The plastic scintillator layers will be segmented in a way such that coarse tracking in 3D can be done on the hit/no-hit information that is available on a nanosecond time scale. Thus a powerful and fast track trigger can be constructed. Different ways of extracting and sensing the light from the scintillator segments will be investigated in order to optimize the solution for cost, performance and segmentation taking advantage of the moderate resolution required. 
The lead-glass will have a good energy resolution, being sensitive over the whole volume. Cherenkov light has the advantage of being direction-sensitive. This gives the possibility to measure the direction of particles or even make it blind to particles in the wrong direction. This would be important to discriminate against fake events of cosmic ray origin. This is an important R\&D topic. The resolution by which the point of impact of gammas from $\pi^0$ decay shall be good in order to obtain good resolution in the reconstruction of $\pi^0$.

This will drive the lateral segmentation of the lead-glass either as lead-glass blocks or a segmentation of the light readout. Whether or not, the readout is direct by photomultipliers or via wavelength shifting  is also an R\&D item.  One could think of segmenting the 25cm of lead-glass in the depth dimension to better measure the range of charged pions. It is questionable if it worth the increased cost of light sensors and electronics. Possibly, the cost of the lead-glass itself is balancing the cost of more readout.  

To conclude, it is considered that this unusual detector application aimed for too high energy for nuclear physics and too low energy for particle physics methods offers many interesting detector R\&D challenges.  It is anticipated that it would be deployed in the ESS test beam described below. The prototype would be supplemented with a scintillator cosmic shield and an inner TPC.  

\subsection{The ESS test beam}%\skippable
%\label{sec:app}

\begin{figure}[H]
  \setlength{\unitlength}{1mm}
  \begin{center}
  \includegraphics[width=1.0\linewidth, angle=0]{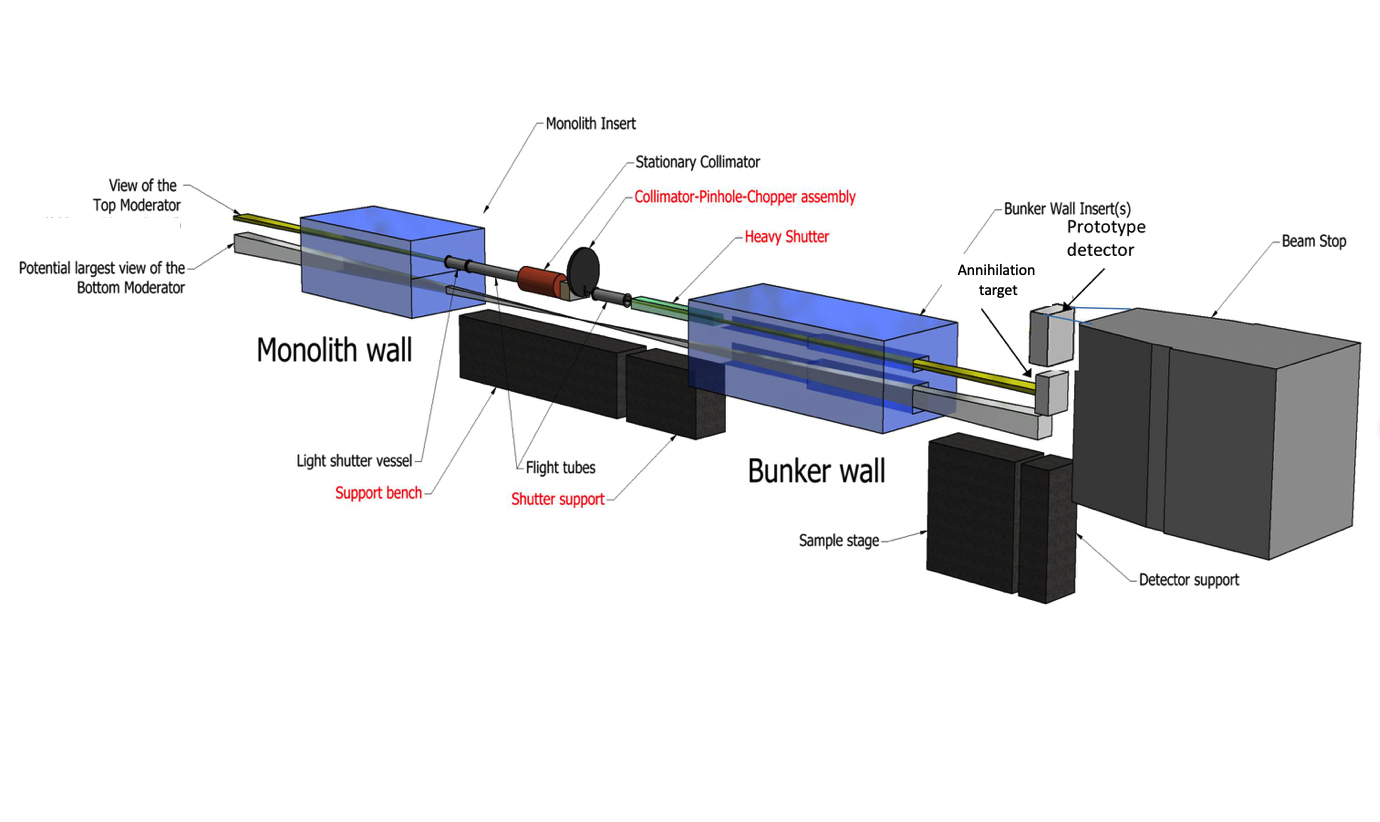}
  \end{center}
  \vspace{-2.25cm}
  \caption{Overview of the planned ESS test beam set-up. A collimated beam would be passed to the carbon target and then to the  detector  prototype  }
  \label{fig:testbeamoverview}
\end{figure}

\begin{figure}[H]
  \setlength{\unitlength}{1mm}
  \begin{center}
  \includegraphics[width=0.75\linewidth,angle=0]{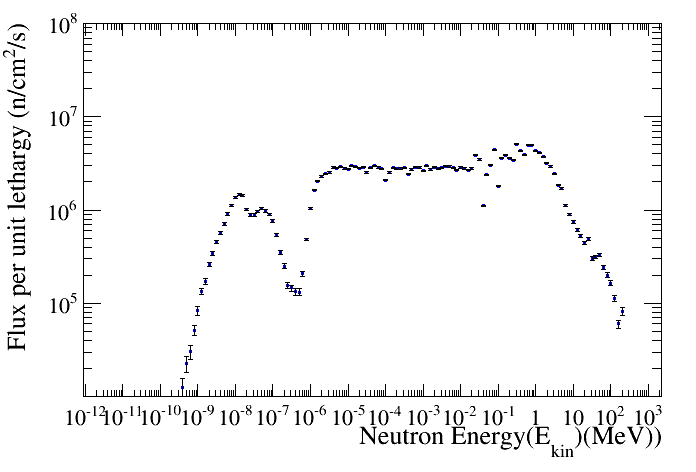}
  \end{center}
  %\vspace{-3.25cm}
  \caption{Neutron energy spectrum at the location of the annihilation target   location of the TBL  }
  \label{fig:ess-test-beam-spectrum}
\end{figure}

A dedicated test beam line (TBL) will be build at ESS to be used initially to verify that the accelerator has successfully delivered beam on target, and to characterize the pulsed neutron beam emitted from the upper moderator (e.g. time structure, spatial distribution, energy dependence etc.), while it can be adjusted to also view a future lower moderator. In the longer term, it will provide supporting measurements for the user program and also serve for the development of key neutron technologies, such as optical components, choppers and detector systems. For the last mentioned purpose, the ESS TBL is an ideal place to perform tests of the prototypes of HIBEAM/NNBAR detectors.
Figure~\ref{fig:testbeamoverview} shows a schematic overview of the TBL.
In its basic configuration, it is a pin-hole (camera obscura) imaging station for viewing the entire width and height of the moderator assembly, with a double-disc chopper at the pin-hole position to provide tuneable wavelength resolution and wavelength band selection. For the moderator characterization, a position sensitive detector (PSD) will be placed at 17 m from the moderator with an adjustable pin-hole at the half-way position, enabling to spatially image the neutrons of different wavelengths emerging from the moderator. The instrument will be in direct line of sight and not be equipped with optics,  but in addition to the double disk chopper will have a series of collimators (one stationary and one adjustable in size), a range of optional beam attenuators and filters and a heavy shutter.
The test for the HIBEAM/NNBAR detectors will be done in combination with the annihilation target. After hitting the annihilation target, the beam will be absorbed in a beam stop that is already part of the TBL. With a direct view of the source, the test beam line represents also an optimal place to make background measurements since it will not only provide cold but also fast and high energy neutrons as can be seen in Figure~\ref{fig:ess-test-beam-spectrum}.  

Planned measurements include:
\begin{itemize}
    \item Test of the full prototype detector 
    \item  Benchmark against Experimental Data of Monte Carlo background simulations 
    \item Gamma background measurements from neutron
    interaction and activation with the annihilation target  and surrounding materials 
    \item Fast neutron background measurements
    \item Cosmics and skyshine background characterization 
\end{itemize}

All these measurements with allow to have a deep understanding of the background and provide a better mitigation strategy implementations both for HIBEAM and the NNBAR experiment.

\end{document}